%% file: main.tex
\documentclass[oneside, letterpaper, 10pt, oldfontcommands]{memoir}

\input{frontstuff/packages.tex}

\settitle{New Physics with PeV Astrophysical Neutrino Beams}
\setauthor{Ibrahim Safa}
\setdepartment{Physics}
\doctors
\setgraddate{2022}
\setdefensedate{August 17, 2022}
\setfoca{Francis Halzen, Professor, Physics} 
\setfocb{Carlos Argüelles, Assistant Professor, Physics}
\setfocc{Ke Fang, Assistant Professor, Physics}
\setfocd{Amy Barger, Professor, Astronomy}
\setfoce{Benjamin Jones, Professor, Physics}

\begin{document}
\frontmatter
\thetitlepage
\clearpage
%\thecopyrightpage
%\cleardoublepage
\clearpage
\setcounter{page}{1}
\input{frontstuff/dedication.tex}
\clearpage
\input{frontstuff/abstract.tex}
\input{frontstuff/acknowledgements.tex}
\clearpage
%\maxtocdepth{subsection}
\SingleSpace
\tableofcontents* % the * means that there isn't an entry for the TOC itself
\clearpage
\DoubleSpacing
\mainmatter
\input{chapters/intro.tex}
\input{chapters/nusources.tex} %source search
\input{chapters/tau_modeling.tex} %tau polarization paper, tau appearance
\input{chapters/taurunner/taurunner.tex}   %software paper
\input{chapters/taus_advantages.tex} %GZK papers, tau neurino transient paper, ANITA search paper
\input{chapters/DMnus.tex} %RMP baby
\input{chapters/conclusion.tex} %goodbye
\clearpage

\listoffigures
\clearpage
\listoftables  

\bibliographystyle{JHEP}
\bibliography{references}
\end{document}

%% file: frontstuff/packages.tex
\usepackage{uwthesis}
\usepackage{graphicx}
\usepackage[dvipsnames]{xcolor}
\usepackage{amssymb}
\usepackage{amsmath}
\usepackage{multirow}
\usepackage{multicol}
\usepackage{makecell}
\usepackage{pdfpages}
\usepackage[colorlinks=true,linkcolor=blue, citecolor=magenta, urlcolor=cyan]{hyperref}
\usepackage[dvipsnames]{xcolor}
\usepackage{booktabs}
\usepackage{listings}
\usepackage{tabularx}
\usepackage{lmodern}
\usepackage{gensymb}
\usepackage[T1]{fontenc} 
\usepackage{placeins}
\usepackage[all]{nowidow}
\usepackage{array}
\usepackage{tikz}
\def\checkmark{\tikz\fill[scale=0.4](0,.35) -- (.25,0) -- (1,.7) -- (.25,.15) -- cycle;} 

\newcommand{\sv}{\langle \sigma v \rangle}
\newcommand{\attrib}[1]{%
\nopagebreak{\raggedleft\footnotesize #1\par}}
\newcommand{\taurunner}{\texttt{TauRunner}}
\newcommand{\Particle}{\texttt{Particle}}
\newcommand{\Track}{\texttt{Track}}
\newcommand{\Body}{\texttt{Body}}
\newcommand{\XS}{\texttt{CrossSection}}

\definecolor{codegreen}{rgb}{0,0.6,0}
\definecolor{codegray}{rgb}{0.5,0.5,0.5}
\definecolor{codepurple}{rgb}{0.58,0,0.82}
\definecolor{backcolour}{rgb}{0.95,0.95,0.92}

\lstdefinestyle{mystyle}{
    backgroundcolor=\color{backcolour},   
    commentstyle=\color{codegray},
    keywordstyle=\color{codegreen},
    numberstyle=\tiny\color{codegray},
    stringstyle=\color{orange},
    basicstyle=\ttfamily\footnotesize,
    breakatwhitespace=false,         
    breaklines=true,                 
    captionpos=b,                    
    keepspaces=true,                 
    numbers=left,                    
    numbersep=5pt,                  
    showspaces=false,                
    showstringspaces=false,
    showtabs=false,                  
    tabsize=2
}

%% file: frontstuff/dedication.tex
\vspace*{\fill}
\settowidth{\versewidth}{To all the people that lived above the buildings that I was hustlin' in front of}
\begin{verse}[\versewidth]
\textit{Yeah, this album is dedicated \\
To all the teachers that told me I'd never amount to nothin' \\
To all the people that lived above the buildings that I was hustlin' in front of \\
Called the police on me when I was just tryin' to make some money to feed my daughter \\ (it's all good).\\}
\end{verse}
\attrib{The Notorious B.I.G (1972 - 1997)}
\vspace*{\fill}

%% file: frontstuff/abstract.tex
\setabstract{Astrophysical neutrinos allow us to access energies and baselines that cannot be reached by human-made accelerators, offering unique probes of new physics phenomena.
This thesis aims to address the challenges currently facing searches for Beyond Standard Model (BSM) physics in the high-energy universe using astrophysical neutrinos, particularly in the contexts of flavor measurements and connections with dark matter. 

The search for new physics with astrophysical neutrinos requires as a prerequisite understanding standard neutrino sources, which remain ambiguous.
We begin by performing a multi-wavelength search for astrophysical neutrino sources using nine years of IceCube data. We find hints of neutrino emission from radio-bright Active Galactic Nuclei (AGN), further supporting recent claims that neutrino emission occurs near the core of AGNs.

Next we turn our attention to BSM searches. 
Accurate flavor measurements of the astrophysical flux provide a smoking gun signature to BSM physics.
This requires a precise measurement of the tau neutrino fraction.
However, tau identification proved a major hurdle in the current generation of observatories.
We confront the problem of astrophysical neutrino flavor measurements by first introducing \taurunner{}, a simulation tool that accurately models the propagation of tau neutrinos including previously neglected effects such as tau lepton energy losses and depolarization in matter.
We show that better modeling of tau neutrino propagation improves IceCube transient point-source sensitivities by more than an order of magnitude at EeV energies, and diffuse flux sensitivities by a factor of two.
Second, we use this software to model IceCube counterparts to anomalous events reported by the ANITA experiment.
After performing an analysis using IceCube data, we show that all Standard Model explanations are ruled out.
Looking ahead to the future of flavor measurements, we also present a study that predicts the production of tau neutrinos via the propagation of electron and muon neutrinos in Earth, finding an irreducible but quantifiable background to next-generation tau neutrino observatories.

\quad Finally, we attempt to address the field's shared ignorance of the origin of neutrino and dark matter masses by exploring potential connections between the two. Specifically, we present an analysis of dark matter annihilation and decay to neutrinos. 
We obtain limits from MeV to ZeV masses using more than a dozen neutrino experiments. Notably, using recent data from the SuperKamiokande experiment, we place the first-ever limit on dark matter annihilation that reaches the thermal relic abundance in the neutrino sector, challenging notions that studies with neutrinos cannot be sensitive enough to make strong claims about the nature of dark matter.}
% An abstract may be required by your department.
\section{Abstract}
\uwabstract
\cleardoublepage

%% file: frontstuff/acknowledgements.tex
\section{Acknowledgements}
I am not just thankful but honestly quite confounded by the chain of circumstances that led to this thesis because only ten years ago I had dropped out of high-school. 
Since then, undoubtedly a lot of things had to work out in my favor and I for one think it's really neat that they did.

First and foremost I want to address Francis. Thank you for always being in my corner, for never questioning why I wasn't making progress on the projects I was supposed to be working on, and for always being open to new ideas no matter how crazy. 
Your compassionate guidance and trust gave me security in myself, my abilities, and the science I am pursuing. 
That is a privilege that not many students have, and it sure made things a whole lot easier. 
I only hope to carry forward a fraction of the passion and vigor with which you approach research, and I promise to continue to ``do things'' rather than reading books.

\noindent Although Francis (pre-covid) was traveling more often than not, he always answered my messages within minutes. 
He also made sure I was surrounded by people who could teach me, hands-on, the grueling details.
Carlos Argüelles, of course, immediately comes to mind. 
You are a fearless scientist. 
You told me that everyone wants glory, but few are willing to mop the floors first. 
You taught me that a scientist needs to know how to ``actually do things''; to write code, to develop new tools, solve pesky problems, think critically especially about concepts that are eerily widely accepted. Only then can we find the anomalies, and that's where the good stuff is. I will always be grateful for these lessons, which you taught me as a mentor, friend, and eventually boss. Another mentor I had the privilege of working with is Ali Kheirandish. Your humane approach to science, idealism, and attention to detail imprinted on me the desire to do not just good science, but honest science.

\noindent I thank Ben Jones for introducing me to neutrinos, for believing in me when I had a lot of enthusiasm but very little knowledge to back it up, even though I proclaimed to be a PMT expert. You are a selfless advisor who is always looking out for the little guy, and you showed me that I can actually make a career out of this thing. 

\noindent I also thank those whose love for the subject sparked in me the interest in Physics: Mohammad Cherri my high-school physics teacher, Greg Sherman at Collin College, and Collin Thomas who invited students to ``come do crazy things'' and then delivered on his promise when he gave me a microwave, some steel wool, and a resilient bacteria he'd found in his backyard to test theories of panspermia. 

\noindent To my grad school companion and lifelong friend Jeff Lazar, thanks dawg it was good to hang. 
I also thank Leslie for stepping up when I needed it most.
My first two years of graduate school were mostly held together by Lisa Everett and Jeff Schmidt. Their commitment to every student's success at UW-Madison was inspiring, I will never forget that.
I thank my good friends Neil, Alex, Zach, Brent, Praful, and Milena for commiserating about graduate school and helping me get through it.
And to the many friends who kept me sane and helped me along the way: Youssef, Jenny, Khalid, Chloe, Ahmad, Joe and Sam, Lindsey.
\noindent Additionally, to those who taught me how to overcome hurdles by putting them smack dab in my way, thank you.

\noindent Most of all, I want to thank my mother, Ghada, my sister Sara, and my brothers Mo and Karim. 
Through all our struggles, you always had my back, and never once questioned my decisions. 
It means the world.

\quad \quad And finally, to Daphne. I'm so very glad I found you.

%% file: chapters/intro.tex
\chapter{Introduction}
\SingleSpace
\epigraph{``Abstruse thought and profound researches I [Nature] prohibit, and will severely punish, by the pensive melancholy which they introduce, by the endless uncertainty in which they involve you, and by the cold reception which your pretended discoveries shall meet with, when communicated.''}{David Hume}
\DoubleSpacing
\noindent In 1930, Pauli hypothesized a neutral particle to explain anomalies in beta decay experiments and save energy conservation~\cite{Pauli:1930pc}.
The neutrino, a particle so elusive that Pauli himself thought would never be detected, was indeed observed near nuclear reactors merely two decades later by Cowan and Reines~\cite{Cowan:1956rrn}.
Not long after their discovery, neutrinos started producing anomalies of their own.

In 1964, John Bahcall predicted a flux of electron neutrinos from the Sun produced by yet unconfirmed nuclear fusion at its core~\cite{Bahcall:1964gx}.
Davis and Bahcall then proposed a chlorine experiment which was sensitive to the capture of electron neutrinos on chlorine atoms.
A few years later Davis, Harmer, and Hoffman would report the first detection of solar neutrinos~\cite{PhysRevLett.20.1205, Cleveland:1998nv}, confirming that nuclear fusion is indeed the fuel of stars;
however, the measured flux from the Sun was only a third of what Bahcall had predicted.
A host of experiments then confirmed this measurement. 
This became known as the solar neutrino problem.
It would not have been a problem, of course, if Bahcall hadn't predicted (accurately) the solar neutrino flux, and then stood by his calculation following the first measurements.

Four decades later Super-Kamiokande would measure a similar effect, namely a deficit of the predicted flux, but this time using muon neutrinos produced in the atmosphere~\cite{Super-Kamiokande:1998kpq}.
When the Sudbury Neutrino Observatory measured the solar flux with all neutrino flavors via neutral current neutrino interactions~\cite{SNO:2002tuh}, it became clear that both Bahcall and the measurements were correct.
The truth was that neutrinos were changing flavor.

What was required to solve this problem was to measure the neutrino flux at several baselines and energies.
These results would then be cemented by measurements of neutral current interactions, which are flavor-independent.
The result meant that neutrinos must have mass and the Standard Model (SM) is incomplete.

We were able to accommodate neutrino oscillations with a parametric extension of the SM. 
Neutrinos would change flavor in-flight, and this flavor change was dependent on the neutrino energy (E) and the distance it traveled (L); 
however, the underlying mechanism which provides neutrinos their mass remains unknown.
In fact, this ad-hoc extension of the SM has already started to break down.
At the turn of the century, the Liquid Scintillator Neutrino Detector~\cite{LSND:2001aii} measured a low-energy electron neutrino excess, followed by a confirmation by the MiniBoone detector, again at different energies and baselines~\cite{MiniBooNE:2007uho}.
This flavor anomaly, aptly named the MiniBoone anomaly, remains unresolved and suggests that the story is not yet over.
What had started as a bold idea put forward by Pauli, followed by an accurate calculation by Bahcall, is now forcing us to rethink our most successful theory of interactions in nature to date.

The recipe is now clear.
In order to understand the underlying neutrino mass mechanism, we must rely on a symbiosis of accurate theoretical predictions and a variety of neutrino measurements at all possible baselines and energies, until new anomalies reveal themselves exposing the incompleteness of our theories.

\begin{figure}[htb!]
    \centering
    \includegraphics[width=0.45\textwidth]{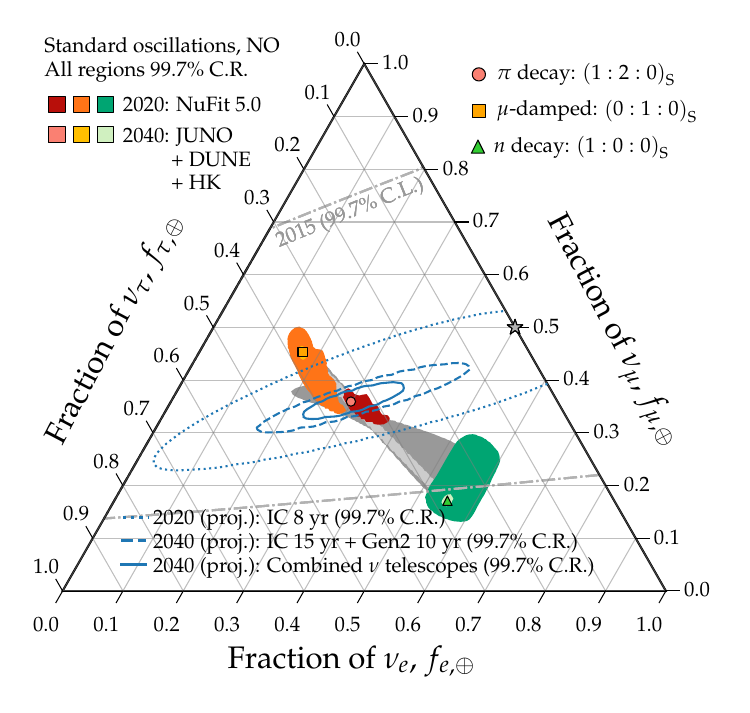}
    \includegraphics[width=0.45\textwidth]{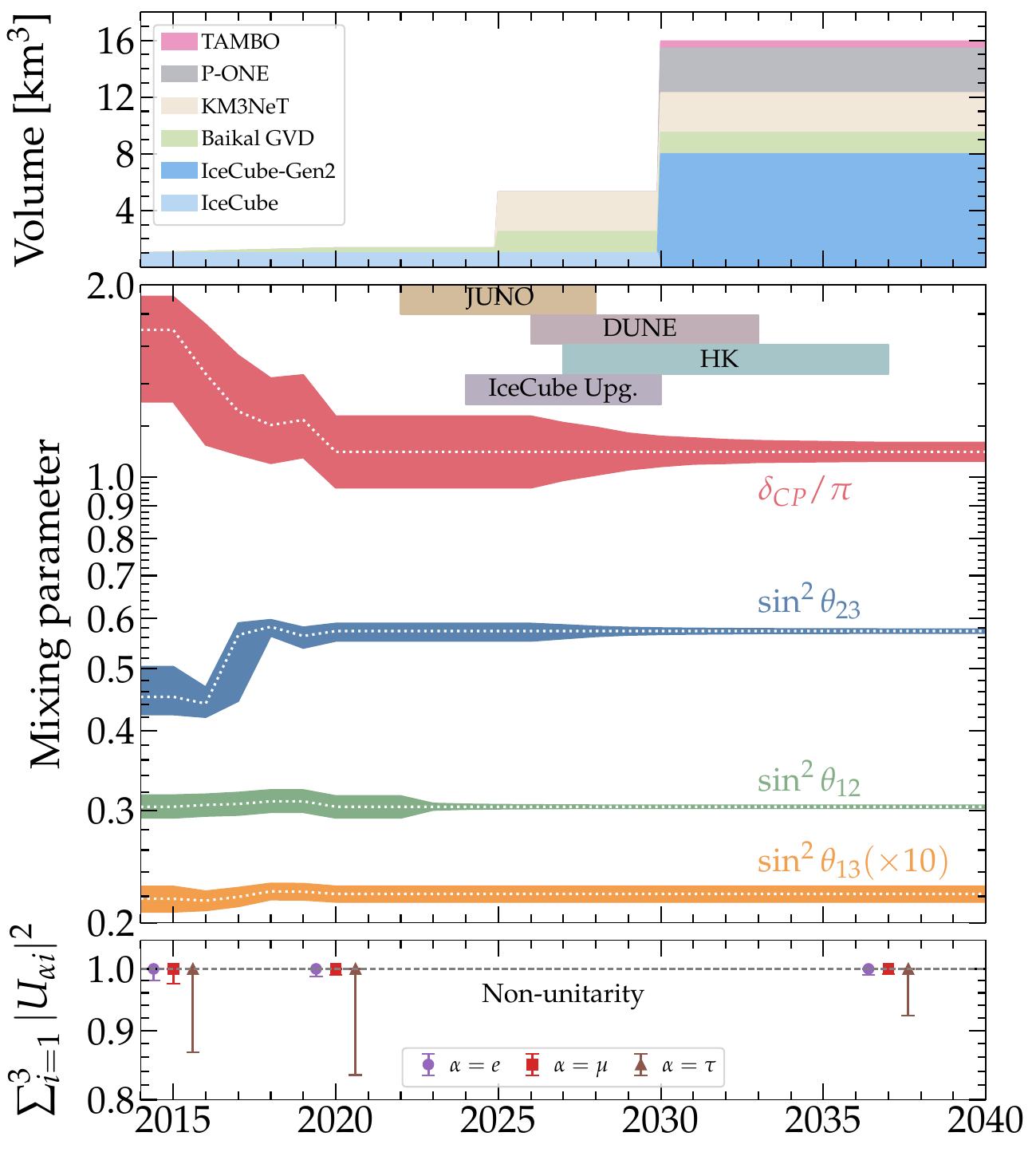}
    \caption[Astrophysical neutrino flavor potential.]{Left: Flavor composition of astrophysical neutrinos at Earth. The triangle includes current measurements and future projections.
    Different colors correspond to different production scenarios.
    Dark hues use the current uncertainties on oscillation parameter measurements, while light hues use the expected uncertainties in 2040.
    Current constraints on the flavor composition are shown as gray lines, while projections are shown in blue.
    Right: Projections for the next two decades that will affect flavor measurements. This includes the evolution of high-energy neutrino telescopes, neutrino mixing angles, and unitarity constraints.
    Top: Stacked effective volume of all high-energy neutrino telescopes.
    Middle: Uncertainties on the neutrino oscillation parameters as a function of time. Experiments relevant for neutrino oscillation parameter measurements are also shown with their proposed timelines.
    Bottom: Constraints on the unitarity of the neutrino mixing matrix for different neutrino flavors.
    Figures are from Ref.~\cite{Song:2020nfh}.
    }
    \label{fig:he}
    %\label{fig:fig_flavor_triangle}
\end{figure}

In 2013, the IceCube Neutrino Observatory reported the first detection of a high-energy astrophysical neutrino flux extending to $\sim10$~PeV energies~\cite{IceCube:2014stg}. 
Interpretations of this discovery gave footholds to address two long-standing problems in physics.
First, the energy content in the Universe is the same in cosmic-rays and neutrinos, suggesting that finding the sources of astrophysical neutrinos will lead to the discovery of cosmic-ray accelerators; this would solve the century-old cosmic-ray mystery.
Second, the astrophysical neutrino flux is predominantly extragalactic~\cite{IceCube:2017trr}, allowing us to access neutrino baselines extending to gigaparsecs, and energies millions of times higher than what can be reached with earth-based neutrino factories.
Fig.~\ref{fig:he} shows expected and measured flavor composition of astrophysical neutrinos at Earth, after oscillations at cosmic scales.
Different neutrino production scenarios are represented by different colors in the triangle. 
These are the three most likely neutrino production scenarios.
However, one can start from any point on this triangle and after oscillation, end up in the gray ``bowtie'' near the center. 
The uncertainties on the final flavor composition are driven by the uncertainties on measured neutrino oscillation parameters.
In the next two decades, experiments such as Juno~\cite{JUNO:2015zny}, Hyper-Kamiokande~\cite{Hyper-Kamiokande:2018ofw}, DUNE~\cite{DUNE:2015lol}, and the IceCube upgrade~\cite{Ishihara:2019aao} will significantly improve our knowledge of oscillation parameters.
The impact is reflected in the flavor triangle as a significant shrinking of the predicted standard oscillation region, shown in the lighter hue colors.
Any measurement that deviates from the expected astrophysical flavor composition will have discovered new physics. 
It is therefore important to measure the astrophysical neutrino flavor composition accurately.
This, today, is hindered by the tau neutrino flavor identification.

Beyond flavor measurements, neutrinos offer another unique probe into one of the most confounding cosmological mysteries; The nature of almost 85\% of mass density in the Universe is unknown.
Overwhelming astrophysical and cosmological evidence shows the bulk of the mass density of the universe cannot be accounted for in the SM.
Observations of the bullet cluster and dwarf galaxies hint at a possible corpuscular nature of this so called dark matter.
If true, this particle must interact at or below the weak scale, and could require an extension to the SM.
Although no signal has yet been detected to confirm a connection between the standard and dark sectors, another chink in the SM armor, neutrinos, may provide a path forward.
Specifically, there is a distinct possibility that the dark sector is only tied to the SM via neutrinos, a scenario motivated by scotogenic neutrino-mass-generation models\cite{Ma:2006km, Blennow:2019fhy}.
In these models, neutrinos acquire some or all of their mass effectively through interactions with a dark field.~\cite{Boehm:2006mi,Farzan:2012sa,Escudero:2016tzx,Escudero:2016ksa,Hagedorn:2018spx,Alvey:2019jzx,Patel:2019zky, Baumholzer19}.
These models introduce heavy neutrino states, sometimes called dark neutrinos, which could also provide a possible explanation of anomalies observed at short-baseline neutrino experiments such as MiniBoone~\cite{Bertuzzo:2018itn,Ballett:2018ynz,Ballett:2019cqp,Ballett:2019pyw}.

Astrophysical neutrinos, therefore, offer exciting potential in the search for new physics which could manifest at production or propagation.
Neutrinos could be produced by dark matter, implying interactions with dark fields that may also explain their mass dynamically.
On their way to Earth, new physics could also manifest in non-standard oscillations, altering their flavor composition.
But above all, understanding their standard sources is a crucial component to answering all such questions.
These three avenues will be explored here. 

But before we begin, a kind warning to the reader. This document is not self-contained, and serves rather as a listing of contributions I made as a graduate student under the supervision of Francis Halzen and Carlos Argüelles.
Familiarity with particle physics at the level of Halzen and Martin \cite{Halzen:1984mc} is assumed, as well as knowledge of standard statistical and experimental techniques as described in the Particle Data Group \cite{ParticleDataGroup:2020ssz}.
Further, I make no attempt to introduce or describe the IceCube experiment as I had no part in designing, building, or operating it; I leave that honor to those who contributed, and point you instead to the excellent detector paper here~\cite{IceCube:2016zyt}.

In this thesis we search for new physics using high-energy astrophysical neutrinos.
As a necessary pre-requisite to new physics searches, however, we begin with an attempt to understand standard neutrino sources. 
In chapter~\ref{ch:sources}, we describe a multi-wavelength search for astrophysical neutrino sources using nine years of IceCube data and find hints of neutrino emission from radio-bright Active Galactic Nuclei (AGN), further supporting recent claims that neutrino emission occurs at the core of AGNs.

We then turn our attention to astrophysical flavor measurements, and focus on improving predictions for one of the more pesky flavors, tau neutrinos, in chapters \ref{ch:taumodeling}-\ref{ch:tau_adv}.
Chapter~\ref{ch:taumodeling} describes several improvements in high energy tau neutrino propagation modeling.
We present the modeling of tau energy losses in matter, and depolarization of taus as a result of multiple interactions.
Further, we present a mechanism by which tau neutrinos are produced in Earth as a result of electron and muon neutrino propagation. 
This has a direct impact on future astrophysical neutrino flavor composition measurements. 
The work presented in this chapter resulted in two publications. The first \cite{Soto:2021vdc} was co-authored with Alfonso Garcia Soto, Pavel Zhelnin, and Carlos Argüelles and published in \textit{Physical Review Letters}. The second \cite{Arguelles:2022bma} was co-authored with Diksha Garg, Mary Hall Reno, Sameer Patel, and Carlos Argüelles and published in \textit{Physical Review D}.

Chapter~\ref{ch:taurunner} presents \taurunner{}, a simulation package that incorporates these effects in a single publicly available software.
The work presented in this chapter was published \cite{Safa:2021ghs} in \textit{Computer Physics Communications} and co-authored with Jeffrey Lazar, Alex Pizzuto, Oswaldo Vasquez, and Carlos Argüelles. 

Chapter~\ref{ch:tau_adv} discusses several advantages that result from accurate modeling of tau neutrinos. 
First, we present an improvement in IceCube's sensitivity to GZK neutrinos by a factor of two. 
Second, we show an improvement in IceCube’s transient source sensitivity by more than an order of magnitude.
Third, we perform an analysis using IceCube data and use tau neutrinos to rule out standard model explanations of anomalous events reported by the ANITA experiment.
Work presented in this chapter resulted in three publications. The first \cite{Arguelles:2022aum} was co-authored with Francis Halzen, Ali Kheirandish, and Carlos Argüelles and published in \textit{ApJ Letters}. The second \cite{Safa:2019ege} was co-authored with Alex Pizzuto, Carlos Argüelles, Francis Halzen, Ali Kheirandish and Justin Vandenbroucke and published in the \textit{Journal of Cosmology and Astroparticle Physics}. The third \cite{Aartsen_2020} was co-authored with Alex Pizzuto and Anastasia Barbano and published together with the IceCube collaboration in \textit{The Astrophysical Journal}.

Then, motivated by scotogenic neutrino mass generation models in which interactions with dark matter effectively provide some or all of the neutrino mass, we present a study of possible connections between dark matter and neutrinos in chapter~\ref{ch:DM}. 
Specifically we derive constraints on dark matter annihilation and decay to neutrinos in our galaxy and from all dark matter halos in the observable universe using more than a dozen neutrino experiments. 
Our limits cross the thermal relic abundance in the neutrino sector for the first time. 
This work \cite{Arguelles:2019ouk} was published in \textit{Reviews of Modern Physics} and co-authored with Carlos Arguelles, Alejandro Diaz, Ali Kheirandish, Andres Olivares Del-Campo, and Aaron Vincent.
Finally, in chapter~\ref{ch:conc}, we conclude.

%% file: chapters/nusources.tex
\chapter{Multiwavelength Search for Neutrino Emitters} %source search
\label{ch:sources}
\epigraph{``See first, think later, then test. But always see first. Otherwise you will only see what you were expecting. Most scientists forget that.''}{Douglas Adams}
\noindent Before our quest for new physics may begin, it is imperative that we understand the standard sources of astrophysical neutrinos for two reasons.
First, subtracting the known standard sources from the total measured astrophysical flux improves our sensitivity to neutrinos produced by exotic sources like dark matter. 
Second, knowing the production location of detected neutrinos opens up a multitude of new physics tests like, for example, neutrino decay, non-standard neutrino interactions, a fourth neutrino state, right-handed neutrinos, lorentz-invariance violation, CPT violation etc.

In this chapter we present an analysis that uses nine years of IceCube data to search for neutrino sources. 
We first summarize IceCube's observations in the last decade in Sec.~\ref{sec:gammas} and motivate a move away from an exclusive reliance on gamma rays; Instead, we advocate for a multi-wavelength approach relying on radio, x-ray, infrared, and optical data to correlate neutrino and electromagnetic sources.
In Sec.~\ref{sec:sourceselection} we describe our chosen catalog and the cuts made based on radio variability. 
We then present the statistical method and event selection in Sec~\ref{sec:anameth}.
The tested hypotheses and analysis sensitivity are then shown in Sec.~\ref{sec:hypoth}.
Finally, in Sec~\ref{sec:sourceresults} we present the results of this search and what it means in the context of other recent results.
\section{Gamma rays are not the whole story}
\label{sec:gammas}
IceCube has detected an astrophysical neutrino flux extending from energies as low as a few TeV to as high as $\sim10$ PeV~\cite{IceCube:2014stg}.
These measurements showed that the energy content in the Universe is the same in neutrinos as it is in cosmic-rays, pointing to a common origin. 
It gave us a foothold into answering the century-old mystery: Where do cosmic-rays come from, and how are they accelerated to such immense energies?

The connection is rather simple: Active galaxies or neutron stars transform some of their gravitational energy via relativistic outflows into particle acceleration. 
Accelerated protons are then stopped by surrounding ambient radiation or dust.
Astrophysical neutrinos in the Standard Model are then created by the decay of mesons or neutrons, which in turn could be created by proton-proton or proton-photon interactions.
A necessary byproduct of all such interactions is also high-energy gamma-rays. 
Therefore neutrino astronomers' first instinct was to search for correlation of high-energy gamma-rays with astrophysical neutrinos.
This led to a multitude of correlation studies between IceCube and GeV-TeV gamma ray sources in the last decade~\cite{IceCube:2021slf,IceCube:2018ndw, Abbasi:2022hdv}.
These studies, however, have indicated that Fermi-detected blazars cannot be the dominant sources of IceCube's astrophysical neutrinos~\cite{IceCube:2016qvd, Huber:2019lrm}.

In 2017, IceCube launched a realtime alert program that informs the astronomy community of any interesting neutrino candidates that have more than $50\%$ chance of being astrophysical in origin, enabling followups by any telescope that has an overlapping field-of-view~\cite{IceCube:2016cqr}.
On September 22nd, 2017, one such alert with estimated $290~$TeV neutrino energy was found to be in spatial coincidence with TXS 0506+056, an active galaxy that is bright in multiple wavelengths of light~\cite{IceCube:2018dnn}.
This IceCube event was coincident with a gamma-ray flare detected by the Fermi telescope. 
A followup search in IceCube's archival data in the direction of TXS 0506+056 found a flare of 13 neutrinos in 100 days, a period spanning mid 2014 to 2015~\cite{IceCube:2018cha}.
This 2014-2015 neutrino flare of TXS-0506+056 was correlated with a rise in its radio flux at 15GHz, and a significant drop in the optical flux; however, no enhancement in the gamma-ray activity was detected~\cite{Fermi-LAT:2019hte}.
Recent studies of this source in fact pointed to hints of gamma-ray suppression at the time of neutrino emission not just in TXS 0506+056, but also in two other gamma-ray sources spatially coincident with high-energy neutrino events~\cite{Kun:2020njy}. 
These observations provided the first clues that sources may not simultaneously be efficient neutrino and gamma-ray emitters.

In hindsight, this may seem quite obvious.
In order to efficiently produce neutrinos, a source must have in their vicinity sufficient target matter or radiation to stop a fraction of the cosmic-rays they accelerate.
A simple comparison of the proton and photon interaction cross sections reveals that material or radiation dense enough to stop protons is more likely to stop photons with equivalent energy.
Of course, the photons are not absorbed, but cascade down in energy until their energy is low enough to escape the source environment.
It is therefore prudent in the selection of sources for our analysis to relax the requirement that neutrino and gamma-ray sources are directly correlated.
It's important to note we do not claim that neutrinos and gamma-rays are unrelated.
Rather we claim that this relationship is more complicated than a simple direct correlation, and we proceed in our source selection without making any assumptions about the gamma-ray connection.

\section{Source Selection}
\label{sec:sourceselection}
We select a special class of AGN for this analysis using radio, infrared, and x-ray data to pick out sources more likely to be active and variable, pointing to an active core.
These scenarios could include but are not limited to galaxy or supermassive black-hole mergers, major matter accretion activity near the core of AGN, and tidal disruption events~\cite{Nellen:1992dw, deBruijn:2020pky, Murase:2020lnu}. 
All such events can cause outbursts in most wavelengths of photons, and manifest as heightened variability at various timescales.
Such signatures are indicators of necessary accelerator and target material to produce neutrinos.
Our main discriminant will be the variability of the 15GHz radio flux.
Previously, two alerts issued by IceCube pointed to possible radio enhancement accompanying high-energy neutrino emission (TXS 0506+056 and PKS 1502+106). 
At the time of the neutrino alerts, these two sources were either close to or at their peak radio emission at 15GHz over a period of more than two decades of continuous observation~\cite{Kun:2020njy}.

The sources are selected from the Candidate Gamma Ray Blazar Survey (CGRaBS) catalog~\cite{Healey:2007gb}. 
CGRaBS is a statistically complete sub-selection of the CRATES catalog (Combined Radio All-Sky Targeted Eight GHz Survey). 
It consists of 1625 radio sources with flat/hard radio spectra.
This catalog was initially created in 2007 to guide Fermi observations, which was set to launch in 2008.
Therefore the sources were selected based on their radio and x-ray spectra to resemble properties of gamma-ray blazars.
Interestingly, only $39\%$ of these sources were seen by Fermi in 10 years. 

To pick out more likely neutrino emitters from this catalog, we first characterize the radio variability of the sources using two measures.
First, the intrinsic modulation index (V)
\begin{equation}
    V=\frac{s}{\bar{I}}=\frac{1}{\bar{I}} \sqrt{\frac{N}{N-1}\left(\overline{I^{2}}-\bar{I}^{2}\right)}
\end{equation}
where $s$ is the standard deviation of flux measurements, and $\bar{I}$ is the mean measured flux density.
This is a simple estimate of the deviation of a set of measurements from the mean measured flux density. 
This does not, however, account for statistical uncertainties in the individual measurements and could therefore be inaccurate, especially for fainter sources.
The second measure we use is $\eta$, a reduced $\chi^2$ test-statistic that quantifies the deviation of individual flux measurements from the weighted mean measured flux density
\begin{equation}
    \eta=\frac{1}{N-1} \sum_{i=1}^{N} \frac{\left(I_{i}-\xi_{I}\right)^{2}}{\sigma_{I, i}^{2}}
\end{equation}
where N is the number of observations, $\sigma_{I, i}$ is the uncertainty on a single measurement, $I_{i}$ is a single flux measurement and
\begin{equation}
    \xi_{I}=\frac{\sum_{i=1}^{N} I_{i} / \sigma_{I, i}^{2}}{\sum_{i=1}^{N} 1 / \sigma_{I, i}^{2}}
\end{equation}
is the weighted mean of flux measurements.

\begin{figure}
    \centering
    \includegraphics[width=0.75\textwidth]{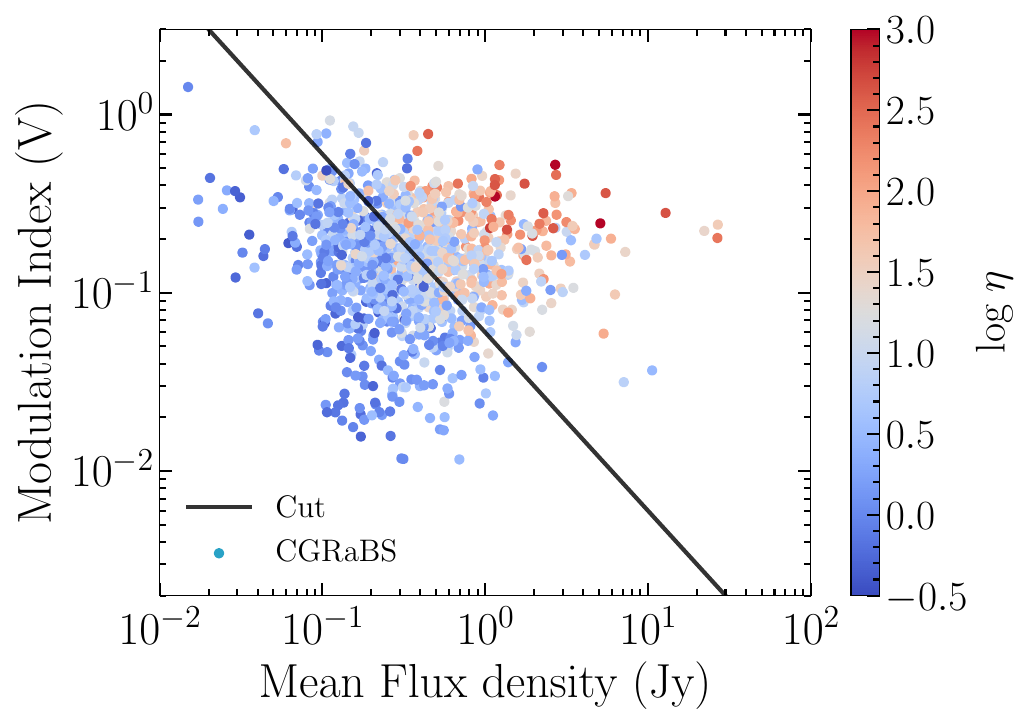}
    \caption[Variability measures for all sources in the CGRaBS catalog above $-20^\circ$ declination.]{Variability measures for all sources in the CGRaBS catalog above $-20^\circ$ declination. Shown is the modulation index (V) as a function of mean radio flux density at 15GHz as measured by the OVRO 40m telescope. The color scale shows the log of the variability test-statistic used to characterize these sources. The black line is where we place the cut to keep more variable sources.}
    \label{fig:variability}
\end{figure}

We use radio data obtained from a fast-cadence 15 GHz radio monitoring program with the 40m telescope at the Owens Valley Radio Observatory (OVRO)~\cite{Richards:2010su}.
The telescope is located in the
Owens Valley, California at an elevation of 1236m. Construction of the telescope was completed in 1966 and has been used for many missions since then.
The blazar monitoring program ran between 2007 and 2020 and collected data at a cadence of 3 days from more than 1500 sources.
Although OVRO has so far only published lightcurves from 2007 till 2016, this provides thousands of measurements per source and is sufficient to characterize their variability using the measures explained above.
A 2-dimensional scatter plot of the modulation index as a function of the mean flux density is shown in Fig.~\ref{fig:variability}. The color scale is the $\chi^2$ test-statistic defined above.
The black line in the figure corresponds to a standard deviation of 65mJy, which is the mean standard deviation of all sources.
We are interested in the most variable sources, and place a cut to keep only those sources whose variability exceeds the mean standard deviation of the population.
This corresponds to the region above the black line, and results in a selection for sources whose test-statistic ($\eta$) is highest.
The final selection, after variability cuts, includes 390 sources.

Finally, we cross-correlate these sources with surveys at other wavelengths, starting with infrared photons.
Infrared photons coming from the dust torus of an AGN are a good candidate for target radiation. 
Therefore we are interested in loud radio AGN that also have associated infrared emission.
However, there are other physical scenarios, such as star-forming regions, where there is significant radio and infrared emission, which doesn't fit the hypothesis we're trying to test.
Since this catalog is composed of blazar candidates, however, we expect the IR emission to be pointing mostly to target radiation associated with the central engine, as described by the blazar sequence~\cite{Murase:2014foa}.
To make our selection, we use a value called $q_{22}$, which is simply the ratio of IR to radio intensities.
Because active central engines are expected to cause an IR flux, this value $q_{22}$ is then supposed to minimize the contamination from sources whose radio emission is not associated with AGN activity.
In other words, IR emission associated with the activity causing radio emission will show a correlation, and this correlation can be captured in the ratio of the two intensities.
We use radio-loud WISE (Widefield Infrared Survey Explorer) blazars detected in all four WISE filters, whose mid-IR colors are similar to the typical colors of confirmed gamma-ray emitting blazars~\cite{DAbrusco:2014nij}. The cross-correlation with CGRaBS yields 238 sources.

We then turn to X-ray data.
The X-ray flux can act as a good proxy to column depth, which in turn is a possible indicator of gamma-ray opaque sources.
We cross-correlate our selection with XMM-Newton~\cite{XMM:2001haf}, an X-ray telescope sensitive to the $0.2-12$ keV flux.
The cross-correlation yields 87 sources.

\begin{figure}[ht!]
   \centering
   \vspace{1cm}
   \subfloat{\includegraphics[width=.47\textwidth] {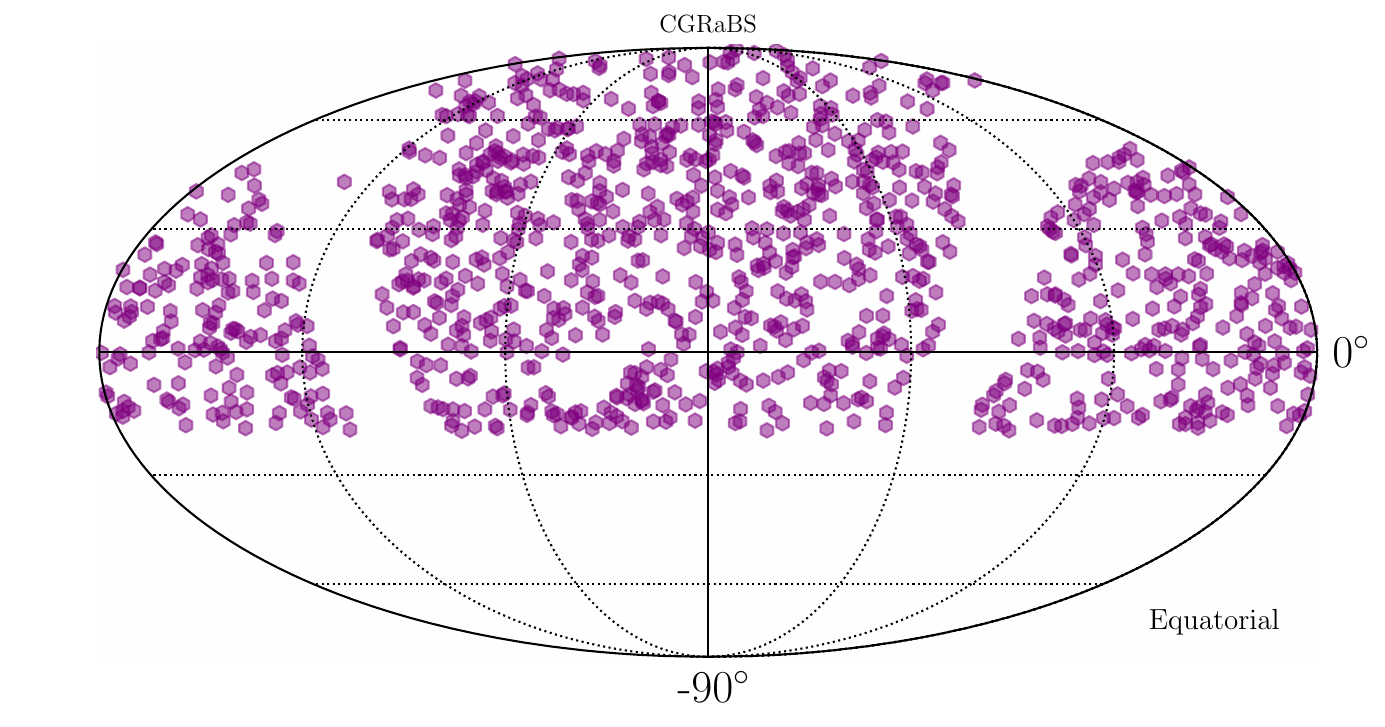}}\quad
   \subfloat{\includegraphics[width=.47\textwidth]{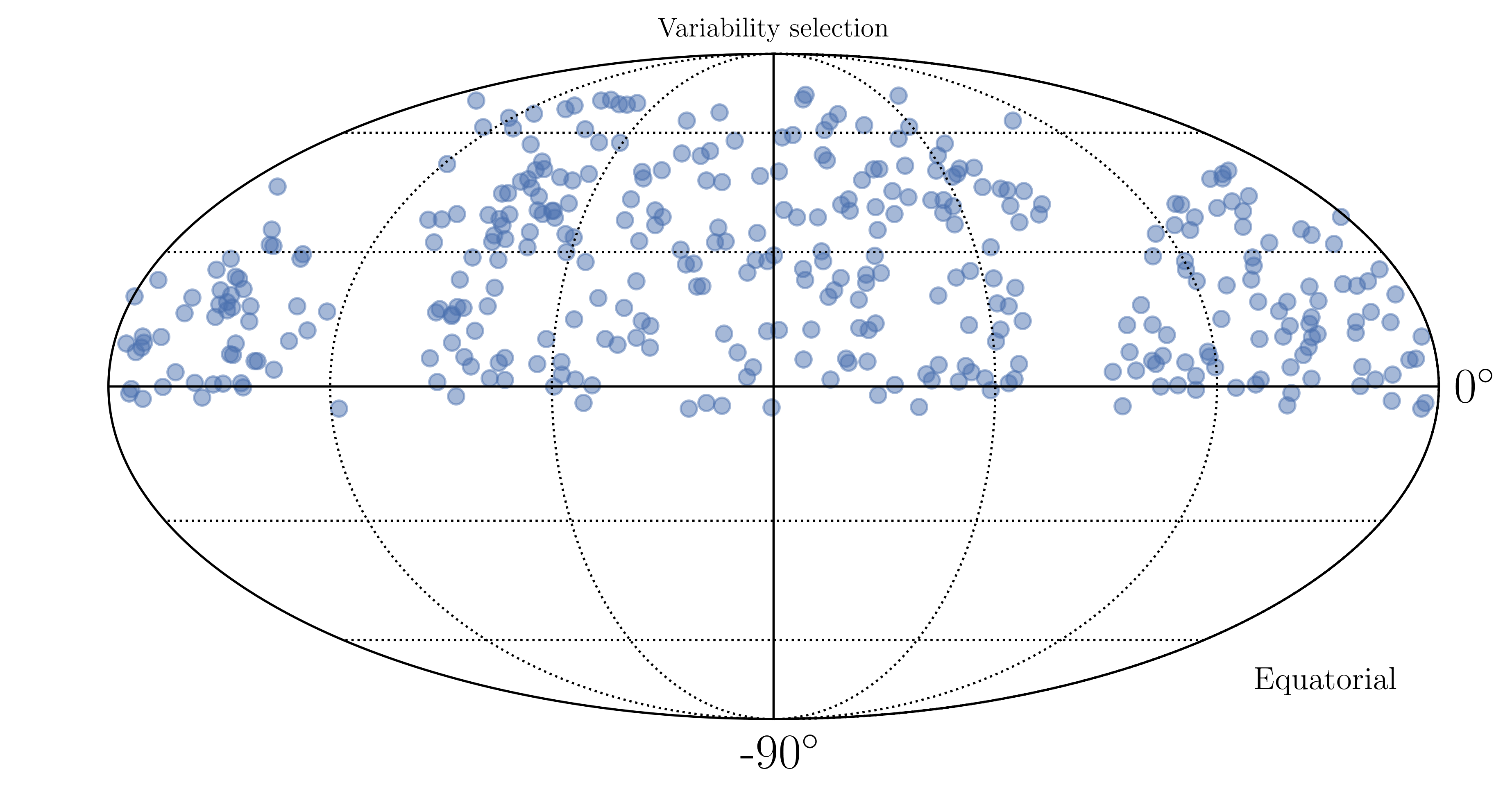}}\\
   \subfloat{\includegraphics[width=.47\textwidth]{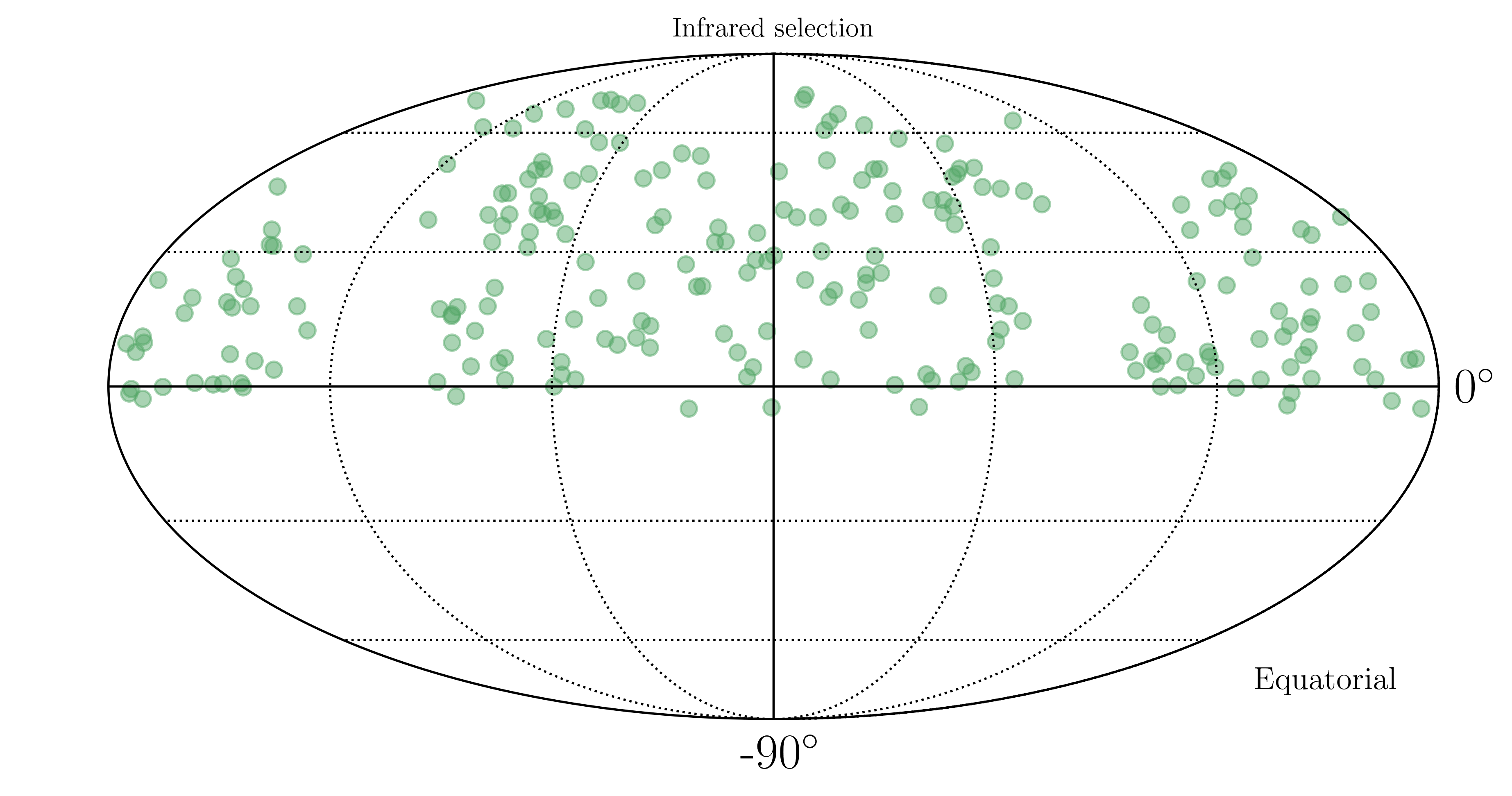}}\quad
   \subfloat{\includegraphics[width=.47\textwidth]{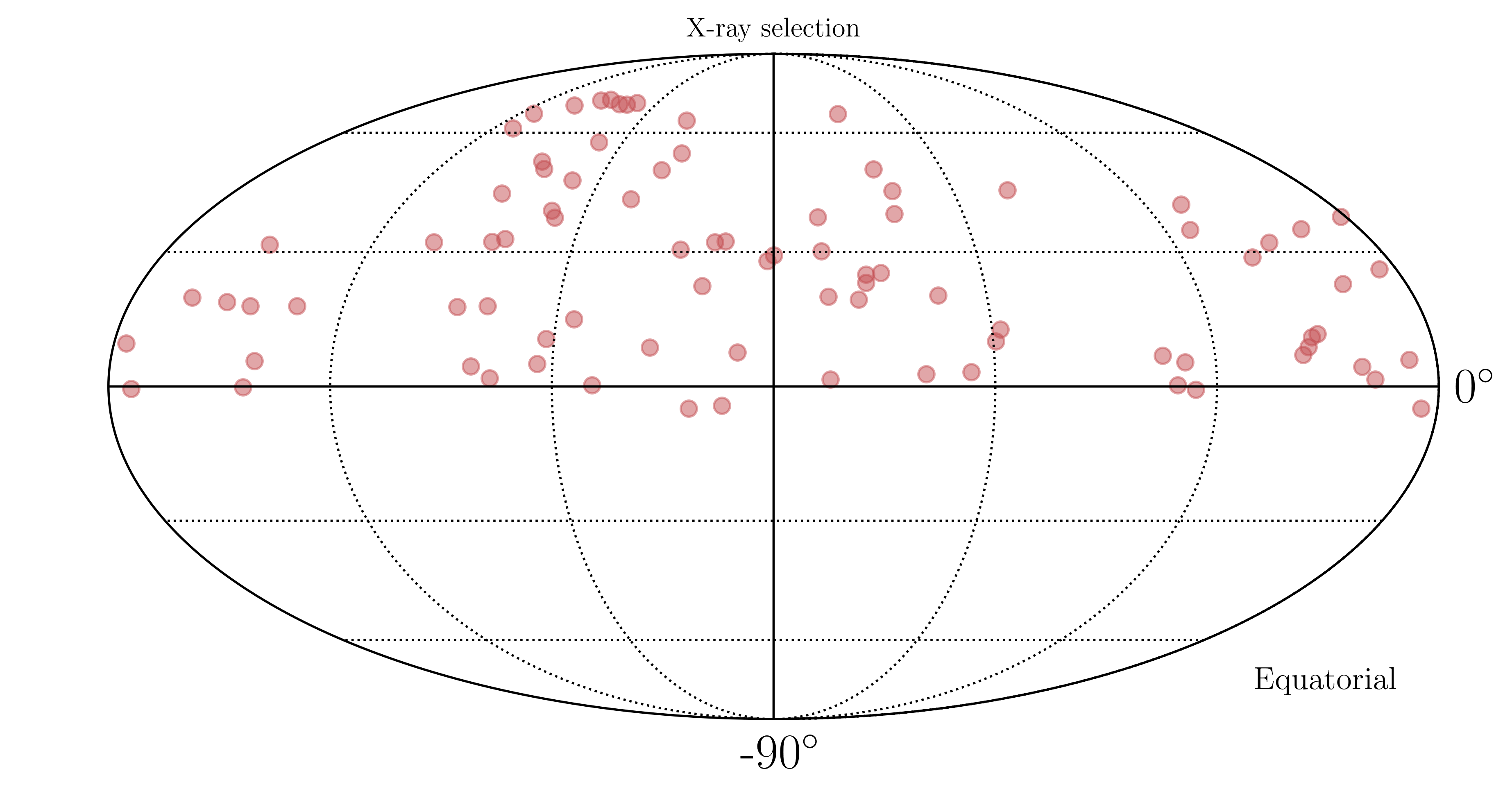}}
   %\vspace{-0.5cm}
   \caption[Skymaps of CGRaBS sources.]{Skymap of CGRaBS sources. Top-left shows all of the sources in the field of view of OVRO. Top-right shows the 390 sources remaining after applying variability cuts, and a declination cut of $-5^\circ$ due to the IceCube event selection used. Bottom-left shows the remaining 283 sources after cross-correlation with the radio-loud WISE catalog, and bottom-right shows the remaining 87 sources after cross-correlation with XMM-Newton.}
   \label{fig:cgrabssky}
\end{figure}

A skymap of all 1158 sources in the CGRaBS catalog above a declination of -20 degrees is shown in Fig.~\ref{fig:cgrabssky}.
This declination band represents the field of view of the OVRO 40m telescope.
We also show skymaps of all cross-correlated selections discussed above, which includes sources after the variability cut, the X-ray cross-correlation, and Infrared cross-correlation.

\section{Analysis Method}
\label{sec:anameth}
Our analysis searches for neutrino spatial and energy correlation from a collection of sources.
We test the hypothesis that all of the sources in a catalog are emitting neutrinos.
This method is called a stacking analysis~\cite{Braun:2008bg}.
Stacking sources together has been shown to improve the sensitivity of source searches.

We use an unbinned maximum-likelihood to search for neutrino emission above the atmospheric and diffuse astrophysical components.
The likelihood is defined as
\begin{equation}
    \mathcal{L}\left(n_{s}, \gamma\right)=\prod_{i=1}^{N}\left[\frac{n_{s}}{N} \mathcal{S}_{\mathrm{stack }}\left(\mathbf{x}_{i},\left\{\mathbf{x}_{S_{k}}\right\}, \sigma_{i}, E_{i} \mid \gamma\right)+\left(1-\frac{n_{s}}{N}\right) \mathcal{B}\left(\delta_{i}, E_{i}\right)\right]
\end{equation}
where $n_{s}$ is the number of signal events and $\gamma$ the spectral index of the sources assuming a simple power-law.
$N$ is the total number of events in the sample, $\mathbf{x}_{i}$ is the sky coordinate of event $i$, $\mathbf{x}_{S_{k}}$ is the sky coordinate of source $S_{k}$, $\sigma$ is the angular uncertainty, $E$ is the energy proxy.
The background pdf, $\mathcal{B}$ is
\begin{equation}
    \mathcal{B}\left(\delta_{i}, E_{i}\right)=\underbrace{\frac{\mathcal{P}_{\mathcal{B}}\left(\delta_{i}\right)}{2 \pi}}_{\text {spatial }} \times \underbrace{\mathcal{E}_{\mathcal{B}}\left(E_{i}, \delta_{i}\right)}_{\text {energy }},
\end{equation}
where $\mathcal{P}_{\mathcal{B}}$ is the background probability of an event and depends only its declination $\delta$ since IceCube is located at the geographic South Pole. 
The stacking signal PDF, $\mathcal{S}_{\mathrm{stack}}$ can be written as
\begin{equation}
\mathcal{S}_{\mathrm{stack}}\left(\mathbf{x}_{i},\left\{\mathbf{x}_{S_{k}}\right\}, \sigma_{i}, E_{i} \mid \gamma\right)=\frac{\sum_{k=1}^{M}\left[W_{k} R_{k}\left(\delta_{S_{k}} \mid \gamma\right) \times \mathcal{S}\left(\mathbf{x}_{i}, \mathbf{x}_{S_{k}}, \sigma_{i}, E_{i} \mid \gamma\right)\right]}{\sum_{k=1}^{M} W_{k} R_{k}\left(\delta_{S_{k}} \mid \gamma\right)}    
\end{equation}
where the sum runs over all $M$ sources, and each individual event signal PDF is weighted by a combination of the detector acceptance at the source location $R_{k}$ and the theoretical source weight $W_{k}$, which will be discussed later on.
Finally, the signal PDF $\mathcal{S}$ is built from simulation and using kernel density estimators to extract the energy dependence of the angular estimators, including intrinsic uncertainties such as the kinematic angle between the neutrino and the muon.

The dataset used for this analysis consists of nine years (2011-2019) of IceCube muon events in the Northern Sky for a total livetime of 3186 days.
We use deep neural networks to reconstruct the muon energies which significantly improves the energy resolution over previous treatments.
A detailed description of the dataset and reconstruction techniques was published in ()

The best estimates for the parameters assuming a signal hypothesis are those which maximize the likelihood function, $\mathcal{L}$. 
This maximized likelihood is then compared against the likelihood of the background only hypothesis such that
\begin{equation}
    \mathcal{TS} = 2 \ln \Bigg( \frac{ \mathcal{L}\left(\hat{n}_{s}, \hat{\gamma}\right)}{ \mathcal{L}\left(n_{s} = 0\right)} \Bigg) \; .
\end{equation}
where variables with~(~$\hat{}$~)~are best-fit parameters. 
This is the test-statistic used for the analysis. 
\section{Hypotheses and Sensitivity}
\label{sec:hypoth}
\begin{figure}[ht!]
   \centering
   \subfloat{\includegraphics[width=.32\textwidth, trim={0.75cm 0.75cm 0.75cm 0.75cm}]{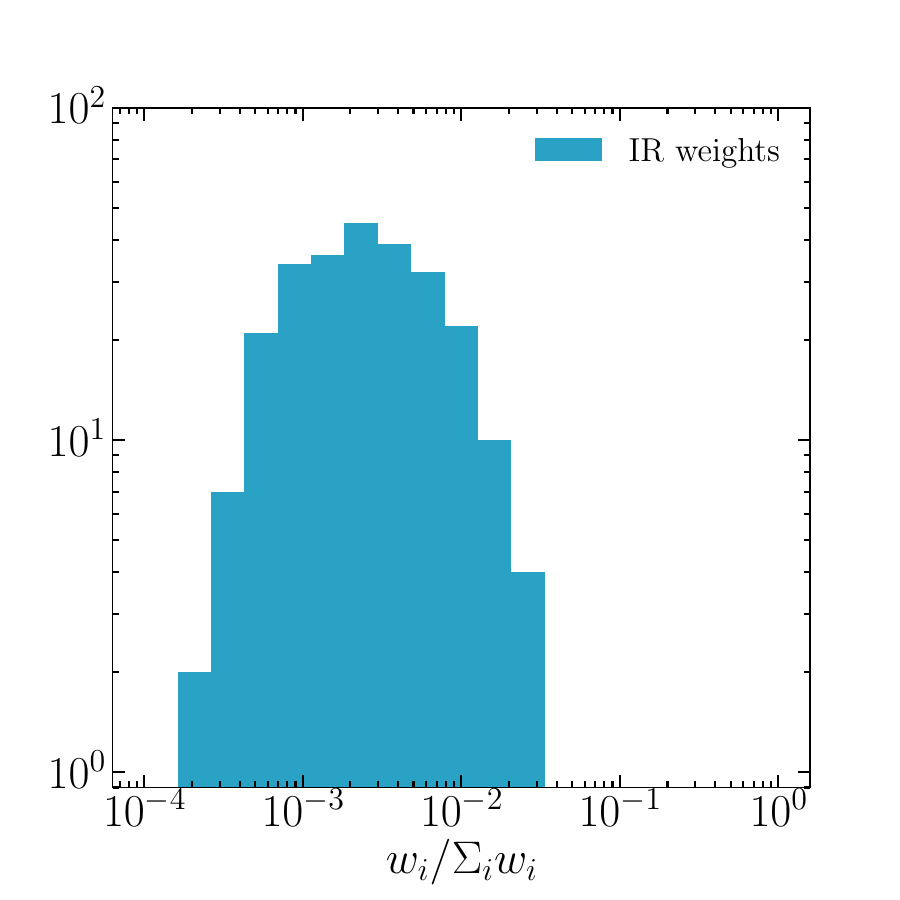}}
   \subfloat{\includegraphics[width=.32\textwidth, trim={0.75cm 0.75cm 0.75cm 0.75cm}]{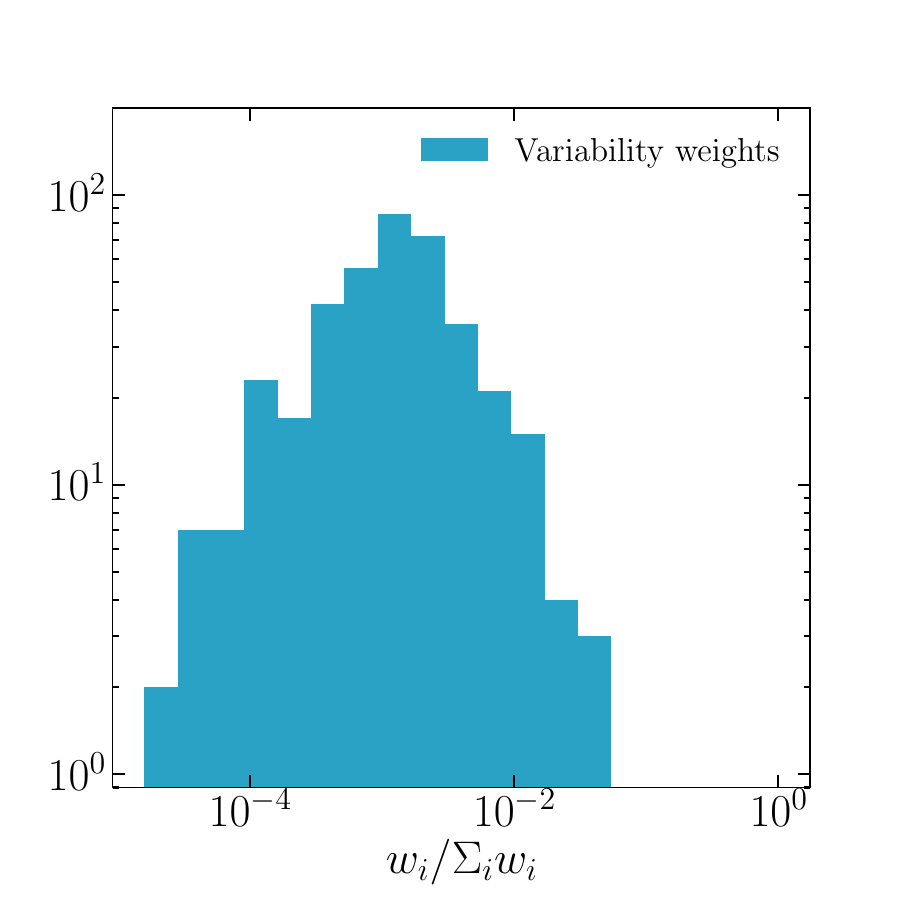}}
   \subfloat{\includegraphics[width=.32\textwidth, trim={0.75cm 0.75cm 0.75cm 0.75cm}]{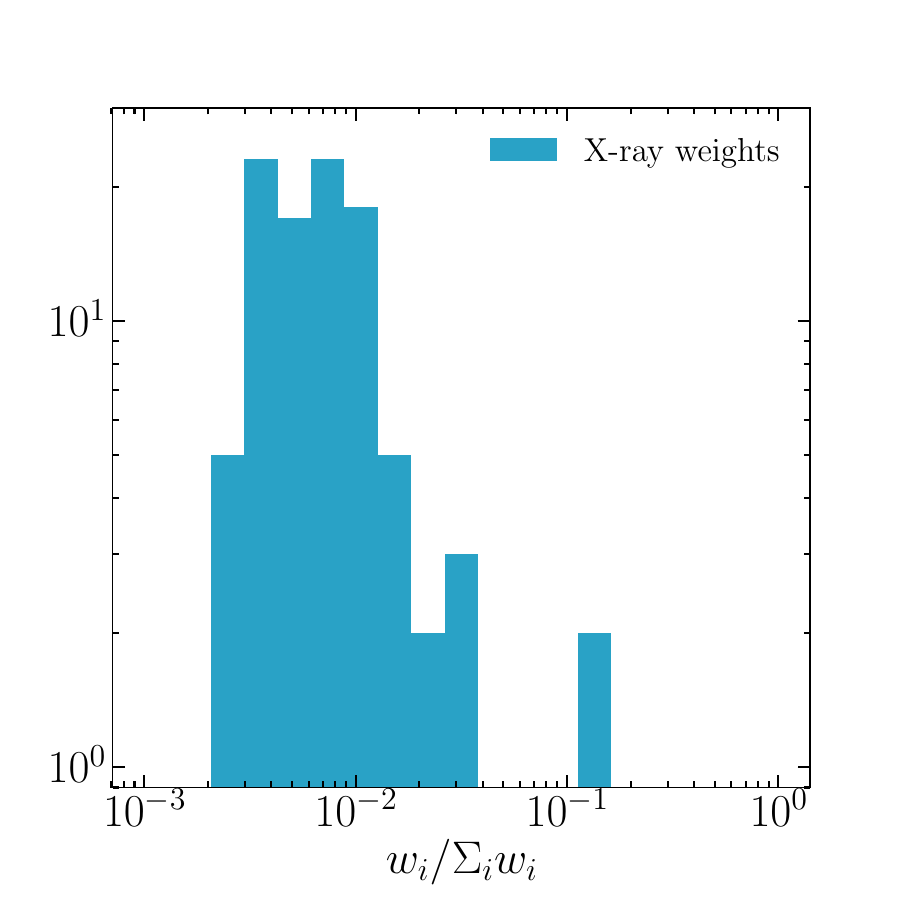}}
   \vspace{0.5cm}
   \caption[Stacking analysis weights.]{Sum-normalized weight distributions shown for all three hypotheses tested. Notably, the X-ray flux weights show two sources that each carry at least 10\% of the weights, whereas the other two hypotheses are more evenly distributed.}
   \label{fig:weights}
\end{figure}

\begin{figure}[ht!]
   \centering
   \vspace{1cm}
   \subfloat{\includegraphics[width=.32\textwidth]{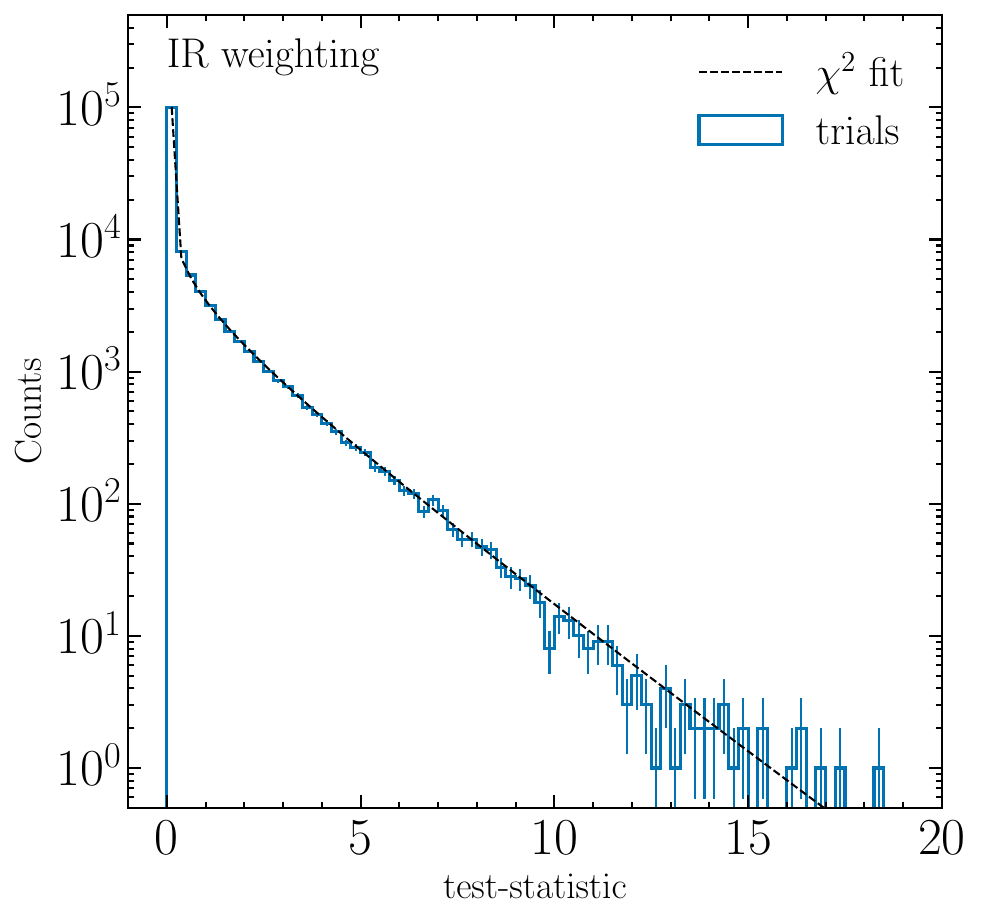}}
   \subfloat{\includegraphics[width=.32\textwidth]{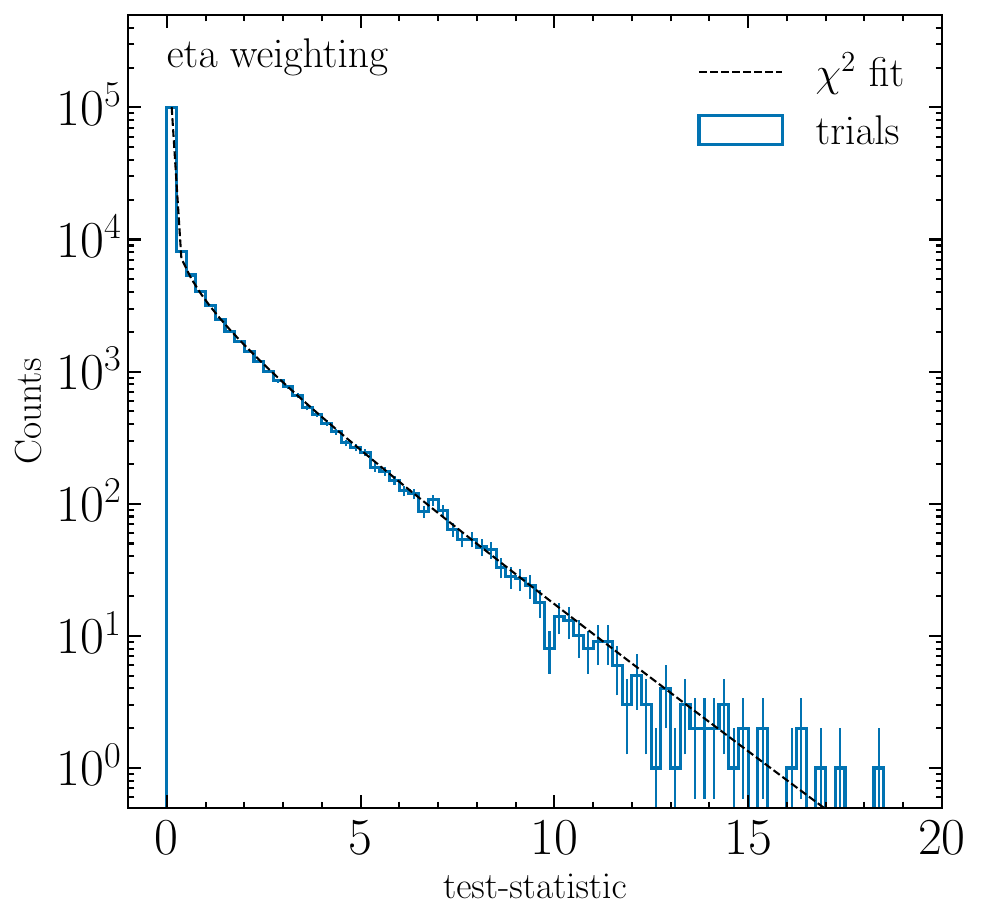}}
   \subfloat{\includegraphics[width=.32\textwidth]{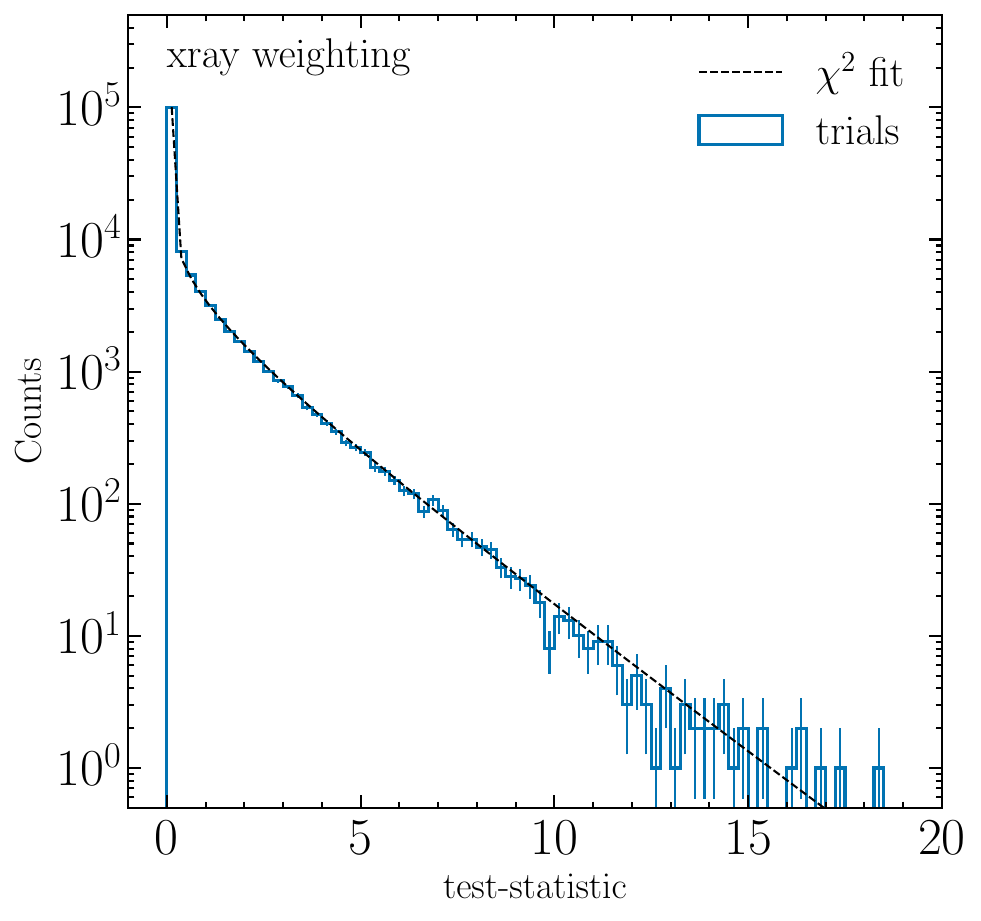}}
   \caption[Background TS distributions]{Distribution of test-statistic values assuming a background-only hypothesis, shown for all three weighting schemes. Overlayed on top of the histogram is a functional fit assuming a $\chi^2$ distribution with a delta function at zero.}
   \label{fig:bgdists}
\end{figure}

\begin{figure}[ht!]
   \centering
   \subfloat{\includegraphics[width=.32\textwidth]{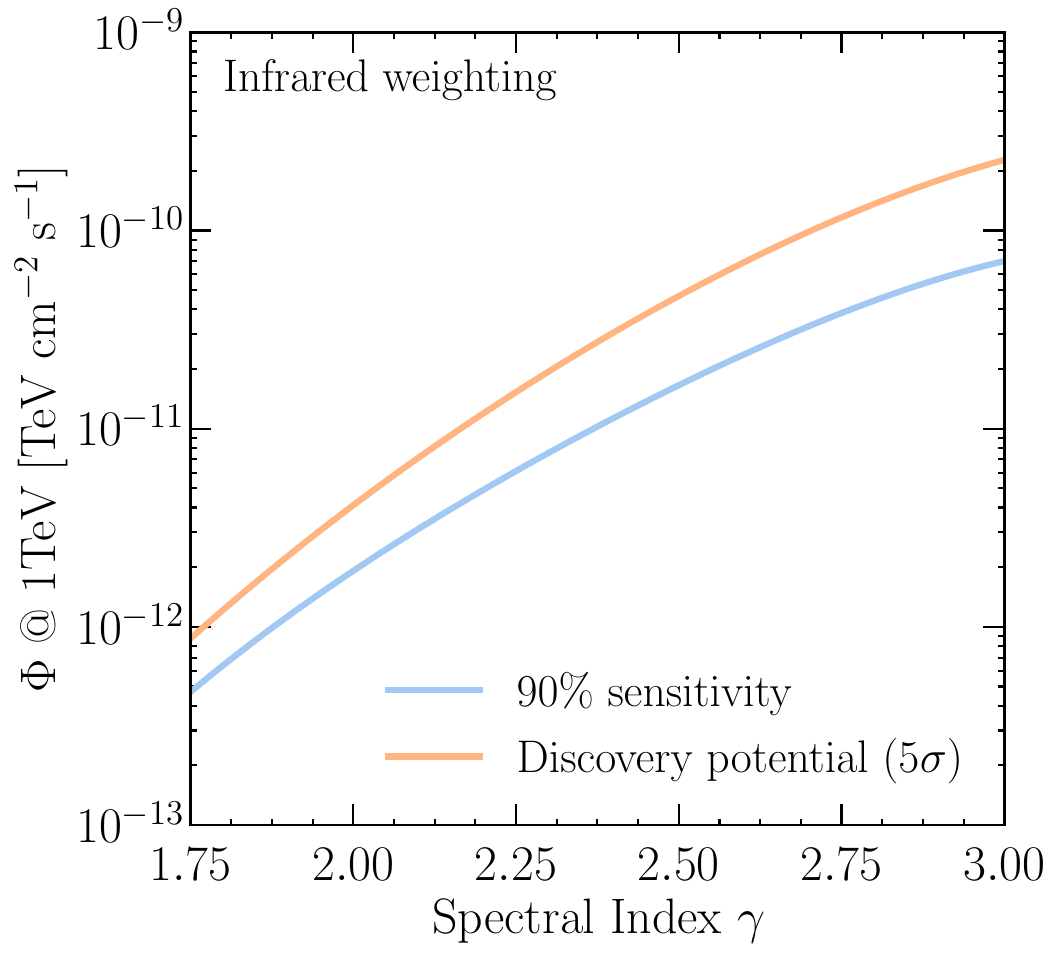}}
   \subfloat{\includegraphics[width=.32\textwidth]{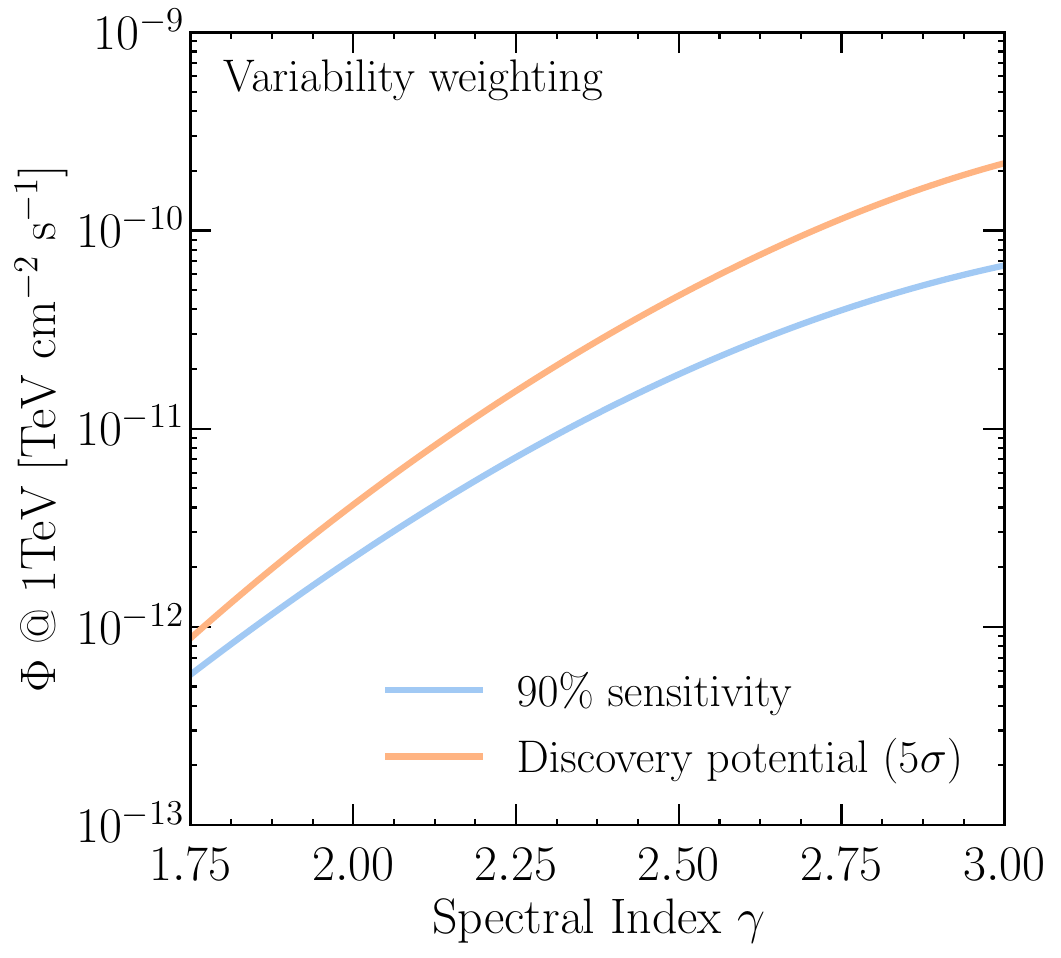}}
   \subfloat{\includegraphics[width=.32\textwidth]{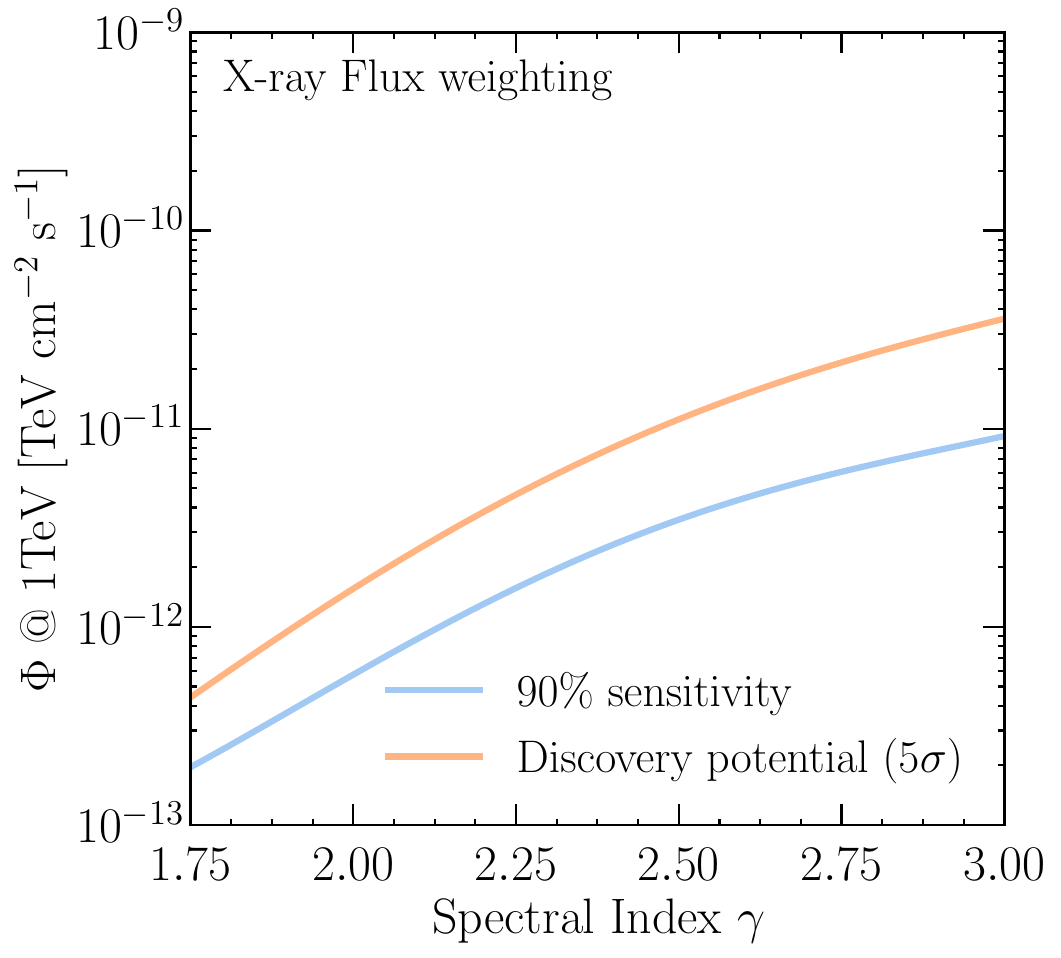}}
   \caption[Analysis sensitivity and discovery potential]{Sensitivity and discovery potential as a function of spectral index, for all three weighting schemes. Sensitivity is defined as the flux where 90\% of signal-injected trials exceed the mean background $\mathcal{TS}$. Discovery potential is defined as the flux needed for 50\% of signal injection trials to exceed the $\mathcal{TS}$ value corresponding to the quoted significance.}
   \label{fig:sensitivities}
\end{figure}

Starting with our source selection after radio variability cuts, we test three hypotheses in this analysis.
These hypotheses use the same methods described above but vary in the final source selection and source weights $W_{k}$. 
The first hypothesis tests for a correlation directly between neutrino emission and radio variability using, as weights, the variability test-statistic $\eta$ described in \ref{sec:sourceselection}.
The second hypothesis uses the correlation between our selected sources and the radio-loud WISE catalog.
We weight this selection by the ratio of radio intensity to infrared intensity at $22$~mm.
Finally, the third hypothesis tests for a correlation with soft X-rays weighting by the X-ray flux measured with XMM Newton. 
Skymaps for these selections can be seen in Fig.~\ref{fig:cgrabssky}, and the distributions of weights is shown in Fig.~\ref{fig:weights}.
We build background distributions for these locations and weights by repeating many trials with randomized data realizations; that is, we assume $n_{s}=0$ and obtain the corresponding $\mathcal{TS}$.
The background distributions obtained for all three hypotheses are shown in Fig.~\ref{fig:bgdists}.
The sensitivity for this analysis is then estimated via injection tests where a signal is inserted in a fake data realization, and the $\mathcal{TS}$ is maximized.
We repeat this test many times for various injected spectral indices and fluxes.
The sensitivity is set at 90\% C.L when 90\% of the trials return a $\mathcal{TS}$ value greater than the mean of background-only trials.
The $5\sigma$ discovery potential is set at the point where 50\% of trials exceed the $5\sigma$ threshold background test-statistic.
Sensitivities and discovery potentials as a function of spectral index are shown in Fig.~\ref{fig:sensitivities} for all three hypotheses. 

\section{Results and Interpretation}
\label{sec:sourceresults}
\begin{figure}
    \centering
    \includegraphics[width=0.5\textwidth]{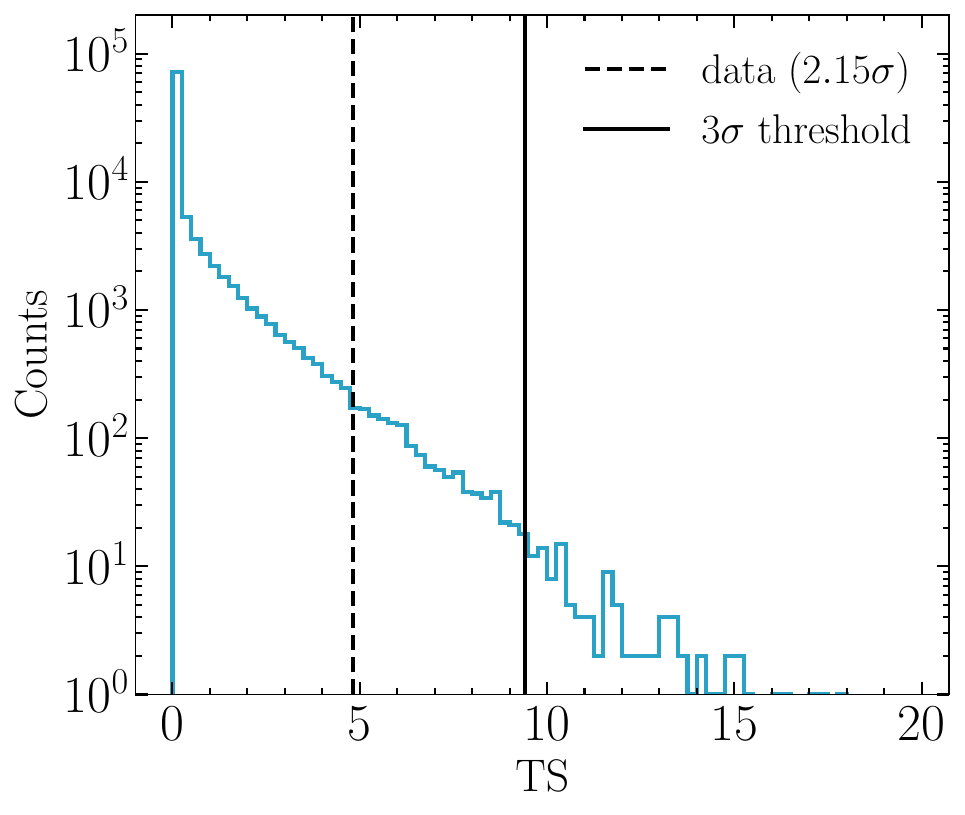}
    \caption[Unblinded test-statistic distribution.]{Unblinded $\mathcal{TS}$ value overlaid on the background distribution for the infrared selection. The p-value corresponds to a one-sided significance of $2.15\sigma$. Also shown is the $3\sigma$ threshold $\mathcal{TS}$ value.}
    \label{fig:unblindedTS}
\end{figure}

\begin{figure}
    \centering
    \includegraphics[width=0.5\textwidth]{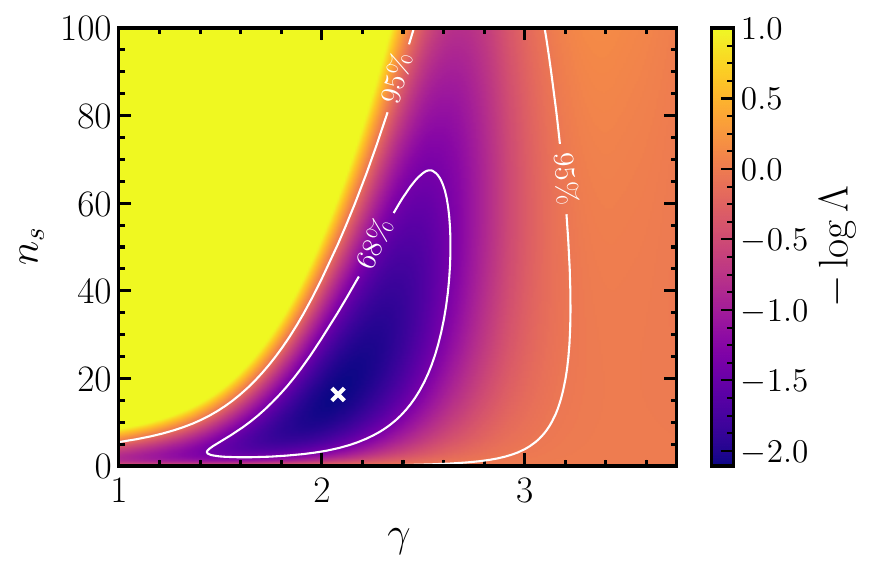}
    \caption[Unblinded log-likelihood scan.]{Unblinded log-likelihood scan showing the best-fit point (white marker) as well as the $1\sigma$ and $2\sigma$ contours. The best-fit point corresponds to 16.33 events with a spectral index of 2.08. $\Lambda$ is the ratio of the signal and background likelihood values.}
    \label{fig:2dscan}
\end{figure}

Our most significant result is the infrared sub-selection. 
The unblinded test-statistic overlayed on the background distribution is shown in Fig.~\ref{fig:unblindedTS}.
We find hints for neutrino emission in the case of our infrared selection at the level of $2.15\sigma$ one-sided significance above background expectation.
We fit $16.33$ events with a spectral index of $2.08$.
The unblinded log-likelihood scan is shown in Fig~\ref{fig:2dscan} where we plot the fitted flux as a function of spectral index. 
The best-fit flux and spectral index as well as the $1$ and $2 \sigma$ contours are shown.
This observation further supports a recent result from the IceCube experiment testing the correlation of neutrinos with AGN cores \cite{IceCube:2021pgw} which showed an excess of $2.6\sigma$ after stacking $32,249$ sources.
We obtain a similar significance but with a factor of $\sim100$ less sources, and more importantly with a negligible overlap between the two source selections.
This gives us increased confidence in the underlying physics scenario being tested. 
Namely, cosmic-rays are likely accelerated near the cores of AGN, and are stopped by the surrounding ambient radiation, producing neutrinos and obscuring the high-energy photons.
Independent groups using IceCube's public alerts with estimated energies $>200$~TeV have also seen a correlation of IceCube neutrinos with radio AGN, and concluded that neutrinos are likely produced by proton interactions with ambient radiation near the core of AGN\cite{Plavin:2020emb}.
Further, an analysis of public icecube muon events in the TeV-PeV energy range found evidence for neutrino emission from compact radio sources~\cite{Plavin:2020mkf}.

These observations, along with the analysis presented here, all point to a coherent picture. 
Astrophysical neutrinos point to cosmic-ray accelerators with significant radio emission coming from an active core.
Relativistic outflows from accreting matter onto that core accelerate cosmic-rays.
These cosmic-rays subsequently interact with ambient radiation near the source where the radiation density is high, making efficient neutrino sources that become gamma-ray opaque. 
The sources are not necessarily gamma-ray quiet, but the gamma-rays likely escape at lower energies, and the observed high-energy gamma rays are produced further downstream from the site of neutrino production. 
Undoubtedly, further years of observation will make this conclusion clear and robust.

%% file: chapters/tau_modeling.tex
\lstdefinestyle{mystyle}{
    backgroundcolor=\color{backcolour},   
    commentstyle=\color{codegray},
    keywordstyle=\color{codegreen},
    numberstyle=\tiny\color{codegray},
    stringstyle=\color{orange},
    basicstyle=\ttfamily\footnotesize,
    breakatwhitespace=false,         
    breaklines=true,                 
    captionpos=b,                    
    keepspaces=true,                 
    numbers=left,                    
    numbersep=5pt,                  
    showspaces=false,                
    showstringspaces=false,
    showtabs=false,                  
    tabsize=2
}

\lstset{style=mystyle}

\chapter{Tau Neutrinos Require Some Care} 
\label{ch:taumodeling}
\begin{figure}[!htb]
    \centering
    \includegraphics[width=0.75\textwidth]{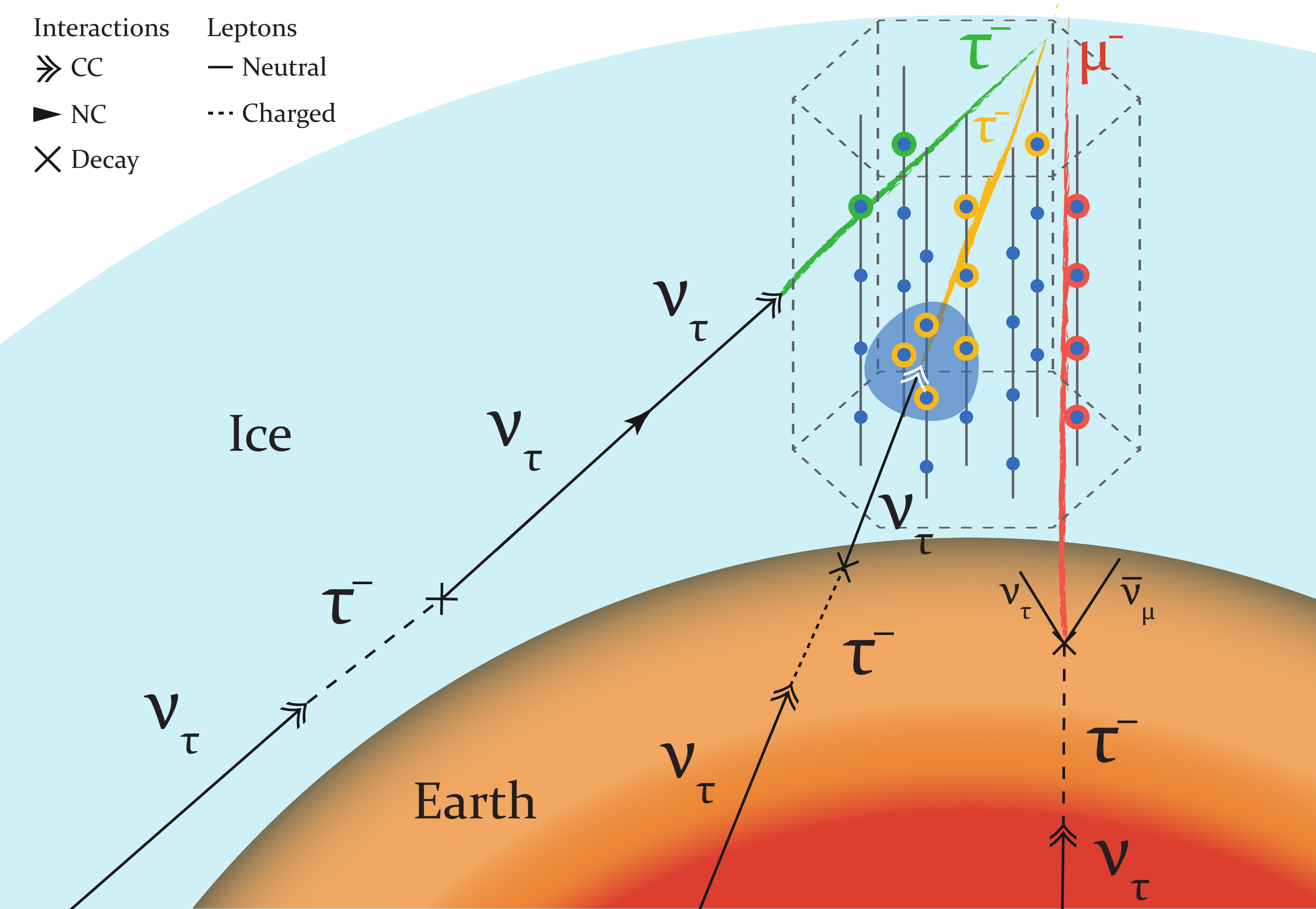}
    \caption[Schematic of lepton propagation through Earth.]{Schematic of lepton propagation through Earth followed by a measurement with the IceCube detector. There are three possible signatures from EeV tau-neutrino secondaries, described here from left to right. Left: A throughgoing tau track, which is possible for taus at or above 10 PeV. Center: The interaction vertex is contained in the fiducial volume of the detector in this case, producing a cascade from the charged-current interaction, along with an outgoing tau track. Right: The tau decays before reaching the detector, producing a muon in ${\sim}18$\% of the cases, which subsequently enters the detector. For clarity, not all particles involved in the interaction are shown. An additional contribution included in the results but not shown here is an NC interaction inside the detection volume.}
    \label{fig:tau_runner_schematic}
\end{figure}

\SingleSpace
\epigraph{``And then there's Ibrahim.. I have no idea what he's working on anymore. I think taus. There's always things to do with the taus''}{Francis Halzen}
\DoubleSpacing
As briefly highlighted in the introduction, measuring the flavor composition of astrophysical neutrinos is a direct test of all physics scenarios that would alter neutrino oscillations. 
Such a test is especially strong with astrophysical neutrinos given the immense baselines these neutrinos traverse, and the energies they possess.
Flavor measurements in this generation are limited by the tau fraction, mainly due to low statistics.
That is why a multitude of experiments are now proposed which focus primarily on detecting tau neutrinos.
These experiments will require accurate modeling of tau neutrino propagation in matter, which is not as straight-forward as other flavors.
In fact we show here that accurate modeling can have significant impact even on the current generation of experiments, namely IceCube.
In this Chapter, we present improvements made to modeling of tau neutrino propagation at high-energies.
As shown in Fig.~\ref{fig:tau_runner_schematic}, tau neutrinos above 10~PeV energies can appear as tracks in detectors.
They may also decay to muons before reaching the detector, thus contaminating muon event selections, if not properly accounted for.
We introduce neutrino interactions in matter is Sec.~\ref{sec:leptons}.
We then describe in Sec.~\ref{sec:lossesdecays} charged tau energy losses in matter, which become relevant above $\mathcal{O}(10)$~PeV energies.
In Sec.~\ref{sec:depol} we derive a treatment for the depolarization of taus in matter.
Finally, in Sec.~\ref{sec:tauappearance} we quantify the effect of tau neutrino production in Earth due to the propagation of muon and electron neutrinos.
We find that the production of tau neutrinos at ultra-high energies is significant and will be the largest irreducible background to future tau neutrino experiments.
These effects are then all discussed in the context of a Monte Carlo software package that is presented in Chapter~\ref{ch:taurunner}.

\section{Tau Neutrinos through Matter\label{sec:leptons}}

The propagation of a flux of neutrinos through a medium can be described by the following cascade equation~\cite{GonzalezGarcia:2005xw}
\begin{equation}
    \frac{d \varphi(E, x)}{d x}=-\sigma(E) \varphi(E, x)+\int_{E}^{\infty} d \tilde{E} ~  f(\tilde{E}, E) \varphi(\tilde{E}, x),\label{eq:transport}
\end{equation}
where $E$ is the neutrino energy, $x$ is the target column density, $\sigma(E)$ the total neutrino cross section per target nucleon, $f(\tilde{E}, E)$ is a function that encodes the migration from larger to smaller neutrino energies, and $\varphi(E, x)$ is the neutrino spectrum.
The first term on the right hand side accounts for the loss of flux at energy $E$ due to charged-current (CC) and neutral-current (NC) interactions, whereas the second term is the added contribution from neutrinos at higher energy, $\tilde{E}$, to $E$ through NC interactions of $\nu_{e, \mu, \tau}$ and CC interactions in the $\nu_{\tau}$ channel.
In Earth, where the column density can be on the order of $10^{33}$ nucleons/cm$^2$, neutrino attenuation is important for 100 TeV energies and higher~\cite{Gandhi:1998ri}. In the Sun, where the column density is, on average, 10$^{37}$ nucleons/cm$^2$, attenuation is relevant for energies greater than $100~$GeV~\cite{Cirelli:2005gh,Blennow:2007tw}. For dense astrophysical environments capable of producing high-energy neutrinos---one of the denser ones being radio galaxies---the column density is around $10^{24}$ nucleons/cm$^2$~\cite{Tjus:2014dna}. Given that the column density required for a neutrino to undergo a single interaction is, on average, $10^{29}$ nucleons/cm$^2$, astrophysical sources are thus transparent even to neutrinos of EeV energies and higher.
In this work, the secondaries produced in CC interactions of other flavors are neglected due to the fact that the electrons and muons lose energy rapidly.
On the other hand, taus produced in CC tau neutrino interactions have a much higher probability of decaying yielding high-energy neutrinos.
This is due to the fact that weak decays scale as $m^5$ and that the tau mass is significantly larger than that of the muon, allowing for more decay modes, which results in a ratio of lifetimes between muons and taus of approximately $10^7$.
While the lifetimes are drastically different, the energy losses above $\sim$1 PeV, where stochastic losses are dominant, are only a factor of 10 smaller for taus than for muons.
These two facts set the critical energy in ice---the energy at which the decay and interaction lengths are equal---to be approximately ${\sim} 10^9$ GeV for taus, while for muons it is ${\sim}10$ GeV~\cite{Koehne:2013gpa}. 
This implies that tau energy losses can be safely neglected below 10 PeV and the decay-on-the-spot approximation is a good one, see e.g.,~\cite{Vincent:2017svp}.
However, in this work we consider neutrino propagation at EeV energies and higher, where this approximation no longer holds and careful treatment of tau energy losses is required; see~\cite{Alvarez-Muniz:2017mpk, Reno:2019jtr} for recent implementations and discussions. 

Measurements of neutrino cross sections have been performed at energies below GeV up to a few PeV~\cite{Patrignani:2016xqp}.
This includes a multitude of results using human-made neutrinos in accelerator~\cite{AguilarArevalo:2010zc,Tzanov:2005kr} and reactor~\cite{Vogel:1999zy,Kurylov:2002vj} experiments as well as natural sources such as solar~\cite{Agostini:2018uly}, atmospheric~\cite{Li:2017dbe}, and astrophysical neutrinos~\cite{Aartsen:2017kpd,Yuan:2019wil}; for recent reviews, see~\cite{Formaggio:2013kya,Katori:2016yel}.
In the future, measurements of high-energy neutrinos from collider experiments will be available in the TeV range~\cite{Feng:2017uoz,Abreu:2019yak}.

\begin{figure}[!htb]
    \centering
    \includegraphics[width=0.85\linewidth]{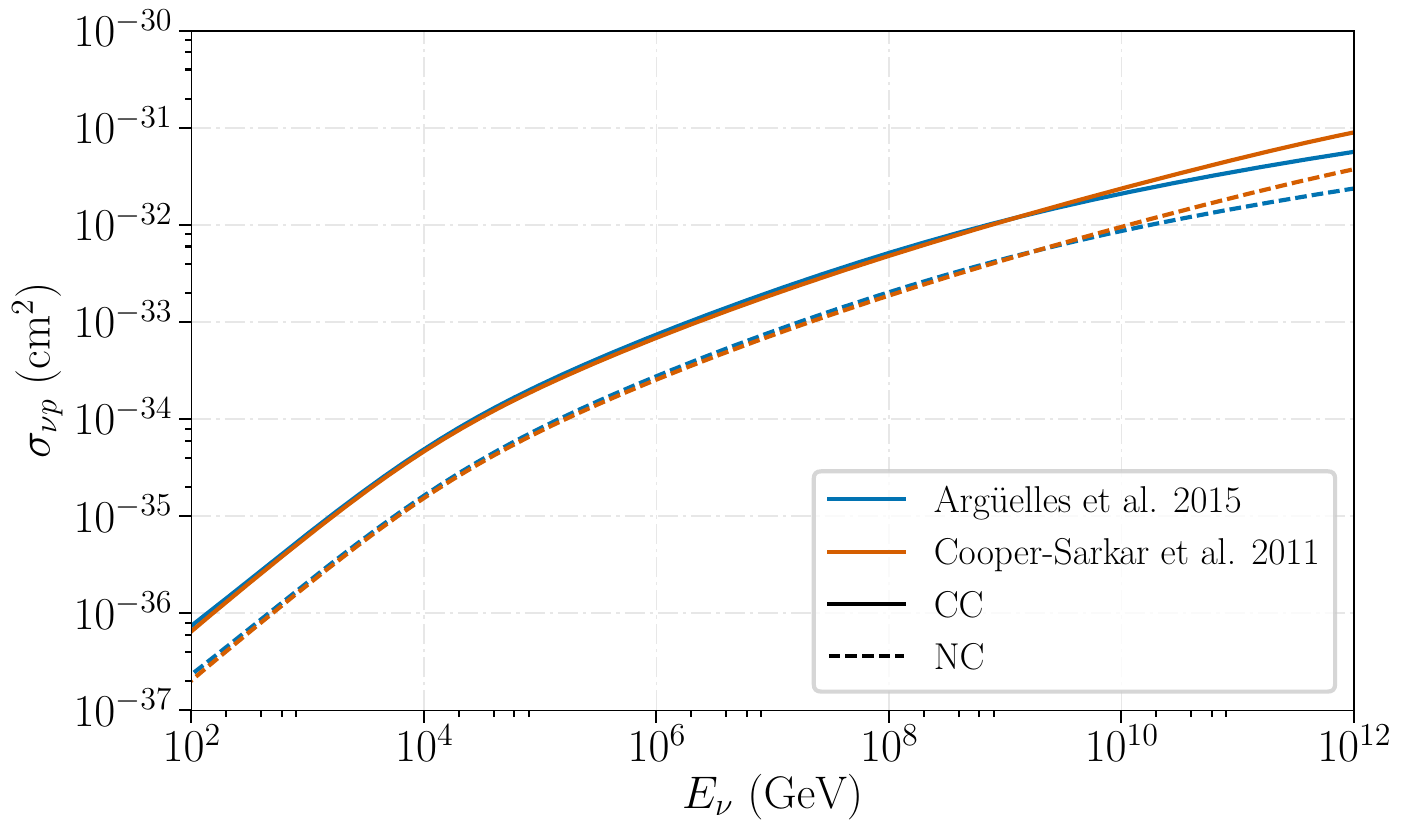}
    \caption[The neutrino-proton cross section as a function of the neutrino energy.]{The neutrino-proton cross section as a function of the neutrino energy. Solid (dashed) lines correspond to charged-current (neutral-current) cross sections. Blue lines~\cite{Arguelles:2015wba} correspond to the model used for this work. Orange lines~\cite{CooperSarkar:2011pa} are also implemented in the software and can be chosen by the user.}
    \label{fig:nucross}
\end{figure}

Unfortunately, these measurements stop short of the region of interest for this work where the fractional momenta, $x_{\rm{Bjorken}}$, of the quarks probed by the neutrino can reach $x_{\rm{Bjorken}}\ll10^{-8}$.
The nucleon structure function is not measured at such low $x_{\rm{Bjorken}}$ and is extrapolated in cross section calculations~\cite{CooperSarkar:2011pa,Garcia:2020jwr}. 
Such extrapolations neglect gluon color screening making perturbative QCD calculations of the neutrino cross section grow faster than allowed by unitarity at extremely high energies~\cite{Froissart:1961ux}.
Phenomenological approaches to include gluon screening parameterize the extremely small $x_{\rm{Bjorken}}$ behavior using a dipole model~\cite{Arguelles:2015wba} of the nucleon so as to result in a $\ln^2(s)$ dependence of the cross section at extremely high energies~\cite{Block:2011vz}.
Calculations using this approach were shown to be in good agreement with the total proton-proton cross section measurements from Auger~\cite{Collaboration:2012wt} and TOTEM at LHC~\cite{Antchev:2013gaa} data.
This ultimately results in a difference of a factor $\sim2$ at $10^{12}$ GeV; see Fig.~\ref{fig:nucross} for a comparison of two PDF models: a perturbative QCD calculation~\cite{CooperSarkar:2011pa}, and a dipole model~\cite{Arguelles:2015wba}.

\section{Tau Energy Losses and Decays}\label{sec:lossesdecays}

\begin{figure}[!ht]
   \centering
   \vspace{1cm}
   \subfloat[][]{\includegraphics[width=.4\textwidth, trim={1.5cm 1.75cm 2cm 2cm}]{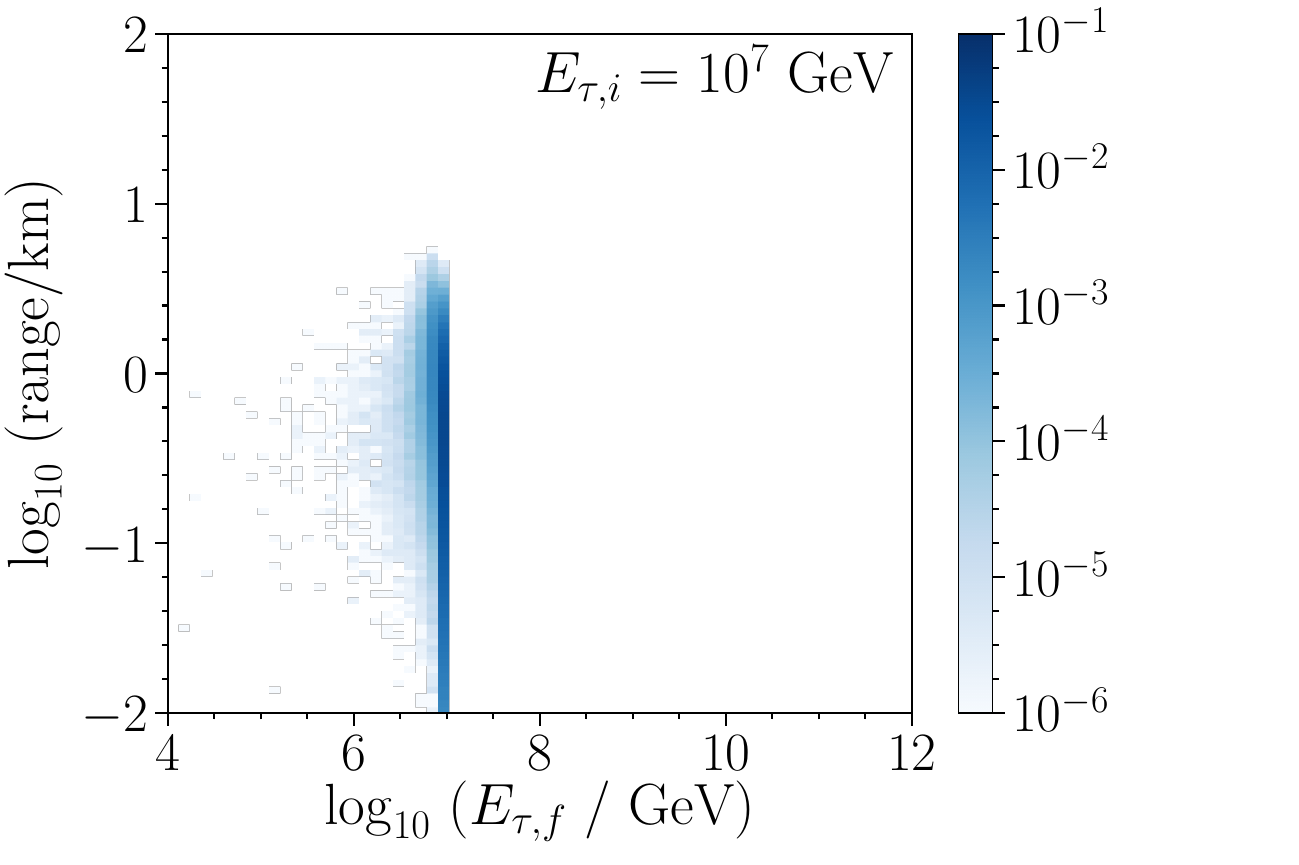}}\quad
   \subfloat[][]{\includegraphics[width=.4\textwidth, trim={1.5cm 1.75cm 2cm 2cm}]{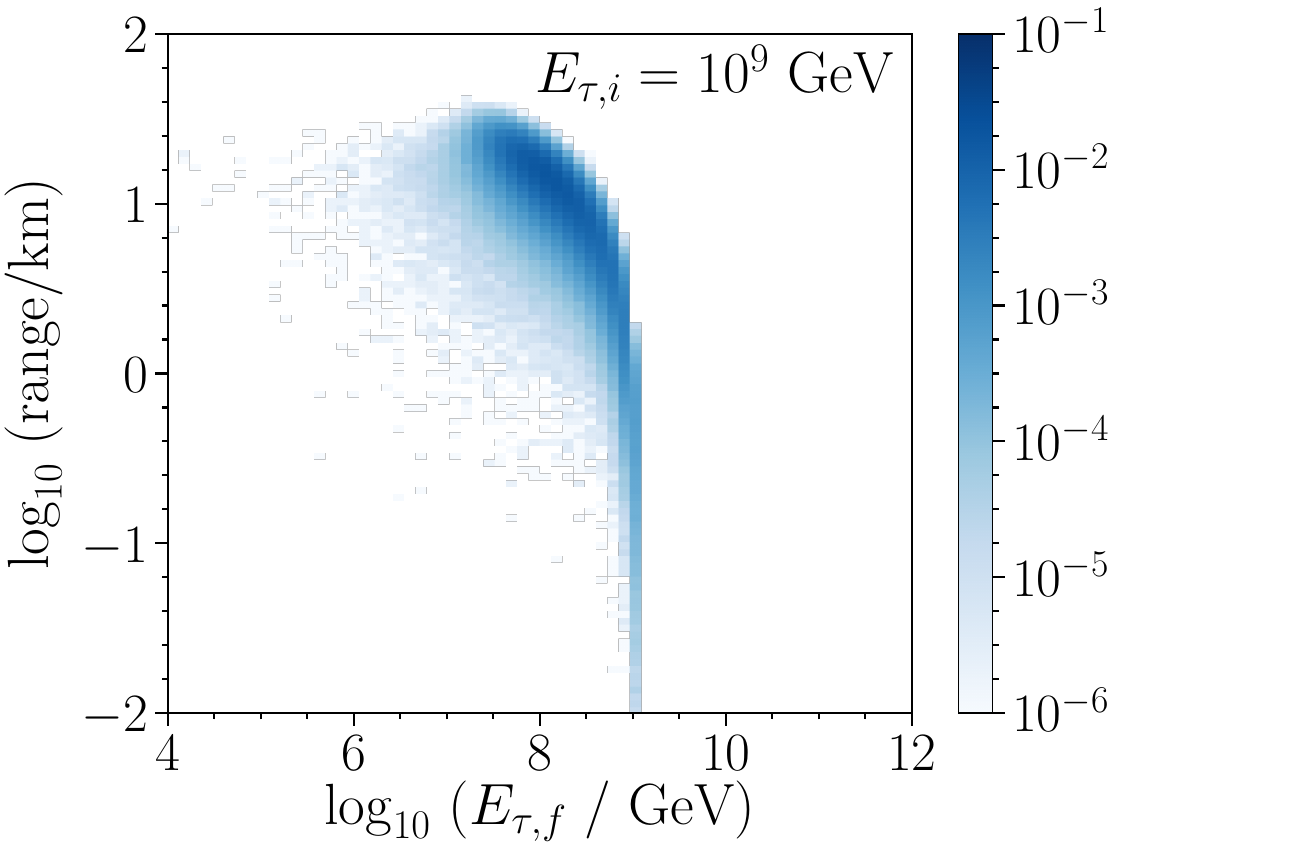}}\\
   \subfloat[][]{\includegraphics[width=.4\textwidth, trim={1.5cm 2cm 2cm 1cm}]{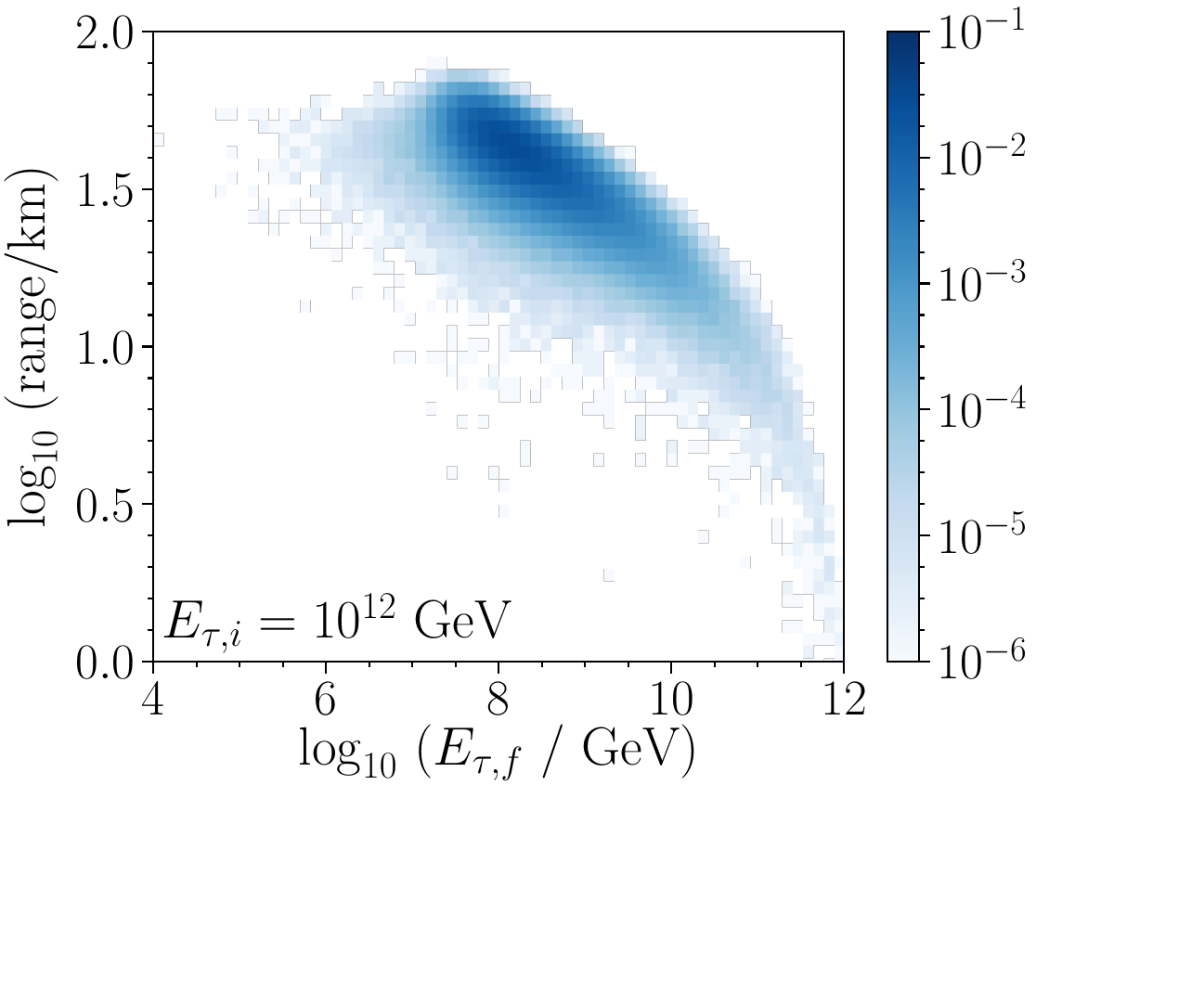}}\quad
   \subfloat[][]{\includegraphics[width=.4\textwidth, trim={1.5cm 2cm 2cm 1cm}]{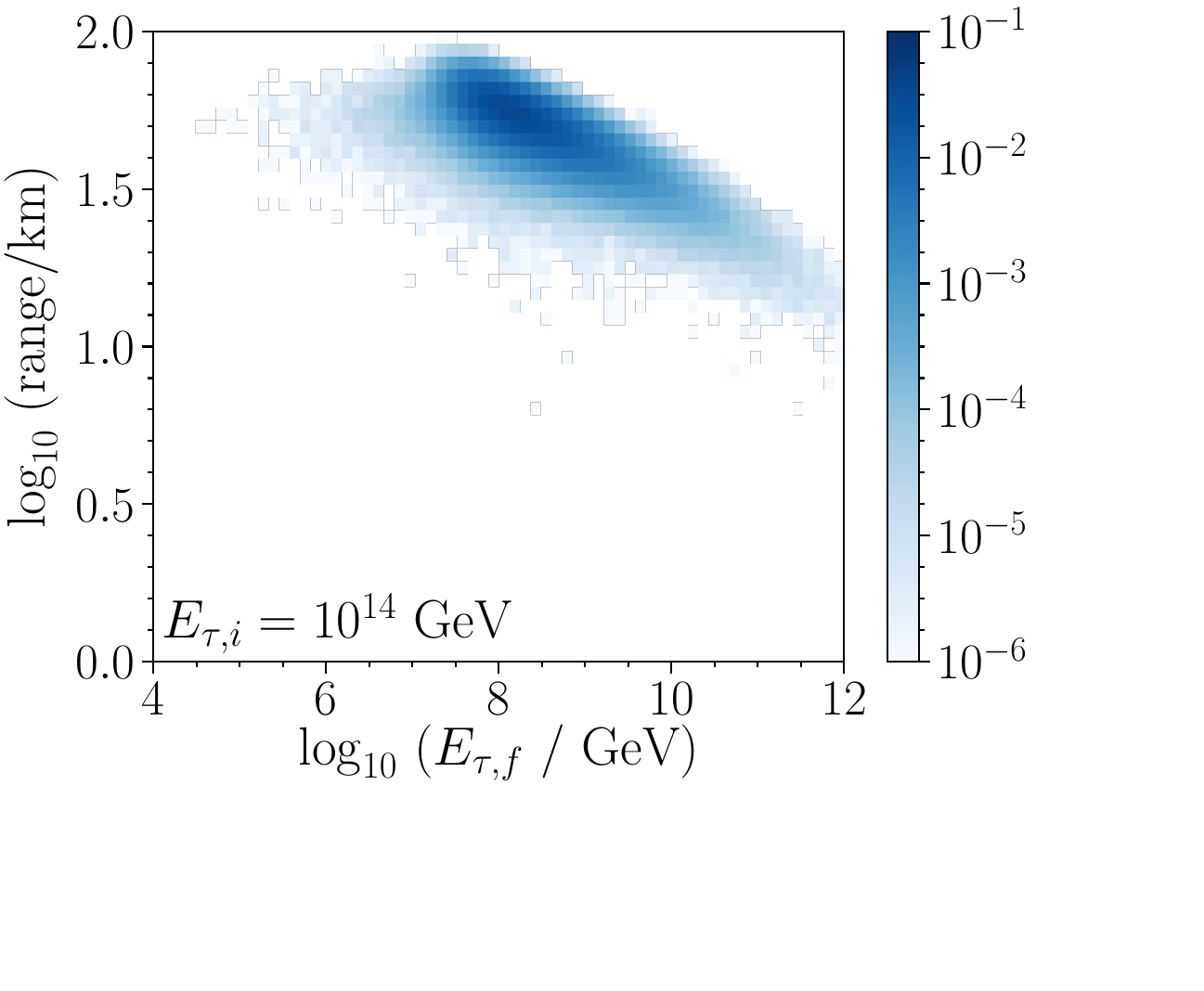}}
   \vspace{-0.5cm}
   \caption[Distribution of final tau energies and total distance traveled before decay for several initial tau energies.]{Distribution of final tau energies and total distance traveled before decay for several initial tau energies. At 10 PeV (upper left) and below, taus lose little energy before decay, while at 1 EeV (upper right) taus reach the critical energy and losses become appreciable. In this regime, the median range increases linearly as the tau becomes more boosted. At 1 ZeV (bottom left) and above (bottom right), these distributions show asymptotic behavior, with taus decaying around 100 PeV and traveling, on average, tens of kilometers.}
   \label{fig:tau_losses}
\end{figure}

When neutrinos undergo CC interactions, they convert to their charged partners through the exchange of a $W$ boson.
Charged particles lose energy in dense media through many processes, and the relative importance of each process depends on the lepton's mass and its energy~\cite{ParticleDataGroup:1994kdp}.
At lower energies, a charged lepton can ionize atoms as it traverses the medium.
This process is described by the Bethe-Bloche equation, and at higher energies scales logarithmically and becomes sub-dominant for all flavors.
A charged lepton can also interact with the electric field of a nucleus, losing energy in the process through the emission of a photon.
This process, called bremsstraahlung, scales like the inverse-sqaured mass of the lepton, and is therefore the dominant energy loss mechanism for electrons.
Another possible interaction with the field of a nucleus leads to the production of electron-positron pairs.
This process scales like the inverse of the lepton mass, and is one of the leading energy-loss mechanisms for $\mu$ and $\tau$.
Finally, the leptons can also lose energy by exchanging a photon with a nucleon, in what is referred to as a photonuclear interaction.
This process dominates tau energy losses at the highest energies ($\geq 10^{9}$~GeV).
This cross section depends on the nucleon structure function, and thus it has the same source of uncertainty as the neutrino-nucleon cross section. Fig.~\ref{fig:tau_losses} shows distributions of final tau energies and total distance traveled before decay for several initial tau energies, obtained by incorporating the discussed models by into a lepton propagation tool, PROPOSAL \cite{koehne2013proposal}. 
We observe that the tau range and minimum energy is bounded. For the range, it grows logarithmically above $10$ kilometers, mainly due to the growing interaction cross section. For the energy at decay, it's bounded around $100$~PeV below which the decay length is much shorter than the interaction length. 

Apart from interacting, $\mu$ and taus can also undergo weak decays.
This process scales like the mass of the lepton to the fifth power, and is therefore the most likely outcome for taus propagating in Earth up to $10^{9}$~GeV.
Above this energy, the total interaction length for other processes becomes shorter than the decay length.
$\mu$, on the other hand, are much more likely to lose all of their energy before decaying at rest, or getting absorbed by a nucleus.
The decay distribution of taus is well described by parametrizations in~\cite{Dutta:2002zc}, which we use in this work.

\section{Tau depolarization in matter}\label{sec:depol}

Tau depolarization in electromagnetic interactions is dominated by scatterings in which the incoming tau loses a substantial fraction of its initial energy.
As discussed in Sec.~\ref{sec:lossesdecays} bremsstrahlung is suppressed relative to $e^+e^-$ pair production and photonuclear interactions for taus~\cite{Dutta:2000hh}. 
Pair production that yields a change in tau energy of more than 10\% are rare.
For example, for $10^9$ GeV tau neutrinos propagating through Earth's mantle, only 0.02\% of the pair production tau scatterings have significant energy loss.
On the other hand, photonuclear interactions have more frequent hard scatterings, in which the final tau energies are less than 90\% of their initial energies.
Thus, here we focus on tau photonuclear energy loss.
We begin by describing the formalism of tau spin polarization, and then move to connect that to the photonuclear interaction cross section. 

The spin polarization three-vector of a tau produced in a CC neutrino interaction, in its own rest frame, can be written as~\cite{Hagiwara:2003di}
\begin{equation}
 \vec{s} = (s_x,s_y,s_z) = \frac{P}{2}(\sin\theta_P\cos\phi_P,\sin\theta_P\sin\phi_P,\cos\theta_P) = \frac{P}{2}(\Lambda\sin\theta_P,0,\cos\theta_P)\,
\label{eq:spinvector}
\end{equation}
where the spin direction is relative to the final state tau momentum direction in the lab frame, taken to be the $z$-axis.
In~Eq. (\ref{eq:spinvector}), $\theta_P$ and $\phi_P$ are the polar and azimuthal angle of the spin vector in the $\tau$ rest frame, and $P$ is the degree of polarization. 
The polarization vector lies in the scattering plane~\cite{Hagiwara:2003di}, thus $\phi_P = 0$ or $\pi$.
Therefore in~Eq. (\ref{eq:spinvector}), $\Lambda$ takes value $+1$ or $-1$ according to the azimuthal angle. 
In what follows, we define ${\cal P}_{z} \equiv 2 s_{z}$.
For a single scattering, we denote the polarization as ${\cal P}_{\tau,z}$.
We are interested in the final, or net, polarization of the tau just before it decays, which we denote as ${\cal P}_z$.
This quantity ${\cal P}_{z}$ enters into the energy distribution of the $\nu_\tau$ from tau decay, which can be described as a function of $z_\nu\equiv E_\nu/E_\tau$ in the form
\begin{equation} \label{eq:dGammadx2}
    \frac{1}{\Gamma}\frac{d\Gamma(\tau\to \nu_\tau)}{dz_\nu}= \sum_i B_i\Bigl(g_0^i(z_\nu)+{\cal P}_{z}g_1^i(z_\nu) \Bigr)\,,
\end{equation}
where $g_0^i(z_\nu)$ and $g_1^i(z_\nu)$ depend on $i$ decay channels with branching fraction $B_i$. 
The full decay width can be modeled as the sum of $\nu_\tau$ plus $e\bar{\nu}_e$, $\mu\bar{\nu}_\mu$, $\pi$, $\rho$, $a_1$, and $4\pi$ final states~\cite{Pasquali:1998xf,Bhattacharya:2016jce}.
We note that for $\bar\tau\equiv\tau^+$ decays, the differential decay distribution is 
\begin{equation} \label{eq:dGammadx2bar}
    \frac{1}{\Gamma}\frac{d\Gamma(\bar\tau\to \bar\nu_\tau)}{dz_{\bar\nu}}= \sum_i B_i\Bigl(g_0^i(z_{\bar\nu})-{\cal P}_{z}g_1^i(z_{\bar\nu}) \Bigr)\,,
\end{equation}
for $z_{\bar\nu}\equiv E_{\bar\nu}/E_{\bar\tau}$. Therefore, CC production of 
a right-handed (RH) $\bar\tau$ will yield the same decay distribution in $z_{\bar\nu}$ as the $z_\nu$ decay distribution of LH $\tau$.
\textbf{\textit{Scattering kinematics}}-- We first define the kinematic variables for the EM interactions.
For an incoming $\tau$ momentum for EM interaction ($k$), target nucleon momentum ($p$) and outgoing tau momentum ($k'$) in the laboratory frame, we write
\begin{equation}
k^\mu = (E_i,0,0,p_i), ~~ p^\mu = (M,0,0,0),~~ k^{\prime\, \mu} = (E_\tau,p_\tau \sin\theta,0,p_\tau\cos\theta).
\end{equation}
Here $E_i$ and $E_\tau$ are the incoming tau and outgoing tau energies in the laboratory frame, respectively, and $k^2=m_\tau^2$.
The Lorentz invariant variables, in terms of energy and angles in the lab frame where the target is at rest, are given by
\begin{eqnarray}
\nu &=& p\cdot q/M = E_i y = (E_i-E_\tau), \nonumber\\
\label{eq:xyq2}
Q^2&=&-q^2=-(k-k')^2 = 2(E_i E_\tau-p_i p_\tau\cos\theta)-k^2-k'^2, \\
x &=& Q^2/(2p\cdot q)\nonumber\,.
\end{eqnarray}
\begin{figure}[t]
    \centering
    \includegraphics[width=0.65\textwidth]{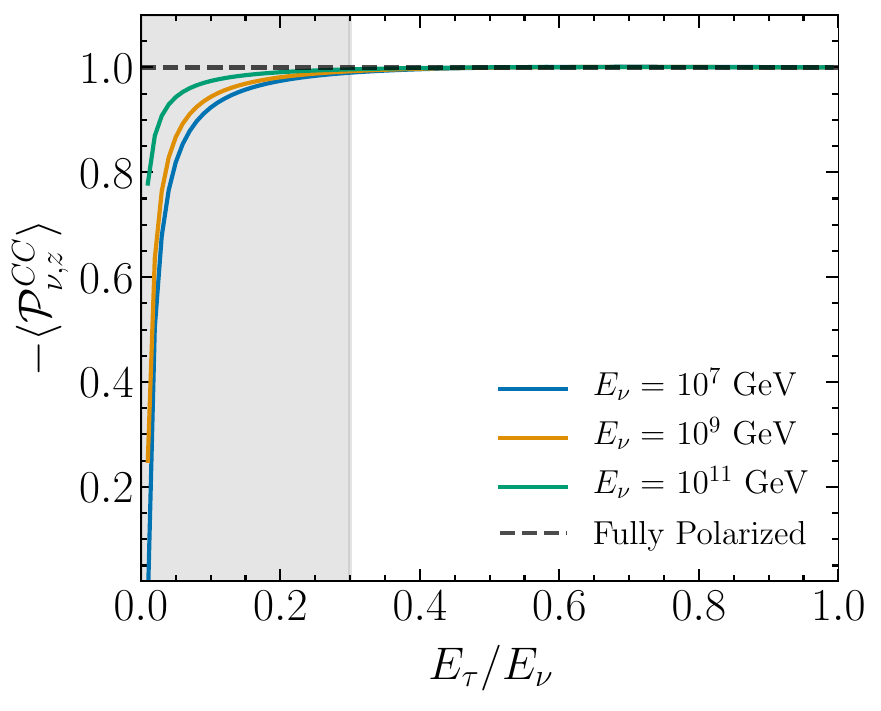}
    \caption[Initial polarization of taus produced from charged-current scattering.]{Initial polarization of taus produced from charged-current scattering.
    The average value of $\langle{{\cal P}_{\nu,z}^{CC}}(y)\rangle$ as a function of tau energy fraction for $\nu_\tau$ CC scattering for different incident neutrino energies.
   The shaded region shows where 10\% or less of the taus emerge with this energy fraction.
    In the massless lepton limit, $-\langle {\cal P} _{\nu,z}^{CC}(y)\rangle = 1$ for all $y$,  corresponding to left-handed out-going taus.
    }
    \label{fig:neutrinopola}
\end{figure}
\textbf{\textit{Tau photonuclear scattering}}-- Fig.~\ref{fig:neutrinopola} shows the average values of the $z$ component of the initial spin polarization vector, $-\langle{{\cal P}_{\nu,z}}^{ CC}\rangle$, as a function of $E_\tau/E_\nu = (1-y)$.
It can be observed that a tau produced by neutrino CC interaction is almost fully polarized for high incident neutrino energies. 
The shaded band shows where 10\% or less of the $\tau$ emerge with this energy fraction.
Thus, it is a good approximation to assume $\nu_\tau\to\tau$ produces left-handed taus in agreement with results found in ref.~\cite{Payet:2008yu}. Therefore in what follows, we assume that taus produced via CC neutrino interactions begin their journey as fully polarized, and proceed to
derive the spin polarization vector for $\tau\to\tau$ electromagnetic scattering in terms of the structure functions.
\begin{figure*}[t]
    \centering
    \includegraphics[width=0.9\textwidth]{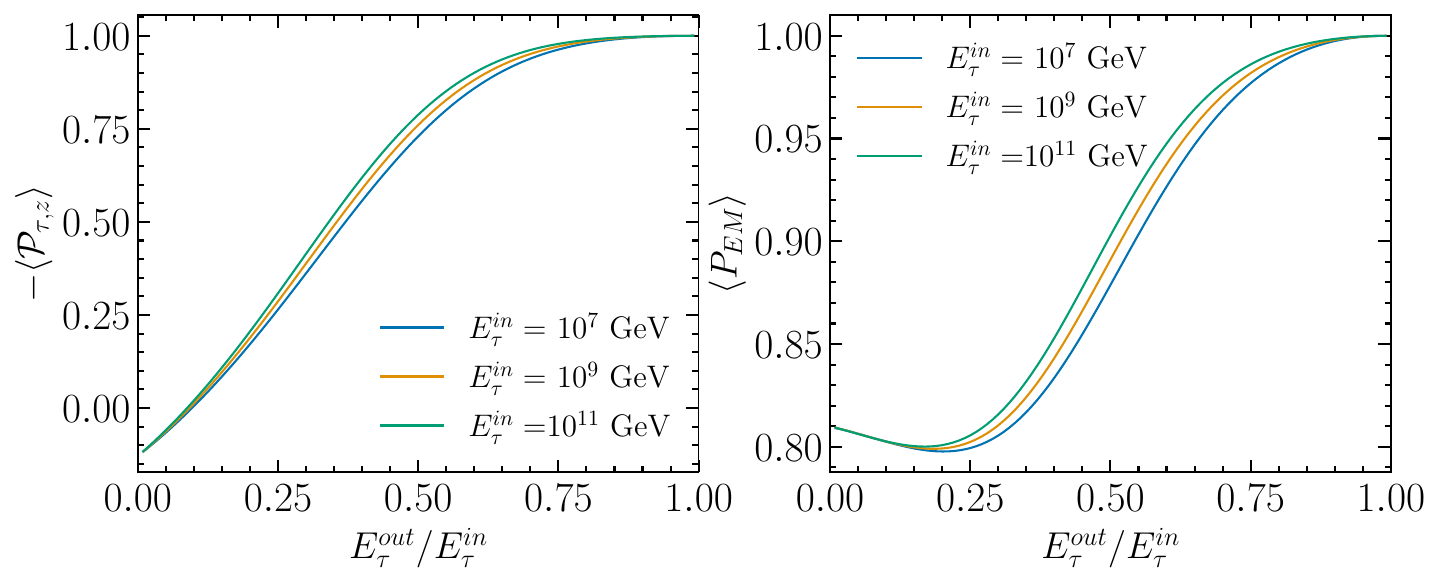}
    \caption[Average polarization of outgoing tau after a single interaction.]{Average polarization of outgoing tau after a single scattering.
    The average polarization about $z$-axis $\langle{{\cal P}_{\tau,z}}\rangle$ (left) and the average total polarization $\langle{P_{EM}}\rangle$ (right) as a function of outgoing tau energy fraction for single
    photonuclear electromagnetic scattering of $\tau$'s with rock for different incident tau energies. Here, $\langle{P^{EM}}\rangle=(\langle{{\cal P}^{EM}_{\tau,z}}\rangle^2$+$\langle{{\cal P}^{EM}_{\tau,x}}\rangle^2)^{\frac{1}{2}}$    
    }
    \label{fig:EMpol}
\end{figure*}
For tau electromagnetic interactions of initially left-polarized tau, the EM leptonic current is
\begin{equation}
    j_\lambda^\mu(\tau\,, EM)= \bar{u}_\tau(k',\lambda)
    \gamma^\mu u_\tau(k,-)\,
\end{equation}
The hadronic tensor for EM case will only have structure functions $W_1$, $W_2$ and reduces to
\begin{eqnarray}
 W_{\mu\nu}&=& -g_{\mu\nu}W_1+\frac{p_\mu p_\nu}{M^2}W_2,,
\end{eqnarray}
by taking into account gauge invariance. 
The differential cross section for tau photonuclear scattering  is 
\begin{equation}
    \frac{d^2\sigma_{\rm EM}}{dy\, dQ^2} = \frac{2\pi\alpha^2}{Q^4}\frac{1}{E_i}F_\tau\,,
\end{equation}
where
\begin{equation}
\nonumber F_\tau = 2W_1(E_i E_\tau-p_ip_\tau\cos\theta -2 m_\tau^2) + W_2(E_i E_\tau + p_ip_\tau\cos\theta + m_\tau^2)\,
\end{equation}
is obtained by contracting the hadronic and leptonic tensors.
The structure functions $W_1$ and $W_2$ are more commonly written in terms of $F_1$ and $F_2$ as
\begin{equation}
W_1 = F_1/M, ~~ W_2 = F_2/\nu \,
\end{equation}
with 
\begin{equation}
     F_1=\frac{1}{2x(1+R)}\Biggl(1+\frac{4M^2x^2}{Q^2}\Biggr) F_2\,.
     \label{eq:f1}
\end{equation}
The differential cross section translates to 
\begin{equation}
\nonumber
    \frac{d^2\sigma_{\rm EM}}{dy\,dQ^2} = \frac{4\pi\alpha^2}{Q^4}\frac{F_2(x,Q^2)}{y}
    \Biggl[ 1-y-\frac{Q^2}{4E_i^2} + \Biggl( 1-\frac{2m_\tau^2}{Q^2}\Biggr)
    \frac{y^2(1+4M^2 x^2/Q^2)}{2[1+R(x,Q^2)]}\Biggr]\,
\end{equation}
The quantity $R(x,Q^2)$, implicitly defined in eq. (\ref{eq:f1}), can be written in terms of the longitudinal structure function $F_L\simeq F_2(x,Q^2)-2xF_1(s,Q^2)$ (in the small $x$ limit) and $F_1(x,Q^2)$.
This gives $R(x,Q^2)\equiv F_L(x,Q^2)/(2xF_1(x,Q^2))$.
At high energies, we can take $R(x,Q^2) = 0$.
The tau spin polarization vector can be related to the tau EM scattering cross section as follows: 
The spin density matrix gives the relation~\cite{Hagiwara:2003di}
\begin{equation}
    dR_{\lambda\lambda'}\sim L_{\lambda\lambda'}^{\mu\nu}W_{\mu\nu}\,.
    \label{densitymatrix}
    \end{equation}
We can then relate the elements of the spin polarization vector to the lepton-hadron contractions.
Up to an overall normalization factor $N$,
\begin{eqnarray}
dR_{++}+dR_{--} &=& d\sigma_{\rm sum} = N (L_{++}^{\mu\nu}+L_{--}^{\mu\nu})W_{\mu \nu},\label{densityA}\\
dR_{+-} &=& s_x d\sigma_{\rm sum} = \frac{L_{+-}^{\mu \nu} W_{\mu\nu}}{(L_{++}^{\mu\nu}+L_{--}^{\mu\nu})W_{\mu\nu}} d\sigma_{\rm sum}, \label{densityB}\\
\frac{dR_{++}-dR_{--}}{2} &=& s_z d\sigma_{\rm sum} =  \frac{(L_{++}^{\mu\nu}-L_{--}^{\mu\nu})W_{\mu \nu}/2}
{(L_{++}^{\mu\nu}+L_{--}^{\mu\nu})W_{\mu\nu}}d\sigma_{\rm sum}\label{densityC}.
\end{eqnarray}
Finally, using Eqs. \eqref{eq:spinvector}, \eqref{densityA}, \eqref{densityB}, and \eqref{densityC} we can evaluate the spin polarization vector components for tau EM scattering: 
\begin{eqnarray}
\nonumber
s_x &=& -\frac{m_\tau}{2}\sin\theta \Bigl[2E_i\, W_1 - {W_2}(E_i +E_\tau)\Bigr]/F_\tau \label{eq:sx} \\
s_y &=& 0 \\
s_z &=& -\frac{1}{2}\Bigl[2W_1(p_i p_\tau - E_i E_\tau\cos\theta)
+ W_2(p_i p_\tau+E_i E_\tau\cos\theta) + W_2 m_\tau^2\cos\theta \Bigr]/F_\tau.
\label{eq:sz}
\end{eqnarray}
The energy distribution of $\nu_\tau$ from $\tau$ decays depend on
${\cal {P}}_{\tau,z}$ (Eq.~\ref{eq:dGammadx2}), so only the $s_z$ component of the spin polarization vector is relevant.
The average value of ${\cal {P}}^{EM}_{\tau,z}$ as a function of $y$, integrated over $Q^2$ is:
\begin{eqnarray}
\label{eq:calPEM}
\langle{{\cal {P}}^{EM}_{\tau,z}}(y)\rangle&\equiv &\frac{d\langle P\cos\theta_P\rangle_{\rm EM}}{dy} = \int dQ^2 \, 2s_z \frac{d^2\sigma_{\rm EM}}{dy\, dQ^2} 
\Biggl(\frac{d\sigma_{\rm EM}}{dy}\Biggr)^{-1} .
\end{eqnarray}
In Fig.~\ref{fig:EMpol} (left), the average value of $-\langle{{\cal P}_{\tau,z}^{EM}}(y)\rangle$ after a single scattering, as a function of $E_\tau^{out}/E_\tau^{in}$, is shown for three incident tau energies.
A high-energy tau passing through rock can get partially depolarized.
The depolarization effect becomes significant for $y \gtrsim 0.2$, namely, for $E_\tau^{out}/E_\tau^{in}\lesssim 0.8$. 
The value of $\langle {\cal P}_{\tau,z}^{EM}\rangle$ depends on the quantities $P$ and $\theta_P$. 
We calculate the average of the polarization, $\langle{P^{EM}}\rangle=(\langle{{\cal P}^{EM}_{\tau,z}}\rangle^2$+$\langle{{\cal P}^{EM}_{\tau,x}}\rangle^2)^{\frac{1}{2}}$, where $\langle{{\cal P}^{EM}_{\tau,x}}\rangle$ is evaluated according to~eq. (\ref{eq:calPEM}) with $s_z\to s_x$. 
In Fig.~\ref{fig:EMpol} (right), we show the average total polarization for different incident tau energies.
We see that  $\langle{P^{EM}}\rangle$ lies between 1 and 0.8 throughout the range of $E_\tau^{out}/E_\tau^{in}$.
The main contribution to depolarization observed in Fig.~\ref{fig:EMpol} (left) is from the polar angle $\theta_P$. 

\section{Production of Tau Neutrinos in Earth}
\label{sec:tauappearance}
\begin{figure}[!ht]
\centering
\includegraphics[width=0.55\textwidth]{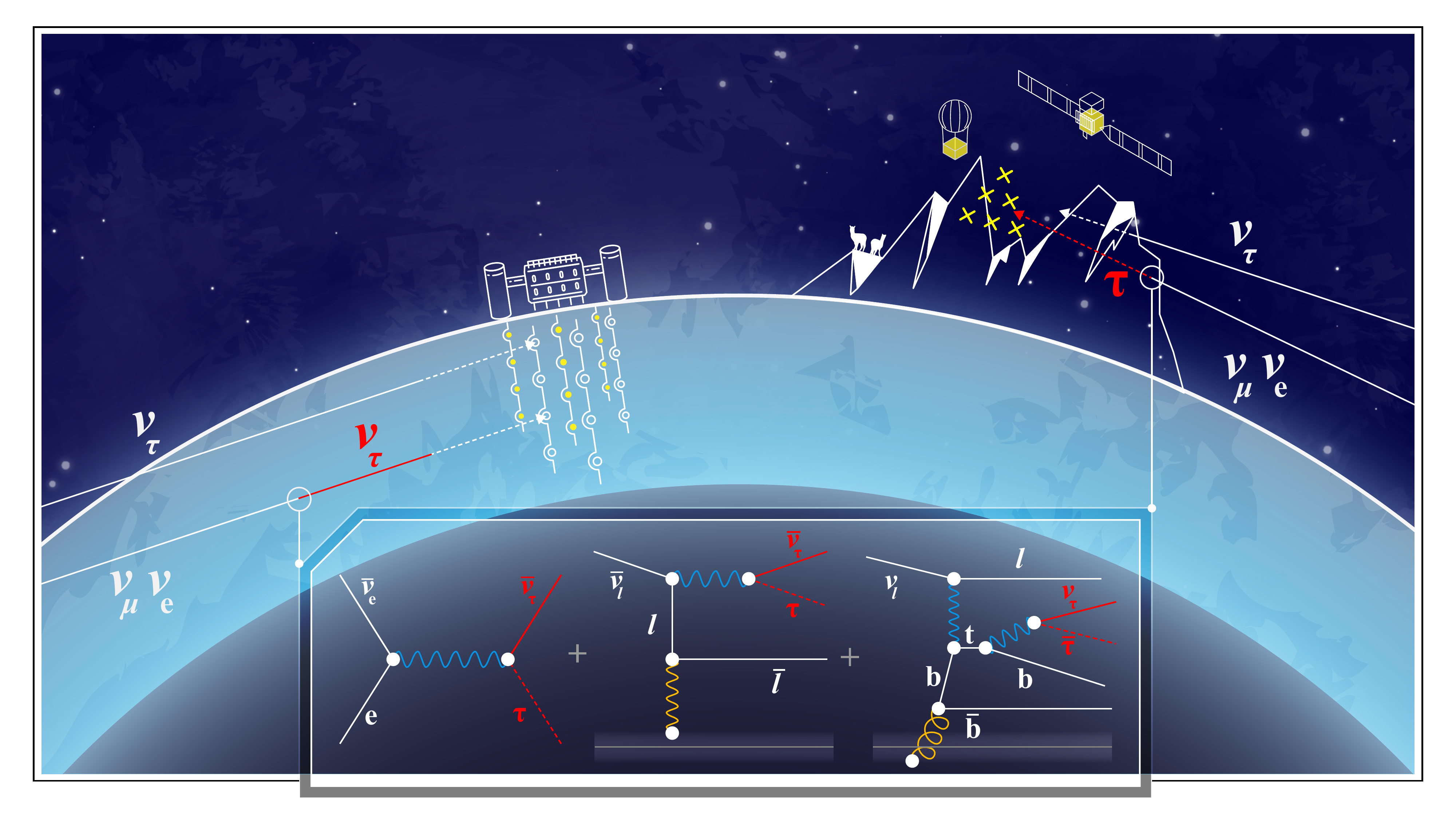}
\caption[Artistic rendition of relevant interactions for tau appearance.]{Artistic rendition of relevant interactions for tau appearance.
The Feynman diagrams summarize the muon- and electron-neutrino interactions that would produce a pair $\tau+\nu_\tau$: Glashow resonance (left), $W$-boson production (middle) and top-quark production (right). 
Ice- and water-Cherenkov detectors are represented on the center left. Earth-skimming and space-based observatories are shown on the center right.
}
\label{fig:scheme}
\end{figure}
\begin{figure}[!ht]
\centering
\includegraphics[width=0.65\textwidth]{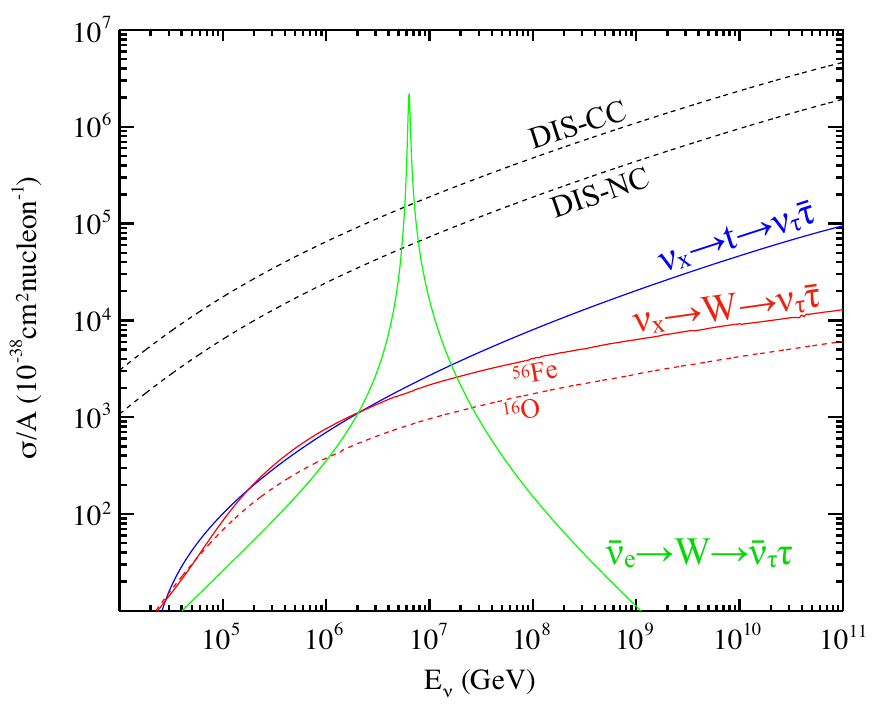}
\caption[Neutrino-nucleus cross sections per nucleon as a function of the neutrino energy assuming an isoscalar target.]{Neutrino-nucleus cross sections per nucleon as a function of the neutrino energy assuming an isoscalar target.
Continuous lines represent the interaction channels in which a $\nu_\tau+\tau$ pair is produced: Glashow resonance (green), $W$-boson production (red), and top-quark production (blue).
The deep inelastic scattering contribution is shown with a dashed line (summing CC and NC).
The $W$-boson production is shown for two different nuclei (O and Fe) in order to illustrate the $Z^2/A$ scaling effect.}
\label{fig:xsec}
\end{figure}
\begin{figure}[!ht]
\centering
\includegraphics[width=1\textwidth]{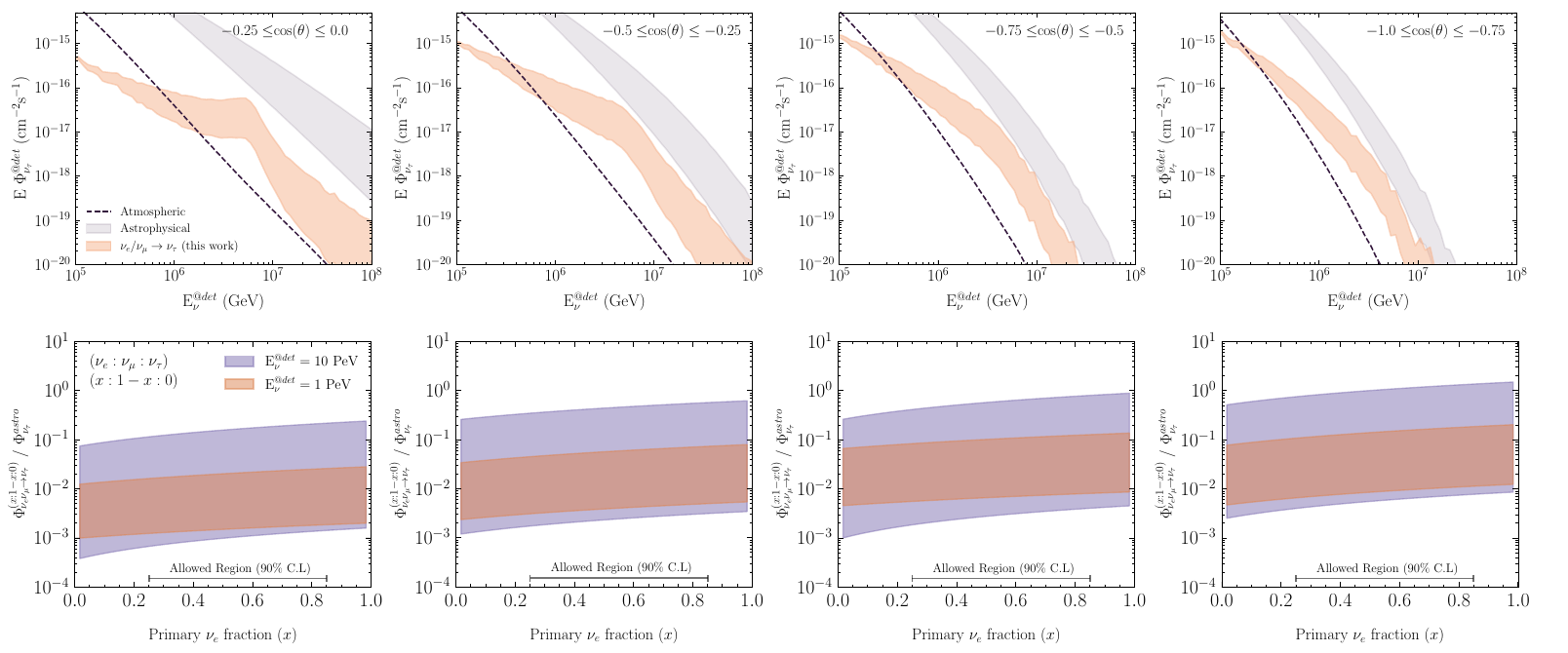}
\caption[Components of tau-neutrino flux at the detector integrated over different angular regions.]{Top: Components of tau-neutrino flux at the detector integrated over different angular regions.
Grey shaded band shows the expected flux between the best-fit spectra from the HESE and throughgoing muons analyses~\protect\cite{IceCube:2020wum,Stettner:2019tok}.
Dashed line shows the atmospheric flux using \texttt{MCeq} (H3a-SIBYLL23C)~\protect\cite{Fedynitch:2015zma}.
The red shaded regions show the secondary tau neutrinos contribution produced by the propagation of muon and electron neutrinos assuming the best fits from IceCube's analysis~\protect\cite{IceCube:2020wum,Stettner:2019tok} and $\left(2:1:0\right)$ flavor composition.
Bottom: Ratio of  the secondary flux assuming different muon- and electron-neutrino fractions to the HESE's best-fit flux assuming $\left(1:1:1\right)$ at $1$~PeV (red) and $10$~PeV (blue).
The bands represent the uncertainty in the primary spectrum.
The black, horizontal line indicate the combined 90\% C.L. from different flavor composition measurements~\cite{IceCube:2018pgc,IceCube:2015gsk}.}
\label{fig:suppl2}
\end{figure}

We show in this section that high-energy electron or muon neutrinos can also produce tau neutrinos
and prove that this effective tau appearance phenomenon has a significant and direct impact on the inferred astrophysical neutrino spectra and flavor measurements. 
For the first time, we identify channels that would yield a significant contribution of secondary tau neutrinos as shown in Fig.~\ref{fig:scheme}.
The secondary contribution arises from channels that produce on-shell $W$ bosons, which decay $10\%$ of the time to $\nu_{\tau}+\tau$~\cite{ParticleDataGroup:2020ssz}.
The dominant interactions are neutrino-nucleus~\cite{Seckel:1997kk,Alikhanov:2015kla,Barger:2016deu,Zhou:2019vxt,Zhou:2019frk} or neutrino-electron~\cite{Glashow:1960zz,Gauld:2019pgt,Huang:2019hgs} $W$-boson production, which become relevant above PeV energies.
In the neutrino-nucleus interaction, when the energy transferred from a neutrino to a nucleon is above the top-quark mass, a $W$ boson can also be produced from the decay of this heavy quark to a $b+W$ pair~\cite{Barger:2016deu}.
The latter process is intrinsically related to the parton distribution functions (PDF) of the sea bottom in the nucleons.
Depending on the PDFs and mass scheme models, top-quark production accounts for $5-15\%$ of the total deep-inelastic-scattering (DIS) cross sections above $100$~TeV~\cite{Bertone:2018dse,Cooper-Sarkar:2011jtt,Cooper-Sarkar:2007zsa,Gluck:2010rw,Connolly:2011vc,Albacete:2015zra,Arguelles:2015wba,Goncalves:2010ay}.
The structure functions for neutrino-nucleon deep inelastic scattering have been implemented in GENIE.
In this study, the HEDIS-CSMS model (\texttt{GHE19\_00b}) is used.
This calculation uses the NLO HERA1.5 PDF set~\cite{Cooper-Sarkar:2010yul} and uses expressions for the massless coefficient functions at NLO.
This is the preferred model in IceCube's analyses.
Additionally, the $W$-boson production and Glashow resonance are computed using the differential cross sections as described in~\cite{Garcia:2020jwr}.
Fig.~\ref{fig:xsec} shows the total cross section for channels in which a tau neutrino is produced from $10$~TeV to $100$~EeV.
For comparison, the deep inelastic scattering contribution is also shown.
The flux at the detector of produced tau neutrinos in different angular regions is shown in Fig.~\ref{fig:suppl2}.
In addition, the relative fractions of secondary tau neutrinos to the astrophysical component at 1~PeV and 100~PeV are shown.

\textbf{\textit{Diffuse cosmic neutrino fluxes ---}} 
The astrophysical neutrino flux has been characterized by the IceCube Collaboration using multiple complementary channels~\cite{Aartsen:2013jdh,Aartsen:2014gkd,Aartsen:2014muf,Aartsen:2015zva,Aartsen:2017mau,IceCube:2018pgc,IceCube:2020acn,IceCube:2020wum,Abbasi:2021viw}. 
As a benchmark scenario, motivated by the Fermi acceleration mechanism~\cite{gaisser2016cosmic}, these analyses model the astrophysical neutrino flux as an unbroken power law in energy.
The results of these analyses agree within errors, and the best-fit spectral index ranges from $2.37 \to 2.89$.
The softest spectral index is measured using high-energy starting events (HESE) with deposited energy above $60$~TeV~\cite{IceCube:2020wum}, while the hardest spectrum is measured when selecting for throughgoing muons in the northern sky~\cite{Stettner:2019tok,Abbasi:2021viw}.
Both analyses set constraints on the fraction of muon neutrinos, whereas similar signatures of electrons and taus in the detector give rise to a degeneracy between the other two flavors.
The smoking-gun signature of tau neutrinos is two separated energy depositions, known as a {\it double bang}. 
These have recently been detected~\cite{stachurska_juliana_2018_1301122}; still, the sample size does not yield significant evidence of a tau-neutrino component~\cite{IceCube:2020abv}.

\begin{figure}[!ht]
\centering
\includegraphics[width=0.4\textwidth]{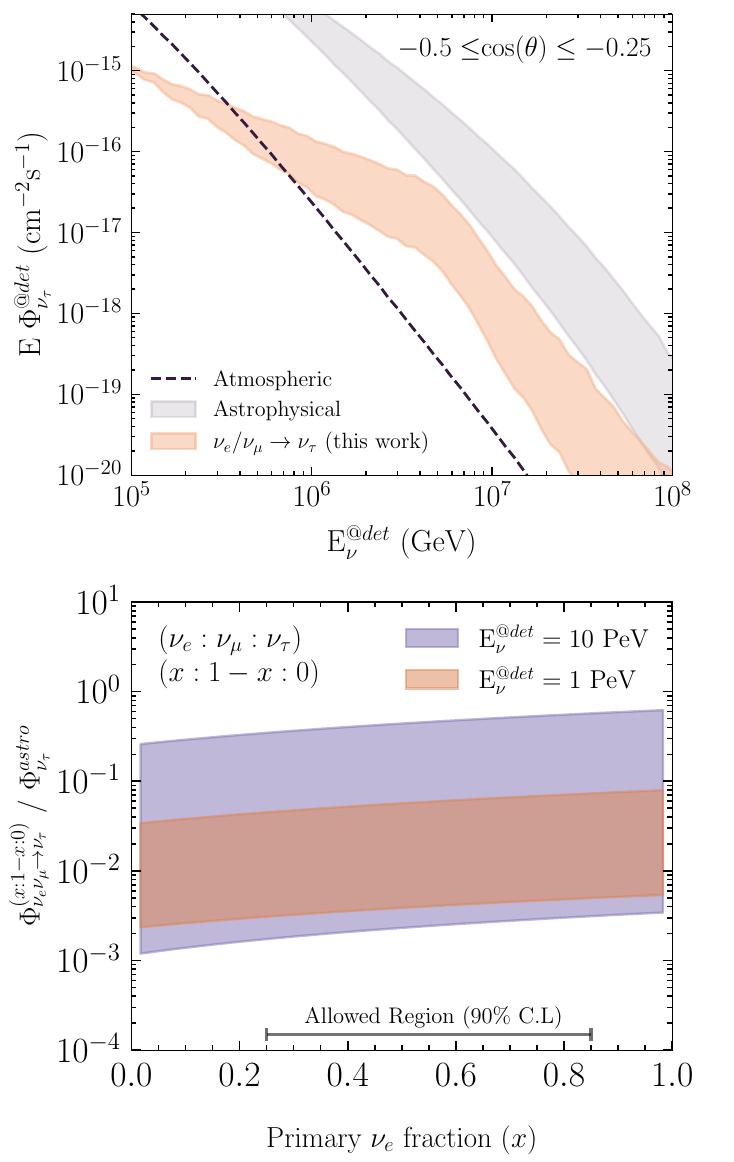}
\caption[Primary neutrino fluxes and appearing tau-neutrino component.]{Primary neutrino fluxes and appearing tau-neutrino component.
Top: Different components of tau-neutrino flux at the detector integrated over $-0.5<\cos\theta<-0.25$.
Grey shaded band shows the expected flux between the best-fit spectra from the HESE and throughgoing muons analyses~\protect\cite{IceCube:2020wum,Stettner:2019tok}.
Dashed line shows the atmospheric flux using \texttt{MCeq} (H3a-SIBYLL23C)~\protect\cite{Fedynitch:2015zma}.
The red shaded region represents tau neutrinos appearing from the propagation of muon and electron neutrinos, assuming $(2:1:0)$ flavor composition.
Bottom: Ratio of the secondary flux assuming different muon- and electron-neutrino fractions to the astrophysical flux assuming $\left(1:1:1\right)$ at $1$~PeV (salmon) and $10$~PeV (lavender).
The bands represent the uncertainty in the primary spectrum.
The black, horizontal line indicates the 90\% C.L. allowed flavor composition~\cite{IceCube:2018pgc,IceCube:2015gsk}.}
\label{fig:astro}
\end{figure}

Here we estimate the expected flux of tau neutrinos at the detector under two different hypotheses.
First, we assume the best-fit results from HESE and throughgoing muons under the canonical $\left(1:1:1\right)$ flavor composition.
Second, we use the best-fit results assuming a zero tau-neutrino component in the primary flux.
Fig.~\ref{fig:astro} shows the expected tau-neutrino flux at the detector for events below the horizon ($-0.5<\cos\theta <-0.25$).
Tau neutrinos produced from muon and electron neutrinos alone can contribute as much as $50\%$ of the primary astrophysical flux for energies above $10$~PeV.
This fraction mainly depends on the primary spectrum but not on the relative composition of muon and electron neutrinos as shown in Fig.~\ref{fig:astro}.
At these energies the yield of secondary tau neutrinos arising from top-quark and W-boson productions is similar, while Glashow resonance dominates at $\sim 6$~PeV.
Additionally, the secondary tau neutrino flux has an angular dependence since it varies with the column depths that neutrinos travel.
Hence, the relative fraction will be higher for steeper angles.
Consequently, even in a scenario where the astrophysical neutrino flux is only made of muon and electron neutrinos, there is a guaranteed flux of tau neutrinos in the detector, which is larger than the atmospheric component (from the prompt decay of heavy mesons) above $\mathcal{O}$($100$TeV).

\begin{figure}[!ht]
\centering
\includegraphics[width=0.4\textwidth]{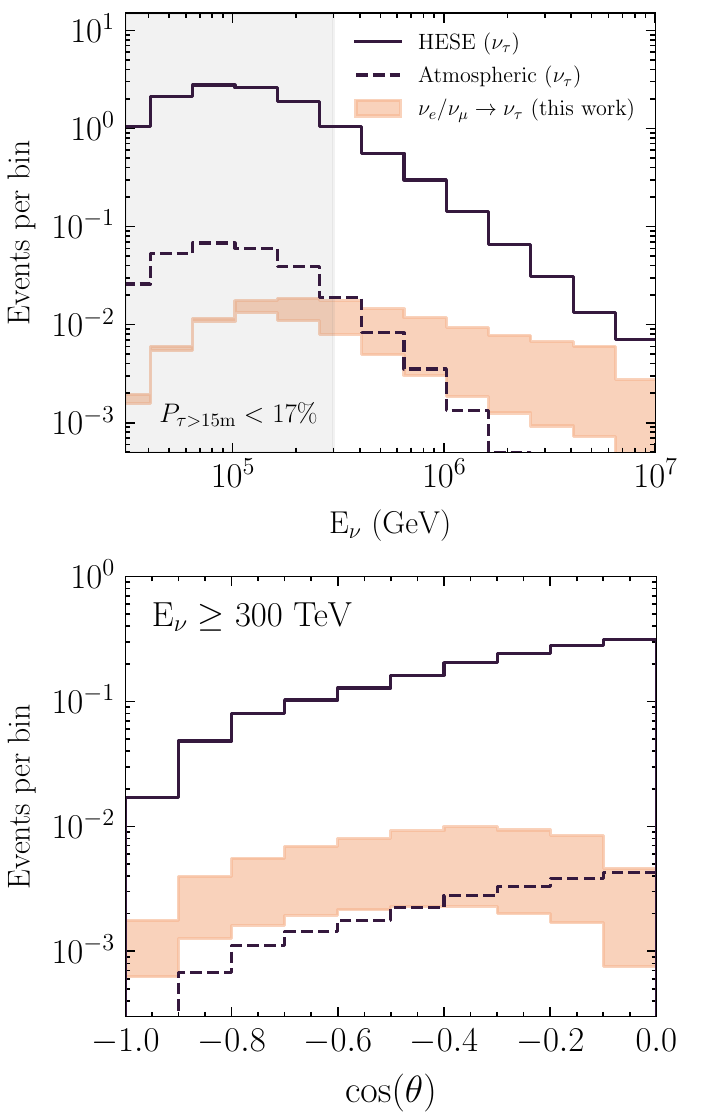}
\caption[Rates of tau-neutrino events in ten years of HESE.]{Rates of tau-neutrino events in ten years of HESE.
Energy (top) and angular (bottom) distribution of tau-neutrino events that pass the HESE selection.
The angular distribution is shown for events with energies above $300$~TeV.
Lines and shaded regions correspond to the fluxes described in Fig.~\ref{fig:astro}.
The grey area represents the region where tau-neutrino identification with current reconstruction methods is challenging.}
\label{fig:rates}
\end{figure}

\textbf{\textit{Impact of tau appearance on discovery of cosmic tau neutrinos ---}}
We now examine how this additional component manifests in the HESE analysis.
To do so, we combine the expected fluxes at the detector with the publicly available effective areas, assuming the best-fit HESE flux and a democratic primary flavor composition.
Fig.~\ref{fig:rates} shows the expected energy and angular distribution of up going tau neutrinos in HESE after ten years of data taking.
Above $300$~TeV, we expect 1.6 upgoing tau neutrinos, but only 0.02 events of atmospheric origin.
The expected rate from secondary tau neutrinos is twice that of the prompt atmospheric component.
Although the expected number of events is small, this secondary component is the dominant irreducible background for any astrophysical tau-neutrino search.
Tab.~\ref{tab:significane} summarizes the capabilities for rejecting the non-tau cosmic component, assuming that all upgoing tau-neutrino events in the HESE sample above an energy threshold are identified. 
We conclude that this additional tau-neutrino contribution reduces the discovery potential for all energy threshold criteria considered.
A thorough breakdown using realistic tau-neutrino identification criteria as well as a likelihood approach that includes the energy and angular information must be performed by neutrino observatories to fully quantify the impact of this intrinsic background.

\begin{table}[!ht]
\centering
\begin{tabular}
{l r@{\hskip 0.15in}c@{\hskip 0.15in}c@{\hskip 0.2in}c}
\hline
\hline
$E_{th}$ & $P_{\tau>15m}$ & HESE & Atmos. & $\nu_\mu/\nu_e\rightarrow\nu_\tau$ \\
\hline
100 TeV &  1\% & 6.63 & 0.13 (6.3) & 0.05-0.11 (6.0-5.7) \\
200 TeV &  9\% & 3.00 & 0.05 (4.3) & 0.03-0.09 (4.1-3.7) \\
300 TeV & 17\% & 1.57 & 0.02 (3.2) & 0.02-0.07 (2.9-2.5) \\
400 TeV & 23\% & 1.12 & 0.01 (2.7) & 0.01-0.06 (2.4-2.1) \\
\hline
\hline
\end{tabular}
\caption[Impact of appearing tau-neutrino component on cosmic tau-neutrino discovery.]{Impact of appearing tau-neutrino component on cosmic tau-neutrino discovery.
Each row shows a different assumed threshold, as indicated in the first column, to identify tau neutrinos in the HESE analysis.
The second column shows the percentage of tau-neutrino interactions with $E_{\nu}=E_{th}$ producing a tau lepton that travels more than $15$m~\protect\cite{IceCube:2020abv}.
The third column shows the number of upgoing tau-neutrino events in HESE in ten years.
The significance ($\sigma$) to reject the non-tau cosmic component for different background hypothesis is shown in parenthesis in the fourth and fifth columns when considering atmospheric and appearing-tau backgrounds.}
\label{tab:significane}
\end{table}

Currently, flavor composition measurements from IceCube are statistically limited due to the size of the detector.
However, the IceCube-Gen2 optical array will be able to detect a sizable amount of PeV neutrinos~\cite{IceCube-Gen2:2020qha}.
The rate of secondary neutrinos due to $W$-boson production, Glashow resonance, and top-quark production is $5-20\%$ above $10$~PeV, as shown in Fig.~\ref{fig:rates}.
In fact, the fraction of any flavor is affected by these new channels since $W$ bosons can also decay to muon and electron neutrinos.
Therefore, any analysis studying the composition of the astrophysical flux using upgoing events must account for these secondary neutrinos.
Thus, our conclusions are generic and extend beyond the HESE analysis.

\textbf{\textit{Ultra-high-energy neutrinos ---}}
Currently, IceCube measurements extend to $10$~PeV in neutrino energy~\cite{IceCube:2020wum,Abbasi:2021viw,IceCube:2021rpz}. 
At higher energies, many experiments have placed upper limits on the neutrino flux~\cite{IceCube:2018fhm,PierreAuger:2019ens,ANITA:2019wyx,ARA:2015wxq}; however, their effective volumes and Earth's opacity for UHE neutrinos limit their capabilities.
One of the most promising ideas to detect EeV neutrinos is to look for Earth-skimming tau leptons~\cite{Sasaki:2017zwd,GRAND:2018iaj,Wissel:2020sec,Romero-Wolf:2020pzh,POEMMA:2020ykm,Wang:2021zkm}.
Almost horizontal tau neutrinos at energies above $10$~PeV will interact in Earth's crust, producing an energetic tau-lepton, which can emerge in the atmosphere and decay.
Subsequently, electromagnetic radiation from upgoing EASs can be detected at optical and radio wavelengths.
Using this technique, experiments like POEMMA~\cite{Venters:2019xwi}, PUEO~\cite{PUEO:2020bnn}, GRAND~\cite{GRAND:2018iaj}, Trinity~\cite{Wang:2021zkm} and Beacon~\cite{Wissel:2020sec} claim to have significantly better sensitivities than current experiments.
In fact, the projected performance of the aforementioned experiments shows that they would be able to discover the cosmogenic neutrino flux for most UHECRs scenarios.

\begin{figure}[!ht]
\centering
\includegraphics[width=0.5\textwidth]{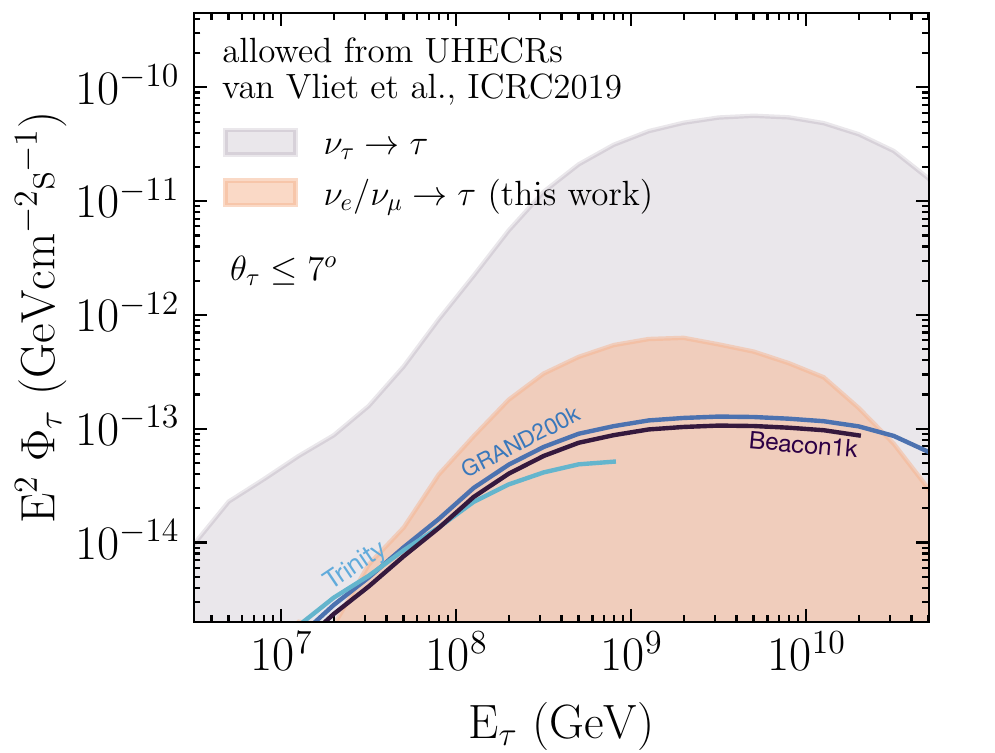}
\caption[Tau-lepton fluxes arriving at Earth-skimming experiments.]{Tau-lepton fluxes arriving at Earth-skimming experiments.
Expected flux of emerging tau for Earth-skimming neutrinos ($\theta_{\tau}<7\degree$).
Shaded curves represent the allowed regions derived from a model fit to Auger data~\protect\cite{vanVliet:2019cpl} with (gray) and without (salmon) tau neutrinos in the primary flux.
Solid lines indicate the flux for which Earth-skimming experiments would collect more than 2.44 events in ten years assuming a $E^{-2}$ primary neutrino spectrum.
}
\label{fig:gzk}
\end{figure}

Fig.~\ref{fig:gzk} shows the predicted energy spectrum of tau leptons emerging almost horizontally from Earth's surface.
In this calculation, we have assumed a cosmogenic model that includes more than a single UHECR source population~\cite{vanVliet:2019cpl}.
The expected secondary flux coming from muon- and electron-neutrino interactions --- as illustrated in Fig.~\ref{fig:scheme} --- is also shown together with the $E^{-2}$ flux for which GRAND, Trinity, and Beacon will observe more than 2.44 events (90\% C.L. upper limit) after ten years of data taking.
At energies above $10$~PeV, the dominant mechanism for the secondary flux is via top-quark production. 
In the case of ultra-high-energy neutrinos, unlike the lower energy astrophysical component, the angular distribution of the primary and secondary components are very similar.
Two main conclusions can be derived from the predictions shown in Fig.~\ref{fig:gzk}.
First, we expect a non-negligible contribution of emerging taus from a primary flux of muon and electron neutrinos. 
This holds true even if the flavor composition of the cosmogenic flux has no primary tau component.
Thus for optimistic models, GRAND, Trinity, and Beacon will detect taus regardless of primary flavor composition.
Second, when they observe emerging taus, they cannot infer the type of neutrino that produced them.
Therefore, these experiments will not set strong constraints on the normalization of the cosmogenic flux without making assumptions about its flavor composition; \textit{i.e.}, the yield of taus is degenerated between small $\left(1:1:1\right)$ and large $\left(1:1:0\right)$ primary fluxes.

There are some nascent reconstruction techniques to detect the optical Cherenkov emission of EASs induced by muons with detectors like POEMMA or Trinity~\cite{Cummings:2020ycz}.
These novel methods would make Earth-skimming experiments sensitive to multiple flavors, allowing them to break the degeneracy previously described.
Similarly, experiments that can detect geo-synchrotron and Askaryan radiation~\cite{Askaryan:1961pfb} --- \textit{e.g.}, IceCube-Gen2~\cite{IceCube-Gen2:2020qha}, RNO-G~\cite{RNO-G:2020rmc} or PUEO~\cite{PUEO:2020bnn} --- would not be affected by this degeneracy because they are equally sensitive to all neutrino flavors.
Therefore, a combined analysis between the mentioned experiments will be fundamental to constrain the normalization and flavor composition of the cosmogenic neutrino flux.

We have shown that the flux of secondary tau-neutrinos is larger than the prompt atmospheric component above a few hundred TeV.
It is the dominant irreducible background for astrophysical tau-neutrino searches in current and future neutrino telescopes. 
Hence, future flavor composition measurements of the cosmic-neutrino flux must account for it.
The relative impact of this background highly depends on the primary neutrino spectrum, so it is fundamental to constrain its shape and normalization further.
We have proved that Earth-skimming experiments are sensitive to cosmogenic fluxes with multiple source populations of UHECRs, independent of the primary flavor composition.
Even in extreme scenarios without primary tau-neutrinos, there will be a distinguishable flux of emerging tau leptons coming from muon- and electron-neutrino interactions.
Nevertheless, experiments sensitive to a single flavor cannot set constraints on the normalization of the flux since there is a degeneracy between primary and secondary contributions.
We therefore conclude that a symbiotic ecosystem of neutrino telescopes with different flavor identification capabilities is needed to understand the origin of UHE neutrinos.

%% file: chapters/taurunner/taurunner.tex
\chapter{\taurunner{}: A Comprehensive Solution to Lepton Propagation}
\label{ch:taurunner}

Now that we have described all the nuances of tau neutrino propagation, we aim to incorporate all these effects into a single, comprehensive, customizable and user-friendly package.
Enter \taurunner, a public \texttt{Python}-based package designed to have minimal dependencies, to allow the user to construct arbitrary neutrino trajectories and propagation media, and to provide interfaces to modify physics inputs such as neutrino cross sections easily.
This chapter describes the software and provides benchmarks and comparisons to other packages that have similar aims.
We expect this software to be useful for next-generation neutrino detectors operating in liquid water (P-ONE~\cite{P-ONE:2020ljt}), solid water (IceCube-Gen2~\cite{IceCube-Gen2:2020qha}), mountains (Ashra NTA~\cite{Sasaki:2017zwd}, TAMBO~\cite{Romero-Wolf:2020pzh}), and outer space (POEMMA~\cite{POEMMA:2020ykm}).

\section{Algorithm Overview}

\begin{figure}[t]
    \centering
    \includegraphics[width=0.68\textwidth]{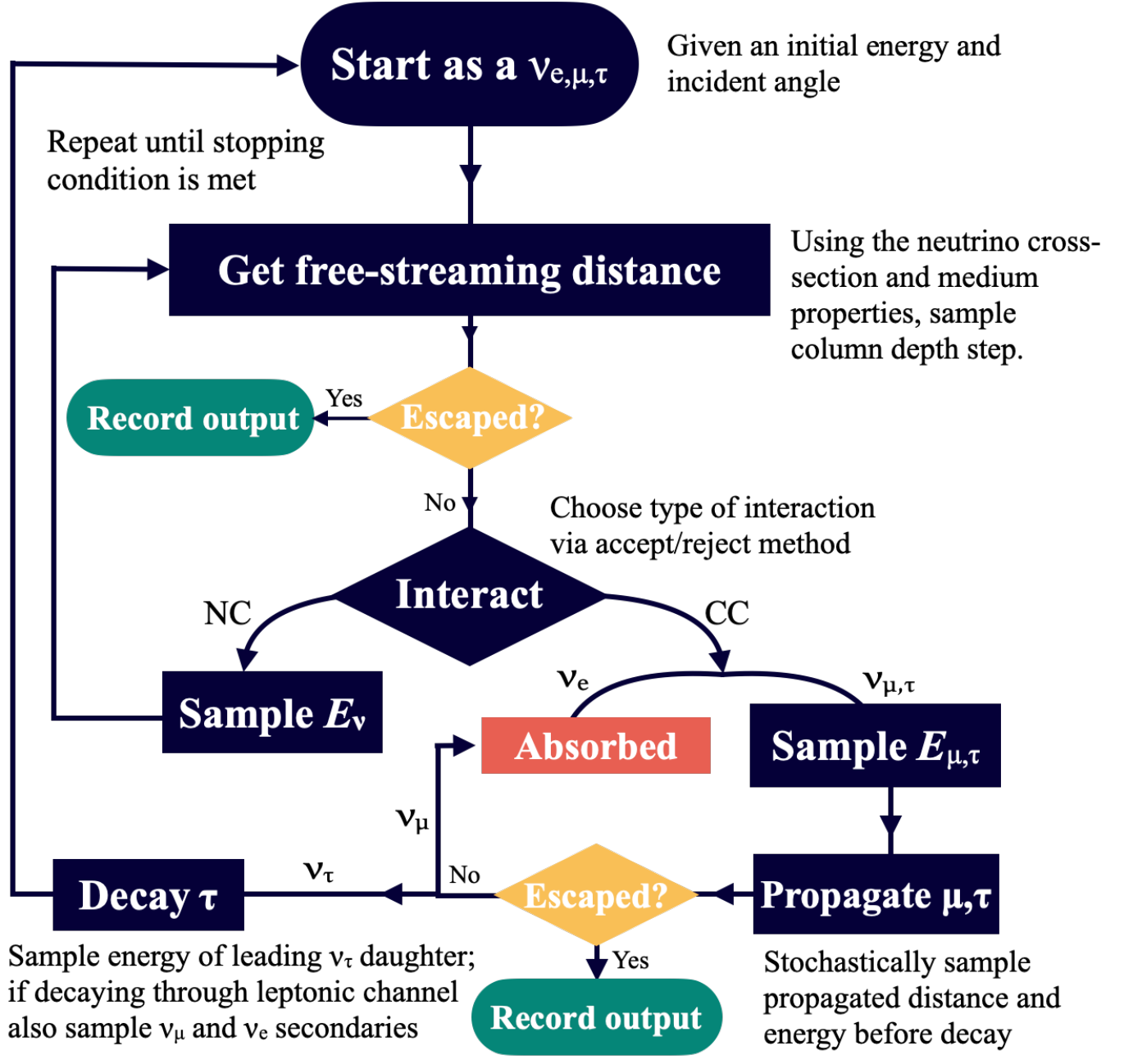}
    \caption[Flowchart of the \taurunner{} propagation algorithm]{Flowchart of the \taurunner{} propagation algorithm.
    Square boxes indicate actions performed by the software. 
    Diamond boxes indicate decision-making stopping points.
    Rounded-corner squared boxes indicate beginning and end of the algorithm.
    Note that users can select also charged leptons as the initial state, in which case
    the algorithm skips straight to the charged particle propagation step.
    }
    \label{fig:flowchart}
\end{figure}

\taurunner{} solves the neutrino transport equation Eq.~\eqref{eq:transport} using a Monte-Carlo approach.
A flowchart of the \taurunner{} Monte-Carlo algorithm is shown in Fig.~\ref{fig:flowchart}. 
Given an initial neutrino type, energy, and incident angle, it begins by calculating the mean interaction column depth, $\lambda_{\textrm{int}}$, which depends on the medium properties and neutrino cross section.
A column depth is then randomly sampled from an exponential distribution with parameter $\lambda_{\textrm{int}}$, and the neutrino advances the corresponding free-streaming distance. 
If the neutrino does not escape the medium, either an NC or CC interaction is chosen via the accept/reject method.
In the case of an NC interaction, the neutrino energy loss is sampled from the differential cross section, and the process repeats.
In the case of a CC interaction, a charged lepton is created with energy sampled from the neutrino differential cross section.

The treatment of the charged lepton then varies according to the initial neutrino flavor.
Electrons are assumed to be absorbed and the propagation stops there. 
$\mu$ and $\tau$, however, are recorded and passed to \texttt{PROPOSAL}~\cite{Koehne:2013gpa} to be propagated through the same medium.
$\mu$ that do not escape will either decay at rest resulting in neutrinos that are below the energies supported by \taurunner{}, or get absorbed.
Therefore a $\mu$ that does not escape is not tracked further. 
Finally, $\tau$s can either escape or decay.
In the latter case, a secondary $\nu_{\tau}$ is created whose energy is sampled from tau decay distributions provided in~\cite{Dutta:2002zc}.
Additionally, if the $\tau$ decays leptonically, $\nu_{e}$ or $\nu_{\mu}$ will be created.
When this happens, the properties of the resulting secondaries are recorded and added to a basket which stores all secondary particles to be propagated together after the primary particle propagation is complete.

\section{Structure of the code}
\label{sec:code}

\taurunner{} may be run either from the command line by running \texttt{main.py} or may be imported to run within another script or \texttt{Jupyter} notebook.
To run from the command line, the user must minimally specify the initial energy, the incident nadir angle, and the number of events to be simulated.
These can be specified with the \texttt{-e}, \texttt{-t}, and \texttt{-n} command line flags respectively.
This will run the \taurunner{} algorithm in Earth with a chord geometry.
The \taurunner{} output will be printed in the terminal unless an output file is specified with the \texttt{--save} flag.
If this option is specified, \taurunner{} will save both a \texttt{numpy} array and a \texttt{json} file with the configuration parameters at the specified location.
In order to ensure reproducibility, the user may specify a seed for the random number generator with the \texttt{-s} flag.
By default, \texttt{main.py} propagates an initial $\nu_{\tau}$ flux, but a user may specify other initial particle types by using the \texttt{--flavor} flag.
Additional options that may be specified by the user can be found in the \texttt{initialize\_args} function of \texttt{main.py} or by running \texttt{main.py} with the \texttt{-h} flag.

To run within another script or \texttt{Jupyter} notebook the user must import the \texttt{run\_MC} function from \texttt{main.py}.
In this latter case one must also create a \taurunner{} \Particle{}, \Track{}, \Body{}, \XS{} objects and a \texttt{PROPOSAL} propagator.
The \Particle{} class contains the particle properties as well as methods for particle propagation.
The \Track{} class parametrizes the geometry of the particle trajectories.
The \Body{} class defines the medium in which the propagation is to occur.
The \XS{} class defines neutrino cross section model.
Additionally, \taurunner{} provides a convenience function for constructing \texttt{PROPOSAL} propagators, \texttt{make\_propagator}, which can be imported from the \texttt{utils} module.
\texttt{Casino.py} combines these classes according to the logic outlined in Fig.~\ref{fig:flowchart}.
After discussing the package broadly, we will discuss conventions and describe \taurunner{}'s output.

\quad \quad In what follows we briefly describe each module:

\textbf{\textit{\Particle{}}}--- A \Particle{} instance contains the structure of a \taurunner{} event.
This includes, among other quantities, the particle's initial and current energies, particle type, and position.
Additionally, it has a number of methods for particle decay and interaction as well as charged lepton propagation.
Finally, the $\tau$ decay parametrization is contained in \texttt{particle/utils.py}.

The user may propagate $\nu_{e}$, $\nu_{\mu}$, $\nu_{\tau}$, $\mu^{-}$, $\tau^{-}$, or any of the corresponding anti-particles in \taurunner{}.
To do this, the user should initialize the the \Particle{} object with the corresponding Particle Data Group Monte Carlo number~\cite{ParticleDataGroup:1994kdp}.
It should be noted that the user may create an $e^{\pm}$, but the internal logic of \taurunner{} assumes all $e^{\pm}$ are immediately absorbed and thus no propagation occurs; see Fig.~\ref{fig:flowchart}.

\begin{figure}[t]
    \centering
    \subfloat[a]{\includegraphics[width=.35\textwidth]{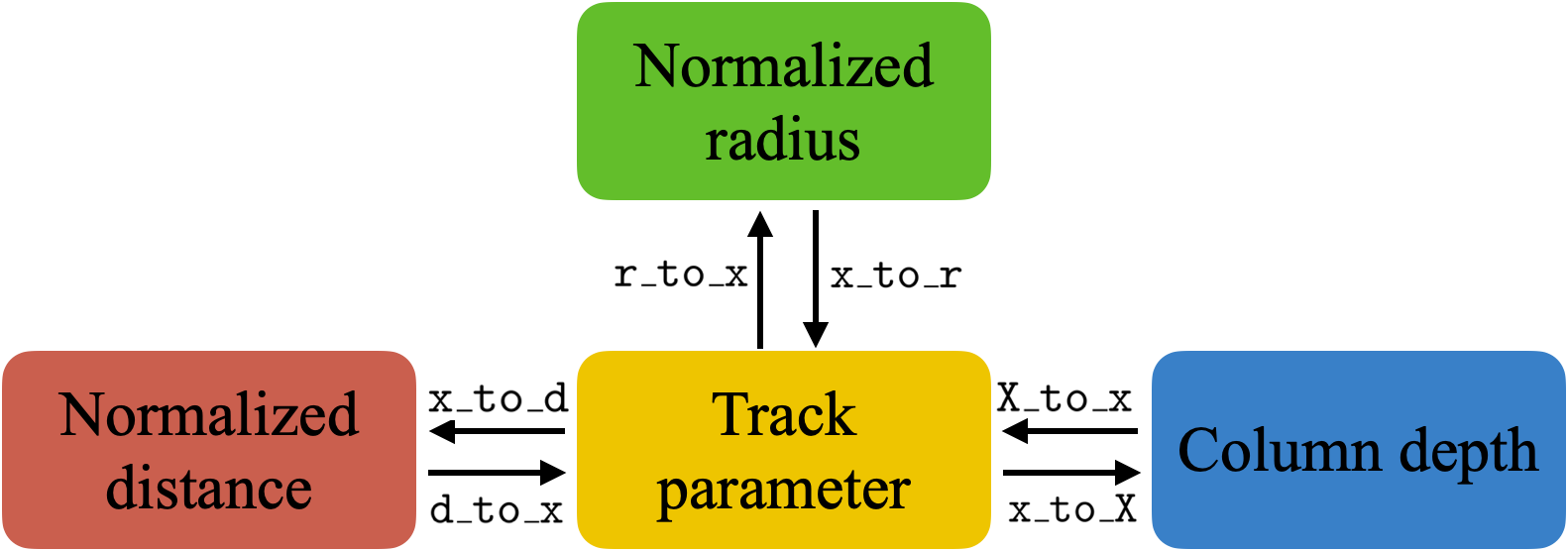}}\\
    \subfloat[b]{\includegraphics[width=0.35\textwidth]{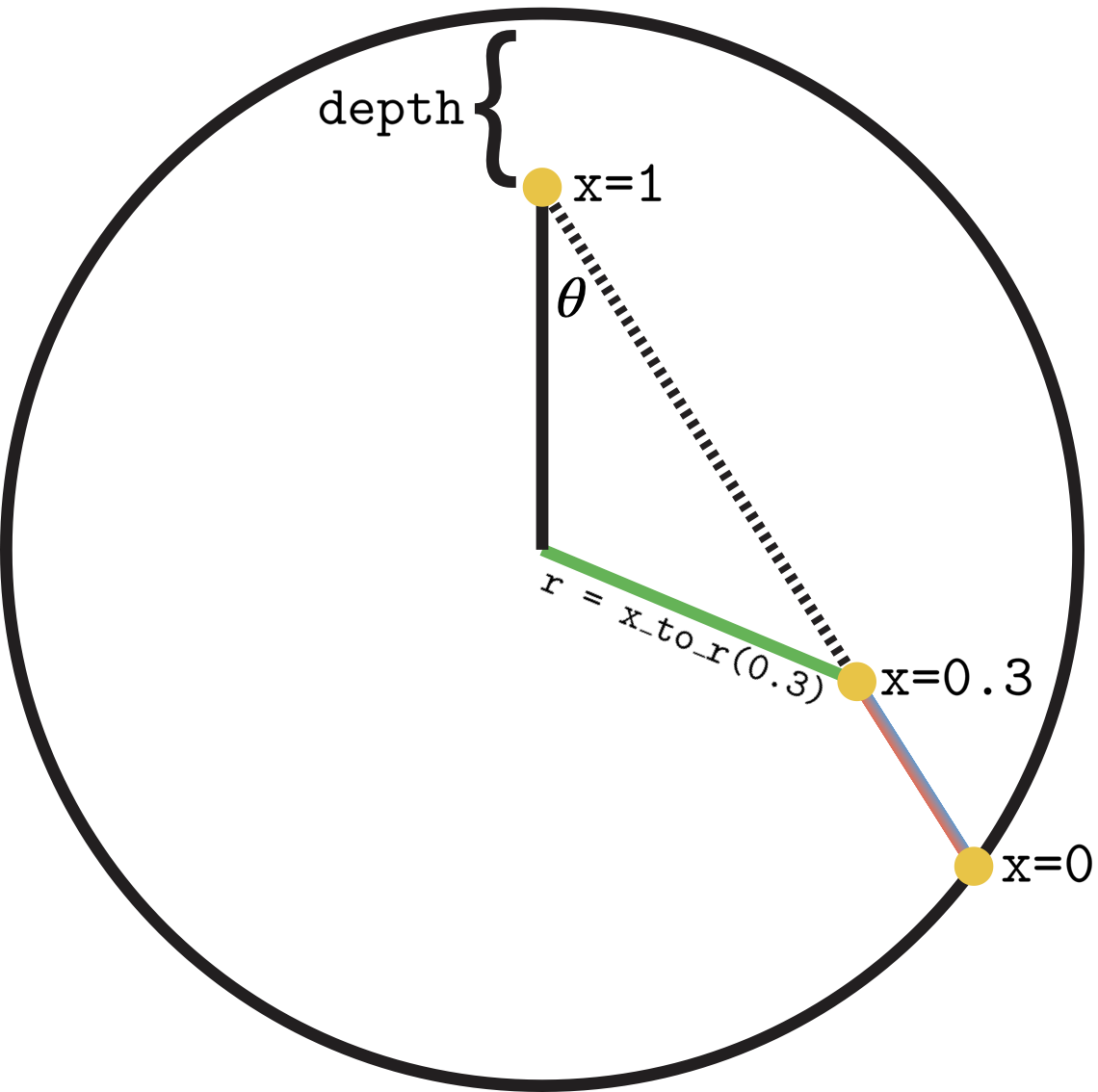}}\\
    \caption[Schematic of \taurunner{} geometry as contained within the \Track{} class.]{Schematic of \taurunner{} geometry as contained within the \Track{} class. (a) shows the relation between the physical quantities relevant to propagation and the affine parameter that parametrizes the \Track{}. The arrows connecting these quantities are labeled with the functions used to convert between them in \taurunner{}. Specifically, these are the functions a user must define in order to specify a custom \Track{} geometry. All distances are normalized with respect to the radius of the body in which the track sits. (b) shows a diagram of these parameters  within a spherical \taurunner{} body. Colors correspond to the boxes in (a). Additionally, it illustrates the \texttt{depth} parameter which intuitively gives the depth of the detector.}
    \label{fig:track_diagram}
\end{figure}

\textbf{\textit{\Track{}}}--- The \Track{} class contains the geometrical information about the particle's trajectory.
A track is parametrized by an affine parameter which defines the position along the trajectory: 0 is the beginning of the trajectory, and 1 is the end.
Almost all of the methods of the \Track{} class are mappings between the affine parameter and physically relevant quantities, \textit{e.g.} radius, distance traveled, and column depth.
The only argument which is generic to the \Track{} class is \texttt{depth} which specifies the distance below the surface of the body at which to stop propagation.
This may intuitively be thought of as the depth of the detector to which the particles are propagated.
An illustration of the \taurunner{} geometry and a diagram of the functional relation of physical quantities to the affine parameter is shown in Fig.~\ref{fig:track_diagram}

The \Track{} class allows the user to make custom trajectories.
The user need only specify mappings between the affine parameter and these variables.
Different trajectories may require additional arguments from the user, depending on the nature of the trajectory.
To illustrate this point, we can look at the two tracks which are implemented by default, the \texttt{Chord} and \texttt{Radial} trajectories.
The former is used for paths which originate outside the \Body{} and cross a section of \Body{}.
The latter is used for paths which originate at the center of the \Body{}.
The former \texttt{Track} describes neutrinos coming from space and passing through Earth on the way to a detector, as in the case of Earth-skimming $\tau$ searches, while the latter gives the trajectory of a neutrino originating in the center of the planet, relevant for searches for neutrinos from gravitationally trapped dark matter.
Clearly, an incoming angle needs to be specified for the \texttt{Chord} trajectory.
Thus, we can see that the necessary arguments for specifying a \Track{} may vary from one geometry to another.

%listing track constructor example
\textbf{\textit{\Body{}}}--- The \Body{} class specifies the medium in which the \Particle{} is to be propagated.
In \taurunner{}, we require that all bodies be spherically symmetric, and so a \Body{} may be minimally specified by a physical radius, and a density profile.
The density profile may be a positive scalar, a unary function which returns a positive scalar, or a potentially-mixed list of positive scalars and such functions.
The sole argument of the functions used to specify the density should be the radius at which the density is to be given, in units of the radius of the body, \textit{i.e.} the domains should be $[0,1]$.
In this system $r=0$ is the center of the body and $r=1$ the surface.
If the user wishes to make a layered body, \textit{i.e.} one where a list specifies the density profile, they must pass a list of tuple with the length of this list equal to the number of layers.
The first element of each tuple should be the scalar or function which gives the density, and the second element should be the right hand boundary of the layer in units of the radius.
The last right hand boundary should always be 1 since $r=1$ is the outer edge of the body.
Lastly, all densities should be specified in $\textrm{g}/\textrm{cm}^{3}$.

In addition to a radius and a density profile, the user may also provide the \texttt{proton\_fraction} argument to specify the fraction of protons to total nucleons in the body.
By default, we assume that the propagation medium is isoscalar, \textit{i.e.} we set the proton fraction to 0.5 throughout the entire body.
As in the case of the density profile, this argument may be a scalar, a function, or a list of function-boundary tuples.
The domains of any functions provided must be [0, 1], and the ranges must be in this same interval.

While the user can construct bodies themselves, there are five bodies implemented by default in \taurunner{}: the Earth, a high-metallicity Sun, and low-metallicity Sun, the moon, a constant density slab.
We use the PREM parametrization to model the densities of Earth~\cite{Dziewonski:1981xy}.
For the Sun, we use fits provided by~\cite{gwensun}.
To instantiate the \texttt{Earth} object, one calls the \texttt{construct\_earth} function, which returns an \texttt{Earth} object.
Additionally, this function allows one to pass in a list of additional layers which will be placed radially outward from the edge of the PREM Earth.
This functionality may be useful for \textit{e.g.} adding a layer of water or ice or adding the atmosphere for simulating atmospheric air showers.
To initialize the Sun, one can use the \texttt{construct\_sun} function.
With this function, the user may specify `HZ\_Sun' or `LZ\_Sun' to use the high- and low-metallicity \taurunner{} suns respectively, or a path to a user defined solar model.

\textbf{\textit{\XS}}--- The \taurunner{} cross sections module defines the neutrino interactions.
Internally, \taurunner{} assumes that cross sections are equal for all neutrino flavors.
Additionally, \taurunner{} uses the isoscalar approximation by default, \textit{i.e.} it assumes a medium is made of equal parts $p^{+}$ and $n$; however, this assumption may be changed by altering the \texttt{proton\_fraction} of the \Body{} object.
The software includes both CSMS~\cite{CooperSarkar:2011pa} and dipole~\cite{Block2014ConnectionOT} cross sections implemented by default; however, it is straightforward for the user to implement other cross section models by providing \texttt{scipy} splines in the appropriate format.
For the total neutrino cross section these splines are \texttt{scipy.interpolate.UnivariateSpline} objects whose $x$-axis is the $\log_{10}$ of the neutrino energy in eV and whose $y$-axis is the $\log_{10}$ of cross section in $\textrm{cm}^{2}$.
The differential cross section splines are \texttt{scipy.interpolate.RectBivariateSpline} objects whose $x$-axis is the $\log_{10}$ of the neutrino energy in eV, whose $y$-axis is a convenience variable which combines the incoming and outgoing neutrino energies, $E_{\textrm{in}}$ and $E_{\textrm{out}}$, given by
$$\eta=\frac{E_{\textrm{out}}-10^{9}~{\textrm{eV}}}{E_{\textrm{in}}-10^{9}~{\textrm{eV}}},$$
and whose $z$-axis is the $\log_{10}$ of incoming energy times the differential cross section in $\textrm{cm}^{2}$.

\textbf{\textit{\texttt{PROPOSAL}}}--- To propagate charged leptons, \taurunner{} relies on \texttt{PROPOSAL}, an open source C++ program with \texttt{Python} bindings.
A utility module to interface with \texttt{PROPOSAL}, \texttt{utils/make\_propagator.py}, is provided with \taurunner{}.
This function instantiates \texttt{PROPOSAL} particle and geometry objects, which are then used to create a propagator instance.
Since \texttt{PROPOSAL} does not support variable density geometries, the \texttt{segment\_body} function is used to segment the \taurunner{} body into a number of constant density layers.
The number of layers is determined by solving for points in the body where fractional change in the density is equal to a constant factor, called \texttt{granularity}.
This argument may be specified by the user, and by default is set to $0.5$.
A single propagator object is created for all $\tau^{\pm}$ and, if needed, for all $\mu^{\pm}$.
Since \taurunner{} assumes $e^{\pm}$ are always absorbed, a propagator will never be made for these.
Whenever a new geometry is used, \texttt{PROPOSAL} creates energy loss tables which are saved in \texttt{resources/proposal\_tables}.
The tables require a few minutes to generate, resulting in an overhead for new configurations, but subsequent simulations with the same geometry will not suffer any slow down.

\subsection{Conventions}
\label{sec:convent}
\taurunner{} uses a natural unit system in which $\hbar=c=\textrm{eV}=1$.
As a consequence of this system, any energy passed to \taurunner{} must be in $\textrm{eV}$.
\taurunner{} includes a \texttt{units} package to easily convert common units to the units \taurunner{} expects.
This may be imported from the \texttt{utils} module, and its usage is demonstrated in several examples.
Additionally, since \taurunner{} assumes that propagation occurs in a spherical body, the radius of this body establishes a natural length scale.
Thus all distances are expressed as a fraction of this radius.

\subsection{Output}
\label{sec:output}
The \texttt{run\_MC} function, which carries out the logic of \taurunner{}, returns a \texttt{numpy.recarray}.
This array may be set to a variable if running \taurunner{} from a script of notebook, or printed or saved if running \taurunner{} from the command line.

In this paragraph, we will describe the fields of this output.
The \texttt{"Eini"} field reports the initial energy of the lepton in $\textrm{eV}$.
The \texttt{"Eout"} field reports the energy of the particle when propagation has stopped in $\textrm{eV}$.
In the case that the particle was absorbed, this field will always read \texttt{0.0}.
The \texttt{"theta"} field reports the incident angle of the lepton in degrees.
The \texttt{"nCC"} and \texttt{"nNC"} fields report the number of charged and neutral current interactions the particle underwent in its propagation.
The \texttt{"PDG\_Encoding"} field reports the particle type, using the Particle Data Group MC numbering scheme.
The \texttt{"event\_ID"} is a number carried byfield reports which initial lepton the particle comes from.
The \texttt{"final\_position"} field reports the track parameter when the propagation was ended.
This may be used to physical quantities of a particle when it was absorbed, or when a user-defined stopping condition was met

\section{Performance}
\label{sec:performance}
For a given primary spectrum and medium through which to propagate, there are a variety of related factors that determine the runtime of the program, including, but not limited to: (1) the initial energy of the neutrinos, (2) the total column depth of the path, (3) the settings for computing energy losses, and (4) which particles are being tracked.

\begin{figure}[!ht]
    \centering
    \includegraphics[width=0.48\textwidth]{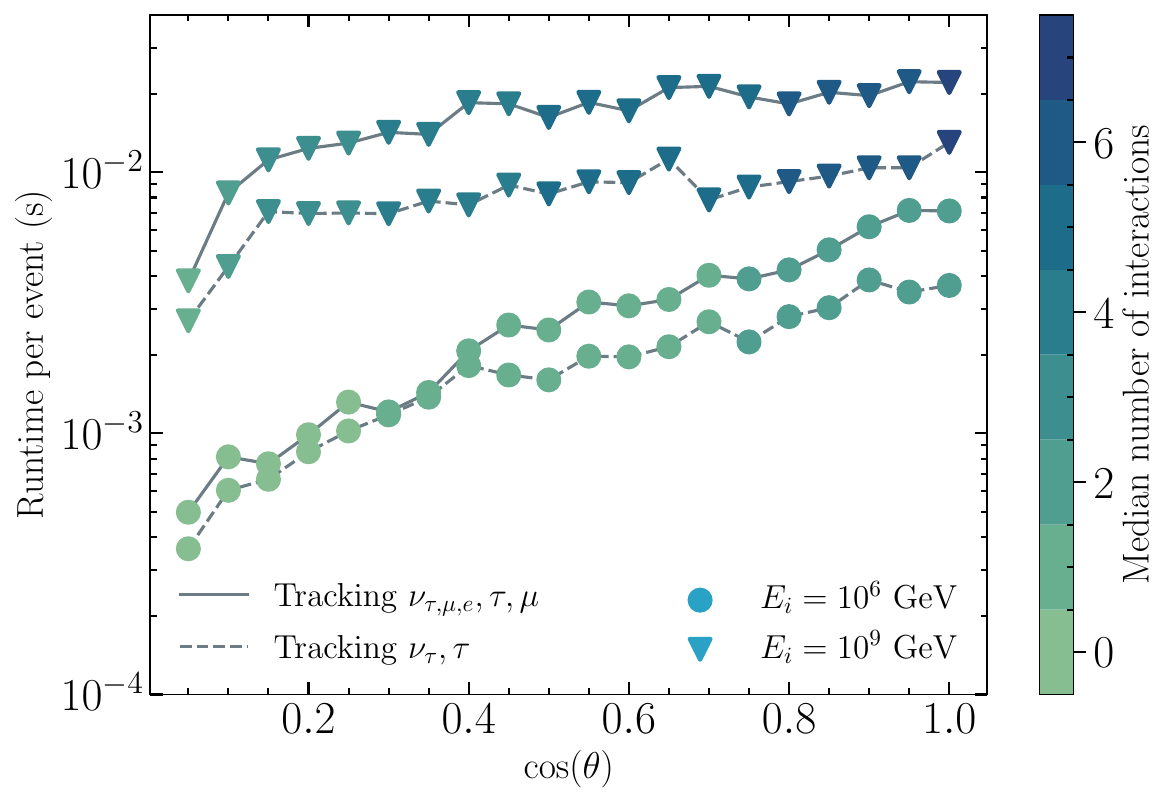}
    \caption[Runtime per $\nu_{\tau}$ event in \taurunner{}.]{Runtime per $\nu_{\tau}$ event. Average runtime per event for various monochromatic fluxes of neutrinos through the Earth, as a function of nadir angle, $\theta$ for incident $\nu_{\tau}$ with energies of 1~PeV (circles) and 1~EeV (triangles). In general, runtime scales with the average number of interactions, which is a function of the energy of the particles and the column depth through which they propagate. The colorbar indicates the median number of NC+CC interactions that the initial beam of $\nu_{\tau}$ undergo. Tracking secondary particles (solid lines) created in $\nu_{\tau}$ CC interactions increases the runtime as the number of particles to propagate increases. Each point represents the average runtime from a simulation including $10^6$ events on a single CPU.}
    \label{fig:performance}
\end{figure}

We show example runtimes for a few different use cases in Fig.~\ref{fig:performance}.
For a fixed \Track{} propagating through Earth, neutrinos with higher initial energy take longer to propagate as they undergo more interactions and as a result experience more stochastic energy losses.
Additionally, those particles that are only being propagated through Earth-skimming trajectories ($\cos(\theta)\approx 0$) can be simulated much quicker than those with large column depths.
This is especially advantageous for proposed Earth-skimming next generation neutrino observatories, \textit{e.g.}~\cite{POEMMA:2020ykm,Neronov:2016zou,GRAND:2018iaj,Aguilar:2019jay,Romero-Wolf:2020pzh}.

By default, all secondary particles that are created as a result of interactions are recorded, meaning that every $\nu_{\tau}$ CC interaction has a chance to increase the number of particles that need to be simulated.
If the user is only interested in outgoing $\nu_{\tau}$ and $\tau$ lepton distributions, this option can be disabled with by setting \texttt{no\_secondaries=True}, which can improve the overall runtime by as much as a factor of two.

Runtime can further be reduced depending on the treatment of energy losses of charged leptons.
By default, energy losses are handled by \texttt{PROPOSAL}~\cite{Koehne:2013gpa}, which treats them stochastically. The user has the choice to ignore energy losses completely, with the setting \texttt{no\_losses=True}, which can improve the runtime by as much as 40\%, although this approximation can only be used in certain scenarios, such as when the initial tau lepton energy is small enough that the interaction length becomes much smaller than the decay length.
This has potential applications for recently proposed indirect searches of ultra-high-energy neutrinos by looking for PeV neutrinos through the Earth~\cite{Safa:2019ege} using large current and next-generation ice or water Cherenkov detectors, such as IceCube-Gen2~\cite{IceCube-Gen2:2020qha}.
Within \texttt{PROPOSAL}, there is also an option to treat energy losses that are below a certain threshold continuously.
We find that setting this parameter to \texttt{vcut=1e-3}, meaning all energy losses that represent less than that fraction of the initial particle energy are treated without stochasticity, achieves an optimal runtime while not neglecting any of the important features that are a result of treating energy losses stochastically.

The first time that a user runs the code, there may be additional overhead while \texttt{PROPOSAL} calculates energy loss distributions for charged leptons.
However, these tables are stored so that future iterations can run more efficiently.
Once the user has run the code at least once and the \texttt{PROPOSAL} energy loss tables are stored, then current runtimes allow users to propagate approximately one million initial EeV $\nu_{\tau}$ through Earth's diameter in approximately eight hours with one CPU.
For an initial energy of one PeV, one million $\nu_{\tau}$ take approximately one hour, depending on the incident angle.
We also found that this runtime varied marginally from machine to machine, and the runtimes in Figure~\ref{fig:performance} and the numbers quoted thus far were all found using a heterogeneous distributed cluster of Linux machines.
The code was also tested on a machine running MacOS with the Apple M1 chip, where the runtimes were found to extremely comparable to those presented above.
For example, $10^4$ $\nu_{\tau}$ with initial energy of one EeV and $\theta=0^{\circ}$ with no secondaries took 0.0127~s per event, on average, and those in the figure above took 0.0124~s per event, on average.

In terms of memory, \taurunner{} can be run on most modern machines, requiring only a few GB of RAM to run.
For example, propagating $10^4$ $\nu_{\tau}$ through the Earth with initial energies of an EeV requires only approximately 1~GB of memory when tracking only $\nu_{\tau}$ and $\tau$, and approximately 3~GB when tracking all particles.
The vast majority of this memory is allocated for calculating energy losses with PROPOSAL, \textit{e.g.} for various trajectories through the Earth and for various initial energies, we found that $\sim 50 - 90\%$ of the memory usage was due to \texttt{PROPOSAL}.
Because most of the memory is due to overhead from the energy losses, there is only a marginal increase in memory usage from propagating many more particles, \textit{e.g.} two sample iterations of the code both took between 2.5~GB and 3.0~GB when propagating $10^4$ or $10^6$ $\nu_{\tau}$ through the Earth with the same initial energies and angles. 

\section{Results and Comparisons}
\label{sec:tauresults}
\begin{figure}[!ht]
    \centering
    \includegraphics[width=0.7\textwidth,trim={0cm 0cm 0cm 0cm}]{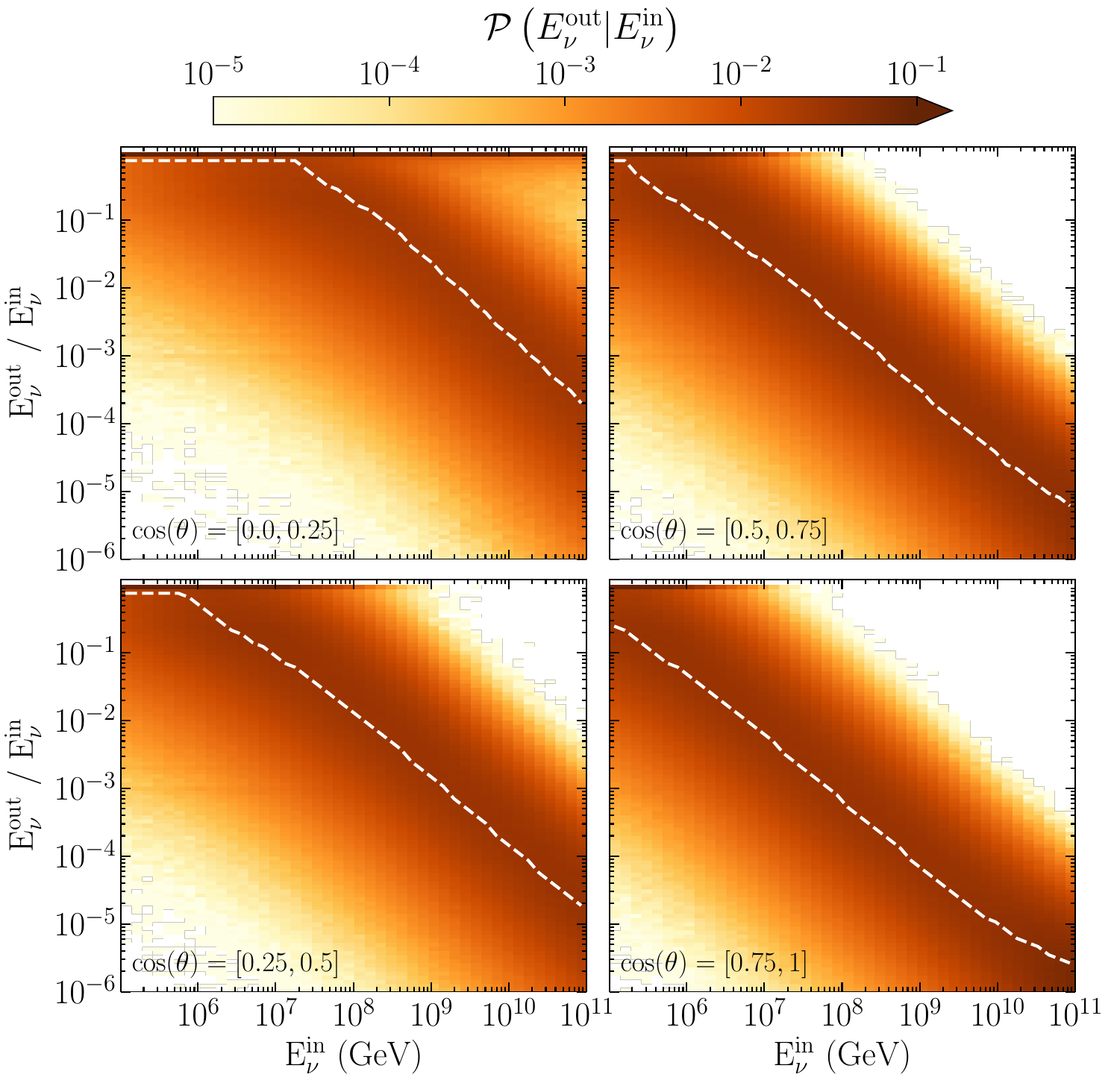}
    \caption[Outgoing $\nu_{\tau}$ distributions for an incident $E^{-1}$ power-law flux.]{Outgoing $\nu_{\tau}$ distributions for an incident $E^{-1}$ power-law flux. Shown are the outgoing tau neutrino energy fraction as a function of the primary tau-neutrino neutrino flux injected as an E$^{-1}$ power-law from 100 TeV to 10 EeV, shown in slices of equal solid angle in the Northern Sky. The dashed line indicates the median outgoing energy}
    \label{fig:outgoing_grid}
\end{figure}

\begin{figure}[!ht]
    \centering
    \includegraphics[width=0.65\textwidth,trim={0cm 0cm 0cm 0cm}]{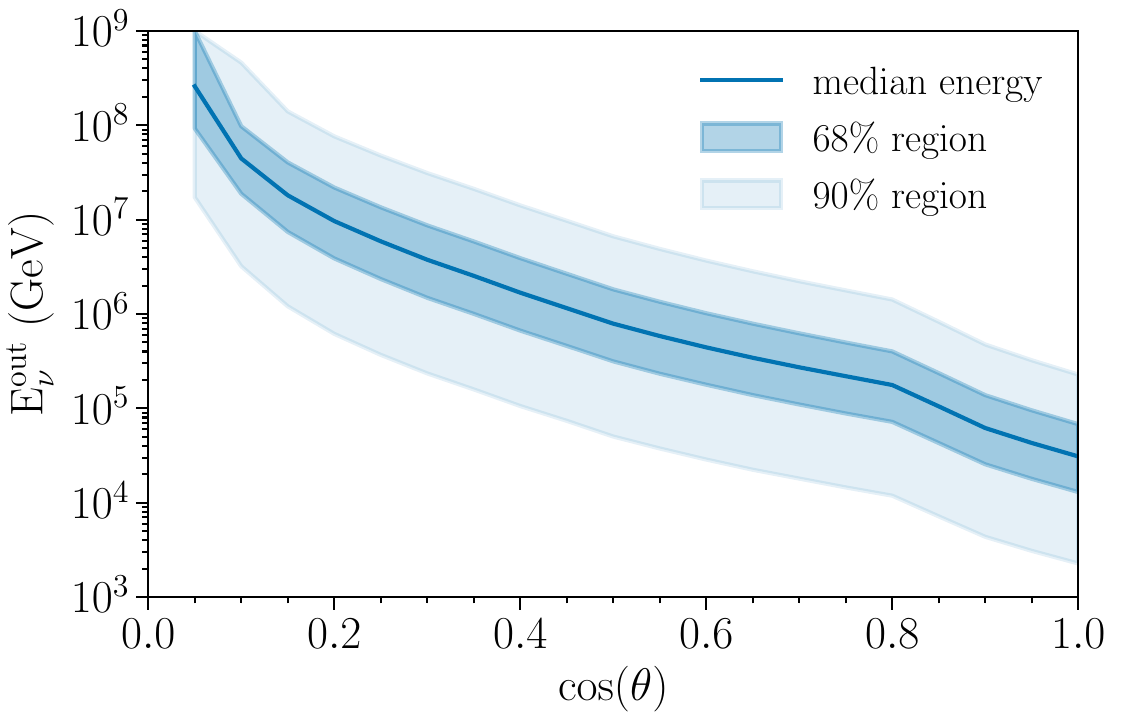}
    \caption[Outgoing energies of secondary tau neutrinos as a function of nadir angle for EeV neutrinos]{EeV tau neutrinos in Earth. 
    Median outgoing energies of secondary tau neutrinos shown as a function of nadir angle. Also, $68\%$ and $90\%$ probability contours for outgoing energies are included. The feature at approximately $\cos{\theta}$ of 0.8 is caused by the core.}
    \label{fig:outgoing_theta}
\end{figure}

The results of several tau neutrino simulation sets are illustrated in this section. 
Fig.~\ref{fig:outgoing_grid} shows column-normalized distributions of outgoing neutrino energy fraction as a function of initial neutrino energy.
Interestingly, the dashed line showing the median outgoing tau neutrino energy fraction varies with a constant slope, corresponding to the energy at which Earth becomes transparent. 
That energy is roughly $10$ PeV at the horizon (top left), $\mathcal{O}(1)$ PeV in the mantle (top right and bottom left), and $\mathcal{O}(10)$ TeV through the core (bottom right). 
This means that for a large fraction of the Northern Sky, tau neutrinos pile-up and escape at energies where the atmospheric neutrino background is relatively low. 
This idea is also made clear when illustrated for a monochromatic flux. In Fig.~\ref{fig:outgoing_theta}, EeV tau neutrinos are propagated and the outgoing energies are plotted as a function of nadir angle. A similar feature can be seen, where a majority of neutrinos in this simulation escape with energy above $100$ TeV.

\begin{figure}[!ht]
    \centering
    \includegraphics[width=0.75\textwidth,trim={0cm 0cm 0cm 0cm}]{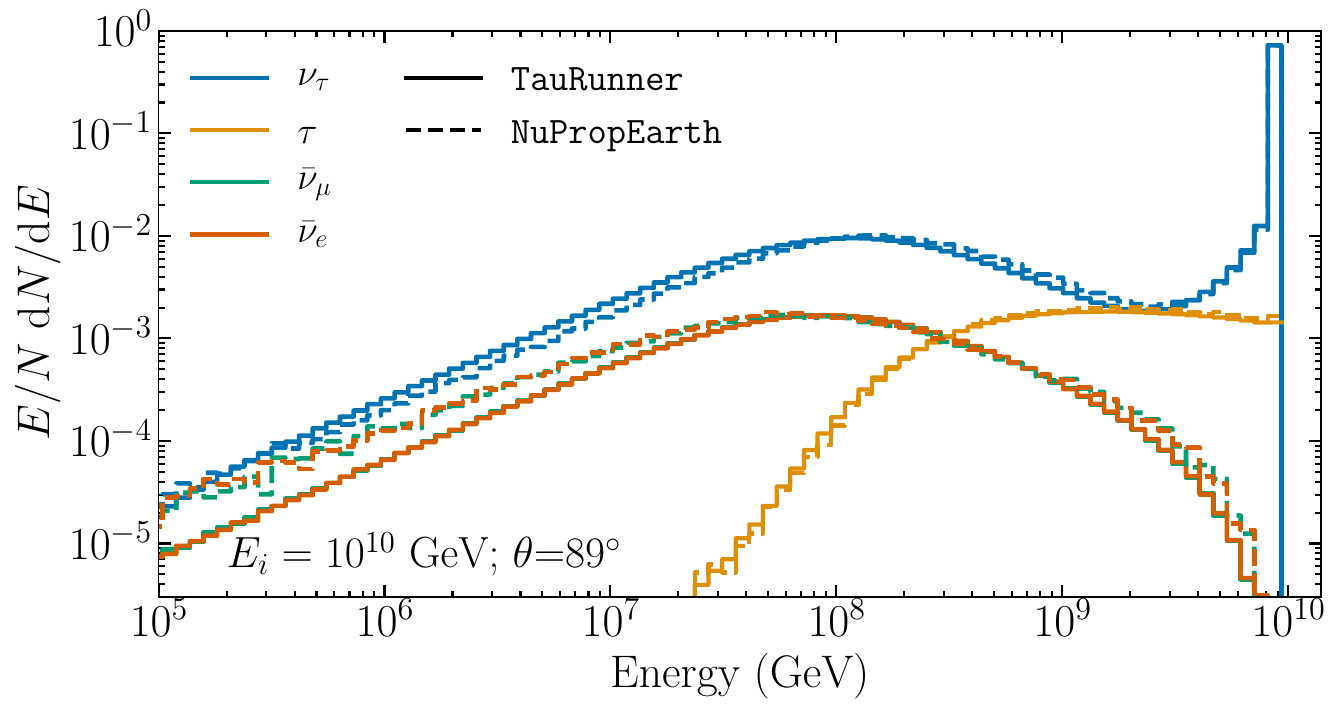}
    \caption[Comparison of \taurunner{} and \texttt{NuPropEarth} output.]{A monochromatic flux of tau neutrinos.
    Outgoing particle energy distributions for a fixed angle and energy. We include secondary anti-electron and -muon neutrinos, as well as charged taus. \taurunner{} shows good agreement with \texttt{NuPropEarth}. This set assumes Earth as a body with a 4km layer of water.
    }
    \label{fig:outgoing}
\end{figure}

\taurunner{} has also been compared to several publicly available packages that perform similar tasks.
A summary of the various tested packages and their features is shown in Tab.~\ref{tab:softable}. 
Besides \taurunner{}, only \texttt{NuPropEarth} offers a full solution in the case of tau neutrinos.
To illustrate this, we show in Fig.~\ref{fig:outgoing} the output of both packages for an injected monochromatic flux of tau neutrinos at $10^{10}$ GeV and one degree below the horizon.
For secondary taus and tau neutrinos, the two packages show excellent agreement. We note that comparisons with \texttt{NuPropEarth} use the trunk version of the code, which has a new treatment for charged particle propagation using \texttt{PROPOSAL} instead of \texttt{TAUSIC}.
Secondary anti-muon and -electron neutrino distributions show slight disagreement in the tails, likely due to different tau polarization treatments.
These differences are still being investigated, and will be addressed in an upcoming work.

\begin{figure}[!ht]
    \centering
    \includegraphics[width=0.55\textwidth,trim={0cm 0cm 0cm 0cm}]{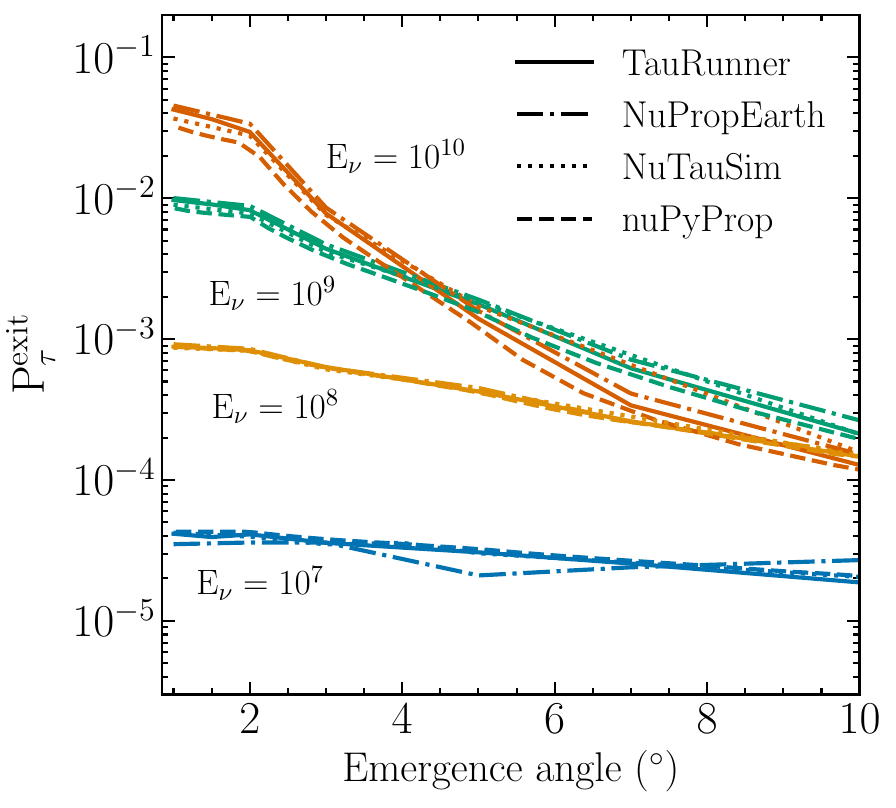}
    \caption[Tau exit probability with \taurunner{} compared with other software packages.]{Charged tau lepton exit probability.
    Different colors correspond to four different monochromatic neutrino energies.
    The  emergence angle is measured with respect to horizon.
    The \taurunner{} prediction (solid line) is compared to NuTauSim, NuPropEarth, and nuPyProp, which are shown in different linestyles.}
    \label{fig:pexit_all_softwares}
\end{figure}

Fig.~\ref{fig:pexit_all_softwares} shows a comparison of the charged tau exit probability in Earth as a function of nadir angle. $P_{\textrm{exit}}^{\tau}$ is the probability that an incoming neutrino will exit Earth as a charged tau. This quantity is especially relevant for future neutrino observatories hoping to detect Earth-skimming tau neutrinos. In that scenario, exiting taus make up the bulk of the expected signal. \taurunner{} again shows great agreement overall with other packages.  

\renewcommand{\arraystretch}{1.5}
\begin{table*}[!ht]
    \centering
    \begin{tabularx}{\textwidth}{ l | c | c | c | c | c }
        \hline
        \hline
        \textbf{Software} & \textbf{Language} & \textbf{Input} & \textbf{Output} & \textbf{Medium} & \\
        \hline
        \taurunner{} & \texttt{Python} & $\nu_{\tau,\mu,e}, \tau, \mu$ & $\nu_{\tau,\mu,e}, \tau, \mu$ & Earth/Sun/Moon/Custom & \\
        \hline
        \texttt{NuPropEarth}\cite{Garcia:2020jwr}  & \texttt{C$++$} & $\nu_{\tau,\mu,e} $ & $\nu_{\tau,\mu,e}, \tau$ & Earth/Custom & \\
        \hline
        \texttt{nuPyProp}\cite{NuSpaceSim:2021hgs} & \texttt{Python}/\texttt{FORTRAN} & $\nu_{\tau}$ & $\tau$ & Earth & \\
        \hline
        \texttt{NuTauSim}\cite{Alvarez-Muniz:2018owm} & \texttt{C$++$} & $\nu_{\tau}$ & $\tau$ & Earth & \\
        \hline
        \hline
    \end{tabularx}
    \caption[Software comparison table]{Each row of this table represents a given package. Input and output particles include their not explicitly mentioned antiparticles. 
    Custom medium refers to a user-defined \Body{} in \taurunner{}.
    }
    \label{tab:softable}
\end{table*}

\section{Implementation of tau depolarization}\label{sec:MC_depolarization}

The theory and derivation of tau depolarization were discussed in Sec.~\ref{sec:depol}. Here, we present the implementation of the ideas discussed in Sec~\ref{sec:depol} and focus on tau propagation in rock, first to determine the distribution of the tau polarization just before it decays, then to illustrate impact of depolarization on the energy distribution of the tau neutrinos that come from tau decays.

\begin{figure*}[!ht]
    \centering
    \includegraphics[width=0.7\textwidth]{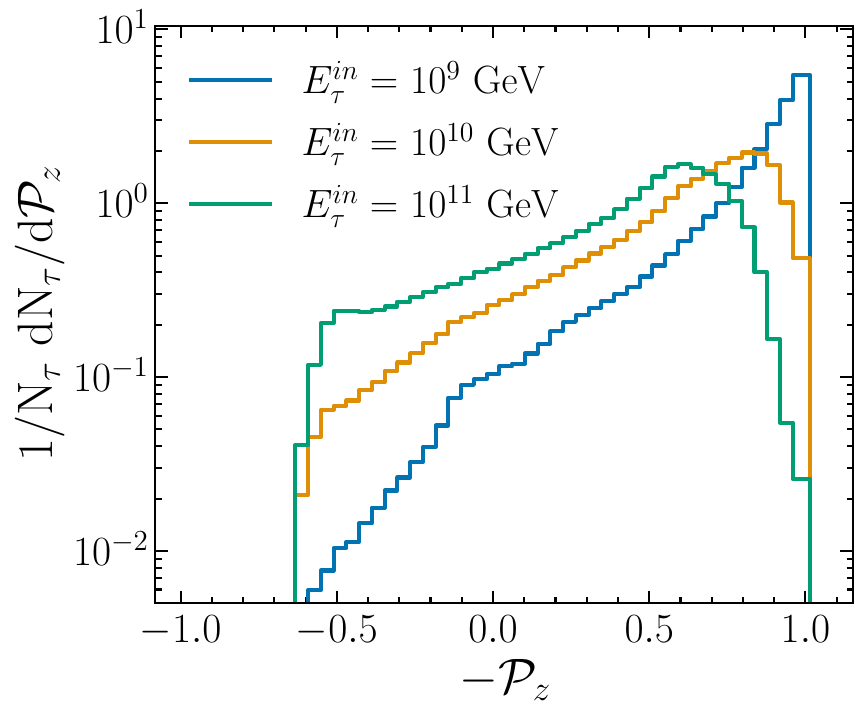}
        \caption[Distribution of final tau polarization before decay, for monoenergetic taus in rock.]{Distribution of final tau polarization before decay, for monoenergetic taus in rock.
        The differential number of taus as a function of final polarization from multiple EM interactions for $10^7$ $\tau$'s propagated through a slab of rock for each initial tau energy.}
    \label{fig:FinalPol}
\end{figure*}

We use \taurunner{} to propagate tau leptons at multiple initial energies through a slab of 200~km.w.e. of standard rock ($A=22$, $Z=11$, and $\rho=2.65\ {\textrm g/cm^3}$), schematically illustrated in the left panel of Fig.~\ref{fig:FinalPol}.
With this depth of rock, all of the taus decay in the slab.
The simulation records $y$ (inelasticity) for each EM interaction of the tau.
For photonuclear interactions, the corresponding $\langle{{\cal {P}}^{EM}_{\tau,z}}\rangle$ and $\langle{P^{EM}}\rangle$ (see Fig.~\ref{fig:EMpol}) are used to determine $\theta_P$ and $P$ for all photonuclear interactions, combined 
according to:
\begin{eqnarray}
   \label{eq:costhp} \cos\theta_{P} &=& \frac{\langle{{\cal {P}}^{EM}_{\tau,z}}\rangle}{\langle{P^{EM}}\rangle} \\
    \label{eq:thpf} \theta_{P,f} &=& \theta_{P,1} \pm \theta_{P,2} \pm \theta_{P,3} \pm ... \\
    \label{eq:pf} P_{f} &=& P_{1}\cdot P_{2}\cdot P_{3}\cdot ...\,,
\end{eqnarray}
Then, the final polarization about the $z$-axis for each tau is given by
\begin{equation} 
\label{finalPolaeqn}
{\cal P}_{z} = P_{f}\cos\theta_{P,f}\,,
\end{equation}
(For a single interaction of the tau, we have defined its polarization about $z$-axis with the notation ${\mathcal{P}}_{\tau,z}$.)
Using Eq.~\ref{finalPolaeqn} with our simulated data we show our results in the right panel of Fig.~\ref{fig:FinalPol} for three incident tau energies, $E_{\tau}^{in}=10^9$~GeV, $10^{10}$~GeV, and  $10^{11}$~GeV.
It is observed that for $E_{\tau}^{in}=10^9$~GeV, the taus are more polarized and therefore the distribution of $-P_{z}$ peaks closer to one.
On the other hand, for  $E_{\tau}^{in}=10^{11}$~GeV, there is more depolarization and we see a shift in the peak away from one.
This is due to the increase in number of interactions for higher initial tau energy which causes more depolarization. 

\begin{figure*}[ht!]
    \centering
    \includegraphics[width=0.8\textwidth]{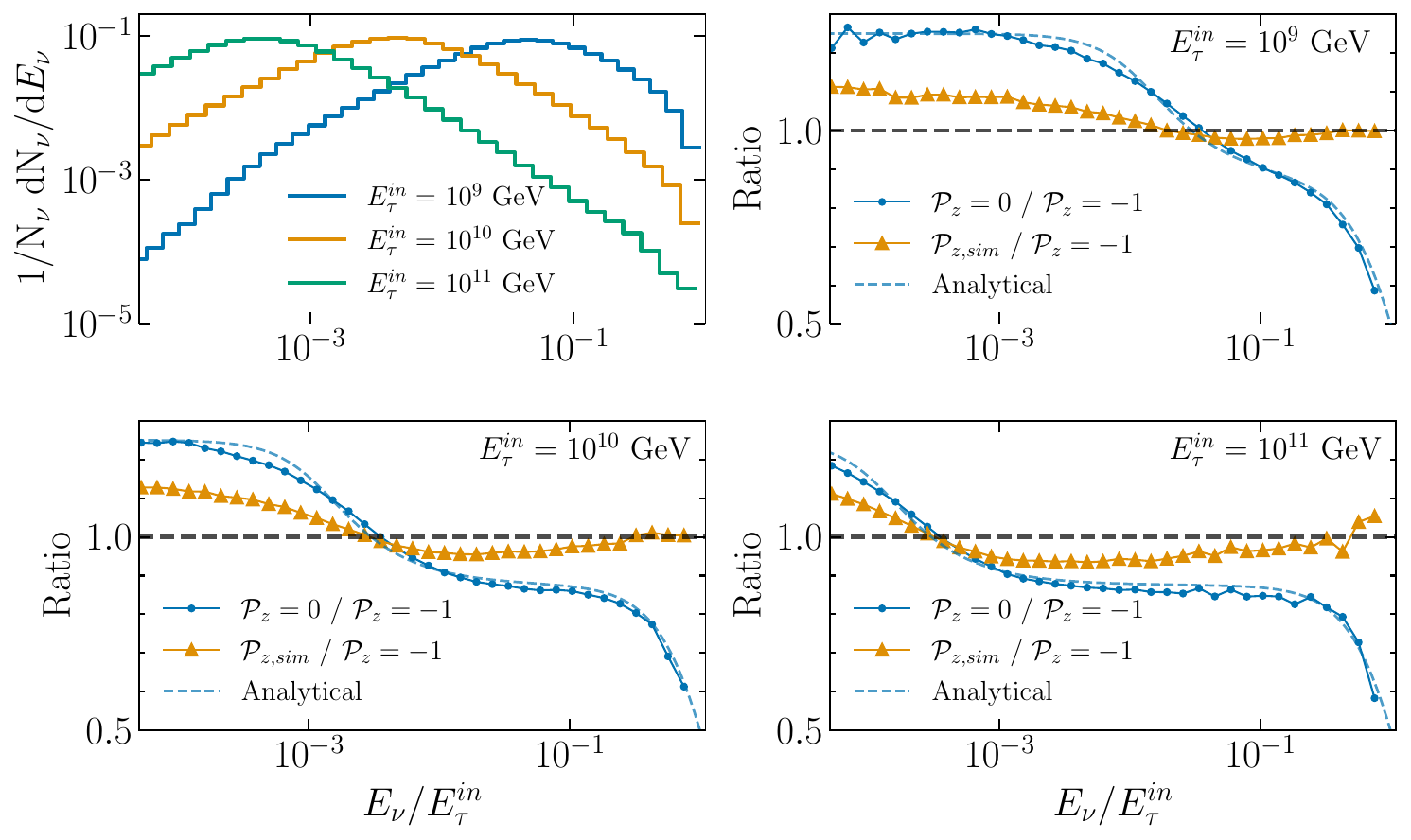}
    \caption[Effect of tau depolarization on outgoing tau neutrinos for multiple starting energies.]{Effect of tau depolarization on outgoing tau neutrinos for multiple starting energies.
    Upper left: Differential number of tau neutrinos as a function of $\nu_\tau$ energy fraction for simulated depolarization taus, shown for three different energies.
    Upper right: The orange (blue) markers show the ratio of the tau neutrino energy distribution with simulated (${\cal P}_z=0$) polarization to the tau neutrino energy distribution for left-handed tau decays as a function of $E_\nu/E_\tau^{in}$ for taus of energy $E_\tau^{in}=10^9$~GeV in rock.
    The dashed blue curve shows the approximate analytic evaluation of the ratio of decay neutrino distributions with ${\cal P}_z=0$ to ${\cal P}_z=-1$.
    The lower plots show these ratios for $E_\tau^{in}=10^{10}$~GeV (left) and $E_\tau^{in}=10^{11}$~GeV (right).}
    \label{Enu_plot}
\end{figure*}
The bump in the right panel of Fig.~\ref{fig:FinalPol} for $E_{\tau}^{in}=10^{11}$~GeV at ${\cal P}_{z} \simeq -0.6$ arises because $\cos\theta_{P,f}$ can have negative values, i.e., $\theta_{P,f}>\pi /2$.
The distribution for $P_f$ peaks at $\sim 0.6$ for $E_{\tau}^{in}=10^{11}$~GeV, which combined with negative values of $\cos\theta_{P,f}$ gives ${\cal P}_{z} \simeq -0.6$ for some fraction of taus.  

We now turn to the neutrino energy distribution from tau decays after propagating in rock.
For each incident tau at a fixed energy, we calculate the energy of the tau neutrino from tau decay including the final tau energy and polarization.
This is done by creating a cumulative distribution function from the neutrino energy distribution equation given in Eq.~\ref{eq:dGammadx2}.

We show the effect of depolarization of taus on $\nu_\tau$ energy distribution in Fig.~\ref{Enu_plot}.
The upper left plot shows the differential number of tau-neutrinos as a function of tau-neutrino energy fraction ($E_\nu/E_{\tau}^{in}$).
Higher energy taus lose more of their initial energy as they propagate farther before their decays.

For three initial tau energies, the remaining  plots in Fig.~\ref{Enu_plot} show ratio of $\nu_\tau$'s produced from unpolarized (${\cal P}_{z}=0$) to fully polarized (${\cal P}_z=-1$) taus (blue markers and curve), and simulated (${\cal P}_{z,sim}$) to fully polarized (${\cal P}_z=-1$) taus (orange markers and curve).
The cross-over points of the ratio plots approximately correspond to the peaks in the upper left plot of Fig.~\ref{Enu_plot}. 

To cross-check our results for the neutrino energy distribution from tau decays given taus incident on rock, we used an approximate analytical equation to get the $\nu_\tau$ spectrum.
We show dashed blue curves with this semianalytic approximation in Fig.~\ref{Enu_plot}.
The ratio of the analytic evaluation of the neutrino energy distribution of unpolarized to fully polarized tau agrees very well with the ratio of distributions for ${\cal P}_z=0$ to ${\cal P}_z=-1$ from the Monte Carlo.

\begin{figure*}[!ht]
    \centering
    \includegraphics[width=0.35\textwidth ]{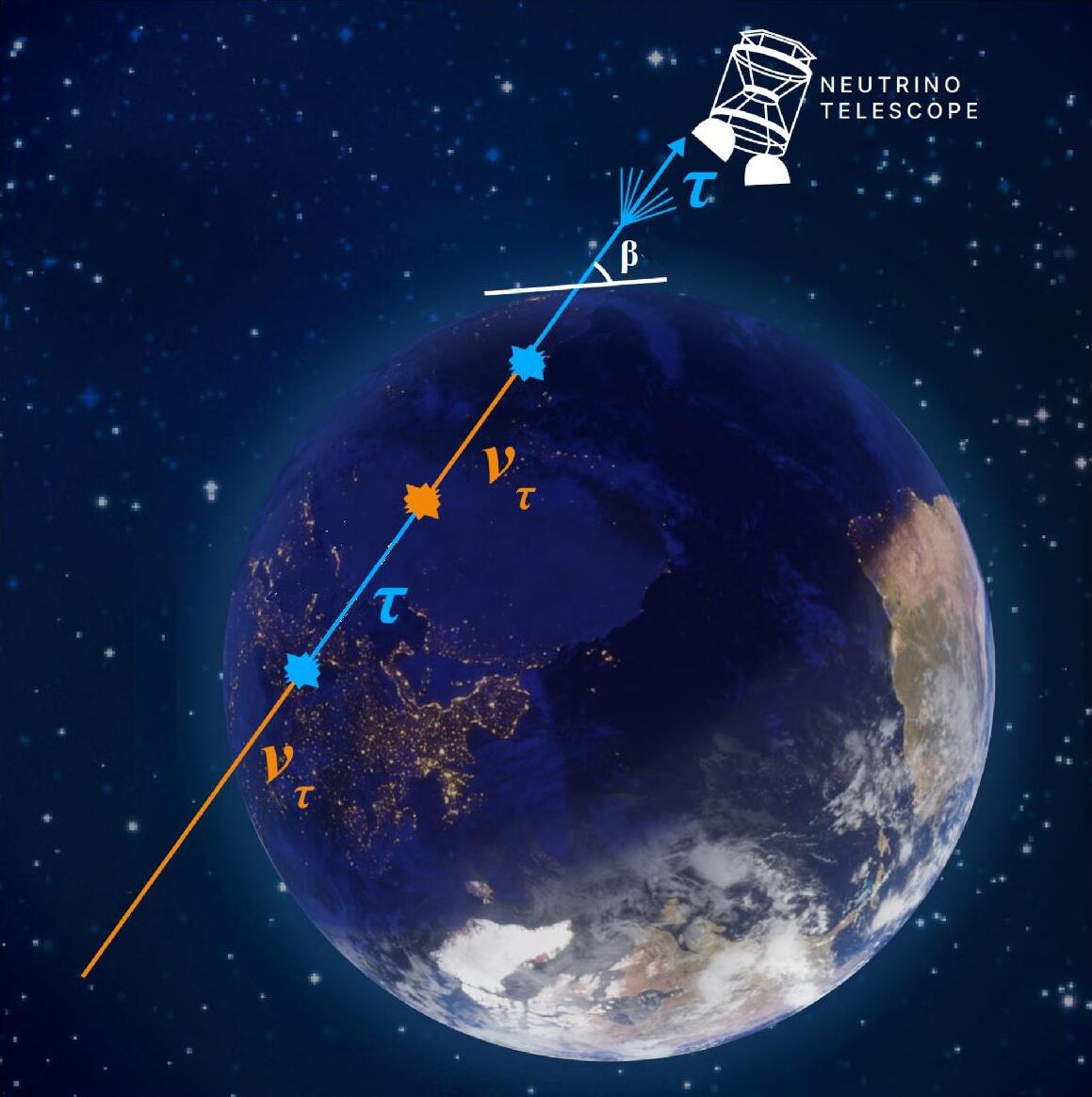} %pdf
    \includegraphics[width=0.6\textwidth]{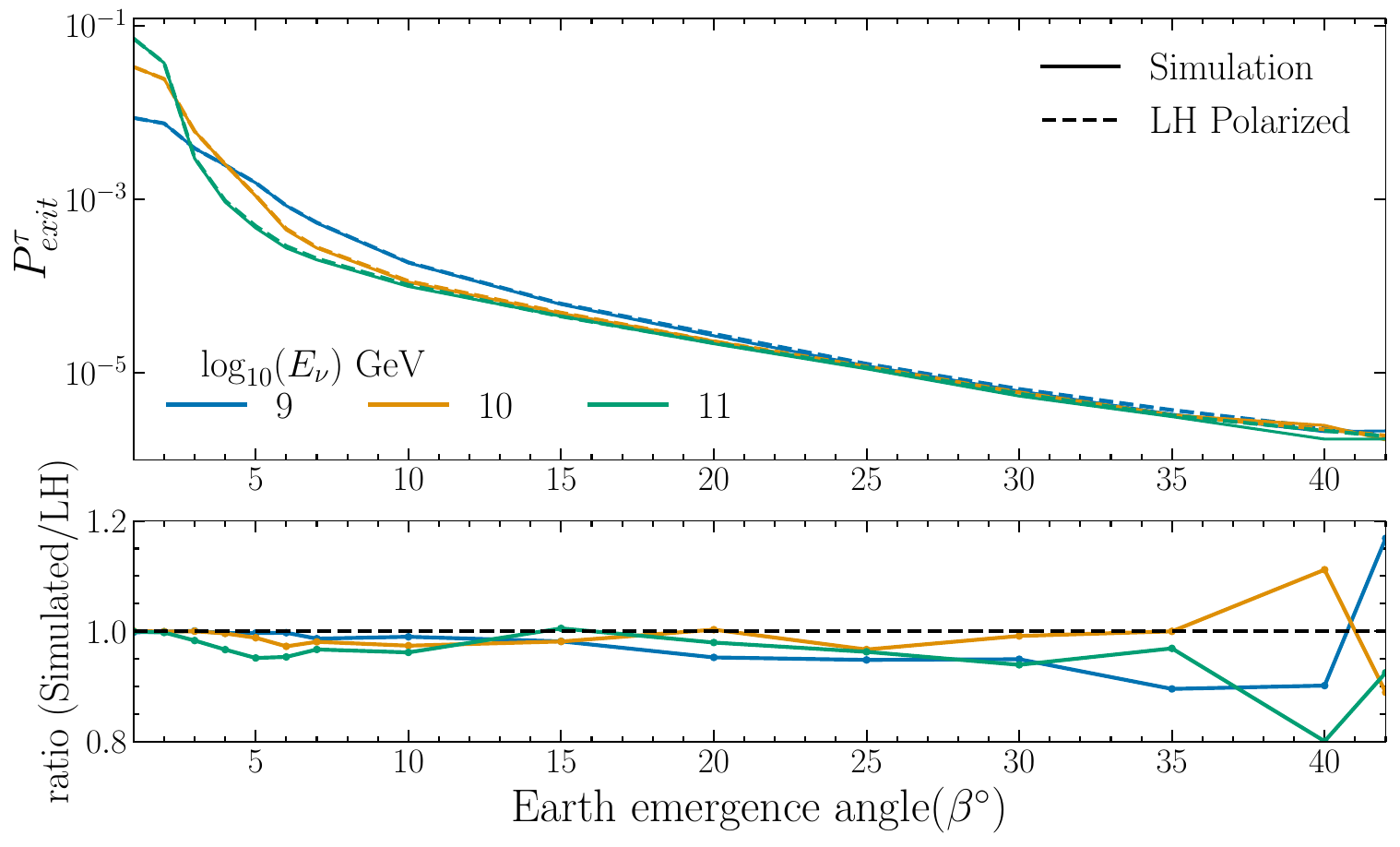}
    \caption[Impact of tau depolarization on the tau exit probability.]{Impact of tau depolarization on the tau exit probability.
    Left: Schematic diagram of a $\nu_\tau$ incident on the Earth that results in an emerging tau after series of CC interactions and tau decays (regeneration). The Earth emergence angle is $\beta$. 
    Right: Exit probability of taus as a function of different Earth emergence angles, for three different initial tau-neutrino energies. It shows a comparison when we consider LH polarization and simulated depolarization for EM interactions of the taus.}
    \label{fig:skimcartoon}
\end{figure*}

\textbf{\textit{Results for Earth-skimming tau neutrinos}}--- Now we implement this depolarization technique into our full neutrino simulation, allowing us to study its effects on neutrino telescopes.
We simulate $\nu_\tau$'s skimming Earth, at different Earth emergence angles ($\beta$), taking into account tau depolarization.
A schematic of Earth-skimming tau neutrino trajectories to produce a tau that emerges from Earth is shown in the left panel of Fig.~\ref{fig:skimcartoon}.

\begin{figure*}[ht]
    \centering
    \includegraphics[width=0.45\textwidth]{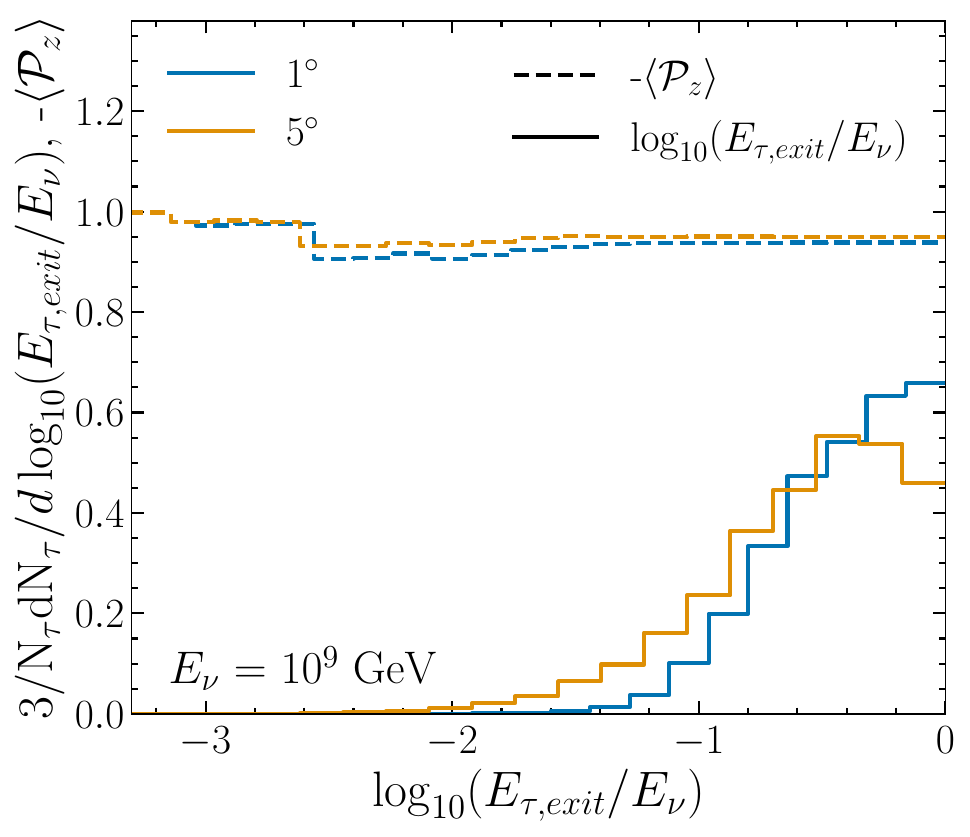}
    \includegraphics[width=0.45\textwidth]{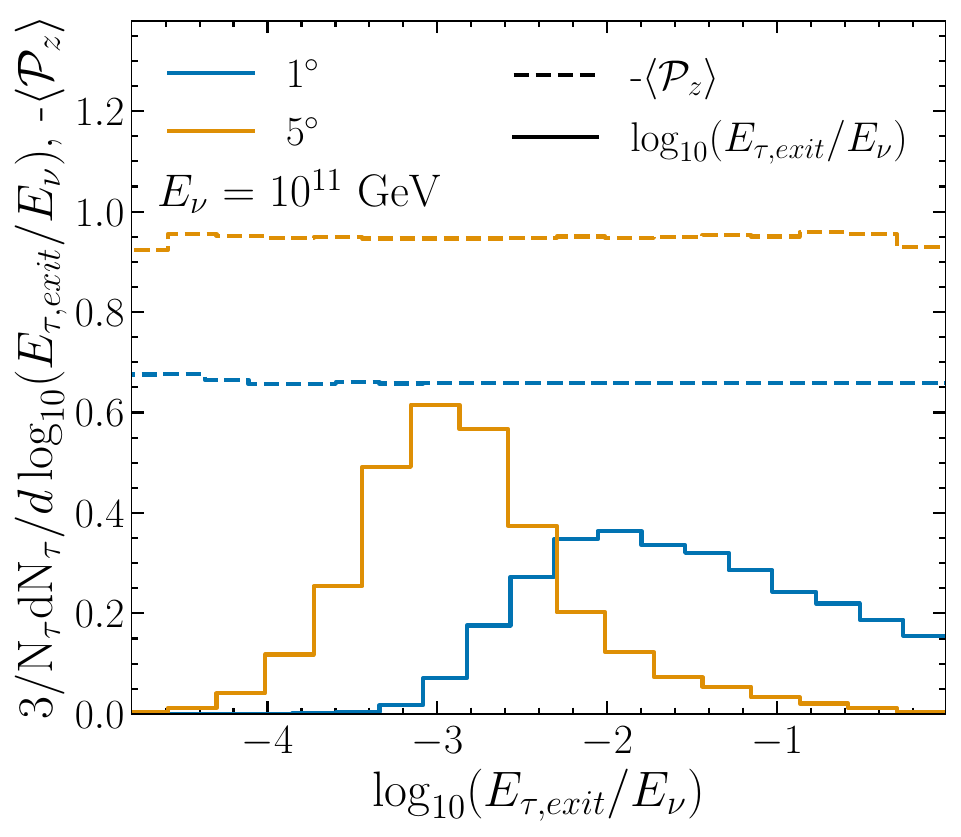}
    \caption[Final energy of Earth-emerging taus and their polarization starting from a neutrino beam.]{Final energy of Earth-emerging taus and their polarization starting from a neutrino beam.
    Differential number of exiting taus and average polarization ($-\langle {\cal P}_z\rangle$) of exiting taus as a function of exiting tau's energy fraction for initial neutrino energies $E_\nu=10^9$~GeV (left) and $10^{11}$~GeV (right) and for $\beta=1^{\circ},5^{\circ}$.
    The energy distribution normalization was chosen so the energy distributions and polarizations can appear in the same figures. Note that the $x$-axes have different ranges in the two panels.}
    \label{Efexit}
\end{figure*}
For Earth-based, suborbital and satellite instruments that detect signals of tau decay-induced extensive air showers, modeling requires the probability that a neutrino produces an exiting tau, the energy of the emerging tau, and its final polarization upon exit.
In the right panel of Fig.~\ref{fig:skimcartoon}, the exit probability of the taus is plotted for different Earth emergence angles, for three different initial tau neutrino energies.
It shows a comparison between fully polarized and simulated depolarized taus.
We observe that the exit probability is changed by 5\% for smaller angles, and 10\% for higher angles, when we consider depolarization in the EM interactions.
This shows us that depolarization has a small impact on the exit probability of the taus.

Fig.~\ref{Efexit} shows us the final energy of the exiting taus with corresponding average polarization.
The energy distributions are normalized so that they appear on the same scale as the polarization curves included in the figures.
The plot on left is for $E_\nu=10^{9}$ GeV for $\beta=1^\circ$ and $5^\circ$. For this energy, the regeneration rate is negligible for both angles. 
The taus that exit the Earth are the ones created from the initial tau neutrinos that interact close to the surface of the Earth. 
Thus there is no significant depolarization and the final energy of the exiting taus is close to the initial tau-neutrino energy. 
The right-hand plot in Fig.~\ref{Efexit} for $E_\nu=10^{11}GeV$,  for $\beta=1^\circ$, the high energy taus that are produced are partially depolarized as they propagate and lose energy on their way to exit Earth. 
For $\beta=5^\circ$, the taus able to exit Earth are created from the regenerated tau neutrinos which interact closer to the surface of Earth.
Because of regeneration, the energy distribution of the emerging taus is lower than for $\beta=1^\circ$.
Since the taus that emerge are produced close to the Earth surface, and with each tau production, its polarization is reset to $-1$, the average polarization for exiting taus is close to $-1$ for $\beta=5^\circ$ for incident neutrinos with $E_\nu=10^{11}$ GeV.
In principle, tau depolarization effects can affect the flux normalization and the energy distributions of tau neutrinos and taus that arrive at underground detectors or taus that emerge to produce upgoing air showers.
We have performed an analysis of the dominant contribution to the depolarization of CC interaction produced left-handed taus as they transit materials. 

The depolarization of taus is not complete for tau energies up to $10^{11}$ GeV. 
With our simulations of tau energy loss in rock, Fig. \ref{Enu_plot} shows that the neutrino energy distributions from tau decays are shifted from LH tau decays by $\sim \pm 10\%$.
We have shown that the tau exit probability is modified at most by $\sim 10\%$ for large Earth emergence angles, where the exit probability is already low. 
The energy distribution of the emerging taus is essentially the same with and without accounting for EM depolarization effects.
Our results for the polarization of Earth-emerging taus show that the average polarization depends on the incident neutrino energy and angle, but it is largely independent of the final tau energy.
Improved modeling of the initial energy of the extensive air shower from tau decays in the atmosphere by including ${\cal P}_z$ is therefore straightforward to implement in Monte Carlo simulations with stochastic energy loss like \texttt{TauRunner}.

%
%\section{Examples}
%\label{sec:examples}

%\quad In this section, we show examples which illustrate many of the capabilities of \taurunner{}.
%\taurunner{} can be run from the command line or imported as a package.
%When a feature can be used via both interfaces, we provide an example for each.

%\textbf{\textit{Installation}}---
%\input{taurunner/examples/installation}

%\textbf{\textit{Monochromatic through Earth}}---
%\input{taurunner/examples/monochromatic}

%\textbf{\textit{Isotropic Flux through Earth with Power Law Distribution}}---
%\input{taurunner/examples/power_law}

%\textbf{\textit{Custom Flux through Earth}}---
%\input{taurunner/examples/custom_flix}

%\textbf{\textit{Radial Trajectory}}---
%\input{taurunner/examples/radial}

%\textbf{\textit{Sun}}---
%\input{taurunner/examples/sun}

%\textbf{\textit{Constant Slab}}---
%\input{taurunner/examples/const_slab}

%\textbf{\textit{Layered Slab}}---
%\input{taurunner/examples/layered_slab}

%% file: chapters/taus_advantages.tex
\chapter{Implications of Accurate Tau Neutrino Modeling}
\label{ch:tau_adv}
This chapter builds on chapters~\ref{ch:taumodeling} and \ref{ch:taurunner} to showcase the physics potential of accurate tau neutrino modeling.
We discuss the impact on GZK neutrino searches for Cherenkov telescopes in Sec.~\ref{sec:GZK} and perform a sensitivity analysis for IceCube, showing that current limits can be improved by up to a factor of two.
We then highlight the promise of detecting ultra-high-energy transients with tau neutrinos in Sec.~\ref{sec:transients}, showing that current IceCube differential sensitivities are in fact orders of magnitude better in the Northern Sky when using muons from tau neutrinos.
Finally, We discuss two followups to anomalous ANITA events using tau neutrinos in Sec.~\ref{sec:ANITA} and show that all standard model explanations are ruled out.
\section{GZK Neutrinos}
\label{sec:GZK}

At energies beyond the so-called Greisen-Zatsepin-Kuzmin (GZK) cutoff ($E \geq 40$ EeV), proton interactions with the CMB restrict the mean free path of cosmic-ray nuclei primaries to less than a few hundred Mpc from sites of cosmic acceleration. A suppression compatible with the GZK cutoff has indeed been observed in cosmic-ray experiments \cite{Abbasi:2007sv, Sokolsky:2008zz, Abraham:2008ru, AbuZayyad:2012ru}. The subsequent decay of the mesons from these interactions leads to an observable, yet currently undetected, flux of GZK neutrinos. 

Although the GZK flux should be isotropic at Earth, searches for this flux have been limited to either half of the sky (downgoing) or often smaller solid angles, specifically looking for Earth-skimming neutrinos, where the probability of detecting a tau in the atmosphere after a single neutrino interaction in Earth is optimized \cite{Abbasi:2010ak, Barwick:2014pca, Aab:2015kma, Aartsen:2016ngq, Allison:2015eky,Allison:2018cxu,Aartsen:2018vtx,Aab:2019auo, Kotera:2010yn}. Here, we show how using the secondary tau neutrino flux will extend this search to the entire sky. Specifically, we look for neutrinos after several interactions in Earth, which emerge at $\mathcal{O}$(PeV) energies. We also show that the rate from Earth-traversing neutrinos is not negligible, and quantify the gain in sensitivity when including them. 

\begin{figure}
    \centering
    \includegraphics[width=0.75\textwidth]{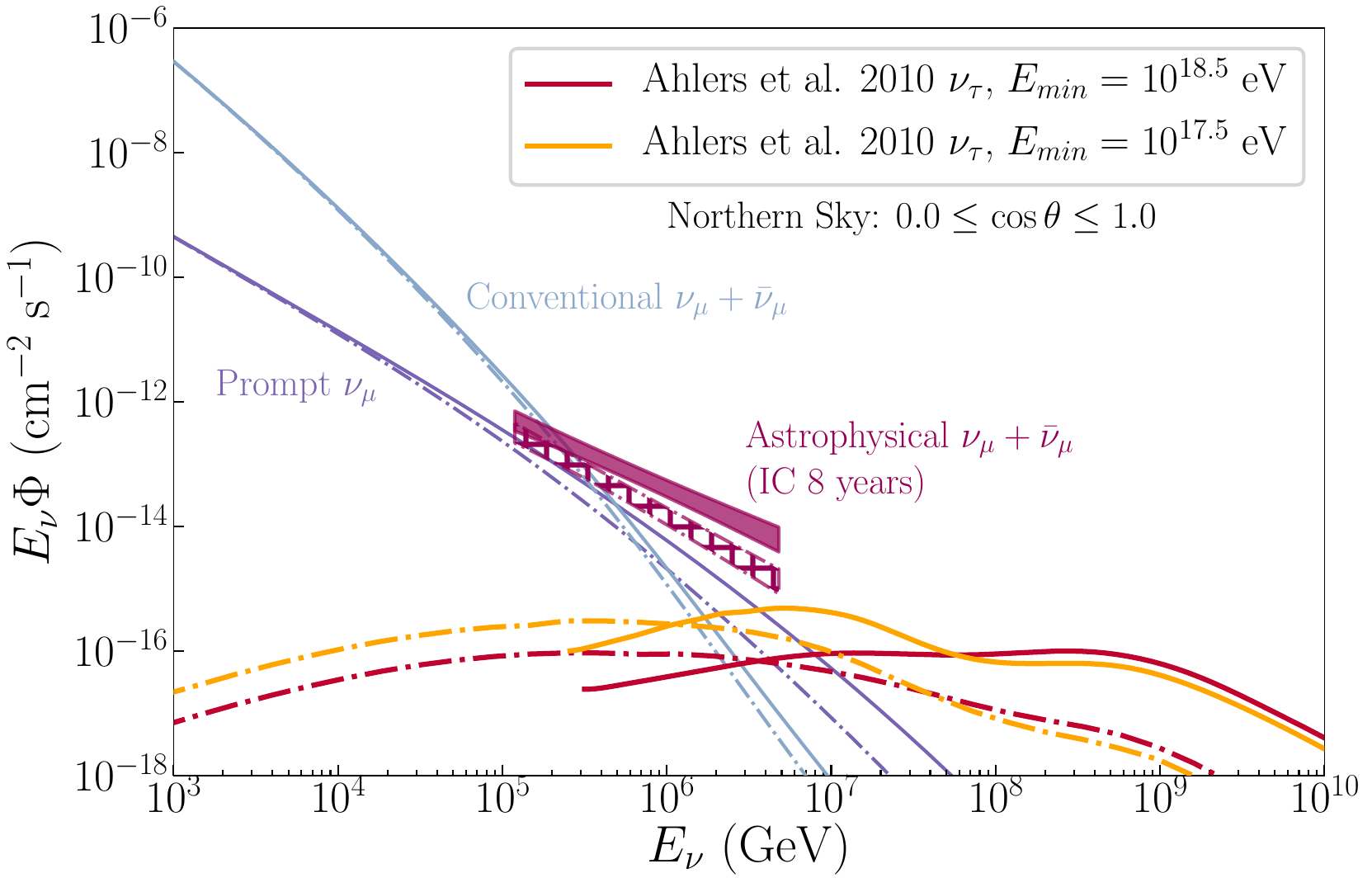}
    \caption[Per flavor neutrino fluxes from 1 TeV to 10 EeV, integrated over the northern sky.]{Per flavor neutrino fluxes from 1 TeV to 10 EeV, integrated over the northern sky. Primary fluxes are shown as solid lines, and fluxes present at the IceCube detector are shown in dashed-dotted lines and hatches. The $\nu_{\tau}$ components for various models of the cosmogenic flux are shown in red and orange~\cite{Ahlers:2010fw}. These spectra are compared to models of both the conventional~\cite{Honda:2006qj} and prompt~\cite{Enberg:2008te} components of the atmospheric flux %\cite{Aartsen:2012uu, Aartsen:2015xup, Abbasi:2010ie}
    as well as measurements of the diffuse astrophysical flux~\cite{Haack:2017dxi}.  Secondary $\nu_{\tau}$ spectra peak at PeV energies, a region of parameter space optimal for neutrino telescopes such as IceCube.}
    \label{fig:gzk_and_friends_wholesky}
\end{figure}

Fig.~\ref{fig:gzk_and_friends_wholesky} displays the secondary tau neutrino flux of GZK neutrinos compared to atmospheric and diffuse astrophysical per-flavor neutrino fluxes in the northern sky. For a representative GZK flux, we choose a model produced from a fit to HiRes data~\cite{Ahlers:2010fw}.
We have also considered other fluxes in the literature~\cite{Globus:2017ehu, Kotera:2010yn, Aloisio:2015ega} and find that the shape of the secondary flux at the detector is roughly consistent and differences manifest mostly in the overall normalization. Predictions of number of events varied from 1 to 18 for the proton-dominated fluxes considered. In the figures, we show our benchmark flux~\cite{Ahlers:2010fw} as an example, as it provides a picture of the average expectation for proton-dominated fluxes.
The conventional component in Fig.~\ref{fig:gzk_and_friends_wholesky} shows the $\nu_{\mu}$ flux produced in cosmic-ray showers in the atmosphere, using the model in~\cite{Honda:2006qj}.
The prompt component is the expected muon neutrino flux arising from atmospheric charm production in cosmic-ray showers; we use the model in~\cite{Enberg:2008te}.
Although there is a predicted tau neutrino component of this prompt flux, predominantly from D-meson decays, the level of this flux is much smaller compared to the prompt electon- and muon-neutrino components.
The astrophysical muon neutrino flux we use is based on eight years of northern sky muon track data from IceCube~\cite{Haack:2017dxi}.
All of these primary fluxes are propagated to the detector.
The fluxes arriving at the detector are then compared to the secondary flux from GZK neutrinos.
The spectrum of the secondary GZK flux is much harder and strongly dependent on declination, providing additional handles to distinguish GZK secondaries from other astrophysical or atmospheric events. 
The tau neutrino component of the astrophysical neutrino flux (produced directly at cosmic accelerators and not from proton interactions with the CMB) could also contribute to the secondary neutrino flux in the northern sky. This contribution, and its relation to the GZK secondaries, depends on the spectral index and maximum energy of the primary astrophysical flux. If the primary astrophysical flux is assumed to be a hard power law (such as that of the through going muon neutrino sample \cite{Stettner:2019tok}) and unbroken out to energies exceeding $10^9$~GeV, then the secondary rate from such a flux would be comparable to or higher than that of the GZK flux up to $\mathcal{O}(10~$PeV). However, if the primary flux is softer (such as the measurement from the high energy starting events sample \cite{Schneider:2019ayi}) or if the energy spectrum cuts off below $10^9$~GeV, then the secondary rate would be lower than that from the GZK flux at energies above $\sim3$~PeV.

\begin{figure}[!ht]
    \centering
    \includegraphics[width=0.9\textwidth,trim={0cm 0cm 0cm 0cm}]{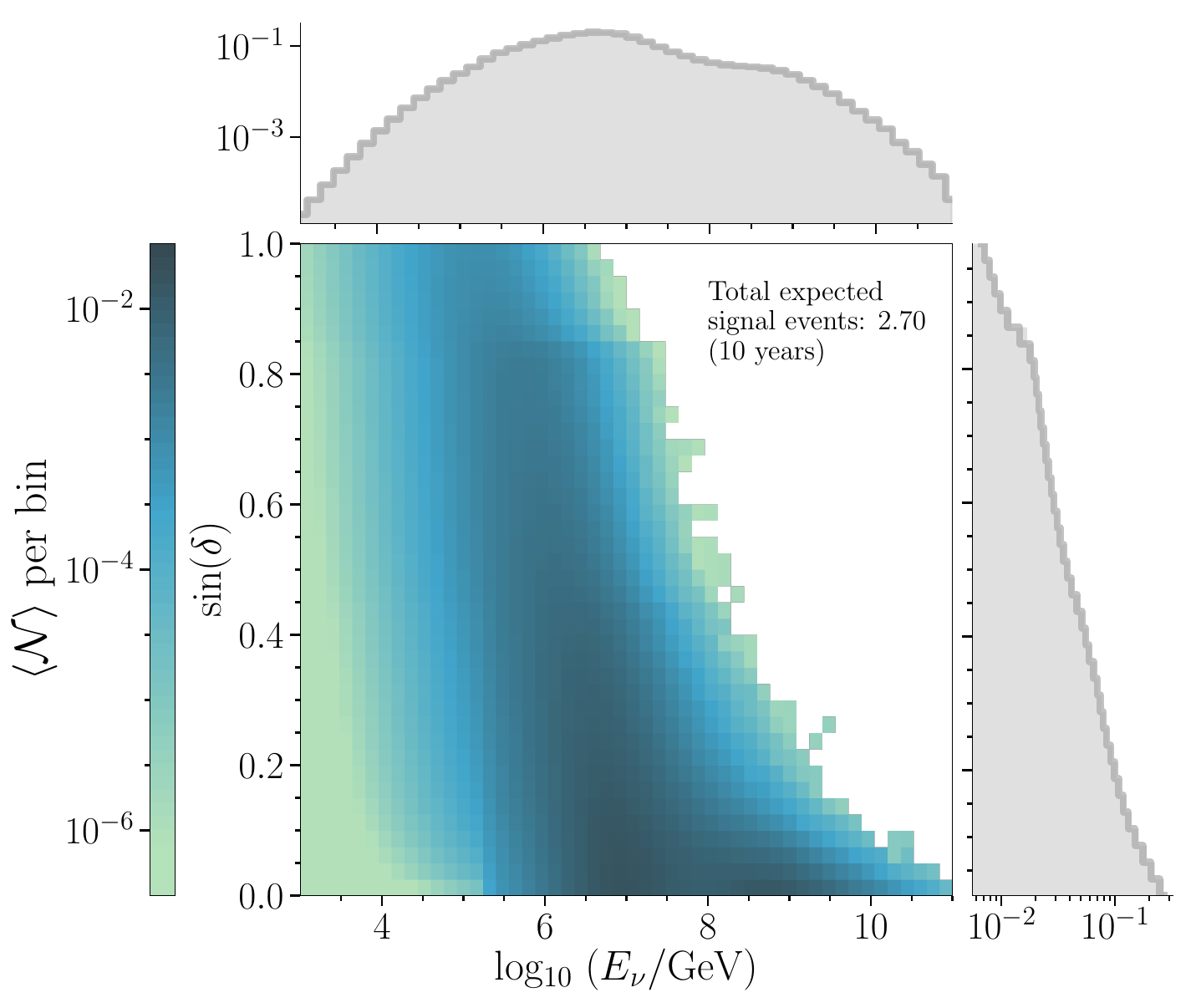}
    \caption[2D signal of GZK neutrinos at IceCube as a function of energy and zenith angle.]{Expected signal of GZK neutrinos at IceCube, assuming the model from \cite{Ahlers:2010fw} and assuming a cosmic-ray composition that is dominated by protons above energies of $10^{17.5}$ eV, after 10 years of data collection. The Earth-skimming contribution represents only about one third of the total expectation, and the majority of events are expected to peak around 10 PeV in true neutrino energy.}
    \label{fig:GZK_signal}
\end{figure}

To further highlight the expected signal shape, we show the resulting expected signal distribution of this benchmark model in Fig.~\ref{fig:GZK_signal}. The number of expected signal events at IceCube is calculated by propagating a $\nu_{\tau}$ flux isotropically over the Northern Hemisphere from incidence on the Earth to a few kilometers away from IceCube. The number of events expected at IceCube is then given by

\begin{equation}
    \mathcal{N}_{\nu}^{\rm{GZK}} ~ = \int dE^{\prime} d\Omega ~ \Phi_{\nu} (E^{\prime}_{\nu}) \Delta T \left[ \sigma_{\nu N}^{CC} (E_{\nu}^{\prime})  \cdot ~ \frac{\Gamma_{\tau \rightarrow \mu}}{\Gamma_{total}} \cdot N^{CC}_{N}(E_{\nu}^{\prime}) + \sigma_{\nu N}^{NC}(E_{\nu}^{\prime}) \cdot N^{NC}_{N} \right], 
\end{equation}
where $ \Phi_{\nu} (E'_{\nu})$ is the emerging flux near the detector, $\sigma_{\nu N}^{CC}$ and $\sigma_{\nu N}^{NC}$ is the neutrino-nucleon isoscalar cross section for CC and NC, respectively, $\Gamma_{\tau \rightarrow \mu} / \Gamma_{total}$ is the tau to muon branching fraction, and $N_{N}$ is the effective number of isoscalar targets. This number is fixed to be $N$ targets in 1 km$^3$ of ice for the NC channel, but has an energy dependence for the CC channel due to the extended muon range, and is given by,

\begin{equation}
    N^{CC}_{N} \left( E_{\nu} \right) = \int d\tilde{E_{\mu}}d\tilde{E}_{\tau}  \frac { d N_{\tau} } { d \tilde{E}_{ \tau} } \left(\tilde{E} _ { \tau } ; E^{\prime}_ { \nu } \right) \frac { d N_{\mu} } { d E_{ \mu} } \left(E_{\mu} ; \tilde{E} _ { \tau } \right) R_{\mu} \left(E_{\mu} \right) A^{geo} \frac{\rho^{ice}}{M_{iso}}.
    \label{eq:ncc}
\end{equation}

\noindent In Eq.~\ref{eq:ncc}, the first and second terms are the tau and muon energy distributions, respectively, $R_{\mu} \left(E_{\mu} \right)$ is the average muon range calculated with MMC,  $A^{geo}$ is the geometrical transverse area (1 km$^2$ in this case), $\rho^{ice}$ is the density of ice, and $M_{iso}$ is the isoscalar nucleon mass. Fig.~\ref{fig:GZK_signal} shows the expected number of events at IceCube binned in true neutrino energy and declination. We find that, assuming a proton-dominated UHECR flux with a minimum crossover energy of 10$^{17.5}$ eV (10$^{18.5}$ eV), IceCube should see 2.70 (1.25) upgoing neutrinos with a hard energy spectrum, peaking at 10 PeV, in ten years of data taking. These events are dominated by the CC channel, with only around 10\% of the signal coming from NC interactions in the fiducial volume of the detector. Of all of the events, we find that only $\sim$0.8 (0.5) would be Earth-skimming, where we have defined Earth-skimming to be up to 5 degrees below the horizon. Therefore, in total, we expect the rate from Earth-traversing neutrinos to be at least twice that from Earth-skimming neutrinos.

\begin{figure}
    \centering
    \subfloat{\includegraphics[width=.46\textwidth]{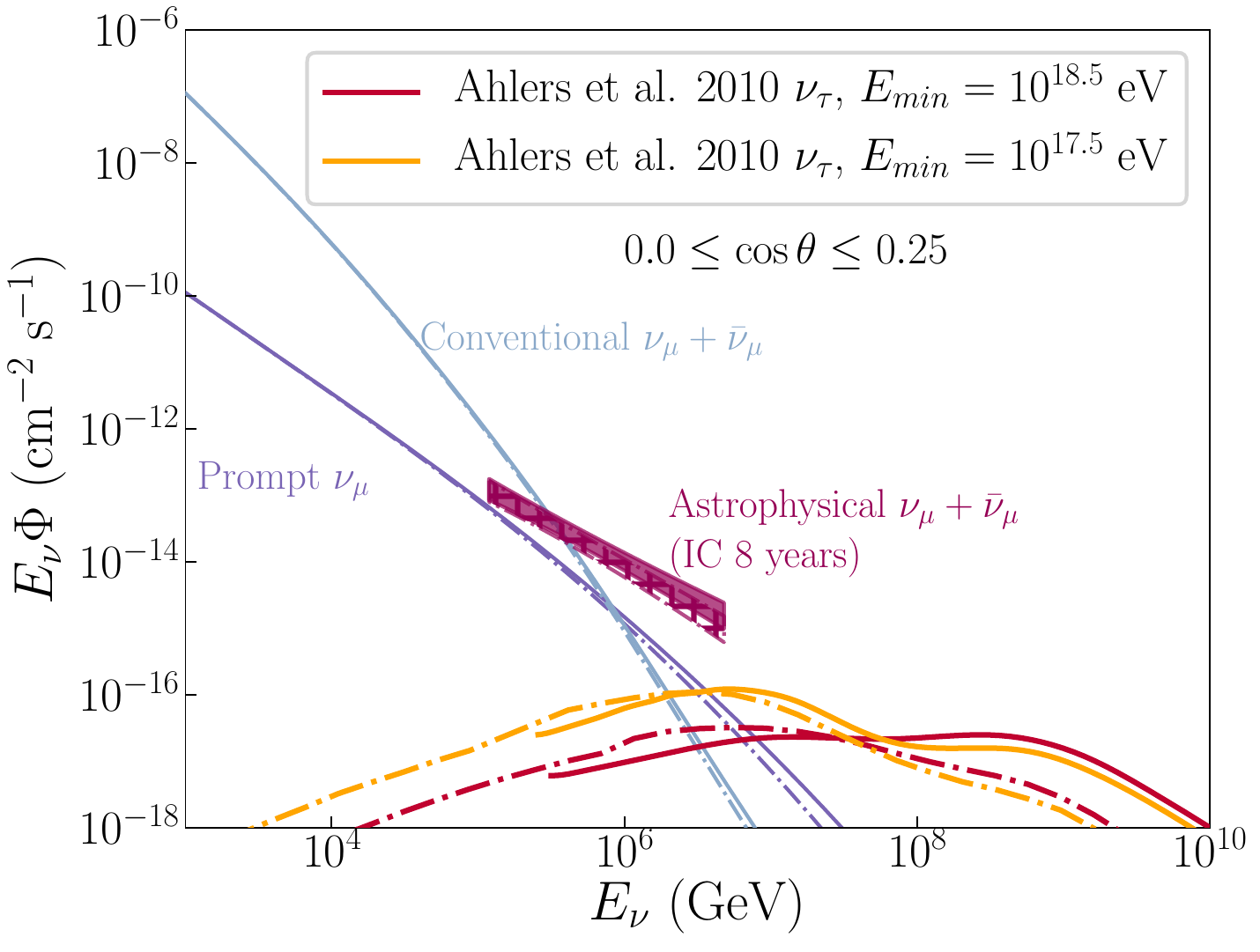}}\quad
    \subfloat{\includegraphics[width=.46\textwidth]{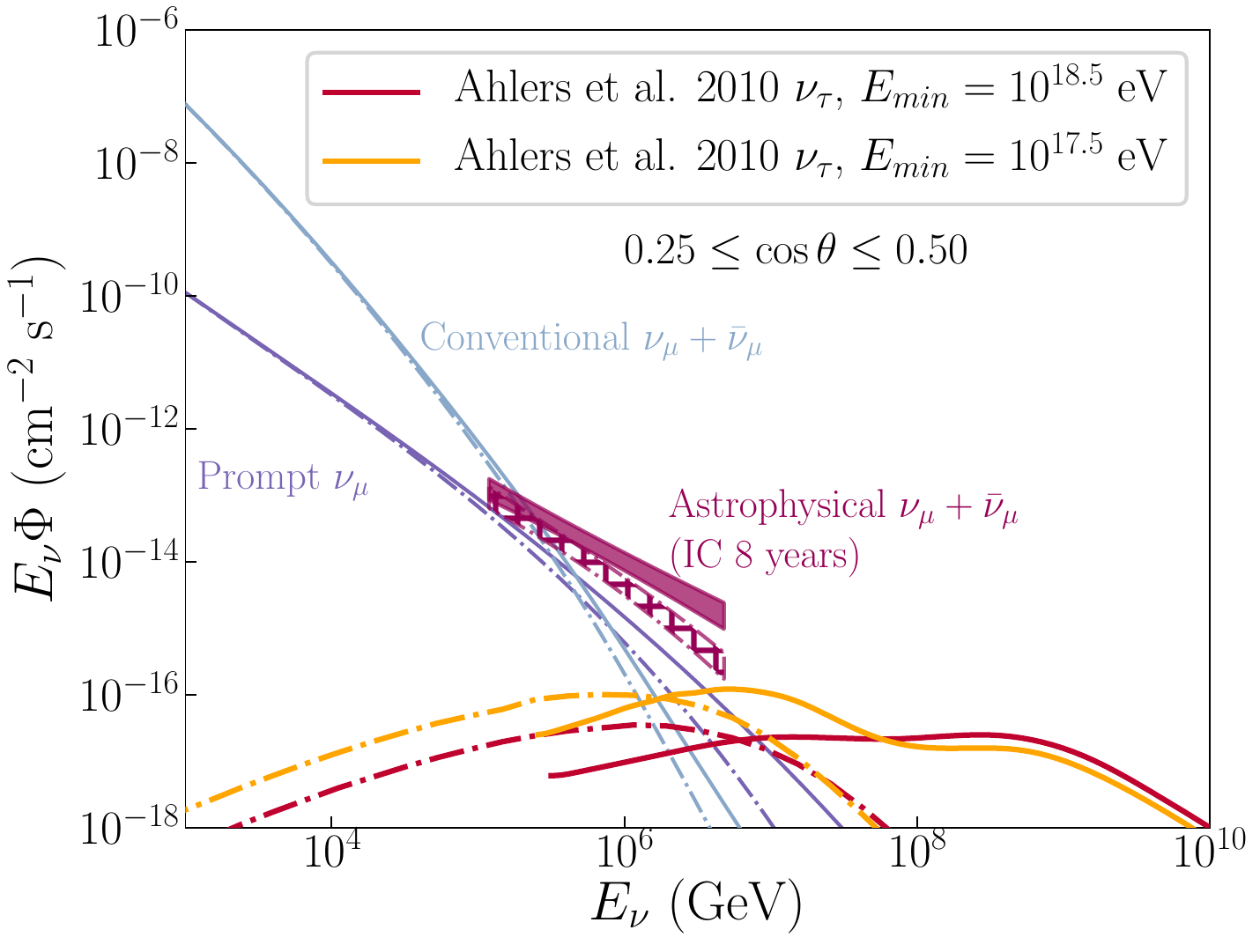}}\\
    \subfloat{\includegraphics[width=.46\textwidth]{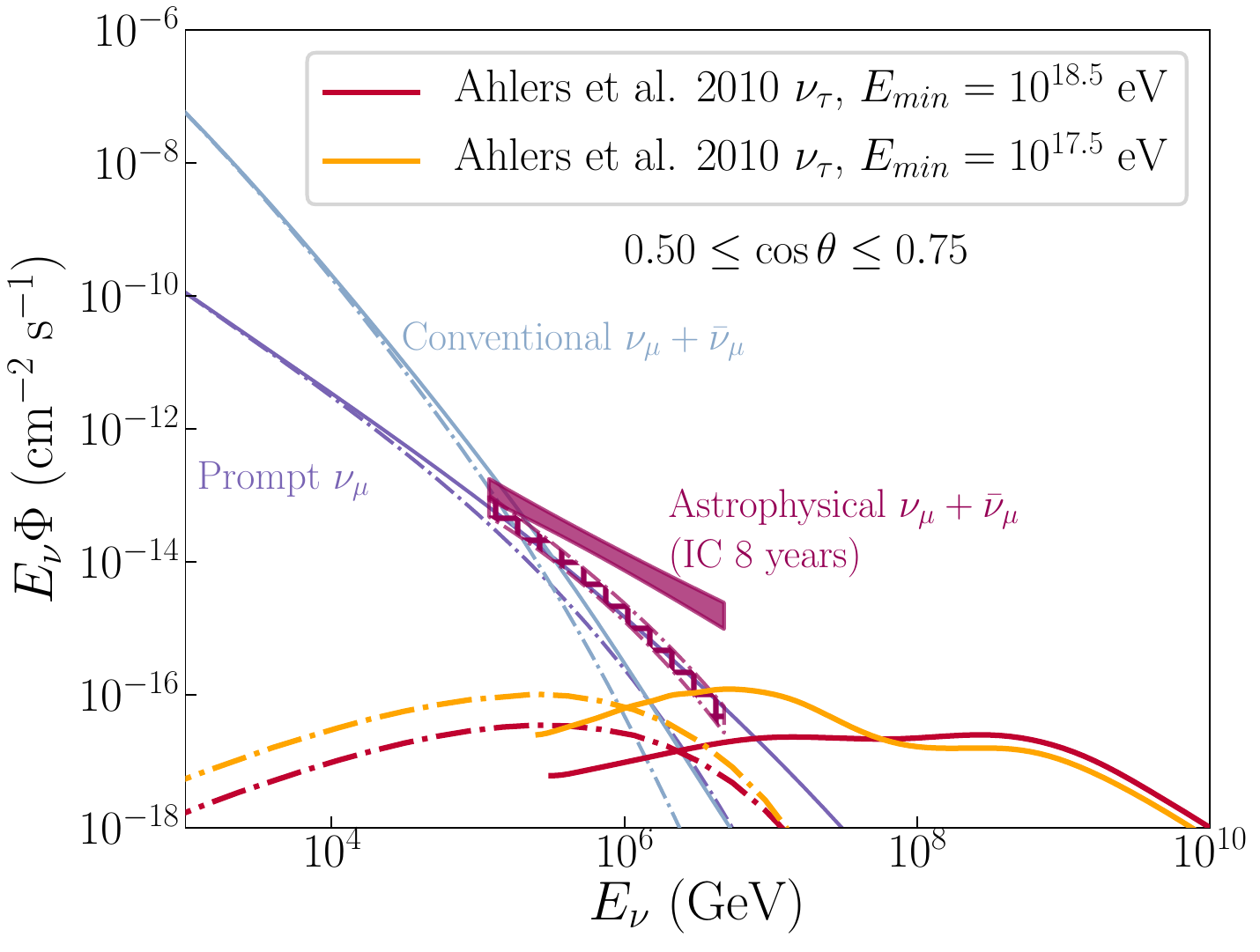}}\quad
    \subfloat{\includegraphics[width=.46\textwidth]{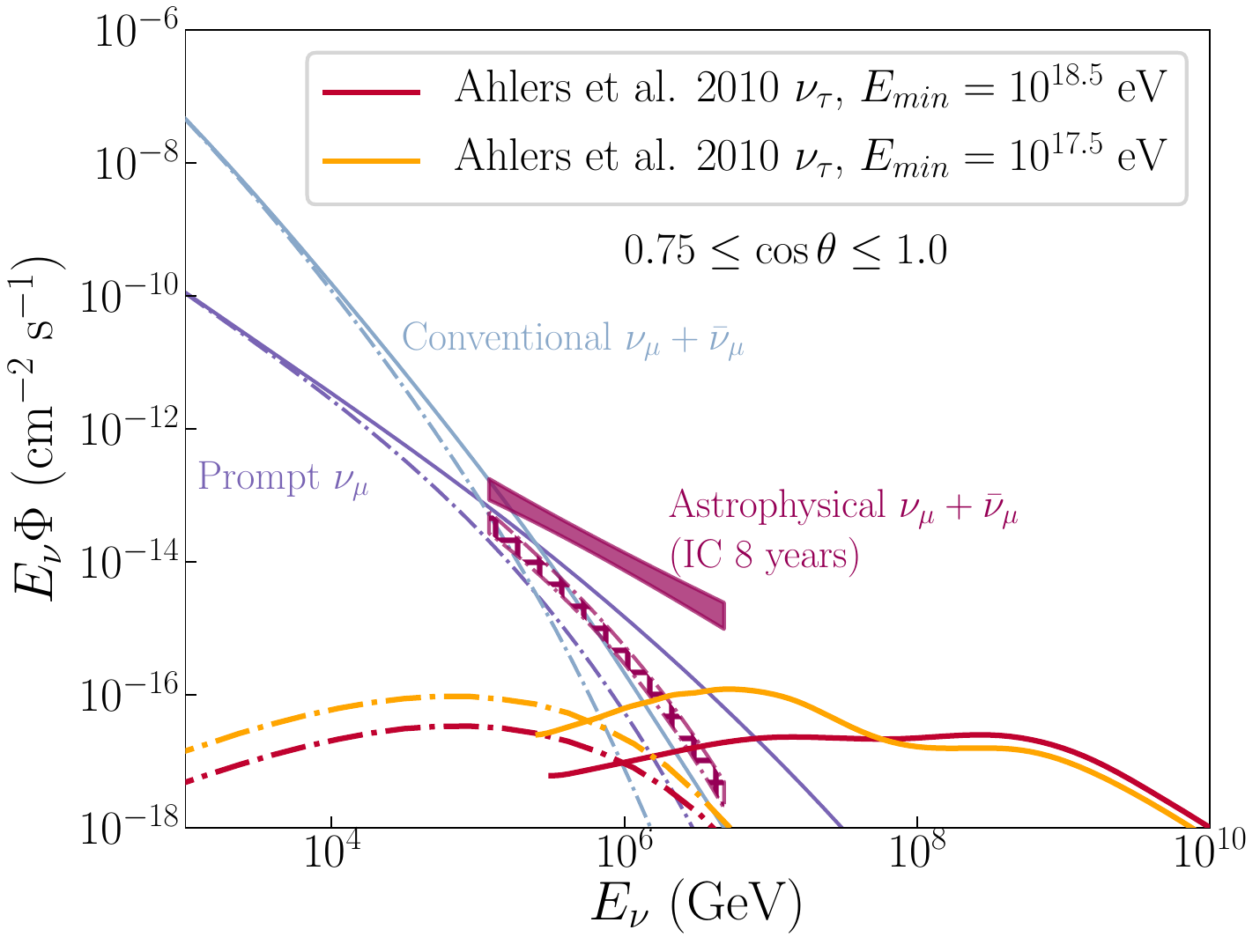}}\\
    \caption[Per flavor neutrino fluxes from 1 TeV to 10 EeV, integrated over various zenith bands in the Northern Sky.]{Per flavor neutrino fluxes from 1 TeV to 10 EeV, integrated over various zenith bands in the Northern Sky. Solid lines are primary fluxes, while secondary fluxes are represented by dashed-dotted lines and hatches. The secondary GZK tau neutrino spectrum is strongly dependent on the incoming zenith angle. For arrival directions towards Earth's core, it contributes equally to the astrophysical flux at IceCube above 2 PeV.}
    \label{fig:gzk_and_friends_bands}
\end{figure}

Fig.~\ref{fig:gzk_and_friends_bands} further demonstrates the declination dependence of this flux through comparison to atmospheric backgrounds, and shows that in certain zenith angle bands with large enough column depth through Earth, the flux arriving at IceCube is higher than the atmospheric background at and above 2 PeV. Given that the event expectation is dominated by the composition of the primary cosmic rays, searching for neutrino secondaries can provide constraints on the composition of primary nuclei, independent of measurements performed by direct cosmic-ray experiments such as the Pierre Auger Observatory and Telescope Array\cite{Aab:2016zth, AbuZayyad:2012ru}.

Next, we illustrate the importance of tau regeneration in searching for GZK neutrinos with a statistical analysis. 
We evaluate the response of IceCube to neutrinos with energy above 100~TeV, where the atmospheric neutrino component is subdominant~\cite{Abbasi:2020jmh}. 
The muon-neutrino effective area is taken from~\cite{Aartsen:2016xlq}. 
We consider two isotropic components: an astrophysical component modeled by a power-law whose parameters have been determined using an independent cascade set of data and a GZK spectrum that we model according to~\cite{Ahlers:2010fw}.
The two components can be differentiated by their energy dependence as well as their angular distribution.
This difference is caused by a contrasting amount of absorption as a function of energy and traversed column depth.

\begin{table}%[!ht]
    \label{tab:model_rejection}
    \makebox[\linewidth]{
    \begin{tabular}{ r c l }
    \hline
    & {\bf Model Rejection} & {\bf Factors} \\
    \hline \hline
        \textbf{Time}  & \textbf{Horizon} & \textbf{Horizon+Upgoing}  \\
         years & $\delta \leq 8^\circ$ &  $\delta \leq 90^\circ$ \\ \hline
        1  & $1.65$ & $0.77$ \\\hline 
        5 & $0.64$ & $0.27$  \\ \hline
        10 & $0.42$ & $0.17$    \\ \hline \hline
        
    \end{tabular}
    }
    % %\internallinenumbers
    \caption[Sensitivity improvement for GZK neutrino searches.]{
    Sensitivity improvement for GZK neutrino searches.
    Each row corresponds to the number of years of data taking for IceCube-Gen2.
    The middle column shows the model rejection factor when measuring neutrinos near the horizon, while the rightmost column lists the corresponding number when measuring both horizon and upgoing events.
    The model rejection factor is defined as the ratio of the baseline GZK flux normalization with respect to the limit obtained in this analysis at 90\%~CL. 
    As a baseline GZK fit we use the flux given by Ahlers \textit{et. al.} 2010 ~\cite{Ahlers:2010fw}.
    }
\end{table}

To quantify the significance of the Earth-traversing contribution, we use a binned Poisson likelihood to compare our prediction to a baseline expectation.
We compute the number of muons produced by the interactions of muon and tau neutrinos in the vicinity of the detector.
We bin the resulting muons linearly using twenty bins in zenith and logarithmically using two bins per decade in reconstructed energy.
We model the detector energy resolution by introducing an 80\% smearing of the initial energy assuming a normal distribution ~\cite{Aartsen:2013vja}.
We compute the 90\% CL. upper limit on the Ahlers-Halzen GZK model normalization, $\phi_0$, using an Asimov data set, i.e., we set the observed number of events to be the expected mean number of events in the absence of a GZK component.
Our results are shown for one, five, and ten years of data taking in Tbl.~\ref{tab:model_rejection}.
The numbers shown are nominally referred to as the model rejection factor (MRF), defined as the ratio of the nominal prediction of the model to the normalization that can be constrained by data at 90\% C.L.
In other words, numbers less than one in Table.~\ref{tab:model_rejection} imply that the baseline model can be detected at greater than 90\% C.L.
When considering only Earth-skimming neutrinos, namely declinations such that $\delta < 10^{\circ} $, we get sensitivities that are comparable to current IceCube constraints~\cite{Aartsen:2013vja}.
However, when we include the Earth-traversing component, the sensitivity improves by approximately a factor of two, with most of the improvement coming from the range between 10$^{\circ}$ to 30$^{\circ}$ below the horizon.

\section{Point Source Transient Limits}
\label{sec:transients}
Astrophysical beam dumps in the vicinity of the sources of UHE cosmic rays provide the opportunity for the production of charged and neutral pions, and therefore, of neutrinos with energies exceeding tens of PeV. Potential sources of UHE neutrinos include binary neutron star (BNS) mergers ~\cite{Kimura:2017kan}, magnetars \cite{Fang:2018hjp,Murase:2009pg, Carpio:2020wzg}, young and fast spinning pulsars~\cite{Fang:2013vla}, supermassive black hole mergers~\cite{Yuan:2020oqg}, gamma-ray burst (GRB) afterglows \cite{Murase:2007yt, Guarini:2021gwh}, and blazars with a luminous dust-torus ~\cite{Murase:2014foa,Petropoulou:2016ujj,Keivani:2018rnh}. The predicted neutrino energy spectrum peaks in the range of 10--100~PeV, energies that are typically below the sensitivity of the current generation of neutrino telescopes.
However, the presence of a flux extending beyond 10~PeV presents an opportunity to probe the sources via the regenerated flux from tau neutrinos.

\begin{figure}[t!]
    \centering
    \includegraphics[width=0.9\textwidth]{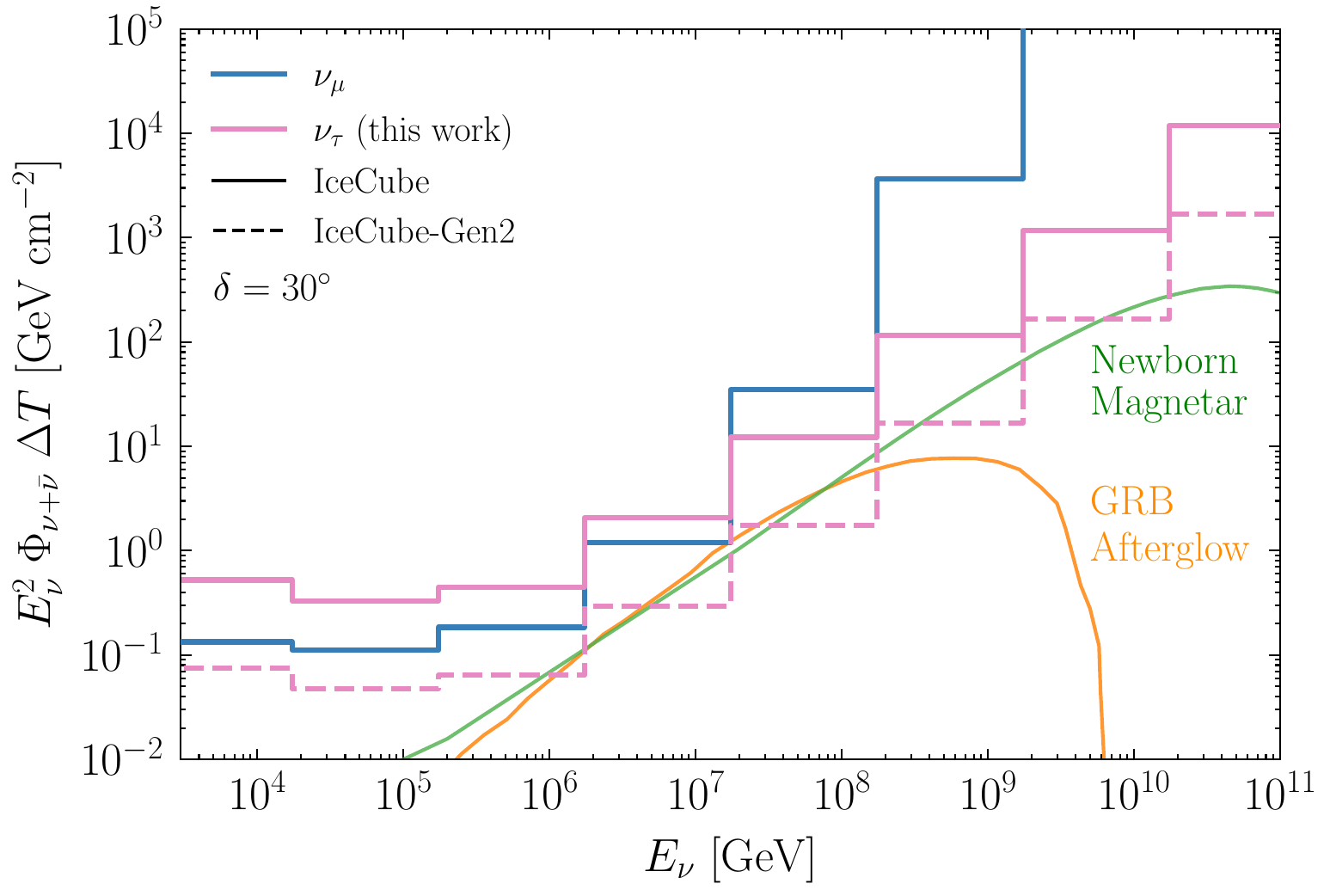}
    \caption[Improved point source differential sensitivity using $\nu_{\tau}$.]{Improved point source sensitivity with $\nu_{\tau}$.
    Differential sensitivity to an $E^{-2}$ time-integrated flux for transient UHE sources at a fixed declination ($\delta = 30^\circ$). The current sensitivity of IceCube is shown in blue, while the sensitivity to muons from tau neutrinos is shown in pastel pink. The projected sensitivity resulting from our method for IceCube-Gen2 (optical-only) is shown as a dashed line.
    To illustrate the power of our method, we compare it to the predicted time-integrated neutrino flux from %binary neutron star mergers \cite{Fang:2017mhl}
    newborn magnetars (green) \cite{Carpio:2020wzg} for a distance of 0.3 Mpc and GRB afterglows (orange) \cite{Murase:2007yt} at 20 Mpc. 
    }
    \label{fig:transients}
\end{figure}

In order to demonstrate the power of the regeneration-based technique for identifying sources of UHE neutrinos, we compute the differential sensitivity of IceCube for a neutrino flux in the range of 1 PeV to 1 ZeV.
Neutrinos with this energy traversing Earth will generate a cascade of muon and tau neutrino fluxes at the detector.
In contrast to the primary all-flavor flux that is attenuated, the regenerated flux results in a substantial intensity of neutrinos of PeV energy, where the sensitivity is optimal for detection by optical Cherenkov detectors.
Here, we utilize the IceCube differential sensitivity for transients computed in ~\cite{IceCube:2020mzw}.
For each energy decade, we inject a flux of tau neutrinos, assuming the spectrum follows a $E^{-2}$ distribution.
To obtain the normalization for each bin, we determine the corresponding flux of secondary neutrinos that reaches the detector.
The sensitivity of IceCube to a flux of muons arriving at the horizon constrains the flux that arrives at the detector in any direction, when excluding Earth effects.
Thus, we use the sensitivity at $\delta = 0$ to constrain the initial flux at higher energies; that is, we constrain the normalization of the injected spectrum at different declinations by folding the sensitivity at the horizon with the expected number of muons from tau decay produced by the propagated secondary neutrinos.
In Fig.~\ref{fig:transients}, we show the corresponding differential sensitivity of IceCube to the time-integrated flux in the energy range of $10^4$~GeV -- $10^{11}$~GeV.
For instance, for a declination of 30$^{\circ}$, the $\nu_\tau$ sensitivity to a transient outperforms the conventional time-dependent search with $\nu_\mu$ for energies greater than $10^7$~GeV.
The improvement grows with energy, exceeding an order of magnitude at $10^9$~GeV.
We also show the corresponding sensitivity for IceCube-Gen2 demonstrating the enhanced reach of time-dependent searches with water-Cherenkov telescopes to sources of UHE neutrinos.
The significance of this improvement is underscored by the power of this methodology to probe neutrinos from GRB afterglows and newborn magnetars, which is out of reach for conventional time-dependent follow-up.
We note that the relative improvement in the sensitivity increases with declination, and reaches more than three orders of magnitude at $10^{9}$ GeV and 60 degrees below the horizon. The method creates new opportunities for the discovery of cosmic accelerators that produce the most energetic particles and were previously inaccessible to Cherenkov detectors.

\section{ANITA Followup deux fa\c cons}
\label{sec:ANITA}
The Antarctic Impulsive Transient Antenna (ANITA) project is a balloon experiment, designed with the primary purpose of detecting the UHE cosmogenic neutrino flux \cite{Gorham:2008dv, Hoover:2010qt, Allison:2018cxu}.
Although this is the project's primary scientific goal, the experiment is sensitive to a wide array of impulsive radio signals, and ANITA's first three flights have resulted in a few interesting detections. 
For a summary of the event properties, see Tab.~\ref{tab:candidates}.
Two of these events had signatures consistent with upgoing air showers produced by a tau \cite{Gorham:2016zah,Gorham:2018ydl}. 
This interpretation requires the decay of a tau (from a tau neutrino CC interaction) to occur in the atmosphere, producing an extensive air shower (EAS). 
This is distinguishable from a reflected EAS initiated by a cosmic ray, in which the radio signal acquires a phase reversal from reflection off of the Antarctic ice, while an upgoing EAS does not display such a phase reversal.
However, this interpretation is problematic as tau neutrinos with energies to which ANITA is sensitive are not likely to travel through the large Earth column depths required for these events.
While it has been noted that these events are unlikely to be caused by an isotropic neutrino flux \cite{Romero-Wolf:2018zxt, Fox:2018syq, Chipman:2019vjm}, discrete-source emission could evade these constraints. 
Beyond the Standard Model (BSM) explanations have also been proposed. 
This includes axion-photon conversion \cite{Esteban:2019hcm}, sterile neutrinos \cite{Chipman:2019vjm,Huang:2018als, Anchordoqui:2018ucj, Cherry:2018rxj}, and heavy SUSY partners or dark matter particle decays \cite{Connolly:2018ewv, Collins:2018jpg, Anchordoqui:2018ssd, Heurtier:2019git, Hooper:2019ytr, Cline:2019snp, Heurtier:2019rkz, Borah:2019ciw, Esmaili:2019pcy}.
Here, we examine the discrete-source emission hypothesis and show that any detection of EeV neutrinos from steep incident angles at ANITA implies a large and detectable flux of TeV-PeV neutrinos with IceCube. 

\begin{table}
\centering
	\caption[Summary of the neutrino candidate event properties from the first three flights of ANITA.]{Properties of the neutrino candidate events from the first three flights of ANITA, from \cite{Allison:2018cxu, Gorham:2016zah, Gorham:2018ydl}. The two anomalous ANITA events (AAE) are those consistent with a steeply upgoing $\nu_{\tau}$ interpretation. Sky coordinates are projections from event arrival angles at ANITA. Localization uncertainty is expressed as major and minor axis standard deviations, position angle. This angle describes the rotation of the major axis relative to the north celestial pole turning positive into right ascension. 
 } \label{tab:candidates}
	\begin{tabular}{ l |c |c |c}
	    \hline
	    \hline
	    & \textbf{AAE-061228} & \textbf{AAE-141220} & \textbf{AAC-150108} \\ \hline \hline
		Event, Flight & 3985267, ANITA-I & 15717147, ANITA-III &  83139414, ANITA-III  \\
		Detection Channel & Geomagnetic & Geomagnetic & Askaryan \\
		Date, Time (UTC) & 2006-12-28, 00:33:20 & 2014-12-20, 08:33:22.5 & 2015-01-08, 19:04:24.237 \\
		RA, Dec (J2000) & 282$^\circ$.14, +20$^\circ$.33 & 50$^\circ$.78, +38$^\circ$.65 & 171$^\circ$.45, +16$^\circ$.30\\
		Localization  & 1$^\circ$.5 $\times$ 1$^\circ$.5, 0$^\circ$.0 & 1$^\circ$.5 $\times$ 1$^\circ$.5, 0$^\circ$.0 & 5$^\circ$.0 $\times$ 1$^\circ$.0, +73$^\circ$.7 \\
		Reco. Energy (EeV) & 0.6 $\pm$ 0.4 & $0.56^{+0.30}_{-0.20}$ & $\geq$ 10\\
		Chord Length (km) & 5740 $\pm$ 60 & 7210 $\pm$ 55 & - \\
        \hline
	\end{tabular}
	\vspace{0.05in}
\end{table}

\begin{figure}[ht!]
    \centering
    \includegraphics[width=0.55\linewidth]{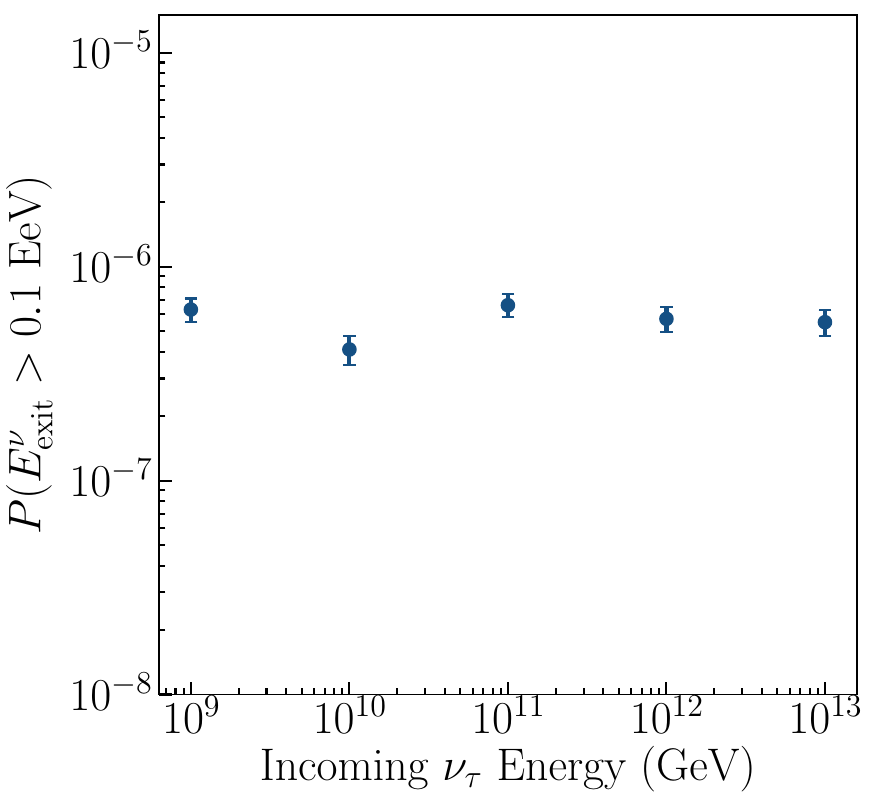}
    \caption[Tau neutrino exit probability at or above ANITA detection threshold.]{Probability for a tau neutrino to exit the Earth with a minimum energy of 0.1 EeV (approximate ANITA threshold), after Earth propagation (based on inferred chord length for AAE141220), assuming $\nu_{\tau}$ incidence at a particular initial energy. Errors are statistical only.}
    \label{fig:tau_prob}
\end{figure}

The number of events detected by ANITA due to tau showers in the atmosphere from a primary neutrino flux, $\Phi \left( E _ { \nu } \right)$, is given by

\begin{equation} \label{eq:anita_exp}
    \mathcal{N} _ { \nu } = \int d E_{ \nu } d E^{\prime} _ { \nu } ~ \Phi \left( E _ { \nu } \right) \frac { d N_{\nu} } { d E' _ { \nu } } \left(E^{\prime} _ { \nu } ; E_ { \nu } \right) \xi _ { a c c } \left( E' _ { \nu } \right) \Delta T \; ,
\end{equation}
where $E_{ \nu }$ and $E^{\prime} _ { \nu }$ are the primary and secondary neutrino energies, respectively, $ d N_{\nu} \left( E' _ { \nu };  E _ { \nu }  \right) /  d E' _ { \nu } $ is the energy distribution of secondary tau neutrinos near the ice surface, $\Delta T$ is the duration of observation, and $\xi _ { a c c } \left( E ' _ { \nu } \right)$ is the ANITA acceptance \cite{Romero-Wolf:2018zxt} in units of cm$^2$sr.
The acceptance incorporates the probability of neutrinos interacting in the ice as well as the probability of a tau decay shower occurring in the atmosphere.
Given that the reported acceptance in  \cite{Romero-Wolf:2018zxt} includes neutrino propagation through Earth, we set the acceptance at all angles to be that near the horizon to remove the Earth absorption effects, which we account for separately with \taurunner{}.
For the incoming flux, we take the minimalistic assumption of a delta function in energy, $\Phi \left( E _ { \nu } \right) = \frac{dN}{dA dt dE_{\nu}} = \Phi _ { 0 } \delta \left( E _ { \nu } - E _ { 0 } \right)$, where $\Phi_{0}$ is the normalization with units $\rm{cm}^{-2} \rm{s}^{-1}$.
Probabilities of tau neutrinos exiting Earth with energies greater than 0.1 EeV are shown in Fig.~\ref{fig:tau_prob} for the chord lengths corresponding to AAE141220. 
For both taus and tau neutrinos, the probability of exiting Earth with an energy larger than 0.1 EeV seems to be fairly independent of energy, for initial tau neutrino energies above 1 EeV.
Therefore, in what follows, we choose $E_0 = 1$ EeV as the primary energy.

\begin{figure}
    \centering
    \includegraphics[width=0.7\linewidth]{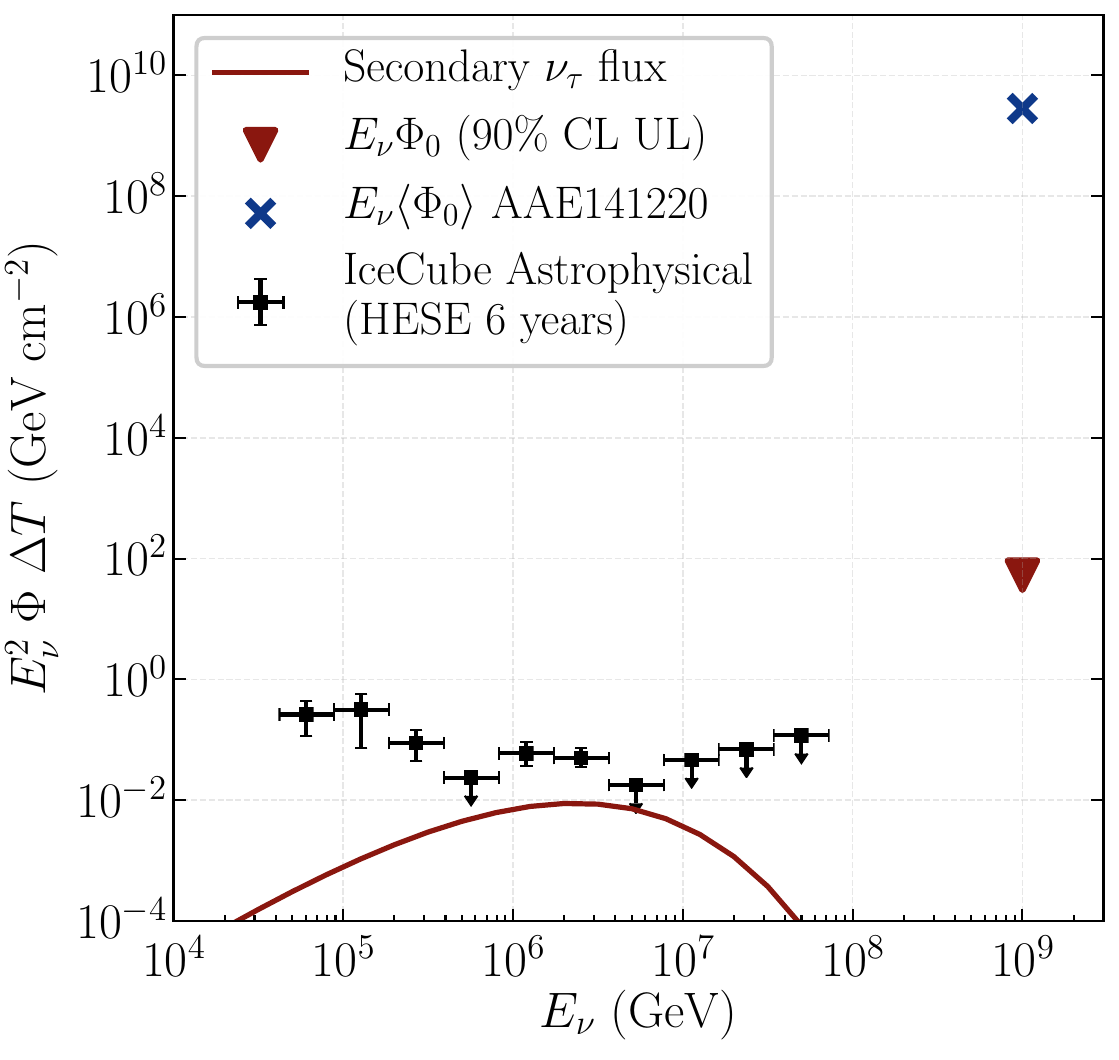}
    \caption[ANITA constraints using HESE.]{Maximum allowed flux of EeV neutrinos (maroon arrow), given an injected mono-energetic neutrino flux at or above the detected ANITA event (AAE141220) energy. The normalization of the secondary flux is set to the maximum that does not exceed IceCube's diffuse astrophysical flux (black bins). The flux needed to produce one event in the third flight of ANITA (blue marker) exceeds the upper limit by many orders of magnitude. We use the published spectrum based on six years of high-energy starting events.}
    \label{fig:ANITA_UL}
\end{figure}

As was discussed above, this primary flux of EeV neutrinos is guaranteed to be associated with a secondary flux of TeV to PeV neutrinos. 
Such a large rate of TeV muons simultaneously crossing the IceCube volume would deposit a large amount of charge.
Large-charge events are promptly reported by IceCube via the EHE and HESE streams\footnote{\url{https://gcn.gsfc.nasa.gov/amon.html}}.
For example, the EHE stream requires three thousand photoelectrons and thirty channels to trigger an alert.
A $\sim$1 PeV muon typically deposits $\sim$200 TeV, which on average results in over four thousand photoelectrons in more than 40 channels, when crossing the full detector \cite{jvs}.
In fact, the largest energy deposit reported in these streams\footnote{\url{https://gcn.gsfc.nasa.gov/gcn3/24028.gcn3}} corresponds to 5 PeV, but it's for a downgoing event; horizontal and upgoing events have not had multi-PeV announcements in these streams.
Thus, we conclude that IceCube has not observed catastrophic events that would be produced by bundles of TeV-neutrino-induced muons.
In what follows, we then take the conservative assumption that a single muon makes it through.
Such events have been observed, and we can compare this expected yield to IceCube's measurement of the high-energy events.
We find the maximum allowed normalization of the incident flux by comparing the secondary neutrino distribution with the measured IceCube astrophysical flux from the high-energy starting event selection (HESE) \cite{Kopper:2017zzm}.
Results for AAE141220 are shown in Fig.~\ref{fig:ANITA_UL}.
The unfolded HESE spectrum is propagated to the detector using \taurunner{}.

The 90\% C.L. upper limit on the EeV primary flux normalization is set by comparing both secondary distributions and requiring that the secondaries produced by the primary EeV flux do not exceed those of HESE at 90\% C.L. 
Given that the time profile of the intrinsic flux is unknown, we place limits on the time-integrated flux.
We take the duration to be 22 days ($\Delta T$ in Eq.~\eqref{eq:anita_exp}), corresponding to the entire ANITA-III flight.
We find the maximum allowed time-integrated flux to be $E^2 \Phi \Delta T \simeq 10^2$ GeV cm$^{-2}$.
Using the maximum allowed time-integrated flux, we calculate the expected number of events from ANITA. 
This yields a maximum expected number of neutrinos of fewer than $ \mathcal{O}\left(10^{-7}\right)$ in 22 days. 
This is illustrated in Fig.~\ref{fig:ANITA_UL} where we show the flux required to produce one event from ANITA as a reference.
It is therefore extremely unlikely for the reported event to be caused by a high-energy tau neutrino. 

It is worth noting, however, that short-timescale transients are allowed to overproduce the measured astrophysical flux as long as they do not overproduce the astrophysical flux integrated over the duration of the measurement.
Therefore in what follows we present a dedicated transient source search using seven years of IceCube data. 
For this analysis, we focus on data collected from 2011 to 2018, containing approximately 900,000 events from 2532 days of operation.

We adopt the procedure described in \cite{Schumacher:2019qdx} to search for counterparts to ANITA events.
Namely, we perform an analysis to test for temporal structure in IceCube data in the location of the ANITA events.
This analysis incorporates the information from the localization of the ANITA events through a joint likelihood.
The sky is divided into grid positions, $\mathbf{x}_s$, and at each point we maximize the likelihood, $\mathcal{L}$, with respect to the expected number of signal events, $n_s$, and other signal parameters contained in the variable $\mathbf{\alpha}$ depending on the different signal hypotheses tested. This likelihood is given by

\begin{eqnarray} \label{eq:general_likelihood}
    \mathcal{L} = \lambda \prod_{i=1}^{N} \Bigg( \frac{n_s}{n_s + n_b}S(\mathbf{x}_i, \mathbf{x}_s, \mathbf{\alpha})  + \frac{n_b}{n_s + n_b}B(\mathbf{x}_i, \mathbf{x}_s) \Bigg) P_A(\mathbf{x}_s) ,
\end{eqnarray}
where $n_b$ is the expected number of observed background events and $N$ is the total number of observed events in the time window.
The vector $\mathbf{x}_i$ contains the event observables such as its reconstructed energy, direction, and reconstruction uncertainty.
$P_A$ is the spatial probability distribution function (PDF) of ANITA events, which are included in Table~\ref{tab:candidates}.
$B$ describes the energy and declination PDF of our background, which is parameterized from data and is the same among all analyses.
Temporal terms in $B$ vary according to the assumed emission time profile.
While the signal PDF $S$ describes the signal hypothesis, the parameter $\lambda$ modifies the likelihood formalism in order to take into account low-statistics problems in some of the analyses. 
In general, the signal PDF, $S$, is defined as
\begin{equation}
    S = S^{space}(\mathbf{x}_i, \mathbf{x}_s, \mathbf{\sigma_i})\cdot S^{energy}(E_i,\delta_i,\gamma)\cdot S^{time} \; .
\end{equation}
These three terms reflect the spatial, energy, and time PDFs, respectively, of our signal hypothesis.
The spatial term, $S^{space}$, expresses the probability for an event
with best-fit reconstructed direction $\mathbf{x}_i$ to originate from a source at the direction $\mathbf{x}_s$, according to a two-dimensional Gaussian function with angular resolution $\sigma_i$. 
The energy PDF, $S^{energy}$, describes the probability of obtaining an event with reconstructed energy $E_i$ given a declination $\delta_i$ under the hypothesis of an $E^{-\gamma}$ power-law energy spectrum, which helps differentiate signal from the known atmospheric backgrounds in our event selection.
The time term, $S^{time}$, describes the time PDF of events observed from the source.

We search for IceCube events in spatial coincidence with the ANITA events in short time windows, $\Delta t$, centered on each ANITA event.
We call this period the \textit{on-time window}.
This is equivalent to setting $S^{time}$ equal to a uniform PDF in this on-time window and to zero for all times outside this window.
To help distinguish potential signals for time windows in which the expected number of background events is small, we set \begin{equation}
    \lambda = \frac{(n_s + n_b)^N}{N!}\cdot {\rm e} ^{-(n_s + n_b)}
\end{equation} 
as in \cite{Aartsen:2017zvw, Aartsen:2014aqy}.
Due to the small statistics for short time windows, the likelihood is only maximized with respect to $n_s$, and the energy dependence in $S^{energy}$ is fixed to an $E^{-2}$ spectrum.
To account for the temperature dependence of atmospheric muon rates \cite{Aartsen:2013jla}, we determine $n_b$ by calculating the rate of events from the surrounding five days of data on either side of our on-time window.
Taking the logarithmic likelihood ratio between the maximum likelihood and that of the null hypothesis results in our test statistic (TS), defined as
\begin{equation} \label{eq:transient_TS}
    \text{TS} = - 2\hat{n_s} +  \sum_{i=1}^{N}2\log \left[ 1 + \frac{\hat{n_s} S(\mathbf{x}_i, \mathbf{x}_s)}{n_b B(\mathbf{x}_i)} \right] + 2 \log \left[ \frac{P_A (\mathbf{x}_s)}{P_A (\mathbf{x}_0)} \right],
\end{equation}
where $\mathbf{x}_0$ is the reported best-fit location of the ANITA event \textcolor{black}{and $\hat{n}_s$ is the value of $n_s$ that maximizes the likelihood}.
TS is calculated for all $\mathbf{x}_s$, and the maximum value is reported.
For this analysis, $P_A$ is a two-dimensional Gaussian assuming the localization uncertainties reported in Table~\ref{tab:candidates}.
As we are not motivated by a specific astrophysical class of objects with characteristic timescales of emission, we consider constant emission over various time windows for each of the ANITA events. 
This technique is similar to previous IceCube searches for gamma-ray bursts and fast radio bursts \cite{Aartsen:2017zvw, Aartsen:2014aqy}.
AAE-061228 is excluded from this analysis because it occurred before IceCube had attained a full detector configuration.
For AAC-150109 we consider three separate time windows: 10 s, $10^3$ s, and $10^5$ s.
During the event time of AAE-141220, IceCube was temporarily not collecting data, due to a run transition that had begun approximately 0.5 seconds before the event and lasting for about one minute.
Because of this, we only investigate hypotheses of constant emission over two time windows ($10^3$ s and $10^5$ s), where the period of time from the run transition is not an appreciable portion of our on-time window.

\begin{table*}%[htb!]
    \caption[Summary of ANITA followup analysis with IceCube.]{Analysis results and upper limits.  Upper limits (90\% C.L) are on the time-integrated $\nu_{\mu} + \bar{\nu}_{\mu}$ power law flux ($E^{-2}$ ) from a point source following the spatial probability distribution provided by ANITA. Limits are set assuming constant emission over a fixed time window.}
    \centering
    \begin{tabular}{l | c| c | c} \hline
     Event & Time Window  & $p$-value & Upper limit (GeV $\cdot$ cm$^{-2}$)  \\ \hline \hline
     \multirow{3}{*}{AAE-141220} & 10 s & - & - \\ 
	 & $10^3$ s & 1.0 & 0.053 \\ 
	 & $10^5$ s & 1.0 & 0.051 \\ \cline{1-4}
	 \multirow{3}{*}{AAC-150108} & 10 s & 1.0 & 0.040 \\ 
	 & $10^3$ s & 1.0 & 0.041 \\ 
	 & $10^5$ s & 1.0 & 0.032 \\ \cline{1-4}
    \hline
    \end{tabular}
    \label{tab:ANITAresults}
\end{table*}

The limits we set on muon neutrinos in the TeV--PeV energy range can constrain generic fluxes of incident tau neutrinos with EeV energies.
As discussd earlier, any incident flux with an EeV $\nu_{\tau}$ component that traverses large Earth chord lengths will result in a secondary flux of lower energy neutrinos, to which IceCube would be sensitive.
We use the same prescription here to constrain a generic point source flux that includes EeV neutrinos.

\begin{figure}[ht!]
    \centering
    \includegraphics[width=0.55\textwidth]{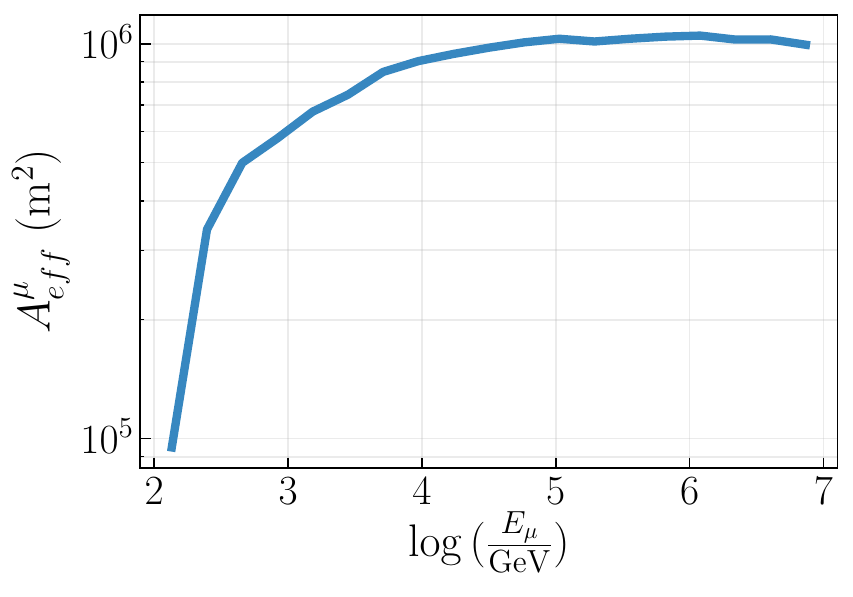}
    \caption[Effective area of the IceCube event selection to muons from the northern sky.]{Effective area of the IceCube event selection to muons from the northern sky, incident on a volume 1.5 km away from the edge of the detector. $E_{\mu}$ is the muon energy incident on this volume.}
    \label{fig:effective_area}
\end{figure}

For any incident flux of neutrinos from the northern sky, $\Phi(E_{\nu}, t)$, the number of expected detected tau-neutrino--induced muon events at IceCube is given by 
\begin{multline}
\langle N _ { \text {IceCube} }^{\mu}\rangle = \int dE _ { \mu } \int dE _ { \tau } \int dE _ { \nu } \Phi \left( E _ { \nu }, t \right) P _ { \tau } ^ { s u r v } \left( E _ { \nu } \right) \frac { dN _ { \tau } \left( E _ { \tau } \right) }{ dE _ { \tau } }  \frac{\Gamma_{\tau\rightarrow\mu}}{\Gamma_{\rm{total}}} \frac { dN _ { \mu } } { dE _ { \mu } } \left( E _ { \tau } , E _ { \mu } \right) A _ { e f f } ^ { \mu } \left( E _ { \mu } \right) \Delta T \\ 
+ \int dE _ { \mu } \int dE _ { \tau } \int dE _ { \nu } ^ { \prime } \int dE _ { \nu } \Phi \left( E _ { \nu }, t \right) P _ { \nu } \left( E _ { \nu } , E _ { \nu } ^ { \prime } \right) \frac { dN _ { \nu } } { dE _ { \nu } ^ { \prime } } \left( E _ { \nu } ^ { \prime } \right)  N^{p} \left( E^{\prime}_{\nu} \right) \\
\times \frac{dN_{\tau}}{dE_{\tau}} \left(E^{\prime}_{\nu};E_{\tau}\right) \frac{\Gamma_{\tau\rightarrow\mu}}{\Gamma_{\rm{total}}} \frac{dN_{\mu}}{dE_{\mu}}\left(E_{\tau};E_{\mu}\right) A_{eff}^{\mu}\left(E_{\mu}\right)\Delta T,
\end{multline}

where the first contribution is from emerging tau-leptons that would decay to muons and then pass an IceCube event selection.
The second contribution is from the remaining $\nu_{\tau}$ flux, the majority of which has cascaded down in energy.
$N^{p} (E_{\nu})$ is the number of targets effectively seen by an incident neutrino with energy $E_{\nu}$.
The effective area of this event selection to muons incident on the detector is displayed in Figure~\ref{fig:effective_area}.
$P _ { \tau } ^ { s u r v } \left( E _ { \nu } \right)$ and $P _ { \nu }(E _ { \nu })$ represent the survival probability of a $\tau$-lepton and $\nu_{\tau}$, given an incident neutrino energy, respectively.
$\Gamma_{\tau\rightarrow\mu}\big/\Gamma_{\rm{total}}$ represents the branching ratio for the tau-decay to muon channel, which is approximately 18\%. 

For ANITA, the number of expected events from upgoing taus can be computed using Eq.~\ref{eq:anita_exp}.
We focus on the non-observation of coincident events in IceCube at $\Delta T = 10^3$~s. A similar procedure can be applied to longer time windows.
Qualitatively, it would result in similar limits up to the lifetime of the ANITA flight.
For longer emission timescales, limits from IceCube become even more constraining as the implied normalization on the ANITA flux would have to increase to compensate for the fraction of time during which ANITA was not taking data.

\begin{figure}
    \centering
    \includegraphics[width=0.55\textwidth]{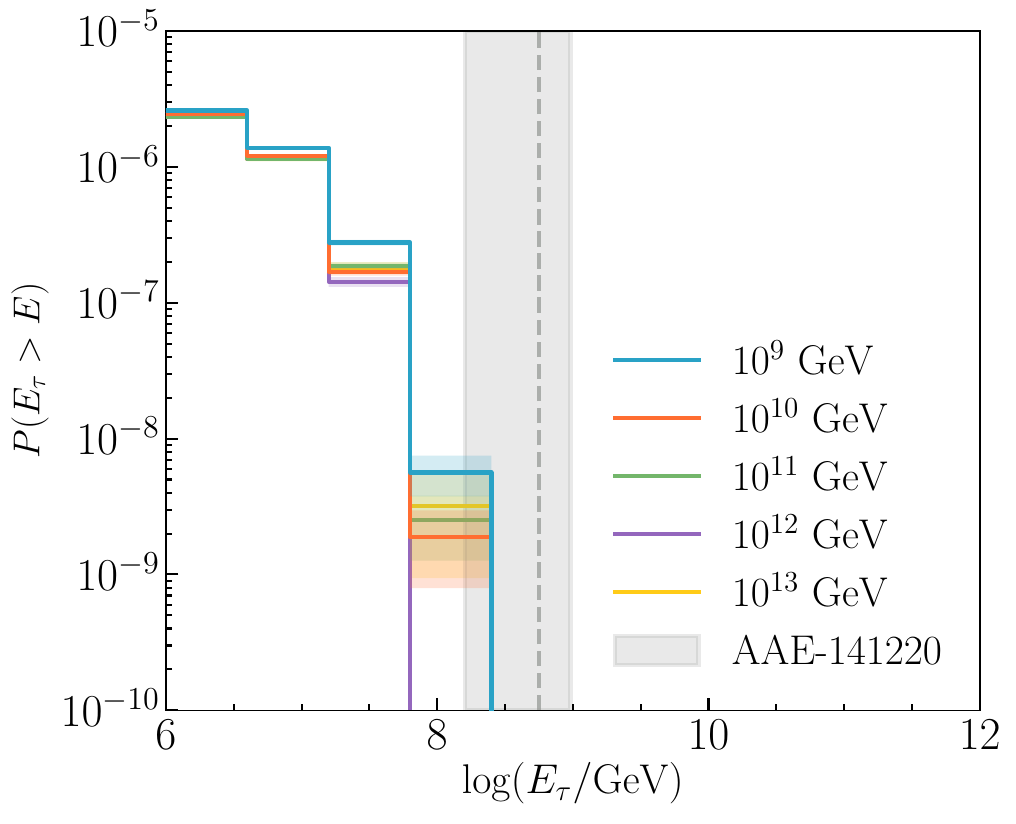}
    \caption[Normalized cumulative distributions for Earth-emerging tau-leptons.]{Normalized cumulative distributions for Earth-emerging tau-leptons. Colors correspond to the incoming tau-neutrino energy, and the gray band is the 95\% containment on the error of the reconstructed shower energy of AAE-141220.}
    \label{fig:tau_exit}
\end{figure}

We inject fluxes described by delta functions in energy,
$\Phi(E_{\nu}, t)= \Phi_0 \delta(E_{\nu} - E_0)$, where now the normalization carries units of $\rm{cm}^{-2}\rm{s}^{-1}$. As explained previously, the secondary tau neutrinos at or above an EeV are degenerate and therefore these limits apply to any assumed initial flux shape. 
After propagating these mono-energetic fluxes, we record what fraction of the incident flux results in a detectable signal at ANITA. 
We repeat this procedure for a variety of injected initial neutrino energies so that we can find the energy that yields the maximum probability of a $\tau$-lepton arriving at ANITA with an energy within the quoted reconstructed energy bounds.
We find that the optimal flux for ANITA corresponds to an injected $\nu_{\tau}$ flux with $E_0 = 1$ EeV.
Normalized cumulative distributions from secondary $\tau$-leptons are shown in Figure~\ref{fig:tau_exit} for injected neutrinos at angles corresponding to the best-fit reconstructed direction of AAE-141220.

\begin{figure*}
\centering
  \includegraphics[width=0.75\textwidth]{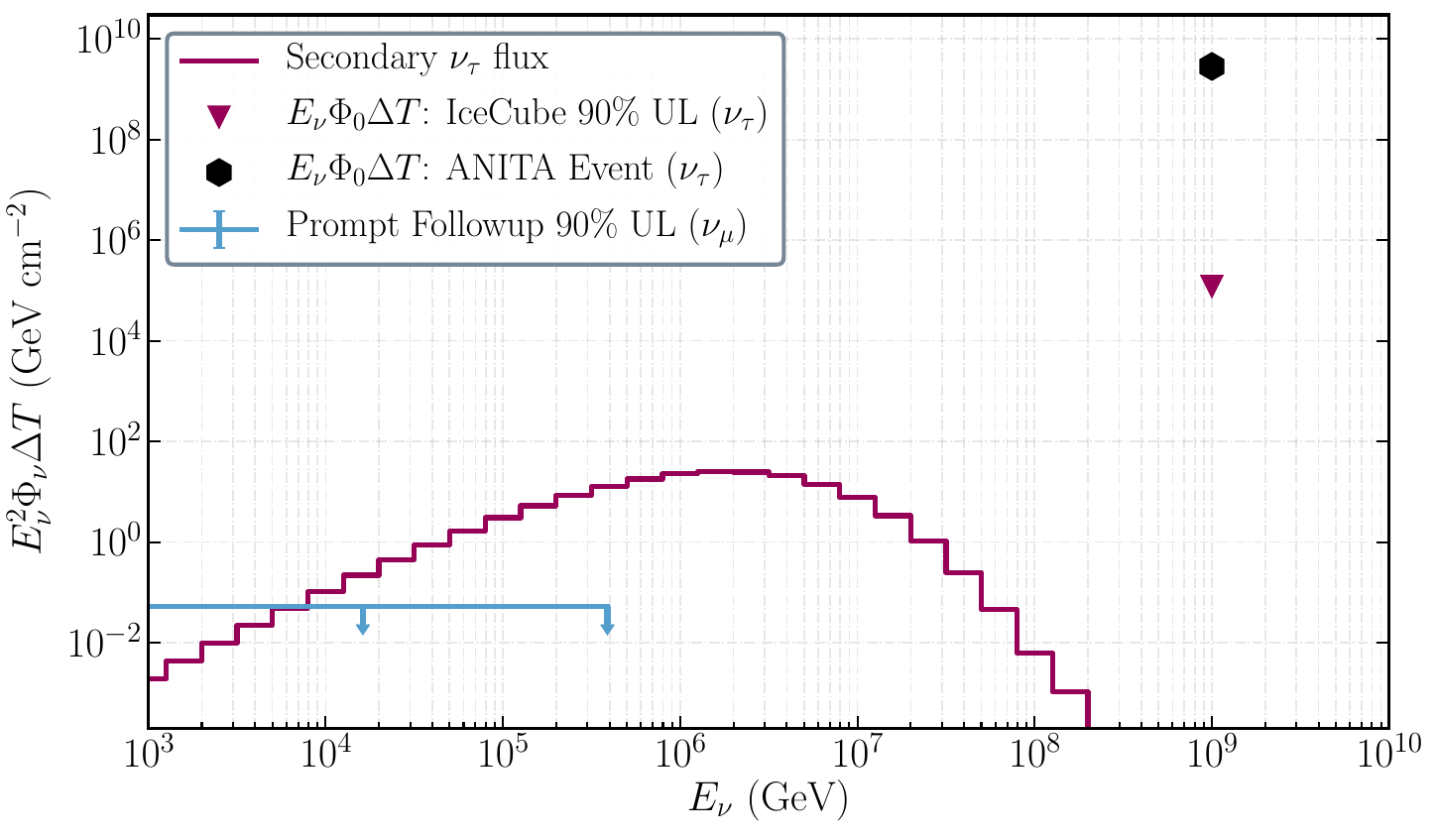}
  \label{fig:taurunner_limits}
  \caption[Transient point-source upper limits on ANITA anomalous events.]{Upper limits (90\% C.L.) placed by calculating the secondary neutrino flux (purple histogram) from an incident flux of EeV neutrinos assuming constant emission over $10^3$ s and comparing to the non-observation of IceCube events for AAE-141220. The flux implied by the ANITA observations (black) overshoots our upper limit (purple arrow) by many orders of magnitude. For comparison, upper limits on the time-integrated muon-neutrino flux from the prompt analysis are shown in blue. All fluxes are per flavor $\nu + \bar{\nu}$.}
\end{figure*}

We next inject a flux of EeV tau neutrinos and find the spectral shape of the secondary $\nu_{\tau}$ flux that would be incident on IceCube.
As we observed zero coincident events in the time window of $10^3$ s around AAE-141220, we calculate the maximum allowed flux normalization at 90\% C.L. on the primary flux that would evade this non-observation. 
The results are displayed in Figure~\ref{fig:taurunner_limits}. 

Although IceCube's sensitivity peaks many orders of magnitude below the reconstructed energies of the ANITA events, the limits set on any potential neutrino source that created AAE-141220 are more constraining by several orders of magnitude than the implied flux by the ANITA observations.
These limits are conservative, and severely constrain any incident spectrum which could produce an observable event at ANITA consistent with the AAE.

%% file: chapters/DMnus.tex
\chapter{Neutrinos and Dark Matter} %RMP baby
\label{ch:DM}
\SingleSpace
\epigraph{``We can, in our conception, join the head of a man to the body of a horse; but it is not in our power to believe that such an animal has ever really existed.''}{David Hume}
\DoubleSpacing
\noindent The goal of this chapter is to compute constraints on the $\chi \chi \rightarrow \nu \bar \nu$ annihilation channel and $\chi \rightarrow \nu \bar \nu$ decay channel from all available extraterrestrial neutrino data.
We focus on the two most promising sources of indirect dark matter signal: 1) the dark matter halo of the Milky Way, in which we are deeply embedded, and 2) the full cosmic flux from the sum of all DM halos within our cosmological horizon.
We cover a mass range from 1~MeV to $10^{15}$~MeV.

The structure of this chapter is as follows: we begin in Sec.~\ref{sec:motivation} with a theoretical overview motivating possible links between the neutrino and dark sectors.
In Sec.~\ref{sec:theory} we review the annihilation signal we are constraining, from the Milky Way halo and from the isotropic background of extragalactic halos.
In Sec.~\ref{sec:pwave}, we detail the calculations needed to extend our analysis to velocity-dependent annihilations, namely $p$-wave and $d$-wave processes.
Sec.~\ref{sec:decay} introduces the flux of neutrinos from dark matter decay in both the Milky Way halo and extragalactic halos.
Sec.~\ref{sec:methods} briefly summarizes the experimental techniques used for neutrino detection in a wide energy range, and describes the statistical methods employed in this work to constrain the neutrino flux from dark matter.
Our results are presented in Sec.~\ref{sec:results}, including results from previous analyses that we recast to be consistent with our halo assumptions, wherever possible.
We then show the results of varying these assumptions in the range allowed by stellar dynamic observations for the Galactic component and simulation results for the extragalactic one.

\section{Motivation}
\label{sec:motivation}
\noindent The Standard Model (SM) of particle physics is the framework that describes matter and its interactions at the most fundamental level.
Despite overwhelming success as a predictive theory, observations indicate that the SM is incomplete.
Neutrinos have nonzero masses, yet the Higgs mechanism that provides masses for the other SM fermions cannot account for the chiral nature of neutrinos and their interactions unless additional particle content is added to the model.
Additionally, overwhelming astrophysical and cosmological evidence points to the existence of a new species of weakly-interacting particles -- dark matter (DM) -- which accounts for $\sim 85\%$ of the mass budget of the Universe.
Local stellar dynamics, galactic rotation curves~\cite{1970ApJ...159..379R,Persic:1995ru}, cluster dynamics~\cite{1937ApJ....86..217Z,1936ApJ....83...23S}, and gravitational lensing \textit{e.g.}\cite{Jee:2008qj,Jee:2007nx} all point to mass-to-light ratios in astrophysical objects that are much higher than could be accounted for by stellar objects and gas; for a historical overview see \cite{Bertone:2016nfn}.
Measured primordial abundances of light elements tell us that Big Bang Nucleosynthesis requires a total baryon density\footnote{By \textit{baryonic} we refer here to stable nonrelativistic matter made of SM particles including neutrons, protons and electrons.} of only $\Omega_b \sim 0.05$, while the Cosmic Microwave Background (CMB) and other probes of large scale structure require the \textit{total} density of nonrelativistic matter to be $\Omega_m \sim 0.3$.
\footnote{More precisely, the baryon density is inferred to be $\Omega_b h^2 = 0.0224 \pm 0.0001$ and the (cold) DM density is $\Omega_c h^2 = 0.120 \pm 0.001$~\cite{Aghanim:2018eyx}, where $\Omega_i$ is the ratio of the density of component $i$ to the critical density, and the Hubble constant is $H_0 \equiv h$100 km s$^{-1}$Mpc$^{-1}$.} 
A leading hypothesis for the nature of this new nonbaryonic component is the Weakly Interacting Massive Particle (WIMP).
The relic abundance of WIMPs today was set as they fell out of equilibrium with the high-temperature plasma of the early Universe.
When the temperature, $T$, fell below the DM mass, $m_\chi$\footnote{We work in natural units where $c = \hbar = k_B = 1$.}, the equilibrium distribution became Botlzmann-suppressed, namely $\sim \exp(-m_\chi/T)$.
At some point, the expansion rate $H(t)$, became larger than the thermally-averaged self-annihilation rate, preventing further annihilation into SM particles, \textit{freezing-out} the relative density of DM particles.
The WIMP scenario predicts the observed relic abundance of DM for values of the thermally-averaged self-annihilation rate $\sv \simeq 3\times 10^{-26}$ cm$^{3}$s$^{-1}$ regardless of the final annihilation channel. 

Thermal production of weakly-interacting DM in the Early Universe implies possible ongoing self-annihilation to SM particles wherever DM exists today.
Significant effort has gone into searches for indirect signatures of DM annihilation.
Annihilation to most SM states yields an abundance of photons with energies on the order of $10\%$ of the DM mass, such that some of the strongest constraints on particle DM models are from the (non) observation of X- and gamma-ray signals from the Milky Way and its satellite galaxies; see \textit{e.g.}~\cite{Fermi-LAT:2016uux,Hoof:2018hyn}.
Cosmic-ray signatures provide similarly constraining limits, reports of excesses notwithstanding; see~\cite{Boudaud:2019efq} and references therein.

As X- and gamma-ray experiments rely, by design, on electromagnetic signals, they are optimal for probing links between the dark sector and quarks or charged leptons, although neutrino detectors can still play a role in these searches~\cite{Cappiello:2019qsw}.

There is a distinct possibility that the principal portal through which the DM interacts with the SM is via the neutrino sector~\cite{Blennow:2019fhy}. 
This naturally arises in ``scotogenic'' models, in which neutrino mass generation occurs through interactions with the dark sector~\cite{Boehm:2006mi,Farzan:2012sa,Escudero:2016tzx,Escudero:2016ksa,Hagedorn:2018spx,Alvey:2019jzx,Patel:2019zky, Baumholzer19}.
These models introduce heavy neutrino states, sometimes called dark neutrinos, which could also provide a possible explanation of the MiniBooNE anomaly~\cite{Bertuzzo:2018itn,Ballett:2018ynz,Ballett:2019cqp,Ballett:2019pyw}. 
``Secret'' neutrino interactions with dark matter have recently become a very active field of investigation, where constraints have been obtained using high-energy astrophysical neutrinos~\cite{Farzan:2014gza,Davis:2015rza,Cherry:2016jol,Arguelles:2017atb,Kelly:2018tyg,Farzan:2018pnk,Pandey:2018wvh,Choi:2019ixb,Capozzi:2018bps,Murase:2019xqi}, solar neutrinos~\cite{Capozzi:2017auw}, cosmology~\cite{Campo:2017nwh,Barenboim:2019tux}, accelerator neutrino experiments~\cite{Aguilar-Arevalo:2017mqx,Arguelles:2018mtc,Hostert:2019iia} and colliders
~\cite{Primulando:2017kxf}.
The past two decades have seen extraordinary progress in the field of extraterrestrial neutrino detection.
Observations span a wide energy range, from the MeV $pp$ solar neutrino flux~\cite{Agostini:2018uly} to the PeV ($10^6$ GeV) high-energy astrophysical neutrinos~\cite{Aartsen:2013jdh,Aartsen:2014gkd,Schneider:2019ayi}.
Furthermore, neutrino flux limits exist all the way up to $\sim$ ZeV ($10^{12}$ GeV) ~\cite{Aab:2015kma,Aartsen:2018vtx}.
With these observations, a multitude of experimental constraints have been derived on the dark matter annihilation cross section to neutrino pairs and decay lifetime, either by experimental collaborations themselves or by independent authors recasting results of previous searches.

Neutrinos are the most weakly interacting stable particles in the SM and, consequently, the hardest to detect.
In the context of indirect detection, this implies that models where DM annihilates predominantly to neutrinos are difficult to rule out.
This makes the study of neutrinos as a final state particle particularly interesting as, so far, all direct and indirect searches for the footprints of DM--SM interactions have come up empty~\cite{Arcadi:2017kky,Tanabashi:2018oca}.

\section{Dark Matter Density Profile}
\label{sec:dmprofile}

To parametrize the galactic DM halo, we use a generalized Navarro-Frenk-White (NFW) profile, which is given by
\begin{equation}
    \rho_{\chi}(r) = \rho_s \frac{2^{3-\gamma}}{\left(\frac{r}{r_s}\right)^\gamma \left(1+ \frac{r}{r_s}\right)^{3-\gamma}}.
    \label{eq:NFW}
\end{equation}
The galactocentric distance is 
\begin{equation}
    r = \sqrt{R^2_0 - 2 \, x \, R_0 \, \cos{\psi} + x^2},
\end{equation}
where $\psi$ is the angle between the Galactic center (GC) and the line of sight, and $R_0$ is the distance from the Sun to the GC. 
We take the Sun to be located $R_0 = 8.127~{\rm kpc}$  from the GC, as determined by recent measurements of the four-telescope interferometric beam-combiner instrument GRAVITY~\cite{Abuter:2018drb}.
We use DM halo parameters compatible with the best-fit values of~\cite{Benito:2019ngh}, {\it i.e.}: a local density\footnote{It is customary to specify $\rho_0 \equiv \rho_\chi(R_0)$ rather than $\rho_s$, as the former can be more directly measured. The two are related by inverting Eq. \eqref{eq:NFW}.} of $\rho_0 = 0.4~{\rm GeV~cm}^{-3}$, a slope parameter $\gamma = 1.2$, and a density $\rho_s$ at  scale radius $r_s = 20~{\rm kpc}$.

Annihilation specifically is sensitive to the clustering of DM, rather than the overall number of particles as it requires the collision of two DM particles. 
Therefore extragalactic neutrino fluxes from DM annihilation are enhanced by clustering. 
Accounting for this enhancement requires integrating the described DM density profile over redshift, accounting for the the mass distribution of halos.
This is often called the ``boost'' factor $G(z)$, and is given by
\begin{equation}
    G(z) = \frac{1}{\Omega_{DM,0}^2\rho_c^2}\frac{1}{(1+z)^6} \int dM \frac{dn (M,z)}{dM}\int dr 4\pi r^2 \rho_{\chi}^2(r).
    \label{eq:overdensity} 
\end{equation}
The first integral is over halo masses $M$ whose distribution is specified by the halo mass function (HMF), $dn/dM$, while the second integral is over the halo overdensities themselves.
We model the latter as self-similar NFW profiles whose densities and radii are specified by a concentration parameter uniquely determined by their mass and redshift.
The parametrization that we employ is based on fits to the MultiDark/BigBolshoi~\cite{2012MNRAS.423.3018P} simulations and can be found in Appendix B  of~\cite{Lopez-Honorez:2013lcm}.

Two uncertainties arise from the integral over $M$.
First is the choice of integration limits, specifically the lower limit, $M_{min}$.
This is because smaller halos are more concentrated, thus contributing more to the injected neutrino energy.
This  means that  choosing arbitrarily low-minimum halo masses results in unrealistic limits.
It is common in the literature to use $M_{min} = 10^{-6} M_\odot$ as a benchmark, although there is no data-driven motivation for this choice.
$M_{min}$ is not well-constrained, and will ultimately depend on model details~\cite{Shoemaker:2013tda,Cornell:2013rza}.
Therefore, in this work we pick $M_{min} = 10^{-3}M_\odot$ as a conservative limit choice.
In section~\ref{sec:haloparams}, we show the effect of varying $M_{min}$ down to $10^{-9} M_\odot$.
The other uncertainty arises from the choice of HMF, $dn/dM$, parametrization. We use the results of the N-body simulation by~\cite{2013MNRAS.433.1230W}, as parametrized in~\cite{Lopez-Honorez:2013lcm,Diamanti:2013bia}.
Several other HMF parametrizations are tested, and the uncertainties due to choice of HMF are quantified in Sec.~\ref{sec:haloparams}. 

\section{Annihilation to neutrinos\label{sec:theory}}

The limits derived on the DM annihilation to neutrinos can be interpreted as an upper bound on the total DM annihilation cross section to SM particles~\cite{Beacom:2006tt,Yuksel:2007ac}, since the latter is larger by definition. 
From a particle physics point of view, the direct annihilation of DM to neutrinos at tree level requires the addition of a neutrino-DM term to the SM Lagrangian that couples them.
Since neutrinos belong to an $SU(2)$ doublet, na\"ive SM gauge invariance implies that coupling neutrinos with DM would also induce an interaction between the DM and the charged leptons, mediated, \textit{e.g.}, by a new $Z$-like particle.
Such interactions are highly constrained, as they lead to production of dijet or dilepton signatures observable at colliders (see \textit{e.g.}~\cite{Carena:2004xs,Lees:2014xha}), fixed target experiments~\cite{Abrahamyan:2011gv}, and direct detection experiments (see \textit{e.g.}~\cite{Blanco:2019hah} and references therein).

Nevertheless, there exist viable models in which the DM phenomenology is dominated by its interactions with neutrinos~\cite{Blennow:2019fhy}.
Coupling only to the heavier lepton generations can strongly mitigate bounds from electron interactions, \textit{e.g.} by introducing a $U(1)_{L_\mu - L_\tau}$ symmetry~\cite{He:1990pn,He:1991qd}.
A more elegant option allows the DM to interact with a sterile neutrino that then mixes with the active neutrinos, leading to direct annihilations of DM to neutrinos if the mass of the sterile neutrino is larger than the DM mass~\cite{Profumo:2017obk,Ballett:2019pyw}.
If the sterile-light mixing is sizable, DM--neutrino interactions will provide the best window to understand such DM models.
A comprehensive review of these scenarios can be found in~\cite{Blennow:2019fhy}.

Finally, we are considering direct annihilation to neutrinos without including electroweak (EW) corrections, which severely complicate the spectral shape computations. These are important at energies above the electroweak scale, and will have two main consequences: 1) the peak of the spectrum will be slightly broadened, and 2) A lower-energy continuum will be produced.
Given the typical energy resolution $\gtrsim 10\%$~\cite{Aartsen:2013vja}  for high-energy neutrino detectors, the former effect is not likely to be important.
The second effect could potentially lead to stronger bounds from the additional flux at  lower energies.
A detailed computation of this effect up to ultra-high-energies has only recently been performed~\cite{Bauer:2020jay}; as these were not available at the time of this analysis we do not include these here.
At sub-TeV energies, these corrections are accurately implemented in numerical codes such as PYTHIA~\cite{Sjostrand:2014zea,Sjostrand:2019zhc}; a comparison between our limits and ones derived using these additional corrections show very little difference [see e.g.~\cite{Liu:2020ckq}].

A more important consequence is the presence of gamma radiation from the decay of EW products, which can potentially provide complementary constraints to dedicated neutrino-line searches~\cite{Murase:2012xs}. 
Using these secondary products, current constraints on the thermally averaged annihilation cross section to neutrinos from Fermi-LAT and HESS hover around $10^{-23}~{\rm cm}^3{\rm s}^{-1}$ in the 300~GeV to 3~TeV mass range~\cite{Queiroz:2016zwd}.
These gamma-ray based constraints are at the same level as current bounds from ANTARES~\cite{Adrian-Martinez:2015wey}, but are expected to be improved by the next generation gamma-ray experiments such as the Cherenkov Telescope Array (CTA)~\cite{Queiroz:2016zwd}.
We will provide an example using these projections for CTA in Sec.~\ref{sec:results}, noting that this only includes prompt gamma rays. 
Inverse-Compton scattering of primary electrons and positrons  with interstellar photons will strengthen the sensitivity of gamma-ray searches.
This effect has been studied for DM decay searches, but not for annihilation $\chi \chi \rightarrow \nu \bar \nu$~\cite{Cohen:2016uyg, Murase:2015gea, Chianese:2019kyl}.

%% Galactic 
\subsection{Galactic contribution\label{sec:galactic}}

We begin by setting limits on DM annihilation to neutrino pairs in the Milky Way (MW) dark matter halo.
The expected flux per flavor of neutrinos plus antineutrinos at Earth, assuming equal flavor composition\footnote{If the flavor composition at the source is not democratic, neutrino oscillation will yield a flavor composition at Earth that is close, but not equal to $(\nu_e:\nu_\mu:\nu_\tau) = (1:1:1)$.
Annihilation to $\nu_e$ only will give $\sim (0.55: 0.25:0.2)$; to $\nu_\mu$: $\sim (0.25: 0.36:0.38)$ and $\nu_\tau$ yields $\sim (0.19: 0.38:0.43)$.}, is given by
\begin{equation}
    \frac{d\Phi_{\nu+ \bar{\nu}}}{dE_\nu} = \frac{1}{4 \pi}
    \frac{\sv}{\kappa m_\chi^2}  \frac{1}{3}  \frac{dN_\nu}{dE_\nu} J(\Omega),
    \label{eq:galaxyAnnRate}
\end{equation}
where $\kappa$ is 2 for Majorana DM and 4 for Dirac DM, $m_\chi$ is the DM mass, and $\sv$ is the thermally averaged self-annihilation cross section into all neutrino flavors.
Going forward we set $\kappa = 2$ (Majorana DM). 
The spectrum in the case of annihilation to two neutrinos is simply ${dN_\nu}/{dE_\nu} = 2\delta(1 - E/m_\chi)m_\chi/E^2$.
$J(\Omega)$ is a three-dimensional integral over the target solid angle in the sky, $d \Omega$, and the distance $dx$ along the line of sight (l.o.s.) of the DM density $\rho_\chi$, namely
\begin{equation}
     J \equiv  \int d\Omega \int_{\mathrm{l.o.s.}}  \rho_\chi^2(x) dx. 
 \label{eq:Jfactordef}
\end{equation}
It is referred to as the $J$-factor and has units of GeV$^2$\,cm$^{-5}$\,sr.\footnote{Another equivalent convention used in the literature is to report the dimensionless quantity $\mathcal{J} = J/\Delta\Omega R_0 \rho_0^2$~\cite{Yuksel:2007ac}.}
In practice, the upper limit of integration can be set at
\begin{equation}
    x_{\rm max} = \sqrt{R^2_{\rm halo} - \sin^2{\psi} R^2_0}+R_{\rm 0}\cos{\psi},
\end{equation}
for some maximum halo radius $R_{\rm halo}$.
Details of the dark matter profile and the values of the variables used in the integration are explained in Sec.~\ref{sec:dmprofile}.

The resulting $J$-factors for $s$, $p$, and $d$-wave annihilation are shown in Tbl.~\ref{tab:Jtable}; the latter cases will be discussed in Sec.~\ref{sec:pwave}.
Some experiments, such as ANITA, AUGER, and GRAND, are only sensitive to a certain region of the sky.
In these cases, the corresponding $J$-factors must be recomputed by converting their respective sensitivity from elevation/azimuth to galactic coordinates, and integrating over the resulting region. 
A value of the $J$-factor is not given for some experiments, where the flux cannot be factored out as in Eq.~\eqref{eq:galaxyAnnRate}. This could be due \textit{e.g}. to an energy-dependent acceptance.
These are also shown in Tbl.~\ref{tab:Jtable}. When the exposure is not a simple declination window, we provide the reference to where it can be obtained.
Recent works~\cite{Pato:2015dua, Benito:2016kyp, Karukes:2019jxv,Benito:2019ngh} have constrained the halo shape and density parameters, using observations of stellar dynamics in the MW.
In Sec.~\ref{sec:haloparams}, we illustrate the effect on the dark matter limits obtained in this work when varying these parameters within those constraints.\\

\begin{table*}[!ht]
    \begin{center}
    \makebox[\linewidth]{
    \begin{tabular}{c | c | c | c | c}
    \hline \hline 
         Experiment& Exposure & $J_s/{10^{23}}$ & $J_p/{10^{17}}$ & $J_d/{10^{11}}$  \\ \hline
        $\heartsuit$ All-sky & All-sky & $2.3$& $2.2$& $3.6$ \\
        $\heartsuit$ GRAND  & Fig. 24 of \cite{Alvarez-Muniz:2018bhp} & $0.28$ & $0.28$  & $0.46$ \\ 
        %$\heartsuit$ Auger & Fig. 1 of \cite{Zas:2017xdj} & $--$  & $-- $ & $--$ \\
        $\heartsuit$ ANITA & dec = $[1.5^{\circ} , 4^{\circ}]$ & $0.018$  & $0.018 $ & $0.028$ \\\hline 
        %DARWIN \cite{McKeen:2018pbb}& All-sky  &  & $0.45$& $2.2$ &$3.6$ \\\hline
        CTA \cite{Queiroz:2016zwd} & Galactic Center \cite{Queiroz:2016zwd} & $0.074$ & $0.12 $ & $0.16$  \\ \hline
        $\heartsuit$ TAMBO & Fig. 3 \& 4 of \cite{Romero-Wolf:2020pzh} & 0.0009 & $-$ & $-$\\ \hline
        $\heartsuit$ Auger & \begin{tabular}{c}${\rm zenith} = [90^{\circ}, 95^{\circ}]$\\${\rm zenith} = [75^{\circ},90^{\circ}]$ \\ ${\rm zenith} = [60^{\circ}, 75^{\circ}]$\end{tabular} & \begin{tabular}{c}0.10\\0.28\\0.27\end{tabular} & $-$ & $-$\\ \hline
        $\heartsuit$ P-ONE & \begin{tabular}{c}$\cos({\rm zenith}) = [-1, -0.5]$\\ $\cos({\rm zenith}) = [-0.5, 0.5]$\\ $\cos({\rm zenith}) = [0.5, 1]$\end{tabular} & \begin{tabular}{c}0.87\\1.2\\0.13\end{tabular} & \begin{tabular}{c}0.85\\ 1.2\\ 0.12\end{tabular} & \begin{tabular}{c}1.4\\ 2.0\\ 0.18\end{tabular}\\ \hline
        \hline
    \end{tabular}
    }
    \end{center}
    % %\internallinenumbers
    \caption[J-factors used to constrain DM annihilation for different experiments and their associated halo parameters.]{J-factors used to constrain DM annihilation for different experiments and their associated halo parameters.
    $J$-factors, given in units of GeV$^2$ cm$^{-5}$ sr, are computed according to Eq.~\eqref{eq:Jfactordef}.
    We use these to find the expected neutrino flux as described in Eq.~\eqref{eq:galaxyAnnRate}.
    Each row corresponds to a different experimental setup given its angular exposure.
    The first column names the experiment; the second column summarizes their angular acceptance; and the last three columns give the $s$-wave, $p$-wave, and $d$-wave $J$-factors, respectively.
    The hearts, $\heartsuit$, indicate new results given in this work.}
    \label{tab:Jtable}
\end{table*}

\subsection{Extragalactic contribution\label{sec:extragalactic}}

In addition to DM annihilation in the MW, annihilation of extragalactic dark matter integrated over all redshifts should provide a diffuse isotropic neutrino signal~\cite{Beacom:2006tt}.
As in the search for extragalactic background light, there are two contributions to this isotropic flux: 1) a ``background'' flux from the diffuse (non-collapsed) distribution of DM, whose rate grows with redshift as $ \Omega_{DM}^2 \sim (1+z)^6$,  and 2) a late-time contribution from the large overdensities in galactic halos. 

In this case, the expected flux of neutrinos plus antineutrinos per flavor at Earth from DM annihilation is given by
\begin{equation}
\frac { d \Phi _ { \nu + \bar{\nu} } } { d E_{\nu} } = \frac {1 } {4 \pi  } \frac { \Omega _ { D M } ^ { 2 } \rho _ { c } ^ { 2 } \sv} { \kappa m _ { x } ^ { 2 } } \frac{1}{3} \int _ { 0 } ^ { z _ { u p } } d z \frac { \left(1 + G ( z )\right) (1+z)^3} { H ( z ) } \frac { d N _ { \nu + \bar{\nu} } \left( E ^ { \prime } \right) } { d E^ { \prime } },
\label{eq:extragalactic_flux}
\end{equation}
where $H ( z ) = H_0\left[ ( 1 + z ) ^ { 3 } \Omega _ {m} + (1+z)^4\Omega_r + \Omega _ { \Lambda } \right] ^ { 1 / 2 } $ is the time-dependent Hubble parameter, $\rho _ {c}$ is the critical density of the Universe,  $\Omega_m$, $\Omega_r$, and $\Omega_\Lambda$ are respectively the fraction of $\rho_c$ made up of matter, radiation and dark energy.
While the upper limit on redshift, $z_{up}$, can in principle be as high as the neutrino decoupling time at $T \sim$~MeV, neutrinos produced at that epoch are redshifted to the point of being invisible to existing detectors.  
${ d N _ { \nu } \left( E ^ { \prime } \right) }/ { d E }$ is the neutrino spectrum at the detector, where $E ^ { \prime }$ ($E$) is the energy at the source (detector). 
The spectrum is related to the source production spectrum via a Jacobian transformation to take cosmological redshift into account, namely
\begin{equation}
\frac { d N _ { \nu + \bar{\nu}} \left( E ^ { \prime } \right) } { d E^ { \prime }  } = 2 \frac{m _ { \chi }}{E^{' 2}} \delta \left( \frac{m _ { \chi }}{E ^ { \prime }} - 1 \right) = \frac { 2 } {  E } \delta \left[ z - \left( \frac { m _ { \chi } } { E } - 1 \right) \right].
\end{equation}
In Eq.~\eqref{eq:extragalactic_flux}, $\sv $ is the thermally averaged cross section.
The first part of the factor $1+ G ( z )$  in the integrand of Eq.~\eqref{eq:extragalactic_flux} represents the isotropic background DM contribution, while $G(z)$ is the halo boost factor at redshift $z$.
It accounts for the enhancement to the annihilation rate in DM clusters and their evolution with redshift, and is defined in Sec.~\ref{sec:dmprofile}.

\begin{figure*}[t!]
%   \beglln{subfigure}
    \includegraphics[width=0.49\linewidth]{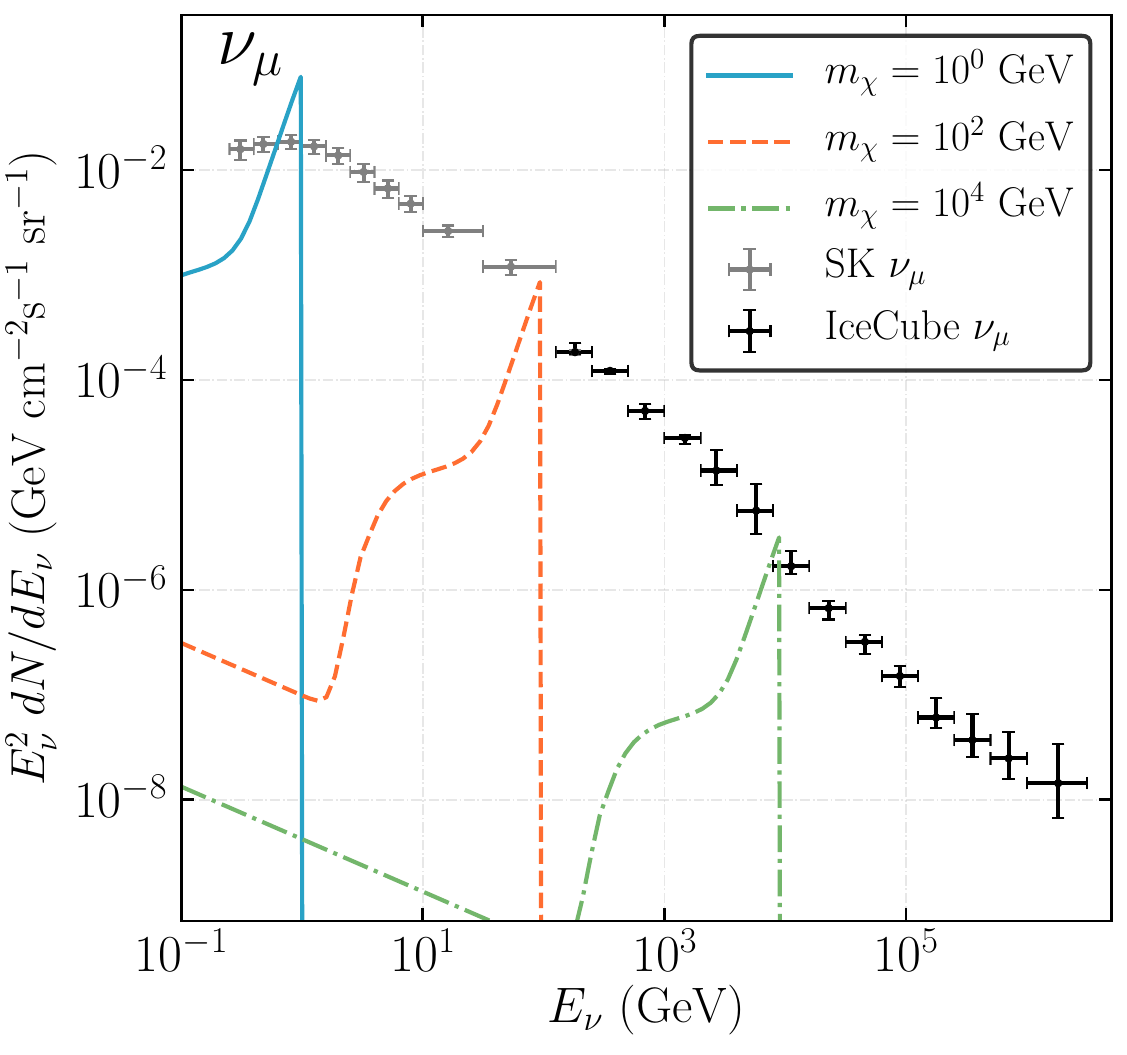}
%   \end{subfigure}
%   \begin{subfigure}
    \includegraphics[width=0.49\linewidth]{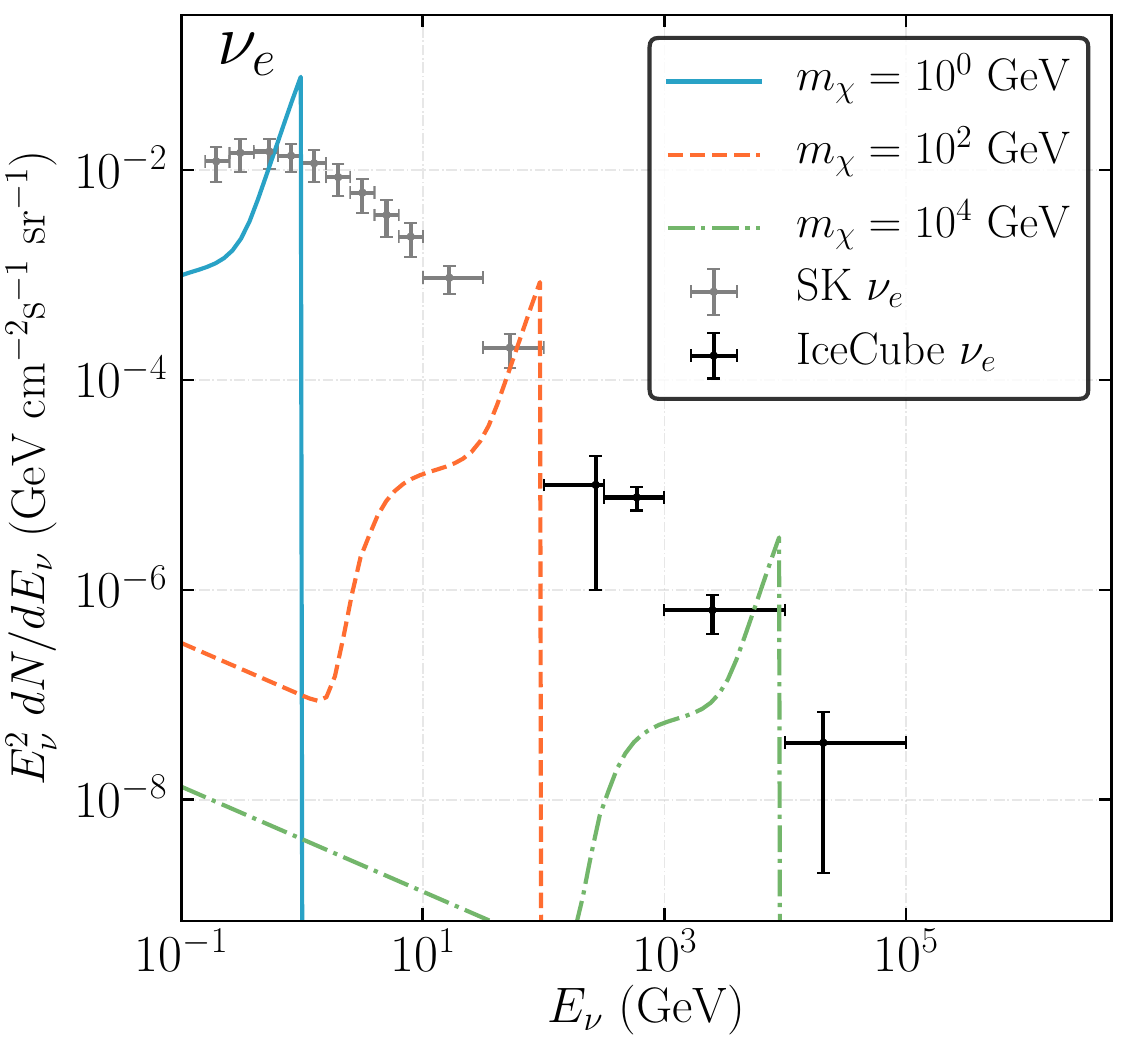}
%   \end{subfigure}
    \begin{center}
    \includegraphics[width=0.51
    \linewidth]{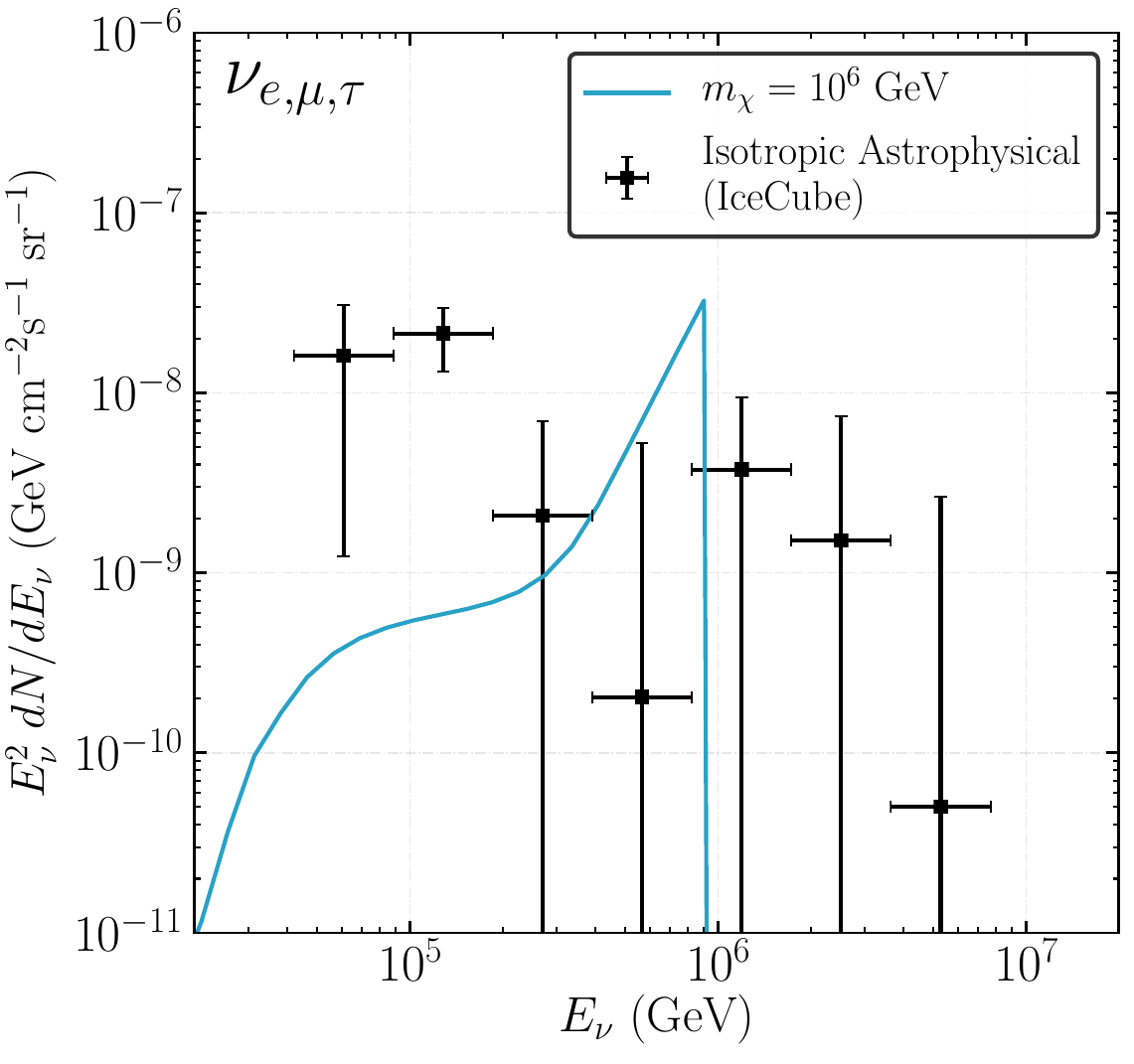}
    \end{center}
%   %\internallinenumbers
  \caption[Examples of neutrino fluxes produced by dark matter annihilation overlayed on the observed neutrino distributions.]{Examples of neutrino fluxes produced by dark matter annihilation overlayed on the observed neutrino distributions.
  Expected flux of neutrinos from extragalactic dark matter annihilation as a function of energy, shown for several dark matter masses.
  Fluxes are computed using the value of the cross section corresponding to the 90\% C.L. limit derived in this work.
  Here, the extragalactic dark matter annihilation fluxes are compared to the unfolded atmospheric fluxes from both Super-Kamiokande~\cite{Richard:2015aua} and IceCube~\cite{Aartsen:2015xup,Aartsen:2016xlq}.
  Top left is the $\nu_{\mu}$ channel; top right is the $\nu_{e}$ channel; the bottom shows a comparison to IceCube's measured per-flavor isotropic Astrophysical flux using 7.5 years of Starting Events~\cite{Abbasi:2020jmh}.
  }
  \label{fig:extragalacticflux}
\end{figure*}

The expected spectrum of DM annihilation to two neutrinos from cosmological sources is shown in Fig.~\ref{fig:extragalacticflux}, for different DM masses.
These are overlaid on the Super-Kamiokande (SK)~\cite{Richard:2015aua} and IceCube~\cite{Aartsen:2015xup,Aartsen:2016xlq} unfolded atmospheric $\nu_e$ and $\nu_\mu$ fluxes as well as the isotropic astrophysical flux~\cite{Abbasi:2020jmh}. 

\section{Velocity-dependent annihilation\label{sec:pwave}}

Certain matrix element vertex structures lead to a suppression of the constant ($s$-wave) part of the self-annihilation cross section.
Expanding in powers of $v/c$, the dominant term may be $p$-wave ($\propto v^2$) or $d$-wave ($\propto v^4$) in the nonrelativistic limit. 
The DM velocity distribution depends on the kinematic details of the structure in which it is bound, as well as its distance from the center of that distribution.
Assuming a normalized Maxwellian distribution, $f(v,r)$, with dispersion $v_0(r)$, the annihilation rate will be proportional to
\begin{equation}
\langle v^n \rangle = \int d^3v f(v,r) v^{n}.
\end{equation}
For $p$- and $d$-wave, this respectively yields 
\begin{eqnarray}
\langle v^2 \rangle &=& 3 v_0^2(r), \\
\langle v^4 \rangle &=& 15 v_0^4(r).
\end{eqnarray}
We obtain the dispersion velocity, $v_0$, by solving the spherical Jeans equation, assuming isotropy. This is given by
\begin{equation}
    \frac{d(\rho(r) v^2_0(r))}{dr}=-\rho(r) \frac{d\phi(r)}{dr} \, ,
    \label{eq:Jean}
\end{equation}
where $\phi(r)$ is the total gravitational potential at radius $r$.
For Galactic constraints, we include not only the contribution of the DM halo to $\phi(r)$, but also follow~\cite{Boddy:2018ike} and include a parametrization of the MW bulge and disk potentials to account for their masses.
These are given by
\begin{eqnarray}
\phi(r)_{\rm bulge}&=&-\frac{G_NM_b}{r+c_b}, \\
\phi(r)_{\rm disk}&=&-\frac{G_NM_d}{r} \left(1-e^{-r/c_d} \right),
\end{eqnarray}
where $G_N$ is Newton's gravitational constant, $M_b=1.5 \times 10^{10}M_{\odot}$, and $c_b =0.6$ kpc are the bulge mass and scale radius, while $M_d=7 \times 10^{10}M_{\odot}$ and $c_d=4$ kpc are the disk mass and scale radius~\cite{Boddy:2018ike}.
Galactic $J$-factors can then be reevaluated via
\begin{eqnarray}
J_{v^n} = \int d\Omega \int_{\mathrm{l.o.s.}}\frac{\left\langle v^{n}(r)\right\rangle}{c^{n}} \rho^2_\chi(r) dx.
\end{eqnarray}

In the case of our extragalactic analysis, we only include the potential from the DM halos themselves. This is conservative, in that the addition of the uncertain baryonic contributions would only strengthen our constraints.
In a similar manner to the Galactic case, Eqs.~\eqref{eq:extragalactic_flux} and \eqref{eq:overdensity} must be modified to include the dependence on $\langle v^n \rangle (r)$.
As long as the annihilation remains a two-to-two process (unlike scenarios in e.g.\cite{Bell:2017irk}), Eq.~\eqref{eq:extragalactic_flux} becomes:
\begin{equation}
\frac { d \Phi _ { \nu } } { d E_{\nu} } = \frac {c } {4 \pi  } \frac { \Omega _ { D M } ^ { 2 } \rho _ { c } ^ { 2 } \sv} { 2 m _ { x } ^ { 2 } } \int _ { 0 } ^ { z _ { u p } } d z \frac { \left(\left[\frac{1+z}{1+z_{\rm KD}} \right]^{n} + G_n ( z )\right) (1+z)^3} { H ( z ) } \frac { d N _ { \nu } \left( E ^ { \prime } \right) } { d E },
\label{eq:extragalactic_flux_vn}
\end{equation}
where the redshift $z_{\rm KD}$ is related to the temperature at kinetic decoupling $T_{\rm KD}$ and the temperature of the CMB today $T_{\rm CMB,0}$ via  $1 + z_{\rm KD} = T_{\rm KD}/T_{\rm CMB,0} \simeq 4.2 \times 10^9 \,  (T_{\rm KD}/\mathrm{MeV})$~\cite{Diamanti:2013bia}.
\cite{Shoemaker:2013tda} obtained a temperature of kinetic decoupling:
\begin{equation}
    T_{\rm KD} \simeq 2.02 \, \mathrm{MeV} \, \left(\frac{m_\chi}{\mathrm{GeV}} \right)^{3/4}.
    \label{eq:TKD}
\end{equation}
In general, kinetic decoupling occurs later than chemical freeze-out and depends on the number of relativistic degrees of freedom $g_\star(T_{\rm KD})$.
At redshifts where the annihilation products are still measurable by earth-based detectors, the factor of $((1+z)/(1+z_{\rm KD}))^n$ still leads to a strong enough suppression that it will always be subdominant to the halo contribution proportional to $G_n(z)$.
The exact value of $T_{\rm KD}$ in Eq.~\eqref{eq:TKD} is thus inconsequential. Eq.~\eqref{eq:overdensity} including velocity dependence is rewritten as follows:
\begin{equation}
    G_n(z) = \frac{1}{\Omega_{DM,0}^2\rho_c^2}\frac{1}{(1+z)^6} \int dM \frac{dn (M,z)}{dM}\int dr 4\pi r^2 \frac{\left\langle v^{n}(r)\right\rangle}{c^{n}} \rho_{\chi}^2(r),
\label{eq:vdepboost}
\end{equation}
where we have used the same HMF as in the velocity-independent case, with the addition of the velocity dispersion, $\left\langle v^{n}(r)\right\rangle$, in the rightmost integral.
~\cite{Diamanti:2013bia} provides the detailed method of solving the Jeans equation to compute $\left\langle v^{n}(r)\right\rangle$ as a function of the DM halo concentration.
For convenience, we provide the following function for the $p-$ and $d-$wave cases:
\begin{equation}
    \ln(G_n) \simeq \sum_{i} c_i~\alpha^i,
\label{eq:polyfit}
\end{equation}
where $c_i$ are the coefficients provided in Tbl.~\ref{tab:coeff}, and $\alpha \equiv \ln(z)$. This parametrization is valid down to redshifts  $\gtrsim 10^{-3}$.

\begin{table}[hb]
\centering
\begin{tabular}{ p{0.33cm} p{2cm} p{2cm} }
 \hline \hline
 & $p-$wave & $d-$wave \\
 \hline
 $c_0$ & $-7.004$   & $-19.88$\\
 $c_1$ & $-1.821$   & $-2.493$\\
 $c_2$ & $-0.5793$  & $-0.804$\\
 $c_3$ & $-0.09559$ & $-0.1636$\\
 $c_4$ & $-0.006148$  & $-0.02101$\\
 $c_5$ & $0$  & $-0.001181$ \\
 \hline \hline
\end{tabular}
%\internallinenumbers
\caption[Coefficients of the polynomial fit to velocity dependent halo boost factors.]{Coefficients of the polynomial fit to velocity dependent halo boost factors.
The coefficients corresponding to Eq.~\eqref{eq:polyfit}, which is a parametrization to the numerical solution of Eq.~\eqref{eq:vdepboost}.}
\label{tab:coeff}
\end{table}

\begin{figure*}[ht]
\centering
\includegraphics[width=\textwidth]{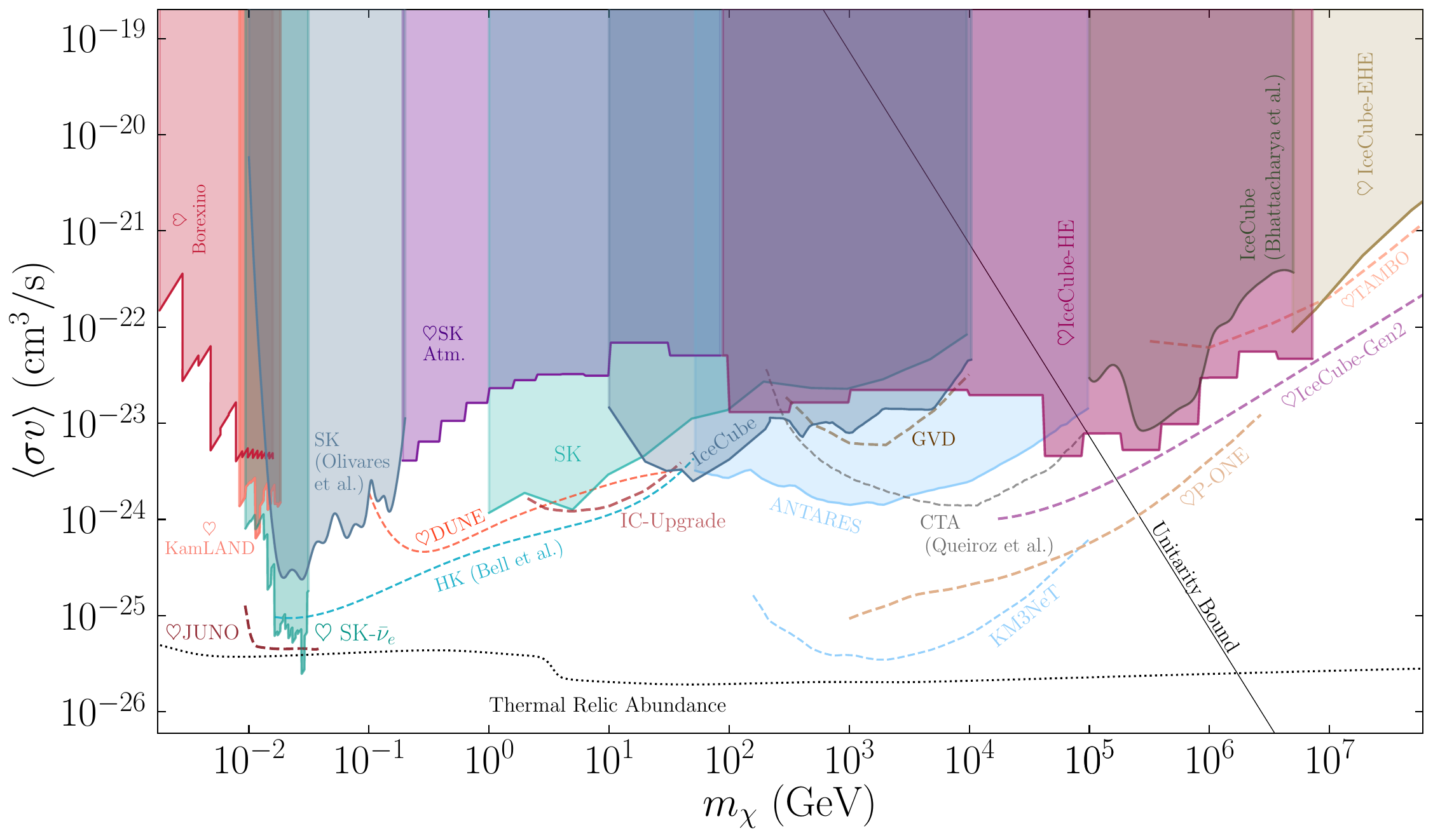}
\caption[The landscape of dark matter annihilation into neutrinos up to $10^8$ GeV.]{The landscape of dark matter annihilation into neutrinos up to $10^8$ GeV.
We show results from this work, as well as previously published limits. Data and corresponding references are detailed in Sec.~\ref{sec:results}. Solid and dashed lines represent 90\% C.L. limits and sensitivities, respectively. Projected sensitivities assume five years of data taking for neutrino experiments and 100 hours of observation for CTA. The dotted line corresponds to the value required to explain the observed abundance via thermal freeze-out. The straight diagonal line, labeled as ``Unitarity Bound,'' gives the maximum allowed  cross section for a non-composite DM particle. These results assume $100\%$ of the dark matter is composed of a given Majorana particle. If instead only a fraction, $f$, is considered these results should be multiplied by $1/f^2$. In the case of Dirac DM, limits would be scaled up by a factor of two. The heart symbols ($\heartsuit$) indicate new results obtained in this work. See Fig. \ref{fig:Indirect_non_unitarity} for constraints and projections up to $10^{11}$ GeV.}
\label{fig:Indirect}
\end{figure*}

\begin{figure*}[ht!]
    \centering
    \includegraphics[width=\textwidth]{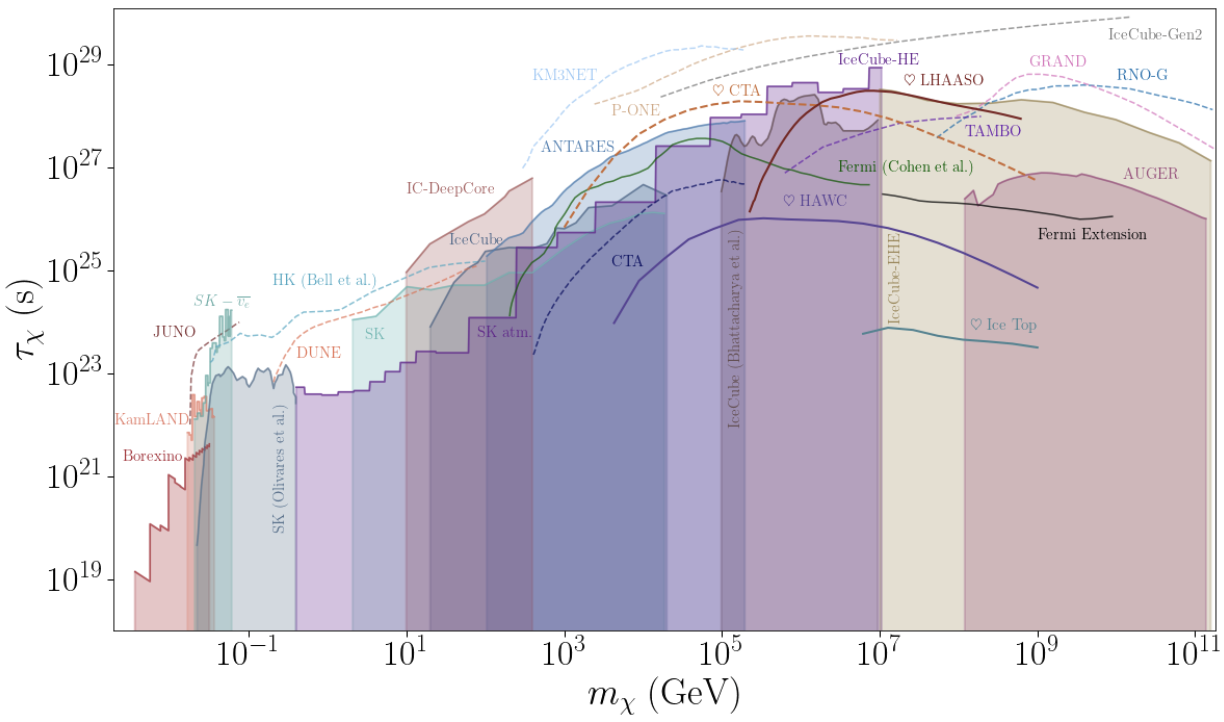}
    \caption[Limits on the lifetime of dark matter decay to neutrino pairs]{Limits on the lifetime of dark matter decay to neutrino pairs $\chi \rightarrow \bar \nu \nu$. DM masses span a few MeV to $10^{14}$~MeV. 
    Limits are especially strong for heavy DM where neutrino observatories such as IceCube and ANTARES dominate.
    Future projections are shown as dashed lines.}
    \label{fig:DecayLimits}
\end{figure*}

\section{Decay to neutrinos}
\label{sec:decay}
Here we focus on direct decay to neutrino pairs $\chi \rightarrow \nu \bar \nu$. 
At masses greater than $\sim$ TeV, electroweak corrections can ``decloak'' the dark matter, producing high-energy photons detectable at current and next-generation gamma-ray observatories. 
We will find, however, that current and future neutrino telescopes retain superior sensitivity across nearly the entire range of masses that we consider.
This is due to three factors: the loop suppression of the gamma-ray production rate, the growth of the electroweak cross section with energy, and the loss of high-energy gamma rays to interactions in the interstellar and intergalactic medium.

\subsection{Galactic contribution\label{sec:galacticdecay}}
The flux of neutrinos and antineutrinos at the detector due to dark matter decay is given by
\begin{equation}
    \frac{d\Phi_{\nu+ \bar{\nu}}}{dE_\nu} = \frac{1}{4 \pi}
    \frac{\Gamma_\chi}{ m_\chi}  \frac{1}{3}  \frac{dN_\nu}{dE_\nu} D(\Omega),
    \label{eq:galaxyDecRate}
\end{equation}
where $\Gamma_\chi = \tau_\chi^{-1}$ is the decay rate of dark matter and $m_\chi$ is the dark matter mass.
A single decay produces two neutrinos each with half the DM mass such that the produced spectrum is
\begin{equation}
    \frac{dN_\nu}{dE_\nu} = 2\delta(E - m_\chi/2).
\end{equation}
The amount of dark matter in the field of view of a given experiment is captured by the factor $D(\Omega)$. 
Sometimes referred to as the $D$-factor, it is given by the line of sight (l.o.s.) integral of dark matter density over the target angle in the sky such that
\begin{equation}
     D \equiv  \int d\Omega \int_{\mathrm{l.o.s.}}  \rho_\chi(x) dx. 
 \label{eq:Dfactordef}
\end{equation}
where $x$ is the l.o.s. distance from Earth and $\rho_\chi$ is the dark matter density, the details of which are described in Sec.~\ref{sec:dmprofile}.
The $D$-factor used for all experiments are shown in Table \ref{tab:Dtable}.
A lot of neutrino experiments are sensitive to the entire sky, those are grouped together. 
For experiments that are only sensitive to a some portion of the sky, the angular bounds of integration used in determining the $D$-factor are calculated by converting their respective sensitivity from elevation/azimuth to galactic coordinates.

\begin{table*}[!ht]
    \begin{center}
    \makebox[\linewidth]{
    \begin{tabular}{c | c | c  }
    \hline \hline 
         Experiment& Exposure & $D/{10^{23}}$  \\ \hline
        $\heartsuit$ All-sky & All-sky & $2.65$  \\
        $\heartsuit$ GRAND  & Fig. 24 of \cite{Alvarez-Muniz:2018bhp} & $0.298$  \\ 
        %$\heartsuit$ Auger & Fig. 1 of \cite{Zas:2017xdj} & $--$  & $-- $ & $--$ \\
        $\heartsuit$ ANITA & dec = $[1.5^{\circ} , 4^{\circ}]$ & $0.052$ \\\hline 
        %DARWIN \cite{McKeen:2018pbb}& All-sky  &  & $0.45$& $2.2$ &$3.6$ \\\hline
        %IceCube (?) & $-$ & $-$ \\ 
        $\heartsuit$ CTA & Galactic Center \cite{Queiroz:2016zwd} & $0.003$ \\ \hline
        $\heartsuit$ TAMBO & Fig. 3 \& 4 of \cite{Romero-Wolf:2020pzh} & 0.001 \\ \hline
        $\heartsuit$ Auger & \begin{tabular}{c}${\rm zenith} = [90^{\circ}, 95^{\circ}]$\\${\rm zenith} = [75^{\circ},90^{\circ}]$ \\ ${\rm zenith} = [60^{\circ}, 75^{\circ}]$\end{tabular} & \begin{tabular}{c}0.11\\0.35\\0.33\end{tabular} \\ \hline
        $\heartsuit$ P-ONE  & \begin{tabular}{c}$\cos({\rm zenith}) = [-1, -0.5]$\\ $\cos({\rm zenith}) = [-0.5, 0.5]$\\ $\cos({\rm zenith}) = [0.5, 1]$\end{tabular} & \begin{tabular}{c}0.83\\1.35\\0.47\end{tabular} \\ \hline
        \hline
    \end{tabular}
    }
    \end{center}
    \caption[D-factors for different experiments used to constrain dark matter decay and their associated halo parameters.]{D-factors for different experiments used to constrain dark matter decay and their associated halo parameters.
    $J$-factors, given in units of GeV cm$^{-2}$ sr, are computed according to Eq.~\eqref{eq:Jfactordef}.
    We use these to find the expected neutrino flux as described in Eq.~\eqref{eq:galaxyAnnRate}.
    Each row corresponds to a different experimental setup given its angular exposure.
    The first column names the experiment; the second column summarizes their angular acceptance; and the last three columns give the $s$-wave, $p$-wave, and $d$-wave $J$-factors, respectively.
    The hearts, $\heartsuit$, indicate new results given in this work.}
    \label{tab:Dtable}
\end{table*}

\subsection{Extragalactic contribution\label{sec:extragalacticdecay}}
DM decay outside the Milky Way also contributes to the observed neutrino flux at Earth.
Extragalactic decays produce a diffuse isotropic signal originating from all redshifts~\cite{Beacom:2006tt}. 
The total expected flux is given by
\begin{equation}
\frac { d \Phi _ { \nu + \bar{\nu} } } { d E_{\nu} } = \frac {1 } {4 \pi  } \frac { \Omega _ { D M }\rho _ { c } } {  m _ { x }\tau_\chi } \frac{1}{3}  
 \int _ { 0 } ^ { z _ { u p } }  \frac {dz} { H ( z ) } \frac { d N _ { \nu + \bar{\nu} } \left( E ^ { \prime } \right) } { d E^ { \prime } },
\label{eq:extragalactic_decayflux}
\end{equation}
where $Omega_{DM}$ is the dark matter fractional energy density, $rho_c$ is the critical density today, $H(z)$ is the Hubble parameter, and $E' = (1+z)E$ is the starting energy of a neutrino accounting for the expansion of the Universe.

\section{Experimental methods\label{sec:methods}}

In this section we will briefly review the different methodologies and technologies used for neutrino detection relevant for the discussion of the experimental results discussed in this chapter.
The results presented in Sec.~\ref{sec:results} rely on our understanding of the backgrounds in the region of interest.
Depending on whether the background flux is known, limits can be either background-agnostic or background informed.
Moreover, the limits highly depend on the systematics that govern neutrino detection, for instance the energy resolution and flavor identification capability.
Below, we first outline the statistical framework for limit-setting, before describing detector physics used over energy ranges considered here, from a few MeV up to $10^{12}$ GeV and beyond.

\subsection{Statistical Methods\label{sec:statistics}}

To contextualize the variety of experimental capabilities, we will first outline the principal statistical treatments used to infer the properties of the flux of neutrinos from dark matter annihilation or decay.
We will explain them in increasing order of complexity and strength.

\subsubsection{Background-agnostic methods} In this method we use the observed data and the detector signal efficiency to constrain the flux of neutrinos from DM.
This method can inform us of the maximum allowed flux, but, by construction, it cannot be used to claim the observation of dark matter.
This technique is predicated on comparing the observed and expected number of events in a given bin, by means of the following likelihood function:
\begin{equation}
   \mathcal{L}(\mu) =  \left\{
                \begin{array}{l r}
                  \mathcal{P}(d | \mu) & (d < \mu),\\
                  1 & (d \geq \mu), \\
                \end{array}
              \right.
\end{equation}
for which the likelihood is less than one only if the predicted number of events $\mu$ is larger than the recorded data, $d$.
The probability distribution $\mathcal{P}$ could be a Poisson or Gaussian distribution depending on the sample size.
Using this likelihood one can construct one-sided confidence upper limits on $\mu$ and, in turn, on the dark matter model parameters.
The strength of this method is determined by experiment exposure, signal efficiency,  and the amplitude of unmodeled backgrounds; these determine the statistical uncertainty and the phase-space over which the bins are defined.
In the case of dark matter, one would ideally bin the events in: energy, direction, and morphology; but often this is either not done due to decreasing statistical power, insufficient Monte Carlo certainty, or increasing difficulty in modeling the systematic uncertainties.

Here we take advantage of this approach in a number of experimental settings.
As examples, we compare the Super-Kamiokande unfolded neutrino energy distribution~\cite{Richard:2015aua} to the dark matter expectation using this technique and perform a similar comparison to the IceCube PeV astrophysical neutrino segmented fit.
We also use this technique when experiments have not seen neutrino events and upper limits are reported, such as the Pierre Auger Observatory's limit on neutrino flux at very high energies.

\subsubsection{Background-informed methods}

A higher statistical power can be achieved by simultaneously modeling the signal -- the event rate due to dark matter -- and background -- any other contribution to the observed rate.
This requires signal and background efficiencies, as well as a model for the background distribution over each observable.
A prototypical likelihood function is:
\begin{equation}
    \mathcal{L}(\theta, \eta) = \mathcal{P}(d | \mu_s(\theta,\eta) + \mu_b(\eta)) \Pi(\eta),
\end{equation}
where $\mu_s(\theta,\eta)$ and $\mu_b(\eta)$ are the expected signal and background counts respectively, $d$ represents the observed counts, and $\theta$ and $\eta$ are the dark matter parameters and nuisance parameters, respectively. 
The latter parameters incorporate the effect of the systematic uncertainties in the signal and background distributions and are often constrained by previous knowledge or \textit{in situ} measurements represented in the function $\Pi(\eta)$.
When the signal and background predictions are well defined, the probability function, $\mathcal{P}$, is taken to be a Poisson function in the small-count regime or a Gaussian function in the large-count regime. 

If the signal or background predictions carry large uncertainties, which is often the case for rare backgrounds or signals that cover very specific parts of phase space such as dark matter lines~\cite{Gainer:2014bta}, stochastic likelihood models can be used~\cite{Glusenkamp:2017rlp,Arguelles:2019izp,Glusenkamp:2019uir}.
For other treatments proposed to tackle this problem see also~\cite{Barlow:1993dm,Bohm:2013gla,Chirkin:2013lya}.

In either case, the treatment of systematic uncertainties is often done by using the profile likelihood method, in which the likelihood function is maximized over the nuisance parameter at each physics parameter point~\cite{doi:10.1146/annurev.nucl.57.090506.123052}.
Alternatively, in Bayesian treatments~(see \textit{e.g}.~\cite{Trotta:2017wnx}) or hybrid frequentist-Bayesian treatments~\cite{Cousins:1991qz} the nuisance parameters are marginalized over by integrating the likelihood function.
In the case that the bin content is large, such that a Gaussian likelihood function is a good approximation, the expectations can be computed accurately.
Often, a multidimensional Gaussian is used where the covariance between bins incorporates both the systematic and statistical uncertainties.
The latter approach does not require additional parameters to incorporate systematic uncertainties into the likelihood, making it computationally advantageous.

With this formalism, background-informed analyses have additional power compared to the background agnostic scenario, provided that experiments are capable of constraining the background size, and separating it from signal.
The ability to constrain background is encapsulated in systematic uncertainties, whereas the separation of background from signal depends on the features of both.
The features in the case of neutrinos from dark matter are a democratic flavor composition, spatial clustering predominantly around the Galactic center, and an energy distribution which is maximal close to the dark matter mass.
Separating dark matter from background using these three features then depends on the experimental direction and energy resolutions, as well as its flavor identification capabilities dictated by the event morphological classification.
The latter is important since natural and anthropogenic sources often have a non-democratic flavor composition.
This is a characteristic of the stronger constraints.
For example, we use the fact that for MeV dark matter one of the main backgrounds are solar neutrinos, which can be efficiently removed by selecting only for antineutrinos in Super-Kamiokande or JUNO; we also rely on this in our predictions of the sensitivities for DUNE and Hyper-Kamiokande in the 100~MeV to 30~GeV energy range, where we use the fact that one can do morphological event analysis to remove muon neutrinos which are the dominant component of the atmospheric flux.

\subsection{Neutrino Detection Methods\label{sec:detectors}}

Because neutrinos only interact via the weak nuclear force, neutrino detection must proceed in at least two steps: first, interaction between a neutrino and a detector electron or nucleus, and second, the detection of the resulting electromagnetic signal. 
Typically, energy from a gamma-ray or electron cascades down via scintillation, additional ionization or Cherenkov radiation and is subsequently measured by optical sensors or charge readout.

The small neutrino detection cross section poses a great challenge in the search for the expected fluxes from dark matter annihilation to neutrinos.
As the dark matter mass increases, larger detectors are necessary to compensate for the smaller flux, which scales as $m_\chi^{-2}$.
Such a scaling can come at the cost of energy and angular resolution, as well as flavor identification, all of which allow differentiation between the dark matter induced neutrinos from other natural or anthropogenic neutrino sources as discussed in the previous section.
In this section, we review the techniques used to detect neutrinos in different energy ranges; see also~\cite{Katori:2016yel,Diaz:2019fwt} for a discussion in the context of neutrino oscillation experiments. Note that the energy ranges detailed here are approximate, and there is naturally some overlap between techniques and physics discussed in each respective subsection.

\subsubsection{Neutrino energies below 10~MeV}

Coherent elastic neutrino-nucleus scattering, namely $\nu A^Z_N \to \nu A^{*Z}_N $, dominates the cross section at the lowest energies \cite{Freedman:1973yd}.
This process, sometimes abbreviated as CE$\nu$NS, has no kinematic threshold and scales quadratically with the atomic number.
However, the maximum recoil energies are very small making its detection difficult; in fact it has only recently been observed using anthropogenic neutrinos in detectors of $\mathcal{O}(10)$~kg of mass~\cite{Akimov:2017ade}.
Future ton-scale dark matter direct detection experiments such as DARWIN~\cite{Aalbers:2016jon} expect to see solar and atmospheric neutrinos via CE$\nu$NS.
Because of the trade-off between detector size and nuclear recoil threshold, they would only be sensitive to DM above $m_\chi \sim 10$~MeV, and provide only marginal improvement over existing dedicated neutrino experiments that use different detection channels.

Neutrino-electron scattering also has no kinematic threshold at detectable energies, and the cross section is predicted without ambiguities that arise from form factors in hadron-neutrino interactions.
This interaction's well-understood kinematics, together with the fact that a single outgoing charged particle is produced, makes it a good channel to use for DM annihilation searches. 
This is because precise energy and directional information can be inferred. 
The angle between the neutrino and the electron is tightly constrained by the kinematics, $E_e \theta_e < 2 m_e$, allowing for an accurate reconstruction of the neutrino direction (it was through this process that in 1998 the Super-Kamiokande experiment made the first image of the Sun in neutrinos~\cite{Fukuda:1998fd}; see also~\cite{Ahmad:2001an,Alimonti:2000xc,Arpesella:2008mt} for subsequent measurements by SNO and Borexino).
Angular information is used to mitigate the $\sim$ 1-10 MeV solar neutrino backgrounds and to search for correlations with the expected angular distribution of DM via $J(\Omega)$.
Unfortunately, the neutrino-electron cross section is approximately $10^{-43}~{\rm cm^2}$ at 5~MeV, which is about a factor of 10 smaller than the dominant neutrino-nucleon process.

The other commonly-used technique to detect sub-10~MeV neutrinos is inverse beta decay (IBD), $\bar\nu_e p \to n e^+$.
This is due to three reasons: first, the large and well-measured IBD cross section, approximately $10^{-42}~{\rm cm^2}$ at 5~MeV~\cite{Vogel:1999zy,Ankowski:2016oyj}, with an uncertainty of $\sim 0.2\%$~\cite{Vogel:1999zy,Kurylov:2002vj}; second, the low-threshold: $E_\nu > 1.806~{\rm MeV}$; and finally, the ability to reduce background by searching for the prompt positron signature followed by the neutron capture.
This detection method is often used with hydrocarbon-based scintillator since it contains a large number of free protons and emits large number of photons, typically $10^4$ per MeV of deposited energy~\cite{leo1994techniques}.
The energy deposited by the prompt signal is the kinetic energy of the positron plus two 511~keV gamma-rays from electron-positron annihilation, and a 2.2~MeV gamma ray from the delayed capture of the neutron on free protons.
In hydrogen-based detectors the neutron capture time is typically 300~$\mu s$.
If the detector is doped with 1\% Gadolinium, this time is reduced to about 20~$\mu s$ and the prompt gamma-ray energy is 8~MeV allowing for an improved background suppression~\cite{Beacom:2003nk}; \textit{e.g.} in the case of Super-Kamiokande a hundredfold background suppression efficiency can be achieved~\cite{Watanabe:2008ru}.
In the search for dark matter this process has the advantage that it is only triggered by $\bar\nu_e$ allowing for very efficient suppression of the solar neutrino flux that dominates the natural backgrounds at sub-10 MeV energies.
In fact, our strongest limit across all dark matter masses comes from an IBD search by Super-Kamiokande; see Fig.~\ref{fig:Indirect}. 
 
\subsubsection{Neutrino energies between 10~MeV and 1~GeV}

Between $\sim 10$~MeV and $\sim 1$~GeV, in Cherenkov detectors the proton is invisible since it is Cherenkov threshold --  approximately 1.3~GeV in mineral oil, 1.4~GeV in water, and can be as low as 1.2~GeV in the Antarctic ice~\cite{2012TCD.....6.4695B}.
This has advantages and disadvantages compared to scintillator detectors, on the one hand it simplifies identification and classification of events since the observed Cherenkov light must be associated with the outgoing charged lepton.
On the other hand, the lack of proton kinematics means that the energy and angular resolution can be greatly degraded. 
The dominant neutrino-nucleon process in this energy range is that of charged-current quasi-elastic (CCQE) scattering, namely $\nu_\alpha N \to \alpha \tilde N$ where $\alpha$ is a charged lepton and $N$ ($\tilde N$) is a proton or neutron.
At high enough energies, muon neutrinos can have CCQE interactions, producing muons which can be identified by the morphology of the Cherenkov ring.
Due to the larger mass, muons tend to preserve their direction as they travel through the detector producing sharper rings than electrons.
Cherenkov detectors can be constructed out of mineral oil, water, or ice.
Although oil-based detectors boast a larger Cherenkov angle and the ability to run without a purification system, they are only utilized in smaller detectors~\cite{Diaz:2019fwt} due to the higher filling cost.
For this reason, multi-kiloton detectors available as of 2020 are all water or ice based. As early as 2022, JUNO will become the first multi-kiloton liquid scintillator detector. 
Since the DM-induced flux is expected to be very small, the larger water or ice Cherenkov detectors currently dominate the constraints over oil-Cherenkov detectors and we will not discuss them further.

\subsubsection{Neutrino energies from 1~GeV to $10^7$~GeV}

Resonant light-meson production is important between approximately 1 and 10~GeV.
Due to the difficulty in cross section modeling, neutrino detection in this range is subject to large uncertainties.
Above 10~GeV the contribution of deep inelastic scattering (DIS), where the neutrino exchanges a $W$ or $Z$ boson with one of the partons inside the nucleon becomes the dominant process.
The production of taus in tau-neutrino charged-current interactions becomes possible above the threshold $m_\tau = 1.777$~GeV, though the cross section is only around 15\% of the charged-current muon-neutrino cross section at 10~GeV, rising to 75~\% at 100~GeV~\cite{Conrad:2010mh}.

Though unsegmented Cherenkov detectors are still used in this energy range, the use of tracking calorimeters, often constructed as segmented scintillators, are popular as they allow for improved reconstruction of outgoing muon tracks, as well as electromagnetic and hadronic showers produced in the interaction vertex.
Notable examples of these types of detectors in contemporary neutrino physics are the NO$\nu$A experiment and the T2K near-detector. 
Sampling calorimeters have also been used to increase the target density, though this comes at the expense of a degraded energy resolution.
In this case a dense material like iron is interleaved with scintillator panels.
This design was used by the MINER$\nu$A experiment~\cite{Aliaga:2013uqz} to perform precision measurements of the neutrino cross section and has been used in the past to measure neutrino oscillations by MINOS~\cite{Sousa:2015bxa}.
In these detectors the morphological features observed in the trackers have been used to identify the different neutrino interaction processes by comparing them to generated event libraries~\cite{osti_925917,Backhouse:2015xva} or convolutional neural networks~\cite{Aurisano_2016,Psihas:2019ksa}.
Given the size of these detectors they are not expected to play a role in the detection of dark matter and are not included in this work.

The newest neutrino detectors in this energy range are the so-called liquid argon time projection chambers (LArTPC)~\cite{Cavanna:2018yfk}.
These detectors consist of an electric field cage filled with liquid argon.
When a charged particle is produced in the argon, it travels through the medium and ionizes the argon atoms, liberating electrons.
An electric field then drifts the electrons to wire planes on one side of the detector, recording a projected footprint of the interaction.
Three dimensional reconstruction is also possible by using the timing of the charge deposition on the wires. 
To localize the event in the third dimension, the drift time of electrons in argon and the initial interaction time need to be known.
The initial interaction time can be known in the case of generic neutrino interactions via the scintillation light produced by the charged particles in argon or, in the case of neutrinos produced in bunches in a beam, by the beam timing.
In the case of dark matter searches, relevant for this work, only the former technique is relevant.
Even though the neutrino-argon cross section is currently poorly understood compared to other  materials conventionally used in neutrino physics, these detectors have the potential for unprecedented particle identification: see \textit{e.g.}~\cite{Acciarri:2017hat,Adams:2018bvi,MicroBooNE:2018mfn}.
Examples of currently operating LArTPC neutrino detectors are MicroBooNE~\cite{Acciarri:2016smi} and ICARUS~\cite{Ali-Mohammadzadeh:2020fbd} at Fermilab.
The next generation experiment in this category is DUNE~\cite{Abi:2020evt}.

At the higher end of this energy range, neutrino telescopes such as ANTARES and IceCube have the largest neutrino collection volumes.
These detectors operate at energies above 10~GeV where DIS is the dominant cross section process~\cite{Gandhi:1995tf}.
These detectors use natural media, such as the Mediterranean water or the Antarctic ice, as targets for the neutrino interaction.
Cherenkov light produced by charged particles by products of these interactions are then observed by photomultiplier tubes (PMTs) arranged on sparse arrays. 
In these detectors the different neutrino interactions map onto different morphologies of the time and spatial distribution of charge in the array.
Neutral-current interactions, charged-current electron-neutrino interactions, and most of the charged-current tau-neutrino interactions produce a morphology known as a cascade.
Because cascades can be contained in the detector, this morphology has the best energy resolution.
Charged-current muon neutrino interactions produce a morphology known as tracks, due to the long travel-time of the muon.
This morphology provides the best directional information.
In water, photons tend to scatter less than in ice, providing more direct light. 
This means that the muon angular resolution in water-based detectors is better than those in ice.
On the other hand, given the longer absorption length of photons in ice compared to water, the effective detector volume is larger for detectors deployed deep in the ice.
Finally, charged-current tau neutrino interactions can produce a variety of morphologies depending on the boost factor of the tau and its decay channel.
For example, around 1~PeV, a tau can travel on average 50~m before decaying producing separated energy depositions known as {\em double bangs}~\cite{Learned:1994wg,Cowen:2007ny}; 
in 2018 IceCube announced the first candidate astrophysical tau events~\cite{stachurska_juliana_2018_1301122,Stachurska:2019wfb}.
Finally, in these detectors one can also observe the electron-neutrino scattering, since at approximately 6.3~PeV an electron antineutrino can resonantly scatter with an atomic electron producing a $W$ on shell~\cite{Glashow:1960zz,Loewy:2014zva}; 
$W$-production of coherent photon scattering can also be important at these energies see~\cite{Seckel:1997kk,Alikhanov:2015kla,Beacom:2019pzs,Zhou:2019vxt,Garcia:2020jwr}. 
The observation of this process provides a unique handle on the ratio of neutrinos to antineutrinos, as well as providing exquisite energy resolution; and in fact, a candidate event has recently been detected~\cite{2019ICRC...36..945L}.

\subsubsection{Neutrino energies above $10^7$~GeV}

At extremely-high energies, the neutrino flux expected from dark matter and other astrophysical sources such as cosmogenic neutrinos is very small, necessitating the construction of detectors with effective volumes much larger than a cubic kilometer.
Neutrino interactions in this energy range occur overwhelmingly via deep inelastic scattering~\cite{Gandhi:1995tf}.
Two main techniques are used to search for neutrinos in this energy range, both of which rely on identifying horizontal or upgoing particles to mitigate the larger cosmic-ray backgrounds.
The first method involves looking for air showers induced by neutrino-nucleus interactions in the atmosphere or just below the surface of the Earth, while the second uses the radio signature produced in very-high-energy neutrino interaction~\cite{Gusev_1984,Markov:1986dx}, known as Askaryan radiation~\cite{Askaryan:1962hbi,Zas:1991jv}.
This former technique can be detected in a number of ways: sparse surface arrays of water Cherenkov tanks are used to identify charged particles from showers as they develop over an area that may span many square-km.
Air fluorescence telescopes and optical air Cherenkov telescopes can also be used alone or in combination with water tanks as is the case for Auger ~\cite{ThePierreAuger:2015rma}. 
The timing, morphology, and amount of light deposition is used to infer the energy of the incoming particle, its direction, and its nature.
In particular, a neutrino will typically travel much deeper into the atmosphere than a cosmic ray or gamma ray before interacting.
Tau neutrinos are particularly promising, as $\tau$ leptons can be produced in a nearby mountain or below the horizon~\cite{Jeong:2017mzv}.
If the tau survives the journey out of the mountain, its decay yields an upgoing air shower~\cite{Reno:2019jtr,Reno:2019qmk}; an EeV $\tau$ typical interaction length is a few kilometers in rock and is shorter than its decay length.
The expected event rate for such processes at cosmic ray observatories like Auger turns out to be higher than from neutrino-induced atmospheric showers, thanks to the high density of rock.
Radio arrays such as GRAND~\cite{Alvarez-Muniz:2018bhp} have been proposed to cover as large an effective area as possible (up to two-hundred thousand square-km) to search for such a signal.

The second method, Askaryan radiation detection, aims to observe neutrinos via the radio emission generated by charge displacement caused by the developing electromagnetic or hadronic shower after DIS scattering.
This emission is distinct from down-going cosmic-ray showers in that the polarization of the radio signal is expected to be different.
This technique has been implemented by using radio antennae either suspended from balloons~\cite{Gorham:2010kv} or buried in the ice~\cite{Anker:2020lre,Allison:2019xtn} in the Antartic continent.
The ability to cover a large area with a single antenna cluster makes this a very scalable and relatively low-cost technique. 

\section{Results\label{sec:results}}

\begin{figure*}[htp!]
\centerline{\includegraphics[width=\linewidth]{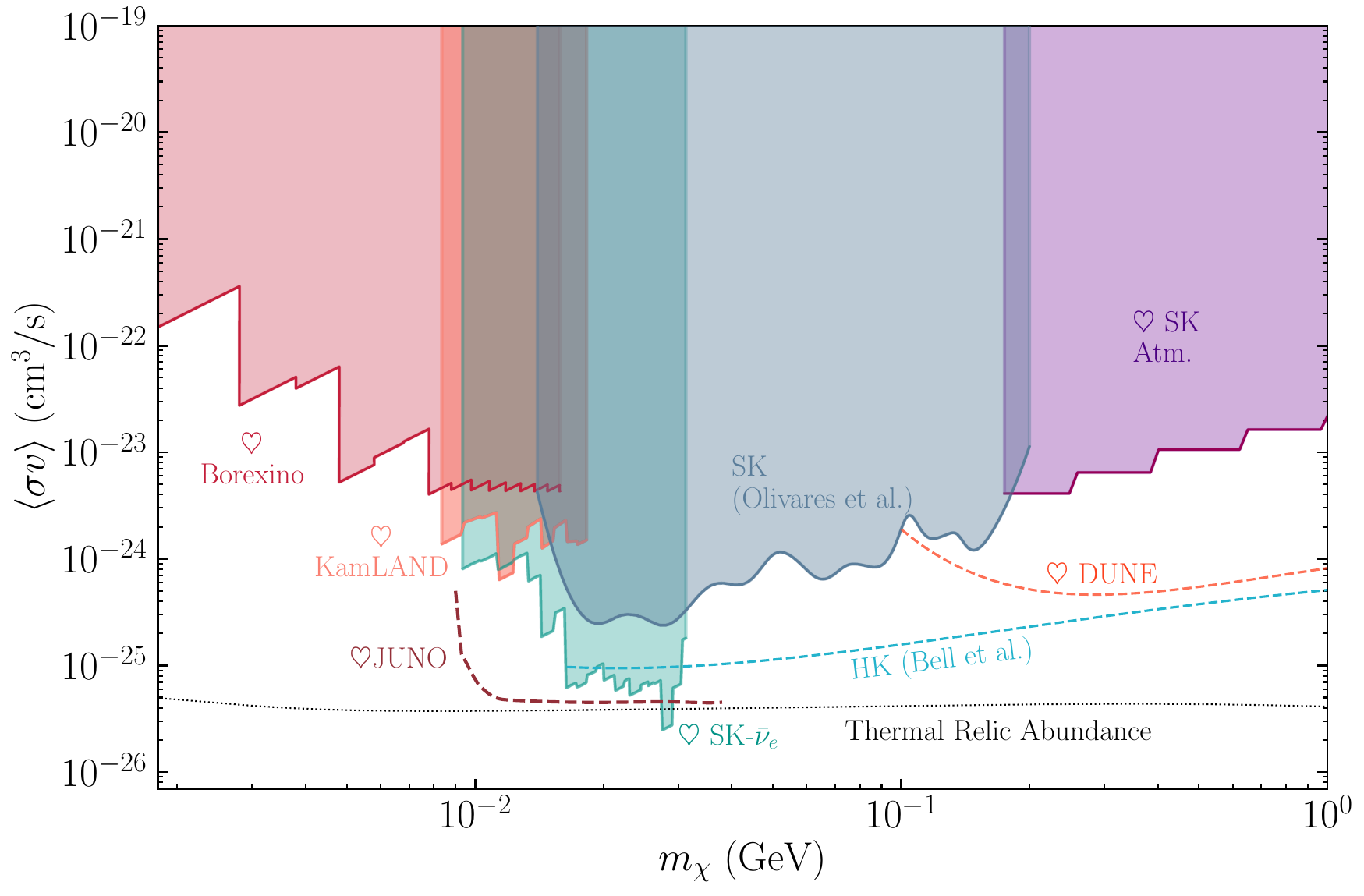}}
%\internallinenumbers
\caption[The landscape of sub-GeV dark matter annihilation into neutrinos.]{The landscape of sub-GeV dark matter annihilation into neutrinos.
Same as Fig.~\ref{fig:Indirect}, but restricted to dark matter masses below one GeV.
}
\label{fig:Indirect_lowe}
\end{figure*}

\begin{figure*}[ht!]
\centerline{\includegraphics[width=\linewidth]{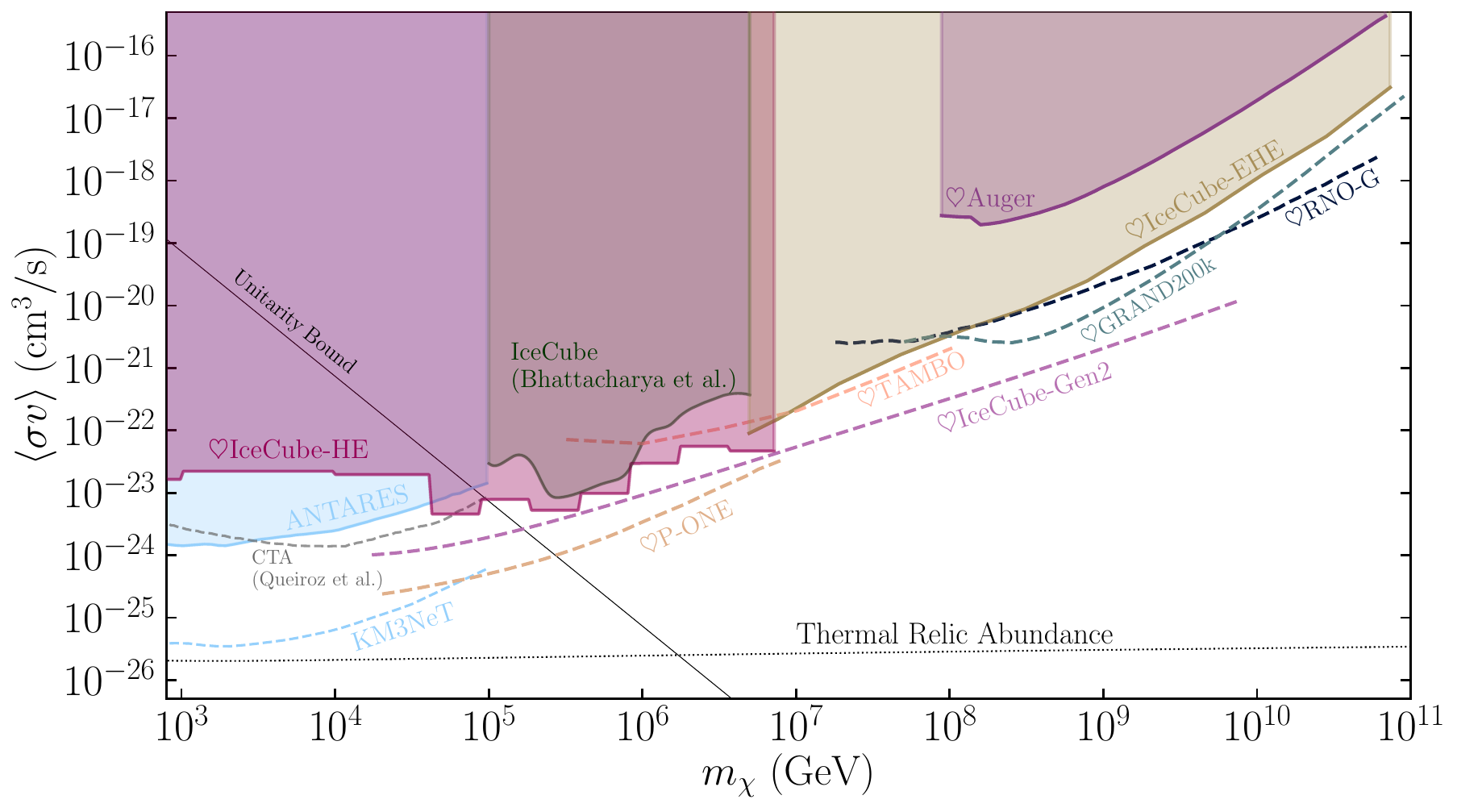}}
%\internallinenumbers
\caption[The landscape of supra-TeV dark matter annihilation into neutrinos.]{The landscape of supra-TeV dark matter annihilation into neutrinos.
Same as Fig.~\ref{fig:Indirect}, but for the high-mass region.
All the experimental constraints in this plot are calculated by converting either the detected flux or the reported upper limit into a conservative upper bound on the DM annihilation cross section.
}
\label{fig:Indirect_non_unitarity}
\end{figure*}

Our main results are shown in Figs.~\ref{fig:Indirect}-\ref{fig:Indirect_dwave}.
Fig.~\ref{fig:Indirect} shows the results derived according to the procedures described in Secs.~\ref{sec:galactic}, in addition to previous results available in the literature.
Fig.~\ref{fig:Indirect_lowe} shows a more detailed view of the low-mass (sub-GeV) range; Fig.~\ref{fig:Indirect_non_unitarity} shows results for the high-mass (10$^3$-10$^{11}$ GeV) region.
Finally, Figs.~\ref{fig:Indirect_pwave}-\ref{fig:Indirect_dwave} provide the constraints and projections in the case of velocity-dependent p-wave and d-wave annihilation, respectively.
Further, Fig.~\ref{fig:DecayLimits} shows results for decay of dark matter according to procedures described in Sec.~\ref{sec:decay}.
We label the results derived specifically for this work with a heart ($\heartsuit$). 

\renewcommand{\arraystretch}{1.5}

In the rest of this section, we describe the data that we used to produce or recast limits according to the procedures outlined in Secs.~\ref{sec:theory} and ~\ref{sec:decay}.
We split the data into three lists: 1) data used to construct constraints in Fig.~\ref{fig:Indirect} and Fig.~\ref{fig:DecayLimits}; 2) previous limits that we have recast; and 3) data used to place limits in the high mass ($m_\chi > 10^3$ GeV) region.
%%%%% BEGIN RESULT DISCUSSIONS

When reporting literature results, where possible, we have rescaled them to use the same halo parameters, \textit{i.e.} consistent $J$- and $D$-factors, as computed in Sec.~\ref{sec:galactic} and Sec.~\ref{sec:decay}.
In this way, we ensure that the constraints we present can be properly compared one with another.
The rescaling could not be done in the case of ANTARES~\cite{Adrian-Martinez:2015wey}, SK~\cite{Frankiewicz:2017trk}, and IceCube~\cite{Aartsen:2016pfc}, since these were event-by-event analyses for which data is not publicly available.
This is unfortunate since the halo parameters used in these studies are no longer preferred  (see discussion in Sec.~\ref{sec:haloparams}).
Shaded regions correspond to experimental limits, whereas dashed lines are projections based on future experimental sensitivity.
Finally, in the case of DM annihilation we include two lines for reference.
First, the dotted black line corresponds to the cross section required to produce the observed relic abundance from thermal freeze-out computed as in \cite{Steigman:2012nb}, and second, the solid black line labeled ``unitarity bound'' corresponds to the perturbative unitarity limit on non-composite WIMP dark matter~\cite{Griest:1989wd}; see~\cite{Smirnov:2019ngs} for a recent discussion.

%%%%%%%%%% Data we used to make limits, from left to right
The limits shown in Figs.~\ref{fig:Indirect} and \ref{fig:DecayLimits}, employing the approach of Secs.~\ref{sec:galactic} and~\ref{sec:decay} use the following data, which we also summarize in Tbl.~\ref{tab:table3}. 
\begin{enumerate}
    \item \textbf{\textit{Borexino:}} Borexino is a large-volume unsegmented liquid scintillator detector located underground at the \textit{Laboratori Nazionali del Gran Sasso} in Italy~\cite{Alimonti:2008gc}. 
    The collaboration has released two event selections: one which has a livetime of 736 days selecting electron-antineutrino candidate events over the entire fiducial volume and another one with 482 days of livetime designed to search for geo-neutrinos~\cite{Bellini:2010hy}.
    These event selections are combined into a single set designed to obtain a pure sample of electron-antineutrinos by means of searching for signatures of inverse beta decay.
    Using this selection, they derive upper limits on the all-sky monochromatic electron-antineutrino flux ranging from $\sim10^5$ to $\sim10^2$ $\bar\nu_e{\rm cm}^{-2}{\rm s}^{-1}$, for energies ranging from $\sim2$ to $17$ MeV, respectively.
    We use the flux upper limits produced by ~\cite{Bellini:2010gn} and recently updated by \cite{Agostini:2019yuq} and compare it with one-sixth of the all-flavor expected flux from dark matter to set our constraints.
    \item \textbf{\textit{KamLAND:}} KamLAND is an unsegmented liquid scintillator detector located in the Kamioka observatory near Toyama, Japan.
    The approximately one kiloton of mineral oil fiducial volume is contained in a 13 meter balloon.
    Beyond its well-known work on reactor neutrinos, KamLAND has measured the $^{8}$B solar spectrum~\cite{Abe:2011em}, searched for geoneutrinos~\cite{Gando:2013nba}, and placed limits on the flux of extraterrestrial neutrinos above $\sim 8.3~{\rm MeV}$~\cite{Collaboration:2011jza} which constrains the supernovae relic neutrino flux. 
    In the latter work, an upper limit on the extraterrestrial flux of $\bar\nu_e$ is derived, which is at the $\mathcal{O}(10)~\bar\nu_e~{\rm cm}^{-2}{\rm s}^{-1}{\rm MeV}^{-1}$ level and is given from 8.3 MeV to 18.3 MeV.
    Using this result, we derive a constraint on the dark matter parameters, shown in salmon.
    Note that in~\cite{Collaboration:2011jza}, the KamLAND collaboration also derives a similar constraint, but with outdated $J$-factors; their result and ours are comparable.
    These are the leading constraints in the $\sim$10 MeV mass range, but we expect that they will be improved by the next-generation liquid scintillator detector in China, JUNO~\cite{An:2015jdp}.
    \item \textbf{\textit{SK:}} Super-Kamiokande (SK) is a 50kt ultrapure water Cherenkov detector located in Kamioka, Japan~\cite{Fukuda:2002uc}. 
    SK can use the morphology of the Cherenkov ring produced by charged particles to perform particle identification, energy measurement, and obtain directional information of the events.
    The unfolded electron- and muon-neutrino fluxes in the sub-GeV to several TeV energy range has been published by SK~\cite{Richard:2015aua}.
    This unfolding uses data from the four stages, SK-I, SK-II, SK-III, and SK-IV, resulting in a total livetime of 4799 days for the fully contained and partially contained event selection and 5103 for the upward-going muon sample. 
    The unfolded fluxes are expected to be dominated by the atmospheric neutrino flux; in fact they are in agreement with model predictions, {\it e.g.} the HKKM model~\cite{Honda:2006qj}, within systematic uncertainties.
    The dominant source of uncertainties on the unfolded fluxes is the neutrino interaction cross section, which introduces an uncertainty of approximately 20\% in the unfolded flux.
    In the case of electron-neutrinos, the second largest uncertainty is due to the small statistics at high energies; which can be up to 10\% in the highest energy bins. 
    For all flavors, all other sources of uncertainty are less than 5\% across all energy bins. 
    We compare the unfolded flux with the expected flux from dark matter to produce limits on Galactic and extragalactic dark matter annihilation and decay. These results are shown in purple in Figs.~\ref{fig:Indirect}, ~\ref{fig:DecayLimits},~\ref{fig:Indirect_pwave}, and ~\ref{fig:hmfparam}, and labeled as \textit{\textbf{$\heartsuit$SK-Atm.}}
    In order to obtain these limits we used a background-agnostic approach as described in Sec.~\ref{sec:statistics}, and a binned truncated Gaussian likelihood in energy with two degrees of freedom.
    Additionally, we perform an analysis using 2853 days of low energy data from SK I/II/III, as well as 2778 days of data from SK phase IV, which led to an upper limit on the relic supernova electron antineutrino ($\bar \nu_e$) flux~\cite{WanLinyan:2018}; labeled \textit{$\heartsuit$\textbf{SK-$\mathbf{\bar \nu_e}$}}.
    The resulting limits on $\sv$ turn out to be the strongest over the entire mass range that we consider, crossing the relic abundance line for masses between 27 and 30~MeV.
    \item \textbf{\textit{IceCube}}: The IceCube Neutrino Observatory is a gigaton ice Cherenkov neutrino detector located at the geographic South Pole~\cite{Aartsen:2016nxy}.
    IceCube has measured the atmospheric neutrino spectrum in the 100~GeV to 100~TeV energy range.
    By separating the events into their observed morphologies (``cascades'' and ``tracks''), the collaboration recently published the unfolded electron- and muon-neutrino flux in this energy range~\cite{Aartsen:2015xup,Aartsen:2016xlq}.
    At energies greater than 60~TeV, using events whose interaction vertex starts in the inner part of the detector~\cite{Aartsen:2013jdh,Schneider:2019ayi}, they have also reported the result of a piece-wise power-law fit to the astrophysical neutrino component using more than six years of data~\cite{Aartsen:2017nbu}.
    We use these to produce background-agnostic limits comparing the produced neutrino flux from DM with the reported unfolding or spectral fits.
    The obtained limits are shown for dark matter masses from 200~GeV to 10~PeV, labeled \textit{\textbf{$\heartsuit$IceCube-HE}} and colored in dark magenta.
    Limits use the same likelihood construction as in the case of the SK limits described above. Note that the muon neutrino atmospheric unfolding reported by IceCube uses northern tracks, which are unfortunately in the wrong hemisphere for the Galactic center. Therefore, for that sample, we only constrain extragalactic emission.
    Dedicated neutrino line searches have not been yet performed by the IceCube collaboration, although sensitivities have been estimated in~\cite{ElAisati:2017ppn,ElAisati:2018vkn} to be stronger than current IceCube constraints in that region.
    We describe the region labeled \textit{\textbf{IceCube-EHE}} below, in the description of the high-mass region.
\end{enumerate}

\noindent Additionally, we use the following previously-published limits on dark matter annihilation and decay obtained by constraining the Galactic flux, rescaled to account for the galactic halo parameters used here unless indicated otherwise:
\begin{enumerate}
    \item \textbf{\textit{Super-Kamiokande diffuse supernovae flux search}}: The gray region labeled \textit{\textbf{SK Olivares et al.}} is an independent analysis of SK all-sky low-energy data which uses SK phases I through III to derive an upper bound on the supernova relic neutrinos~\cite{Hosaka:2005um,Cravens:2008aa,Abe:2010hy}. 
    This analysis covers neutrino energies from 10 MeV to 200 MeV; see~\cite{Li:2014sea} for a recent discussion of backgrounds in the low-energy range.
    The upper limit on supernova relic neutrinos was then converted into dark matter annihilation constraints, and was originally presented in~\cite{Campo:2017nwh,Campo:2018dfh,Olivares-Del-Campo:2019qwe}.
    Recently, SK phase-IV data has placed new constraints on the $\bar\nu_e$ flux in the 10 to 30~MeV energy range~\cite{WanLinyan:2018}.
    These observations improve over KamLAND constraints~\cite{Collaboration:2011jza} by a factor between 3 and 10 in their overlapping energy range.
    Thus these observations dominate the constraints for dark matter masses below $\sim 20~{\rm MeV}$.
    Where they overlap, the \textit{Olivares et al.} limits are not quite as strong as the \textit{SK-$\bar \nu_e$} limits that we have presented, because their background modelling could not use angular information which is not publicly available.
    \item \textbf{\textit{Super-Kamiokande Galactic dark matter search}}: The teal region, labeled \textit{\textbf{SK}}, is from~\cite{Frankiewicz:2015zma}.
    This analysis uses muon-neutrino data in the energy range between 1 GeV and 10 TeV collected by SK over 5325.8 days.
    Since this analysis relies on angular information that is not public, it has not been rescaled to account for our choice of galactic halo parameters.
    \item \textbf{\textit{IceCube/DeepCore Galactic dark matter search}}: The IceCube limits are from~\cite{Aartsen:2016pfc} and use 329 days of IceCube data.
    These  place constraints for masses in between 25~GeV and 10~TeV.
    At the lowest masses, these limits include data from DeepCore, an array of more closely spaced inner strings in  IceCube.
    In addition, we include a limit derived from 3 years of data using primarily tracks to constrain Galactic center emission~\cite{Aartsen:2017ulx}.
    For display purposes, we join these two lines, choosing the best limit at each point, and show it in navy blue, simply labeled as \textit{\textbf{IceCube}}.
    \item\textbf{\textit{IceCube-Bhattacharya et al.}} is taken from ~\cite{Bhattacharya:2019ucd}'s channel-by-channel unbinned likelihood analysis of the High-Energy Starting Event (HESE) data, including energy, angular, and topology information.
    They include both Galactic and extragalactic constraints.
    Constraints that we derive (IceCube-HE) using only spectral information follow these limits quite closely at higher energies since the small sample size prevent angular information from contributing significantly.
    \item \textbf{\textit{ANTARES dedicated Galactic dark matter search}}: The light blue region, labeled ANTARES, is from a Galactic center analysis of nine years of ANTARES muon neutrino and antineutrino data~\cite{Albert:2016emp,Adrian-Martinez:2015wey}. This covers the dark matter mass range from $53$ GeV to $100$ TeV.
    \item \textbf{\textit{Baikal dedicated Galactic dark matter search} (not shown)}: The Baikal underwater neutrino telescope~\cite{Belolaptikov:1997ry,Aynutdinov:2006ca}, NT-200, is a water Cherenkov detector deployed in Lake Baikal, Russia.
    It has an instrumented volume of approximately 100~kt and is comprised of 192 optical modules arranged on eight strings, with a typical distance between strings of 21~m.
    The collaboration performed an analysis looking for dark matter annihilation in the Galactic center into neutrinos using data recorded between April of 1998 to February of 2003~\cite{Avrorin:2015bct}.
    This analysis claimed to place limits on the cross section at the $10^{-22}~{\rm cm}^3 {\rm s}^{-1}$ level for a 1~TeV dark matter mass.
    We do not add this result to our constraint summary because there are stronger results in this mass range, but we do show the projections of the next generation detector at Lake Baikal, GVD.
\end{enumerate}
Finally, Fig.~\ref{fig:Indirect}, includes next-generation sensitivities that can be reached by future experiments. These are shown as dashed lines:
\begin{enumerate}
\item \textbf{\textit{DUNE}}: The Deep Underground Neutrino Experiment (DUNE) far detector will be a 46.4 kiloton liquid argon Time Projection Chamber (TPC)~\cite{Acciarri:2015uup,Abi:2020wmh} constructed at the Sanford Underground Research Facility (SURF) in South Dakota, USA.
Its main advantage in detecting neutrinos from DM annihilation and decay is its improved particle identification, using morphological reconstruction, with respect to Cherenkov detectors like Super-Kamiokande, ANTARES, or IceCube, which \textit{e.g.} can be exploited to make improved measurements of solar neutrinos~\cite{Capozzi:2018dat}.
Thus, a dedicated DUNE analysis utilizing the expected improved directional capability can prove effective in a search for Galactic dark matter annihilation to neutrinos.
We derive projected sensitivities for dark matter masses in the range from 100~MeV to 30~GeV and show them in Figs.~\ref{fig:Indirect} and \ref{fig:DecayLimits} as dashed orange lines.
The dominant background in this energy range is from atmospheric neutrinos.
We use the predictions provided by \cite{PhysRevD.92.023004} at the Homestake gold mine at SURF, taking into account oscillations through the Earth using the nuSQuIDS package~\cite{Delgado:2014kpa,Arguelles:2020hss,nusquids}.
In our analysis, we consider $e$- and $\tau$-flavored charged-current interactions and compare the expected energy distribution; \textit{i.e.} we do not take into account event-by-event directional information.
We use a fractional charged lepton energy resolution of $2\% + 15\%/\sqrt{E/{\rm GeV}}$~\cite{Acciarri:2015uup} and assume the idealized condition of $100\%$ efficiency.
In our analysis, charged-current electron-neutrino interactions are assumed to deposit all their energy in the detector, while tau-neutrino charged-current interactions will deposit less visible energy due to the invisible neutrinos produced in the prompt $\tau$ decay.
Since we expect that DUNE morphological identification will be able to single out muon-neutrino charged-current processes, we choose to remove them from the analysis as they are the primary contributor to the atmospheric neutrino background.
Limits are derived using a binned Poisson likelihood and a background-informed method as described in Sec.~\ref{sec:statistics}.
We note that, due to liquid argon TPC's morphological reconstruction capabilities, a proper Galactic center analysis including directionality would benefit from the inclusion of muon-neutrino charged-current interactions, and thus our projections are conservative.
\item \textbf{\textit{Hyper-Kamiokande}}: Building on SK's technology, a new water Cherenkov detector with a fiducial mass of 187~kton called Hyper-Kamiokande (HK) will be built in Kamioka, Japan~\cite{Abe:2018uyc}.
Due to its larger size, this detector will be able to place stronger limits on the DM annihilation and decay to neutrinos than its predecessor~\cite{Campo:2018dfh}.
In fact, Hyper-Kamiokande is estimated to reach $\sim 10^{-25}~{\rm cm}^3 {\rm s}^{-1}$ for $1~{\rm GeV}$ dark matter and $\sim 10^{-22}~{\rm cm}^3 {\rm s}^{-1}$ at $10^4~{\rm GeV}$ with ten years of data taking~\cite{Migenda:2017tej}.
Furthermore, the possibility of doping both the SK and the HK detectors with gadolinium (Gd) will reduce the dominant background for low-energy analyses by a factor of five and, consequently, improve the constraints on DM parameters~\cite{Horiuchi:2008jz,Laha:2013hva,Bell:2020rkw}.

Similar to our DUNE analysis, we assume that the dominant background in this energy range is due to atmospheric neutrinos, where we use the predictions provided by \cite{PhysRevD.92.023004} at the Kamioka mines, and allow these neutrinos to oscillate through the Earth using the nuSQuIDS package~\cite{Delgado:2014kpa,nusquids}.
We only consider $e$- and $\tau$-flavored charged-current interactions, without taking into account directionality.
We make the same assumptions as our DUNE analysis regarding energy deposition, while using an energy resolution of $1.5\% + 2\%/\sqrt{E / {\rm GeV}}$~\cite{Jiang:2019xwn}. We use total energy rather than lepton (visible) energy, which leads to a sensitivity overestimate of $\sim 40\%$ but simplifies the analysis. In principle, it is be possible to record lepton and proton energy above the proton Cherenkov threshold, [see e.g~\cite{Fechner:2009aa}.
We follow the same statistical procedure as in DUNE and, like DUNE, the sensitivity strength derives primarily from the expected electron- and tau-neutrinos signal.
Taking advantage of this channel explains why our estimates are better than ones presented by \cite{Migenda:2017tej}; see \cite{Beacom:2004jb} for a discussion on ``shower power.''
These projected sensitivities, especially at low energies, are subject to a $\sim 30\%$ uncertainty due to a combination of atmospheric background uncertainties and neutrino cross sections.
\item \textbf{\textit{JUNO}}: The Jiangmen Underground Neutrino Observatory~\cite{An:2015jdp} is a 20~kt unsegmented liquid scintillator detector under deployment in the Guangdong province of China.
The detector has a muon tracker on top of it and is also surrounded by water. 
Both of these systems can be used to veto cosmic-ray muons by either tagging them in the muon tracker or by detecting their Cherenkov light in water.
Due to its large volume and good energy resolution (estimated to be $3\%/\sqrt{E/{\rm MeV}}$) we expect that this experiment will have good sensitivity for neutrino line searches.
We estimate the sensitivity of JUNO to dark matter annihilation to neutrinos in the electron antineutrino channel following the proposal given in \cite{PalomaresRuiz:2007eu}.
We use background estimates derived for diffuse supernova background searches, as presented in \cite{An:2015jdp}.
Below 11 MeV, reactor antineutrinos dominate the background.
Between 11 and 40 MeV, the backgrounds are primarily neutral current interactions from atmospheric neutrinos, with sub-dominant charge current contributions.
According to our projection, JUNO is expected to constrain the velocity-averaged annihilation cross section better than $10^{-25}~{\rm cm}^3 {\rm s}^{-1}$ in the 10 to 40~MeV mass range.
The estimate is shown in dark red in Fig.~\ref{fig:Indirect}, and its equivalent dark matter lifetime is shown in Fig.~\ref{fig:DecayLimits}.

\item \textbf{\textit{IceCube Upgrade}}: The IceCube Upgrade is an extension of the current IceCube/DeepCore array with seven closely-packed strings.
These new strings will be separated by approximately 20 meters and each contain 100 photomultiplier tubes spaced vertically by 3~meters~\cite{Ishihara:2019aao}.
Additionally, a number of calibration devices and sensors will be deployed to improve the modelling of the ice~\cite{Nagai:2019uaz,Ishihara:2019uei}.
In ~\cite{Baur:2019jwm} a preliminary estimation of the IceCube Upgrade sensitivity was performed. It is expected to be better than $10^{-24}~{\rm cm}^3 {\rm s}^{-1}$ for a 10~GeV dark matter mass.
\item \textbf{\textit{IceCube Gen-2}}: The next-generation ice Cherenkov neutrino observatory in Antarctica is a substantial expansion to the current IceCube observatory, aiming at enhancing the detector volume by a factor of ten~\cite{Aartsen:2014njl}.
This increased effective area is expected to provide a better sensitivity to resolve sources of high-energy cosmic neutrinos and identify components of cosmic neutrino flux.
Dark matter limits from IceCube presented here should therefore scale by at least the increased sample size due to the larger effective area.
We have recast the estimates of diffuse flux sensitivity given in~\cite{Aartsen:2019swn} to estimate the sensitivity to dark matter.

\item \textbf{\textit{Baikal-GVD}}: The Baikal Gigaton Volume Detector (GVD) is a planned expansion to the existing NT-200 detector, and is currently being deployed in Lake Baikal, Russia.
The detector has recently reached an effective volume of $\sim0.35$~km$^3$ and has already seen first $\nu$-light~\cite{Avrorin:2019vfc}.
The full array will contain 10,386 optical modules divided among 27 clusters of strings, and is expected to have a final instrumented volume of around $1.5$~km$^3$.
The sensitivity of GVD to Galactic dark matter has been estimated in~\cite{Avrorin:2014vca} and is shown as a dashed brown line labeled \textbf{GVD}.
\item \textbf{\textit{KM3Net}}: The km$^3$-scale water Cherenkov detector currently under construction in the Mediterranean sea is designed to provide high-purity increased effective areas in the Southern Hemisphere.
The larger effective area and improved angular resolution, compared to ANTARES, are expected to provide better constraints on dark matter.
Two separate sites are under construction for low- and high-energy regimes~\cite{Adrian-Martinez:2016fdl}.
The high-energy site, called KM3NeT/ARCA, will consist of two detector array blocks located approximately 100~km offshore from \textit{Porto Palo di Capo Passero}, Sicily, Italy~\cite{Aiello:2018usb}.
Each block is expected to have 115 strings with an average spacing of 90~m.
The low-energy site, called KM3NeT/ORCA, consists of one array block and is under deployment approximately 40~km south of Toulon, France; close to the ANTARES site.
The array is made out of 115 strings with an average horizontal spacing of 20~m.
Each string contains 18 optical modules; in KM3NeT/ARCA they are spaced vertically by 36~m, while in KM3NeT/ORCA they are spaced 9~m.
The horizontal spacing and number of strings are proportional to the effective volume of the experiment, while the vertical spacing is related to the energy threshold~\cite{Halzen:2005qu}.
KM3NeT/ARCA's science program is mainly oriented towards higher-energy (astrophysical) neutrino searches, while KM3NeT/ORCA will measure neutrino oscillations using atmospheric neutrinos.
Assuming an $E^{-2}$ democratic-flavor astrophysical neutrino flux with a normalization of $\sim 1.8\times 10^{-8} {\rm GeV}^{-1}{\rm s}^{-1}{\rm cm}^{-1}{\rm sr}^{-1}$ and an exponential cut-off at 3~PeV they expect to see 11~$\nu_\mu$'s, 41~$\nu_e$'s, and 26~$\nu_\tau$'s in five years of KM3NeT/ARCA operation~\cite{Adrian-Martinez:2016fdl}.
In Figs.~\ref{fig:Indirect} and ~\ref{fig:DecayLimits} we show the KM3NeT/ARCA expected sensitivity to dark matter annihilation to neutrinos in five years of data taking~\cite{gozzini2019search}.
Their sensitivity is within a factor of a few from the expected relic abundance cross section for dark matter masses around a TeV.
\item \textbf{\textit{P-ONE}}: The Pacific-Ocean Neutrino Experiment (P-ONE) is a newly proposed multi-cubic kilometer neutrino detector utilizing sea water as Cherenkov medium~\cite{Agostini:2020aar}.
P-ONE would be deployed in the Cascadia Basin, off the coast of Vancouver island in the Pacific Ocean, taking full advantage of the Ocean Network Canada infrastructure and expertise already in place.
The main goal of the experiment is to explore the origin of the extraterrestrial neutrino flux.
A pair of test strings, named STRAW~\cite{Bedard:2018zml}, has already been successfully deployed and has collected water absorption data.
The first phase of the detector, known as the Pacific Ocean Neutrino \textit{Explorer}, involving ten strings is planned to begin deployment in 2023.
Each string is planned to be equipped with twenty photomultiplier tubes.
The full detector is expected to be complete by 2030 with 70 strings.
Projected limits include backgrounds from atmospheric and diffuse astrophysical neutrinos, and use the exposures shown in~\cite{Agostini:2020aar}.
\item \textbf{\textit{TAMBO}}: The Tau Air-Shower Mountain-Based Observatory is a proposed array of small water-Cherenkov tanks to be deployed on either the Colca Valley or Cotahuasi Canyon in Peru~\cite{Wissel:2019alx,Romero-Wolf:2020pzh}.
These are two of the world's four deepest valleys and their unique geometry allows for efficient detection of Earth-skimming PeV $\nu_\tau$.
Most of the Colca Valley runs along a North-South corridor, though a smaller section of it has an East-West corridor.
If deployed in the East-West corridor of the Colca valley, the declination band covered is $-15.5 \pm 10$ degrees, while in the North-South corridor it would be $-15.5\pm 50$ degrees.
These two provide two extreme configurations in terms of its GC exposure, while a deployment in the Cotahuasi canyon, which has an approximately diagonal corridor, would provide an intermediate exposure. 
TAMBO's effective area is expected to be 10 times larger than IceCube $\nu_\tau$~\cite{Aartsen:2013jdh} at a PeV and 30 times larger at 10 PeV.
The use of the Earth-skimming technique is complementary to very-high-energy Earth-traversing neutrino searches~\cite{Safa:2019ege} and the fact that it relies on the Cherenkov effect, rather than the higher energy threshold Askaryan effect, gives it unique potential to constrain dark matter in the tens of PeV mass range.
Depending on the final geometry of TAMBO its sensitivity to dark matter ranges from $10^{-22}~{\rm cm}^3~{\rm s}^{-1}$ to $4 \times 10^{-21}~{\rm cm}^3~{\rm s}^{-1}$ for a 1 PeV dark matter mass.
Sensitivities shown here are recast from the diffuse flux sensitivity presented by~\cite{Wissel:2019alx}.

\item \textbf{\textit{CTA}}: The Cherenkov Telescope Array is a planned network of 99 air Cherenkov telescopes in the southern hemisphere and 19 in the northern hemisphere that will collectively provide full-sky coverage of the gamma ray sky over an energy range from 20 GeV to 300 TeV~\cite{Acharya:2017ttl}.
Several CTA prototypes have been built and some have already seen first light. The telescopes are projected to have an angular resolution down to 0.1 degrees and a duty cycle of $\sim15\%$.
For high-mass dark matter annihilation into neutrinos, electroweak final-state radiation can also lead to the production of gamma rays, despite a completely ``invisible'' $\nu \bar \nu$ final state, and can thus be constrained by gamma ray observations of the Galactic center with CTA; see Sec.~\ref{sec:theory} for more details.
The expected limits from CTA were computed in~\cite{Queiroz:2016zwd}, and shown as a dashed silver line assuming 100 hours of observation.
\end{enumerate}

We note that the 10 MeV -- 1 GeV range can in principle be covered by future tonne-scale dark matter direct detection experiments such as DARWIN and ARGO~\cite{McKeen:2018pbb}.
However, these are still in their planning phases, meaning that construction is still decades away, and very long ($\gtrsim$ 10 years) exposure times are required to be competitive with HyperK. For this reason we do not show them here. 

Fig.~\ref{fig:Indirect_non_unitarity} shows the extension of available constraints to larger masses, above the ``unitarity bound,'' accessible \textit{e.g.} for composite DM models~\cite{Frigerio:2012uc}.
These bounds are calculated by converting either the detected flux or reported upper limits, from observatories sensitive to these mass range, into a conservative upper bound on the DM annihilation to neutrinos.  The following experiments are sensitive to this regime:
\begin{enumerate}
\item \textbf{\textit{Auger}}: The Pierre Auger Observatory is a hybrid detector consisting of both an array of water Cherenkov surface detectors and atmospheric fluorescence detectors.
Located in Malarg\"ue, Argentina~\cite{ThePierreAuger:2015rma} and operational since 2004, the collaboration has made a multitude of measurements of the highest energy cosmic rays.
This includes measurements of the spectral distribution of cosmic rays beyond the GZK limit, anisotropy searches, as well as fits to their mass composition.
Beyond the extensive cosmic ray program, Auger is able to probe extremely-high-energy neutrinos by searching for showers developing deep in the atmosphere, since showers induced by cosmic rays are likely to develop much earlier.
Another possible detection channel is upgoing tau lepton showers, which are induced by Earth-skimming tau neutrino interactions near Earth's surface.
In 2017, the collaboration reported a limit on the diffuse flux of high energy neutrinos between $10^{8}-10^{11}$ GeV~\cite{Zas:2017xdj} which we use to set a background-agnostic bound on $\sv$ and $\tau$ for such energies (purple line in Figs.~\ref{fig:Indirect_non_unitarity} and ~\ref{fig:DecayLimits}).
\item \textbf{\textit{IceCube-EHE}}: Beyond the astrophysical neutrino flux, IceCube performs searches for GZK neutrinos using a dedicated sample of events that deposit extremely high energies (EHE) in the detector.
The most recent search used nine years of data and set limits on the GZK flux.
We use these limits~\cite{Aartsen:2018vtx} to derive an bounds on the DM annihilation cross section and decay lifetime between $10^{7}-10^{11}$ GeV, represented by a light brown line in Figs.~\ref{fig:Indirect_non_unitarity} and ~\ref{fig:DecayLimits}).
\item \textbf{\textit{GRAND}}: The Giant Radio Array for Neutrino Detection is a proposed large-scale observatory consisting of 200,000 radio antennas covering 200,000 km$^2$ near a mountain range in China.
This experiment plans to use the surrounding mountains as a target for Earth-skimming tau neutrinos.
After the neutrinos interact in the mountain, a tau lepton should be observed exiting the mountain and subsequently decaying in the atmosphere.
The immense coverage will allow GRAND to probe GZK neutrino fluxes that are at least an order of magnitude below current limits~\cite{Alvarez-Muniz:2018bhp}.
We convert their 3-year sensitivity to the GZK neutrino flux between $10^{8}-10^{11}$ GeV into sensitivities on $\sv$ and $\tau$ shown as a dashed navy blue line in Figs.~\ref{fig:Indirect_non_unitarity} and ~\ref{fig:DecayLimits}.
\item \textbf{\textit{RNO-G}}: The Radio Neutrino Observatory in Greenland aims to measure the neutrino flux above $10^{16}$~eV~\cite{Aguilar:2019jay}.
The array of antennas to be deployed in the ice are designed to detect the Askaryan radio emission from extremely high-energy neutrinos traversing the Earth and atmosphere.
The design and deployment of RNO relies upon the experience and expertise obtained in successful deployment and operation of ARA and ARIANNA~\cite{2012APh....35..457A,Barwick:2014pca}.
The plan is to deploy 35 stations such that each station will consists of a surface array and a deep array.
The surface array is going to be used for cosmic-ray detection while the deep array, benefiting from a large effective volume, will detect neutrinos.
\end{enumerate}

\begin{table*}[!ht]
\begin{center}
{
\begin{tabular}{ |c|c|c|c| } 
 \hline
 \textbf{Energy Range} & \textbf{Experimental Analysis} & \textbf{Directionality} & \textbf{Detected Flavor} \\ \hline 
 $2.5-15$ MeV & Borexino \cite{Bellini:2010gn}& $\boldsymbol{\times}$ &  $\bar{\nu}_e$ (IBD) \\ \hline
 $8.3-18.3$ MeV & KamLAND \cite{Collaboration:2011jza} &  \checkmark &  $\bar{\nu}_e$ (IBD) \\\hline
 {$10-40$ MeV} & JUNO \cite{An:2015jdp} &  \checkmark &  $\bar{\nu}_e$ (IBD) \\\hline
  \multirow{2}{*}{ $15-10^3$ MeV} 
  & SK \cite{Campo:2017nwh} & $\boldsymbol{\times}$  &  $\bar{\nu}_e$ (IBD)\\ \cline{2-4}
  & DARWIN \cite{McKeen:2018pbb} & $\boldsymbol{\times}$  &  All Flavors (Coherent)\\\hline
  $0.1-30$ GeV & DUNE \cite{Abi:2020evt} \newline HK \cite{Campo:2018dfh} & $\boldsymbol{\times}$ &  $\nu_e, \bar{\nu}_e, \nu_{\tau}, \bar{\nu}_{\tau}$ (CC)\\\hline
  $1-10^4$ GeV & SK \cite{Frankiewicz:2015zma, Abe:2020sbr} & \checkmark &  All Flavors \\\hline
    $20-10^4$ GeV & IceCube \cite{Aartsen:2016pfc} & \checkmark &  All Flavors\\\hline
     $50-10^5$ GeV & ANTARES \cite{Adrian-Martinez:2015wey} & \checkmark &  $\nu_\mu,\,\bar{\nu}_\mu$ (CC)\\\hline
         $0.2-100$ TeV & CTA \cite{Queiroz:2016zwd} &  \checkmark &  All Flavors (Bremsstrahlung)\\\hline
          %$> 100$ PeV & RNO~\cite{Aguilar:2019jay} &  \checkmark &  All Flavors \\\hline
          $10 - 10^4$ GeV & IC-Upgrade~\cite{Baur:2019jwm} &  \checkmark &  All Flavors \\\hline
          $> 10$ PeV & IC~Gen-2~\cite{Aartsen:2014njl} &  \checkmark &  All Flavors \\\hline
          $10 - 10^4$ TeV & KM3Net~\cite{Adrian-Martinez:2016fdl} &  \checkmark &  All Flavors \\\hline
          $1 - 100$ PeV & TAMBO~\cite{Wissel:2019alx} &  \checkmark &  $\nu_\tau,\,\bar{\nu}_\tau $ (CC) \\\hline
          $> 100$ PeV & GRAND~\cite{Alvarez-Muniz:2018bhp} &  \checkmark &  $\nu_\tau,\,\bar{\nu}_\tau $ (CC) \\\hline
\end{tabular}
}
\caption[Summary of current and future experiments used to derive constraints on $\sv$ and $\tau$ for different energy ranges.]{Summary of current and future experiments used to derive constraints on $\sv$ and $\tau$ for different energy ranges. 
The table also indicates whether the experimental analysis used directional information and which neutrino flavors it relied on.
}
\label{tab:table3}
\end{center}
\end{table*}

\subsection{Velocity-dependent annihilation}

\begin{figure*}[ht!]
\centerline{\includegraphics[width=\linewidth]{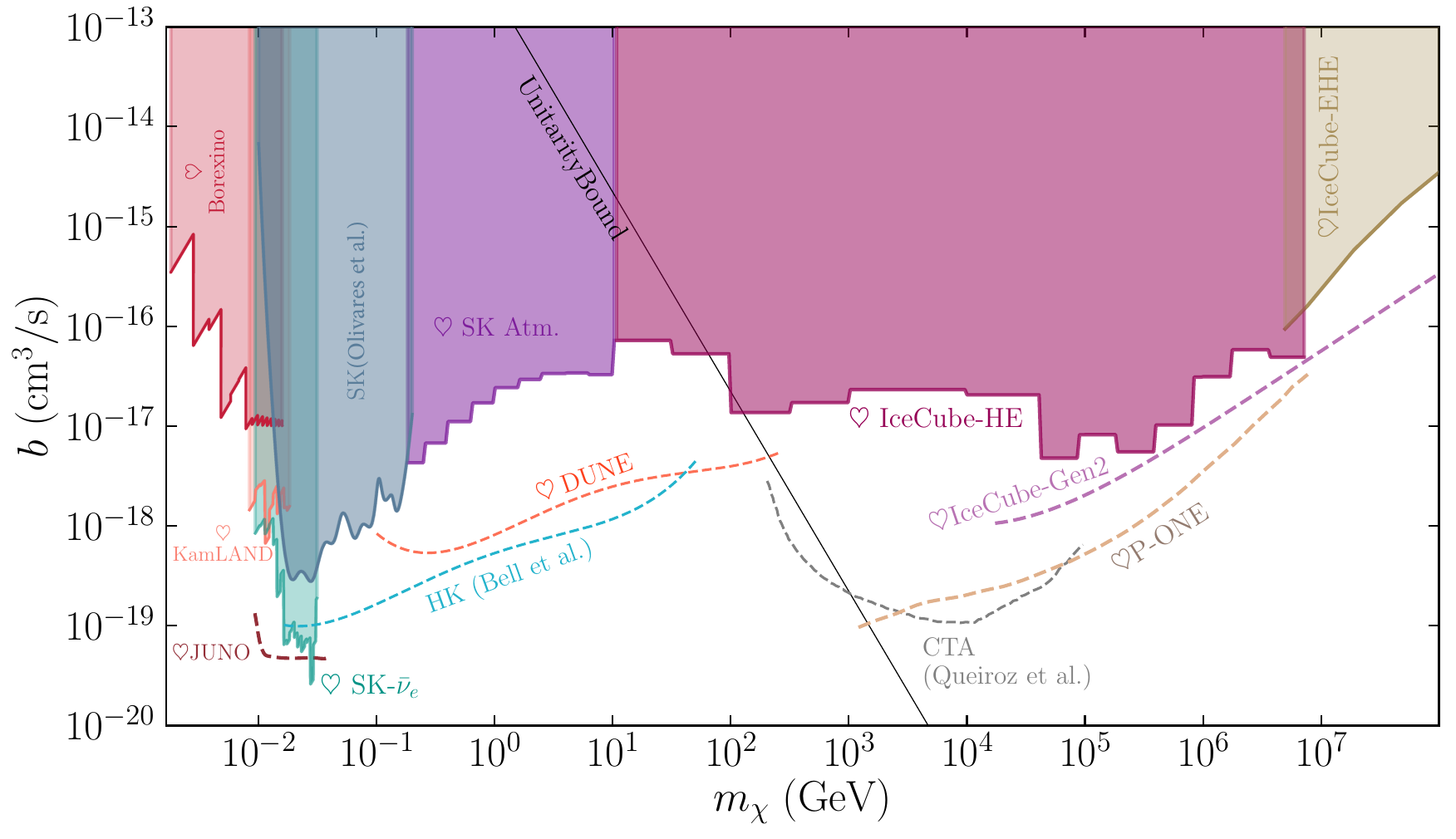}}
%\internallinenumbers
\caption[Limits on $p$-wave, $\sv = b(v/c)^2$, velocity-dependent annihilation cross-section of dark matter to two neutrinos.]{Limits on $p$-wave, $\sv = b(v/c)^2$, velocity-dependent annihilation cross-section of dark matter to two neutrinos. The cross section needed to explain the observed abundance for thermal DM is $\langle \sigma v_r \rangle = 6\times 10^{-26}\; \rm{cm}^3/\rm{s}$.}
\label{fig:Indirect_pwave}
\end{figure*}

\begin{figure*}[hbt!]
\centerline{\includegraphics[width=\linewidth]{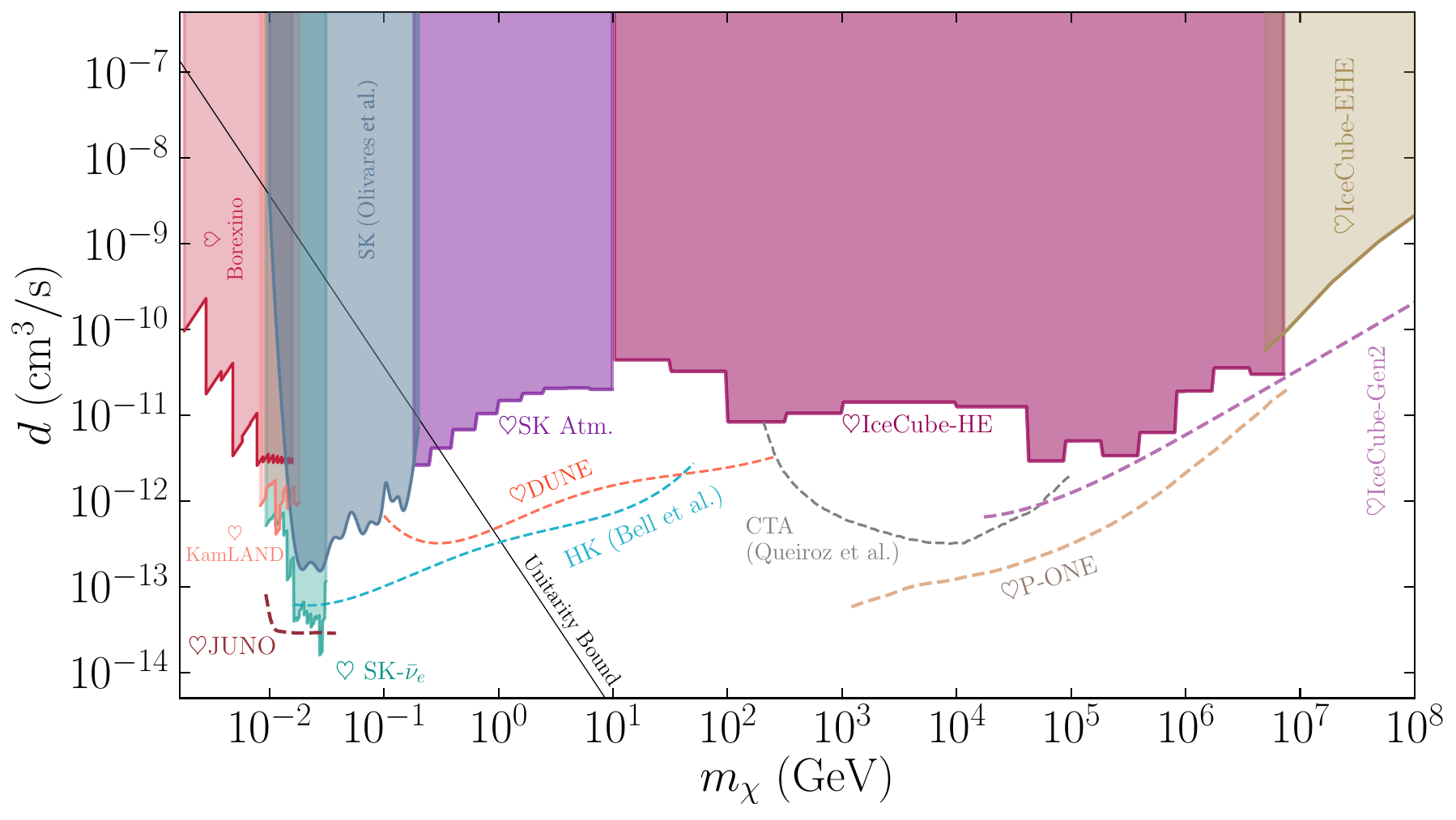}}
%\internallinenumbers
\caption[Limits on the annihilation of neutrinos to dark matter through a $d$-wave process $\langle \sigma v \rangle = d (v/c)^4$.]{
Limits on the annihilation of neutrinos to dark matter through a $d$-wave process $\langle \sigma v \rangle = d (v/c)^4$.
}
\label{fig:Indirect_dwave}
\end{figure*}

Fig.~\ref{fig:Indirect_pwave} shows the corresponding limits for $p$-wave annihilation, and Fig.~\ref{fig:Indirect_dwave} provides limits on $d$-wave annihilation.
In these cases, we follow the procedures outlined in Sec.~\ref{sec:pwave}, to reweight the astrophysical portion of the flux prediction (Eqs.~\eqref{eq:galaxyAnnRate} and~\eqref{eq:overdensity}) to account for the dark matter velocity dispersion.
We do this for all-sky searches since analyses where the angular distribution of the neutrinos has been taken into account are not easily re-scaled when considering the velocity distribution of DM particles within the halo.
Similarly, all the constraints taken from the literature are re-scaled using our choice of halo parameters (see Tbl.~\ref{tab:Jtable} for halo parameters and $J$-factor for the different analyses in the literature). Unsurprisingly, the limits on $\sv$ are much weaker for $p-$ and $d-$wave processes due to the strong velocity suppression. In contrast to the $s-$wave case, where the smallest halos tend to dominate the expected signal, velocity-suppressed annihilation is strongest in the largest DM halos where dispersion velocities are higher. These limits are thus insensitive to the value of the minimum halo mass $M_{min}$. However, the constraints from annihilation in the Milky Way halo remain dominant over the extragalactic contribution. 

\subsection{Dark matter halo uncertainties\label{sec:haloparams}}

As previously mentioned, a major source of uncertainty comes from the spatial dark matter distribution, because of the $n_\chi^2$ dependence in the annihilation signal.
For Galactic constraints, this is mainly reflected by uncertainties in the Milky Way dark matter distribution.
For extragalactic constraints, we focus on the shape of the halo mass function and the minimum dark matter mass, which determines how far down extrapolations of the HMF must go to account for the total DM contribution.

\begin{figure*}[t!]
    \centering
    \includegraphics[width=0.8\linewidth]{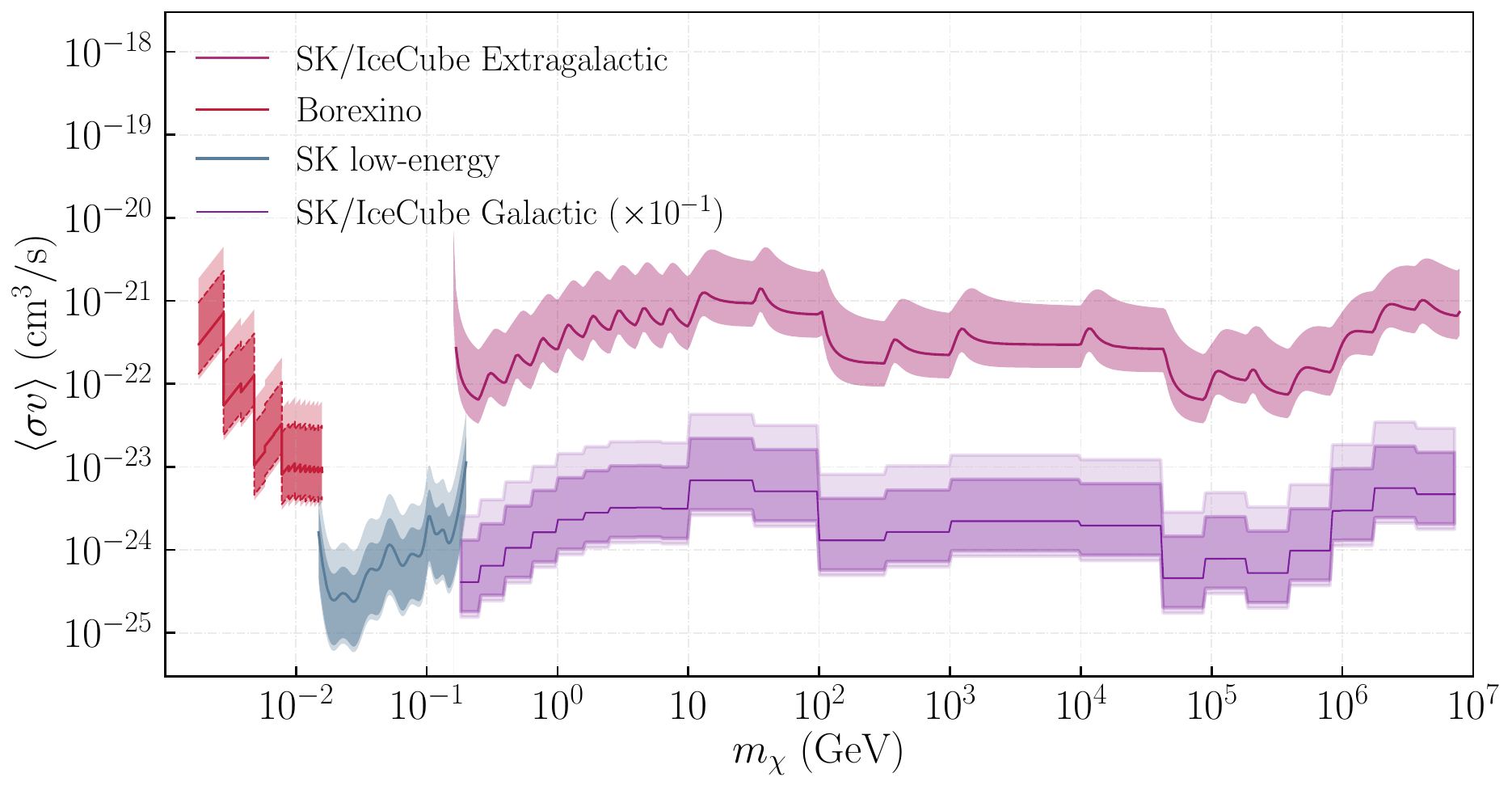}
    \caption[Uncertainties on the $s$-wave annihilation cross section for a subset of our results.]{
    Uncertainties on the $s$-wave annihilation cross section for a subset of our results.
    Solid lines correspond to the limits discussed in Sec.~\ref{sec:results}.
    For all Galactic limits, namely Borexino (red, leftmost), Super-Kamiokande low-energy (grey, scond region from left), Super-Kamiokande and IceCube (lower, rightmost), the 68\% (dark bands) and 95\% (light bands) uncertainties arise from the allowed variation on the dark matter distribution in the Milky Way, assuming a generalized NFW profile.
    The width of the uncertainty band for the extragalactic limits (upper, rightmost), obtained by comparing to the unfolded neutrino flux from IceCube and Super-Kamiokande, is dominated by the choice of the minimum halo mass, $M_{min}$, although it includes the uncertainty in the choice of HMF $dn/dM$, see Fig.~\ref{fig:hmfparam}.
    For our nominal choice of HMF, we choose the value of $M_{min}$ that yields the weakest constraint.
    }
    \label{fig:sigma_uncert}
\end{figure*}

\textit{\textbf{Milky Way halo shape parameters:}} To quantify the effect of the uncertainty on the MW halo shape parameters, we use the code provided by the authors of \cite{Benito:2019ngh}, which computes the log-likelihood as a function of halo shape parameters \{$\rho_0,r_s,R_0,\gamma$\}, given observed stellar kinematics data.
We profile over the 4 degrees of freedom, modifying the code to account for GRAVITY measurements of $R_0$, and obtain 68\% and 95\% C.L. ranges on the $J$-factors which we propagate to a range on $\sv$ for the Borexino, SK, and IceCube analyses. These are shown as dark and light bands, respectively, in Fig.~\ref{fig:sigma_uncert}.

\begin{figure}[ht]
    \centering
    \includegraphics[width=0.55\columnwidth]{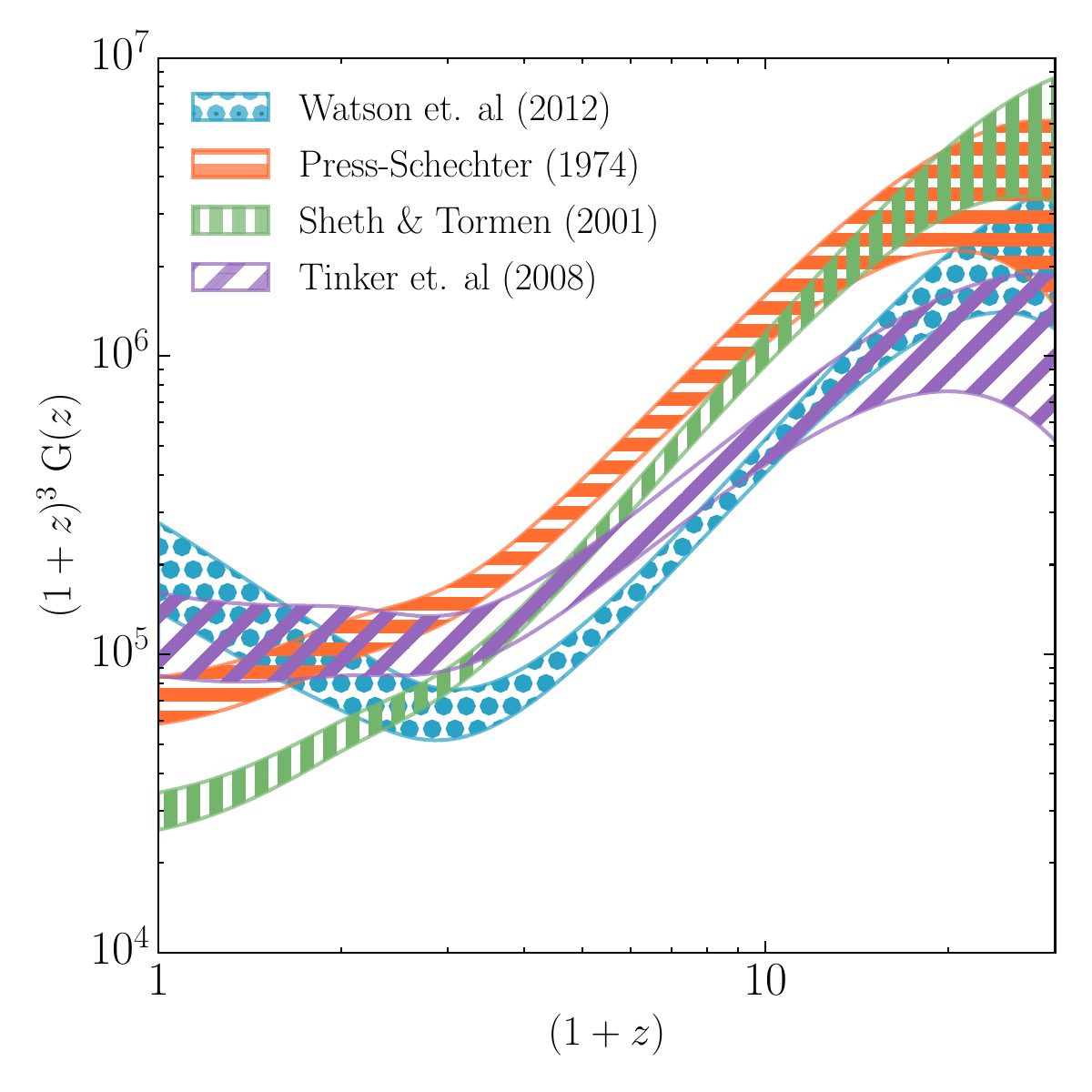}
    \caption[The halo boost factor $G(z)$ as a function of redshift for several parametrizations of the HMF $dn/dM$.]{
    The halo boost factor $G(z)$ as a function of redshift for several parametrizations of the HMF $dn/dM$.
    Our extragalactic constraints use Watson~\textit{et. al}~\cite{2013MNRAS.433.1230W}. The bands represent varying choices of minimum halo mass, from $10^{-3}$ to $10^{-9}$ solar masses. Fig.~\ref{fig:sigma_uncert} shows the effect of choosing a different parametrization on the limits.
    }
    \label{fig:hmfparam}
\end{figure}

\textit{\textbf{Halo Mass Function uncertainties:}} The largest contributions to uncertainties in the cosmological limits come from 1) the choice of HMF parametrization, and 2) the choice of minimum halo mass, $M_{min}$.
In our analyses we have employed the simulation-driven HMF fit by Watson et al.~\cite{2013MNRAS.433.1230W}.
Fig.~\ref{fig:hmfparam} shows the boost factor $G(z)$ defined in Eq.~\eqref{eq:overdensity}, for four different parametrizations from the literature: the analytic Press \& Schechter formalism~\cite{1974ApJ...187..425P,1991ApJ...379..440B}, Sheth \& Tormen~\cite{Sheth:1999mn,Sheth:1999su}, and Tinker~\cite{Tinker:2008ff}.
The width of the bands comes from varying the minimum halo mass from $10^{-3}$ to $10^{-9} \, M_\odot$. The band labeled ``Extragalactic'' in Fig.~\ref{fig:sigma_uncert} shows how this range propagates through to the cross section constraints.
Since there is no way of statistically quantifying the error on the HMF and minimum halo mass, we choose the most conservative scenario $M_{min} =10^{-3} \, M_\odot$ for our choice of HMF, corresponding to the solid magenta line in Fig.~\ref{fig:sigma_uncert}.

%% file: chapters/conclusion.tex
\chapter{Conclusion}
\label{ch:conc}
\SingleSpace
\epigraph{``It's still all good..\\ And if you don't know, now you know''}{The Notorious B.I.G}
\DoubleSpacing
\noindent The work has only just begun, and new physics in the neutrino sector remains elusive. 
But we are getting ever-closer, and discoveries are undoubtedly around the corner.
The mechanism by which neutrinos obtain their mass remains the most promising avenue to discovering physics beyond the standard model, and astrophysical neutrinos offer a unique probe of likely scenarios. 
Flavor composition measurements are a smoking gun signature to BSM effects.
And while the current generation of neutrino observatories lacks the definitive sensitivity, the next generation of astrophysical neutrino detectors will be able to make definitive claims using astrophysical flavor composition.
We have shown in Chapters~\ref{ch:taumodeling}-\ref{ch:tau_adv} that better modeling of the tau flavor significantly improves the sensitivity to transient sources, GZK fluxes, and allows accurate background predictions for the next generation.

The standard sources of astrophysical neutrinos are also becoming clearer, with several hints pointing to gamma-ray obscured AGN cores, as we have shown in Chapter~\ref{ch:sources}.
Identifying the sources will open up a multitude of new physics tests like \textit{e.g.} neutrino decay, non-standard neutrino interactions, a fourth neutrino state, right-handed neutrinos, lorentz-invariance violation, CPT violation and more.

Beyond that, subtracting the known sources from the total measured astrophysical flux improves our sensitivity to neutrinos produced by exotic sources like dark matter.
If dark matter is of a corpuscular nature, it is not an illogical leap to assume that the mechanism responsible for dark matter and neutrino masses could be shared. 
Again, astrophysical neutrinos allow us to ask definitive questions about connections between the two. 
As we have shown in Chapter~\ref{ch:DM}, indirect dark matter searches with neutrinos are promising.
Indeed, we reach the thermal relic abundance cross section in the neutrino sector for the first time at dark matter masses of $\sim~$10 MeV using data from Super-Kamiokande.
We are entering an era of precision neutrino astrophysics, and we look forward to the day when questions asked in this thesis are mundane and the answers are taken for granted, and in retrospect seem quite obvious.